\documentclass[fleqn,usenatbib]{mnras}

\usepackage{newtxtext,newtxmath}

\usepackage[T1]{fontenc}
\usepackage{ae,aecompl}
\usepackage{longtable}
\usepackage{pdflscape}

\usepackage{graphicx}	
\usepackage{amsmath}	
\usepackage{amssymb}	

\usepackage{color}

\title[Magnetospheric effects on the radiation of pulsars]{Evidence for magnetospheric effects on the radiation of radio pulsars}

\author[Ilie \& Johnston \& Weltevrede]
{C.D. Ilie$^{1}$\thanks{email: cristina.ilie@postgrad.manchester.ac.uk},
{S. Johnston$^{2}$} and {P. Weltevrede$^{1}$}
\\
$^{1}$University of Manchester, Jodrell Bank Centre of Astrophysics, Alan Turing Building, Manchester, M13 9PL\\
$^{2}$CSIRO Astronomy and Space Science, Australia Telescope National Facility, PO Box 76, Epping, NSW 1710, Australia\\
}

\date{Accepted 2018 November 29. Received 2018 November 28; in original form 2018 September 17}

\pubyear{2018}

\begin{document}
\label{firstpage}
\pagerange{\pageref{firstpage}--\pageref{lastpage}}
\maketitle

\begin{abstract}
We have conducted the largest investigation to date into the origin of phase resolved, apparent RM variations in the polarized signals of radio pulsars. From a sample of 98 pulsars based on observations at 1.4~GHz with the Parkes radio telescope, we carefully quantified systematic and statistical errors on the measured RMs. A total of 42 pulsars showed significant phase resolved RM variations. We show that both magnetospheric and scattering effects can cause these apparent variations. There is a clear correlation between complex profiles and the degree of RM variability, in addition to deviations from the Faraday law. Therefore, we conclude that scattering cannot be the only cause of RM variations, and show clear examples where magnetospheric effects dominate. It is likely that, given sufficient signal-to-noise, such effects will be present in all radio pulsars. These signatures provide a tool to probe the propagation of the radio emission through the magnetosphere.

\end{abstract}




\section{Introduction}


Soon after the discovery of pulsars, 50 years ago \citep{hbp+68}, it was observed that their radio signals are highly linearly polarized \citep{ls68}, with the position angle (PA) of many pulsars changing across rotational phase in a characteristic S-shape swing, well described by the Rotating Vector Model (RVM) \citep{rc69}. Observed discontinuities in the PA swing in the form of 90$^{\circ}$ jumps have been explained with the co-existence of two orthogonally polarized modes (OPMs) \citep{brc76}. The observed polarized radiation is thus thought to be a superposition of the two OPMs, with the overall degree of linear polarization depending on the relative contribution of each OPM at a specific pulse longitude \citep{scr+84,vt17}.  

When the pulsar radiation propagates through the magnetised interstellar medium (ISM), it is affected by Faraday rotation. This results in a rotation of the orientation of linear polarization ($\Delta$PA) as a function of observing wavelength ($\lambda$), given by the expression 
\begin{equation}
\label{eg:basicRM}
\Delta\text{PA} = \text{RM} \, \lambda ^ 2.
\end{equation}

\noindent Here the constant of proportionality is known as the rotation measure (RM), and is related to the properties of the ISM via

\begin{equation}
\text{RM} = \frac{e^3}{2\pi m_e^2 c^4} \int_{0}^{L} n_e B_{||} \text{d}l ,
\label{eq:rm2}
\end{equation}

\noindent  where $e$ and $m_e$ are the charge and mass of the electron, $c$ is the speed of light in vacuum, $n_e$ is the electron density, $B_{||}$ is the component of the magnetic field parallel to the line of sight, $L$ is the distance to the pulsar and d$l$ is distance element along the line of sight \citep[e.g.][]{lk05}. 
Generally, it is assumed that the radiation from the pulsar does not undergo changes as it traverses the magnetosphere, and therefore that equation~(\ref{eg:basicRM}) represents the contribution from the ISM alone. 
Using combined measurements of RM and dispersion measure (DM), the average magnetic field along the line of sight can be estimated, and thus the structure of the Galactic magnetic field  \citep[e.g.][]{man72,man74,tn80,ls89,hmq99,mwk+03,njk+08,hmv+18}, and it therefore important to test the above assumption. 

If Faraday rotation is the only source of frequency dependence of the PA, we expect the derived RM to be independent of the rotational phase of the pulsar. This can be tested using observations with high time resolution and signal-to-noise (S/N).
The first authors to perform such an analysis were \cite{rbr+04}. They showed that the apparent RM of PSR B2016+28 varied by $30 \,\, \,\text{rad m}^{-2}$ as a function of pulse longitude. More recently, \citet{dgm+15} also saw apparent RM variations in a selection of millisecond pulsars.
\cite{rbr+04} investigated the origin of the frequency dependence of the PA for this pulsar using single pulse analysis and argued that it originated because of the incoherent addition of two non-orthogonal OPMs (quasi-OPMs) which had different spectral indices. The existence of OPMs with different spectral indices was later also observed by \cite{kjm05} and \cite{sse+06}. \cite{nkk+09} concluded that although this effect can explain the apparent RM variations across pulse phase in the case of some specific pulsars, it cannot be generalized to the entire pulsar population. 

\cite{rbr+04} argued that the observed apparent RM variations across pulse phase are not caused by Faraday rotation within the pulsar magnetosphere, since this would lead to significant depolarization. \cite{nkk+09} investigated the possibility that a generalized Faraday effect could be the cause. Following work from \cite{km98}, it was suggested that in this scenario the apparent RM variations should occur there where the circular polarization changes most rapidly with rotational phase.
Although they did not find such correlation, generalized Faraday rotation was not dismissed completely, as the constraints on this theory are not well defined.

Interstellar scattering, which causes a shift of polarized radiation to a later rotational phase in a frequency dependent manner, will cause apparent RM variations.  \citet{kar09} showed, using simulations, how even a small amount of scattering can affect the shape of the PA swing, most notably in the case of intrinsically steep PA swings. OPMs situated at phases close to where the PA swing is changing the fastest were also more likely to be affected by scattering.
\citet{nkk+09} observed the largest RM variations coinciding with the rotational phases where the PA was the steepest, and concluded that scattering was the dominant cause of apparent RM variations. 
More recently, \citet{nsk+15}, using low frequency observations, concluded that the amplitude of the RM variations due to scattering should follow a $\lambda^{-2}$ law.

In this paper, we quantify and investigate the nature of the observed phase-resolved apparent RM variations, RM($\phi$), and whether interstellar scattering is the dominant mechanism responsible. 
We take a statistical approach, using a large sample of pulsars.
It should be stressed that these apparent RM variations quantify changes in $\partial \text{PA}(\lambda,\phi)/ \partial \lambda^2$. Hence, in the presence of other frequency dependent processes, the derived RM is not entirely a measure of the magneto-ionic properties of the ISM. From here onwards, unless otherwise stated, when we refer to RM, we refer to the RM defined in equation~(\ref{eg:basicRM}), rather than the RM from equation~(\ref{eq:rm2}).

In Section~\ref{sect:obs} we outline the details of our observations, while Section~\ref{sect:method} describes the methodology used in this analysis. In Section~\ref{sect:results}, the results are presented and the pulsars which showed significant phase-resolved apparent RM variations are discussed on a case by case basis.
The results related to the sample as a whole are discussed in Section~\ref{sect:discussion} and a summary is given in Section~\ref{sect:summary}.

\section{Observations}
\label{sect:obs}
A sample of the brightest pulsars from \cite{jk18} ranked by S/N were selected for this analysis. 
The data were collected over the period of January 2016 to February 2017, using the Parkes radio telescope, at a frequency of 1.4 GHz and a bandwidth of 512 MHz, using the H-OH receiver. 
Individual observations of each pulsar were summed together in order to increase the S/N. The data were reduced to 32 frequency channels.  
Details of the observations and the calibration scheme used can be found in \cite{jk18}.

\section{Method}
\label{sect:method}

The method we used to measure the RM is based on the most basic form of RM synthesis technique (RMST), which was developed by \cite{bur66} and later extended and implemented by \cite{bd05}. The RMST is based on calculating the complex Faraday dispersion function, $\widetilde{F}(\text{RM})$, using a Discrete Fourier Transform (DFT) given by the equation

\begin{equation}
\widetilde{F}(\text{RM}) = K \sum_{c=1}^{N} \widetilde{P}_c e^{-2i\text{RM}(\lambda_c^2-\lambda_0^2)},
\label{eq:rmst}
\end{equation}

\noindent where $K$ is a normalization constant, $c$ is the frequency channel index, $\widetilde{P}_c$ is the observed linear polarization expressed as a complex number, $Q + iU$, in terms of the Stokes parameters $Q$ and $U$, $\lambda_c$ is the wavelength of channel $c$ and $\lambda_0$ is a reference wavelength \citep[see also][]{g09}. The power spectrum of this function represents the RM spectrum, and $|\widetilde{F}(\text{RM})|^2$ will peak at the RM of the pulsar. Since we are only interested in the shape of $\widetilde{F}(\text{RM})$, we can set $\lambda_0$ to 0 and $K$ to 1, in equation~(\ref{eq:rmst}). Effectively, this method consists of multiplying the complex polarization vector of each individual frequency channel  with a trial RM and $\lambda^2$ dependent complex exponential, therefore it de-Faraday rotates  the linear polarization before summing it over all frequencies. The RM spectrum is produced by taking the square of this function, which is effectively the degree of linear polarization as of function of the trial RM. The peak of this function,  i.e. when the linear polarization is maximized, represents the optimum RM.

To obtain RM($\phi$), the calculation was performed for each pulse longitude bin ($\phi$) in a similar manner. 
The RMST algorithm has been included in the  \textsc{PSRSALSA}\footnote{https://github.com/weltevrede/psrsalsa} software package \citep{wel16}, publicly available at the link provided.

An alternative method for measuring RMs, used by \cite{njk+08,nkk+09}, consists of performing a fit of the PA as a function of $\lambda^2$ to compute the RM. One has to be careful with this method  concerning the non-Gaussianity  of the uncertainties on the PA in the case of low linear polarization signal, hence normally the PAs are computed only for bins where the linear polarization exceeds a certain cut-off, therefore losing sensitivity. In the case of low linear polarization, \cite{njk+08,nkk+09} estimate the uncertainties on PAs from the distribution described in \cite{nc93}. 
In principle, the two methods mentioned are equivalent, however the RMST method, as implemented here, avoids the complexity of non-Gaussian error bars, as it does not require the determination of the PA with associated uncertainties.

Although analytic errors can be determined on the RMs derived using RMST \citep{bd05}, they rely on assumptions which are not necessarily correct. Here, we attributed a statistical uncertainty on each measured RM by adding random white noise with a standard deviation determined from the off-pulse region to the data of each frequency channel, and re-performing the analysis for a large number of iterations, i.e. bootstrapping. Thus, a distribution of RMs was obtained and the standard deviation was taken as the statistical uncertainty.
This provides a robust error determination method.  
No a-priori assumptions have to be made about the underlying signal, and non-Gaussian errors will be properly taken into account.
Assigning an analytic uncertainty on the derived RM is possible \citep{bd05}, but requires assumptions about, for example, the shape of the band-pass \citep[see also][]{sl15}. Furthermore, the spectral shape of the source and scintillation conditions will also affect the shape of the RM spectrum, complicating the determination of an accurate analytical uncertainty.

\begin{figure}
\includegraphics[width=0.47\textwidth]{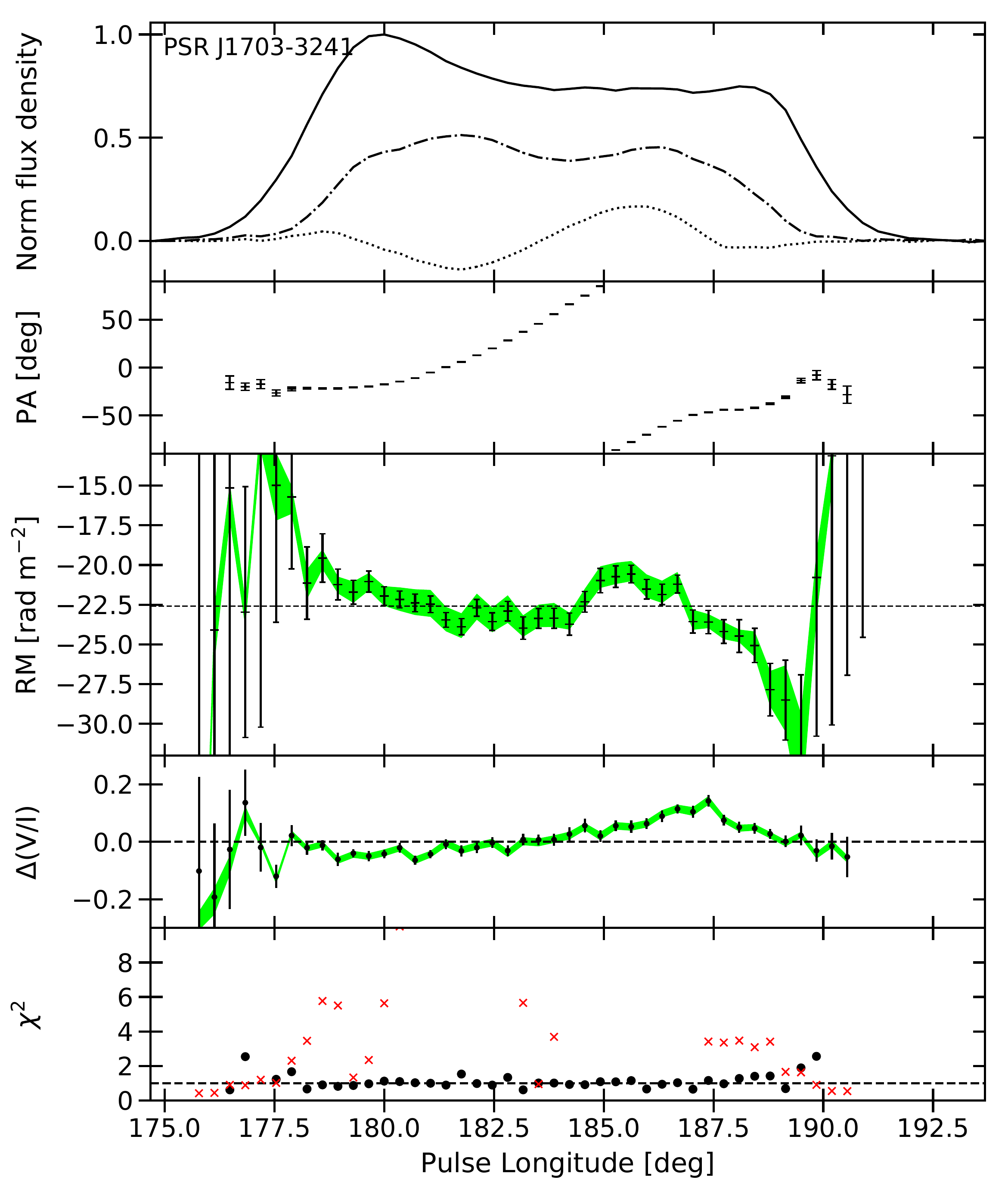}
\caption{PSR~J1703$-$3241.
In the top panel the solid line denotes Stokes $I$, the dashed line shows the linear polarization
and the dotted line the circular polarization. 
The second panel displays the frequency-averaged position angle of the linear polarization. Position angles were only plotted when the linear polarization exceeded 2 sigma. 
The third panel shows RM($\phi$) with associated statistical uncertainties represented by the errorbars. The shaded region (green in the online version) represents the $1\sigma$ systematic uncertainty contour region. The horizontal dotted line plotted is $\left<\text{RM}(\phi)\right>$. 
The fourth panel shows the phase-resolved $\Delta(V/I)$ values with their associated statistical and systematic uncertainties. 
The bottom panel displays the $\chi^2_{\text{PA}\left(\lambda^2,\phi\right)}$, represented by the black circles  and $\chi^2_{\text{V/I}\left(\nu,\phi\right)}$, represented by the red crosses. The horizontal dashed line corresponds to a  reduced $\chi^2 = 1$. The plots for the 98 pulsars are in the online supplementary material, in Fig.~A.1 $-$ Fig.~A.26.}
\label{fig:example}
\end{figure}

RM($\phi$) curves with their associated statistical uncertainties were plotted for each pulsar and the results can be found in the online supplementary material (Fig.~A.1 $-$ Fig.~A.26). An example of a typical plot is shown in Fig.~\ref{fig:example}. In the top panel, the integrated pulse profile is displayed with the solid line denoting Stokes $I$, the dashed line showing the linear polarization, $L$, and the dotted line the circular polarization, Stokes $V$. The second panel shows the frequency averaged PA and in the third panel RM($\phi$), along with associated uncertainties.

In order to assess deviations from Faraday law at a given pulse phase, the PA was computed at those frequencies where the linear polarization exceeded $2\sigma$. The $\lambda^2$ dependence was removed according to equation~(\ref{eg:basicRM}) using the determined RM($\phi$), and the $\chi^2$, $\chi^2_{\text{PA}\left(\lambda^2,\phi\right)}$, of the remaining variability was determined. 
This can be seen for all pulsars as shown in the case of an example pulsar displayed on the left-hand side of Fig.~\ref{fig:example} in the bottom panel.
Note, that when deviations from the Faraday law are observed, the measured RM will not fully quantify $Q$ and $U$ as function of frequency.
Nevertheless, since at least some of these deviations will be absorbed in the RM (as demonstrated by e.g. \citealt{nkk+09} or \citealt{kar09}), variability in the RM($\phi$) curves can be expected, hence it is a good indicator for additional frequency dependent effects.

RM values for the profiles (i.e. non-phase resolved), RM$_{\text{profile}}$, were also determined. The methodology was very similar to the one described above in the case of RM($\phi$). A RM spectrum was computed using equation (\ref{eq:rmst}) for all pulse longitude bins in a selected on-pulse region. The RM power spectra were then summed and the RM determined.  These values, as well as their corresponding statistical uncertainties obtained from bootstrapping, are displayed in Table \ref{table:bigtable}. A similar test to $\chi^2_{\text{PA}\left(\lambda^2,\phi\right)}$ was performed. 
The data were de-Faraday rotated using the determined RM$_{\text{profile}}$, the frequency averaged PA was subtracted for each pulse longitude bin, before averaging the Stokes parameters in pulse longitude. A reduced $\chi^2$ was determined and the results are displayed in Table~\ref{table:bigtable} as $\chi^2_{\text{PA}(\lambda^2)}$.

Scattering will affect the measured RM$_{\text{profile}}$, but \citet{njk+08} avoided this contamination by averaging the Stokes parameters over pulse longitude before measuring the RM. Since scattering does not affect the pulse longitude integrated Stokes parameters, the determined RM is unaffected \citet{kar09}. We will refer to this RM as RM$_{\text{scatt}}$.
Comparison of RM$_{\text{profile}}$ and RM$_{\text{scatt}}$ provides an indication if scattering affected the polarization. The measurement of RM$_{\text{scatt}}$ is less sensitive compared to that of RM$_{\text{profile}}$, as averaging Stokes parameters over pulse longitude leads to depolarization depending on the steepness of the PA. 
Our measured values of RM$_{\text{scatt}}$ can be found in Table \ref{table:bigtable}.

We obtained the fractional circular polarization change across the frequency band, $\Delta(V/I)$, with associated statistical uncertainties, in a similar manner to what is presented in \cite{nkk+09}, for each pulse longitude.  
This is displayed in the fourth panel of the example plot shown on the left-hand side of Fig.~\ref{fig:example}. We quantified the deviation from no variation as $\chi^2_{\text{V/I}}(\lambda^2,\phi)$, shown in the bottom panel of the plot, as red crosses.

\cite{nkk+09} argued that significant $\Delta(V/I)$ variations can be taken as evidence for generalized Faraday rotation in the pulsar magnetosphere.
If this is responsible for the observed RM($\phi$) variations, then we expect the greatest variations to coincide in pulse longitude with the greatest change in $\Delta(V/I)$.
However, we here point out that interpreting $\Delta(V/I)$ in terms of generalized Faraday rotation is complicated by the fact that scattering is also capable of creating $\Delta(V/I)$ variations as a function of pulse longitude.
We expect that if a pulsar is affected by interstellar scattering, then the greatest change in $\Delta(V/I)$ coincides with that part of the profile where Stokes $V$ is changing most rapidly as a function of pulse longitude (see below for a simulation). This is not necessarily where the largest RM($\phi$) variations occur. This therefore potentially allows the distinction between which frequency dependent effect is responsible for the apparent RM variations. 

\begin{figure}
\includegraphics[width=0.47\textwidth]{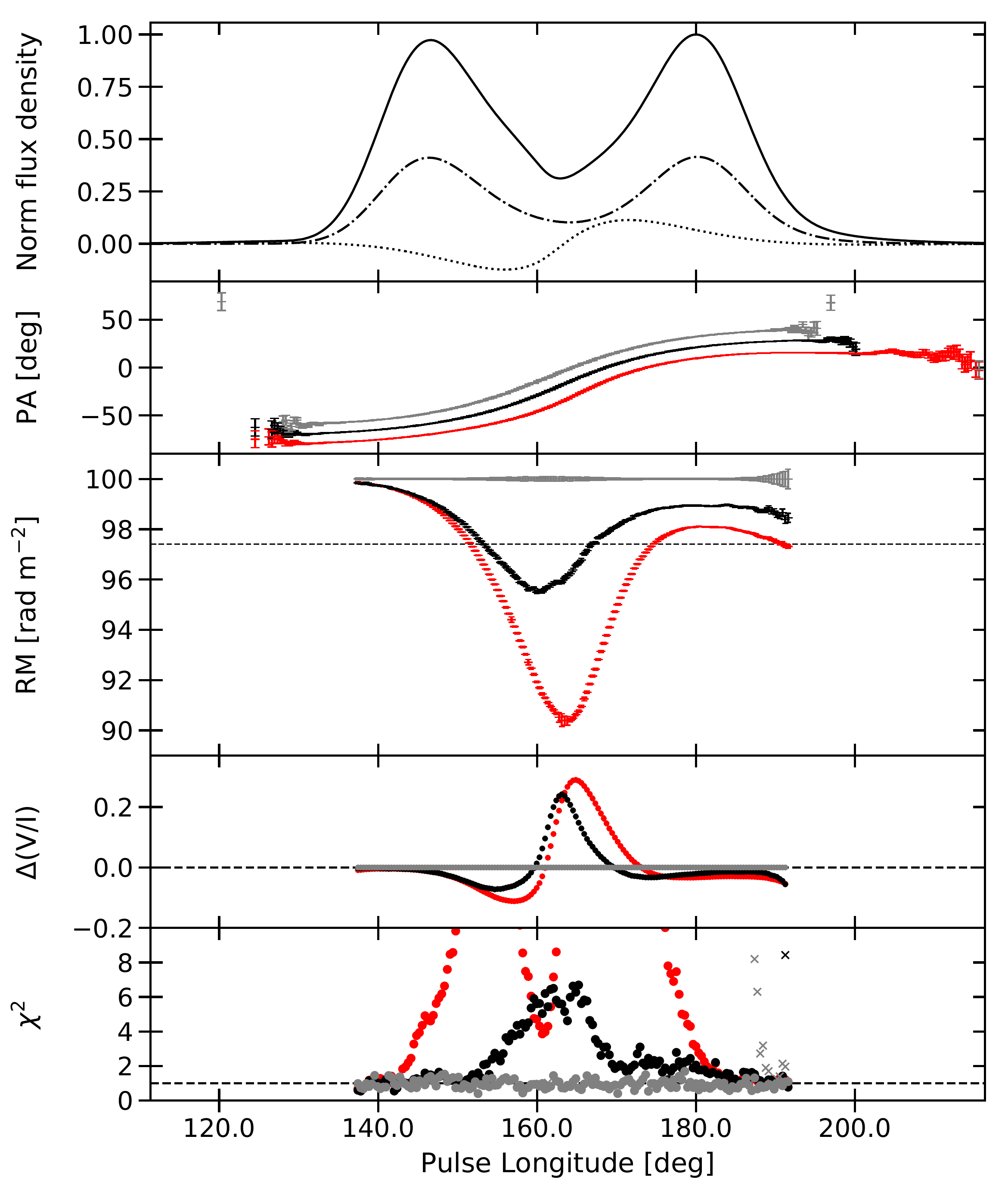}
\caption{Simulations demonstrating the effect of scattering with a timescale of 4 ms (represented in black) and 8 ms (represented in red, in the online version), for a pulse period of 1 sec. The curves without scattering were also represented in grey. The reference frequency is 1.4 GHz and the bandwidth 512 MHz. The panels are as described in Fig.~\ref{fig:example}.
In the second panel, a vertical shift of 10$^{\circ}$ in PA is applied to help distinguishing between the simulations.}
\label{fig:example_sim}
\end{figure}

To demonstrate the effect of scattering, a simulation was performed on a synthetic frequency resolved pulse with varying PA and Stokes $V$ with pulse longitude, and an intrinsic RM of 100 rad m$^{-2}$. Scattering was added to the profile with timescales, $\tau_{\text{scatt}}$, of 4 ms and 8 ms, for a pulse period of 1 sec.
An exponential tail of the form $\exp(-t/\tau_{\text{scatt}})$, where $t$ represents time, was convolved with the Stokes parameters in the modified data in the Fourier domain, similar to the simulation done in \citet{kar09}. We take $\tau_{\text{scatt}}\propto\nu^{-4}$ relative to a reference frequency of 1.4 GHz. A bandwidth of 512 MHz was assumed.
The results are shown in Fig.~\ref{fig:example_sim}. It is clear that scattering is capable of creating  $\Delta(V/I)(\phi)$ variations and these coincide with where Stokes $V$ is changing rapidly. Note that scattering also produces RM variations in a region where the PA swing is steepest, as expected from \citet{kar09}, as well as deviations from Faraday law, as observed in the bottom panel of Fig.~\ref{fig:example_sim}.
Note that the amplitude of variations is larger with larger amounts of scattering, consistent with the findings of \citet{kar09}.

\subsection{Systematic uncertainties}
\label{subsect:sys}

For pulsars with high S/N, the statistical uncertainties can be small enough that systematic effects will dominate. Some of these systematic effects could produce an additional frequency dependence of the polarization, resulting in apparent phase-resolved RM variations. We attempted to determine and quantify a number of systematic effects.

Instrumental effects can produce a peak in the RM spectrum at a value of 0 rad m$^{-2}$, which could lead to erroneous estimates of the RM and its uncertainties (see e.g. \citealt{sbl15}).
All RM spectra were visually inspected for such peaks, and none were observed, hence these effects were not further considered.

An inaccurate DM value introduces a frequency dependent dispersive delay, which will affect the PA as a function of frequency, hence variations in RM($\phi$). We measured DMs using \textsc{tempo2}\footnote{http://www.atnf.csiro.au/research/pulsar/tempo2/} \citep{hem06} for each of the pulsars analysed. However, these measurements are affected by profile variations with frequency.
Hence, in order to correct for such an effect further, we obtained RM($\phi$) after applying 20 trial offsets, DM$_{\text{offset}}$, around the determined value of DM, from $-0.5$ to 0.5 cm$^{-3}$pc in steps of 0.05 cm$^{-3}$pc. For each trial, we computed the  RM$(\phi)$ and the weighted mean, $\left<\text{RM}(\phi)\right>$, and obtained a reduced $\chi^2$, $\chi^2_{\text{RM}(\phi)}$, of the RM$(\phi)$ with respect  $\left<\text{RM}(\phi)\right>$. For each pulsar, the results for the DM$_{\text{offset}}$ which gave the lowest $\chi^2_{\text{RM}(\phi)}$ were displayed in Table \ref{table:bigtable}.
However, by minimising the variability in the RM($\phi$) curves, which could be caused by using an incorrect DM ensures that, if variability is detected, it is not because of this systematic effect.
This does not imply that this is the correct DM. As a consequence, the variations will be underestimated.

A possible systematic effect is the imperfect alignment of individual observations when creating the final datasets. 
The alignment is limited by the the time of arrival (ToA) uncertainties of each pulsar. 
We quantified this effect through simulations. 
The Stokes parameters of the pulsars were first averaged over frequency and duplicated to form 32 frequency channels. This ensured that all RM($\phi$) variations were eliminated, while the shape of the average PA remained unaffected.
Fifty such individual observations were created for each pulsar, with phase offsets sampled from a Gaussian distribution with a standard deviation equal to the respective ToA uncertainty. The ToA uncertainties were smaller than 0.01$\%$ of the pulse period.
These individual observations were added together and RM($\phi$) curves were obtained as described before. 
No significant variations were observed.

Another systematic effect which was quantified is the varying RM of the ionosphere, RM$_{\text{iono}}$, which can change depending on the time, day or season of observation. 
For our observations, we considered RM$_{\text{iono}}$ to vary with $\pm 3$ rad m$^{-2}$ \citep{ssh+13}.
If the alignment of individual observations were done perfectly, we would only expect a constant change in RM with pulse phase in each observation due to RM$_{\text{iono}}$. 
However, if the individual observations are not perfectly aligned, we expect the effects of changing  RM$_{\text{iono}}$ to introduce additional systematic frequency dependent variations. Based on the simulation described earlier, we allowed the RM in each observation to randomly vary within the specified limits. The only pulsar for which we observed this effect to create RM($\phi)$ variations is the Vela pulsar, J0835$-$4510. 
The results are displayed in Fig. B.1, in the online Supplementary material. The contribution of this systematic effect appears to be less than $5\%$, much smaller than the next systematic to be discussed.

Polarization imperfections in the H-OH receiver can be responsible for significant RM($\phi$) variations compared to the statistical ones.
Assuming the distortions to be linear, the transformation from un-calibrated Stokes parameters to intrinsic Stokes parameters can be determined by using a Mueller matrix, a $4\times4$ real matrix  \citep[e.g.][]{mue48, hpn+01}. 
One of the seven independent parameters of this matrix is the overall gain and is not useful for the scope of this paper. 
Two other parameters are the differential phase and gain. These were corrected for by performing a short pulse calibration observation, for each pulsar, which pointed slightly offset from the source right before the actual observation.
The remaining four parameters, which are the ones we are interested in, are referred to as the leakage parameters. These correct for the effect where one of the dipoles records some signal that should have been detected by the other dipole.

We simulated artificial receiver imperfections, by randomizing values for the four leakage parameters. For simplicity, it was assumed that they have a linear frequency dependence. 
When generating these artificial leakage parameters, it was ensured that no off-diagonal elements of the Mueller matrix exceeded 1.5$\%$ conversion between Stokes parameters at any frequency channel, while reaching this maximum percentage in at least one frequency channel. This limit was chosen based on the following test. 
From our sample, we chose all pulsars which did not show any RM($\phi$) variations (e.g. J1048$-$5832 and J1709$-$4429). 
Our assumption was that the imperfections of the receiver could not generate more RM($\phi$) variations than were already observed.
The limit was therefore chosen as the maximum value which did not create additional apparent variations in RM($\phi$).

The systematic uncertainties because of receiver imperfections were quantified for each pulsar by randomly generating 100 receiver imperfections obeying the above description, which then were used to distort the pulsar signal in the process of polarization calibration \footnote{Note that these variations are in general too small to result in a noticeable peak centred at RM=0 rad m$^{-2}$ in the RM power spectrum.}. For each of these 100 different distortions, we calculated values for RM$_{\text{profile}}$, RM$_{\text{scatt}}$, RM$(\phi)$ and $\Delta(V/I)(\phi)$. The standard deviation of these values was quoted as the systematic uncertainty in Table~\ref{table:bigtable}. 
The systematic uncertainties determined for the RM($\phi$) and $\Delta(V/I)$ are displayed for each pulsar, for example in Fig.~\ref{fig:example}, in the third and fourth panels as a $1\sigma$ contour grey shaded region over-plotted over the RM($\phi$) values (green in the online version).

\section{Results}
\label{sect:results}

Pulsars described in Sections~\ref{sect:highRM} and~\ref{sect:noRM} and in Table~\ref{table:bigtable}, for which the phase-resolved RM profile  had never been investigated previously in the literature, are marked with an asterisk (*).

The plots for the 98 pulsars have been included in the online supplementary material, in Fig.~A.1 $-$ Fig.~A.26 and an example is displayed in Fig.~\ref{fig:example}. All plots were aligned so that the total intensity peaked at pulse longitude $180^{\circ}$. For the six pulsars from the sample, which had both a main pulse (MP) and an interpulse (IP), we aligned the MP peak at pulse longitude $90^{\circ}$ and hence, the IP peaked around pulse longitude $270^{\circ}$.
The results from the analysis described in the Section~\ref{sect:method} can be found in Table~\ref{table:bigtable}.

The resulting RM($\phi$) profiles allowed us to classify the pulsars as follows. Pulsars which had $\chi^2_{\text{RM}(\phi)} > 2$ were classified as showing significant RM($\phi$) variations and they are discussed in more detail, on a case to case basis, in Section~\ref{sect:highRM}. 
Six pulsars which were initially classified as showing significant RM($\phi$) variations, were removed from this section based on their high systematic uncertainties. 
A total of 42 pulsars, out of our sample of 98, were classified as showing significant RM($\phi$) variations.

For all cases where we saw RM($\phi$) variations,  we observed deviations from the expected $\lambda^2$ dependence (as quantified by $\chi^2_{\text{PA}\left(\lambda^2,\phi\right)}$, but also by inspecting PA versus $\lambda^2$ directly), implying that Faraday law fails to describe the full frequency dependence of the PA, and there must be another frequency and pulse longitude dependent effect present. Therefore, the results obtained from the panel where $\chi^2_{\text{PA}\left(\lambda^2,\phi\right)}$ is displayed, were not discussed on an individual basis. 

Unless otherwise stated in individual cases, RM$_{\text{profile}}$ and RM$_{\text{scatt}}$ were consistent, providing no indication whether the RM was affected by interstellar scattering or not. 

\subsection{Pulsars with significant RM variations}
\label{sect:highRM}

\textbf{PSR J0034$-$0721}. 
The profile of this pulsar has a central peak and a long tail.
$L$ is low ($<20\%$), with two drops to zero at the longitudes where OPM jumps occur in the PA swing. 
Apparent RM($\phi$) variations occur after pulse longitude $175^{\circ}$, with the largest deviations at the second OPM jump, although \cite{nsk+15} observed no RM($\phi$) variations at 150 MHz. 
There are no significant $\Delta(V/I)$ variations observed.  
If scattering was responsible for the apparent RM variations, we would not necessarily expect to see large $\Delta(V/I)(\phi)$ variations, given Stokes $V$ is relatively constant as function of longitude. Furthermore, we would expect the largest RM variations at longitudes where the PA swing is steep or breaks occur, which is observed, indicating that scattering may be the primary cause for the observed RM variations.

\textbf{PSR J0255$-$5304*}. 
This pulsar has a two component profile, with low $L$ and a complex PA
swing. There is an OPM jump at $\sim179^{\circ}$.  RM($\phi$) variations are as high as $\sim$ 90 rad m$^{-2}$ but are sensitive to the choice of DM. The largest variations can be observed towards the centre of the profile although deviations can be seen at all pulse longitudes.  
We see significant variations in $\Delta(V/I)$ towards the centre of
the profile.  However, the largest variations occur towards the trailing
part of the profile, where Stokes $V$ is changing strongly. We conclude that scattering is likely responsible for the observed RM variations.

\textbf{PSR J0401$-$7608*}. The profile of this pulsar has three blended components, with the central one being the strongest, 
and $L$ is moderately high, especially in the trailing component. The PA swing is flat, except in the central region of the profile, which 
is also where the deviations in RM($\phi$) occur ($\sim 20$ rad m$^{-2}$). 
The lack of significant deviations in $\Delta(V/I)$ indicates we cannot distinguish between scattering and magnetospheric effects being responsible for the apparent RM variations.

\textbf{PSR J0452$-$1759*}. This pulsar displays complex profile and polarization frequency evolution. 
$L$ is low ($\sim 20\%$), with several OPM jumps present in the PA swing. 
The largest RM($\phi$) deviations ($\sim 80$ rad m$^{-2}$) occur at the pulse longitudes of the OPM jumps, and where the PA is the steepest, whereas between  $184^{\circ}$ and $192^{\circ}$, the PA swing and RM($\phi$) are relatively flat. 
The values of RM$_{\text{profile}}$ and RM$_{\text{scatt}}$ are inconsistent, indicating that low-level scattering might affect the pulsar.
Other authors \citep[e.g.][]{kmn+15,lkk15,phs+16} indeed reported finding small amounts of scattering at lower frequencies.
There are significant $\Delta(V/I)$ variations across the whole profile, with the largest where the PA swing is the steepest and the RM($\phi$) curve is also changing the most. In this region, Stokes $V$ is also changing, as is expected for scattering. Therefore all indicators are consistent with  interstellar scattering being the main mechanism responsible for the apparent RM variations.

\textbf{PSR J0536$-$7543*}. This pulsar has a high degree of linear polarization and a steep `S'-shaped PA swing, except for the observed OPM break towards the trailing edge, around pulse longitude $185^{\circ}$.  
The RM$(\phi)$ curve is flat in the leading part of the profile, where the PA swing is also relatively flat. Deviations ($\sim 10$ rad m$^{-2}$) occur starting at longitude $\sim170^{\circ}$, where the PA curve is steepest. 
Significant $\Delta(V/I)(\phi)$ variations can be seen in the same region, which is also where Stokes $V$ changes the most,
hence we cannot distinguish which frequency dependent effect is responsible for the apparent RM variations (see Section~\ref{sect:method}).

\textbf{PSR J0738$-$4042}. 
The PA swing of the pulsar reveals five OPM jumps. At the pulse longitudes where these jumps  occur, significant deviations can be seen in RM($\phi$). Towards the leading edge of the profile, $L$ is weak and the RM($\phi$) uncertainties are high, however a significant dip can be observed in RM$(\phi)$.
\cite{nkk+09} classified this pulsar as having low apparent RM($\phi$) variations, since they only report a slow deviation in the centre of the profile, as we see between longitudes $\sim170^{\circ}$ and $\sim185^{\circ}$. However, we observe much larger deviations with a maximum amplitude of $\sim$ 35 rad m$^{-2}$ where OPM jumps occur. 
Complex intensity and polarization evolution with time and frequency has been reported for this pulsar \citep{krj+11}, explaining the difference in the shape of our profile compared to what was seen in 2004 and 2006. 
The greatest change in $\Delta(V/I)(\phi)$ occurs at pulse longitude $\sim165^{\circ}$, coincident with the largest change in RM($\phi$), and with an OPM transition. However, this is not where Stokes $V$ changes most rapidly. This suggests that scattering may not be the dominant cause for the observed RM variations, and a magnetospheric effect is significant for this pulsar.

\textbf{PSR J0820$-$1350*}. The PA swing is very steep with several kinks around pulse longitudes $179^{\circ}$ and $184^{\circ}$, however there are no OPM jumps.  $L$ is low ($\sim20\%$), and Stokes $V$ has comparable magnitude.
RM($\phi$) variations are present across most pulse longitudes. The highest amplitude variations are located where the two kinks in the PA swing occur. 
The low degree of $L$ and the very steep PA swing means that RM$_{\text{scatt}}$ is not very significant, as reflected in the high systematic and statistical uncertainties.
We see large changes in $\Delta(V/I)(\phi)$ up to pulse longitude $182^{\circ}$, coincident with several changes in Stokes $V$, as expected for 
scattering. It is however curious that the $\Delta(V/I)(\phi)$ variations occur only up to pulse longitude $182^{\circ}$, even if Stokes $V$ is slowly changing until pulse longitude $184^{\circ}$. There might be a direct correlation between the $\Delta(V/I)(\phi)$ and RM($\phi$) curves, if the RM($\phi$) is distorted downwards before pulse longitude $181^{\circ}$. There could be magnetospheric effects affecting the polarization of this pulsar.

\textbf{PSR J0837$-$4135*}. This pulsar has a three component profile, with a bright central component and a weaker post and pre-cursor. $L$ is low ($\sim$$20\%$) and is comparable with Stokes $V$. 
The shape of the PA swing is complex: up to pulse longitude $175^{\circ}$ it is flat with a slight upwards gradient; in this region, the RM($\phi$) curve is flat. 
The highest amplitude apparent variations in RM($\phi$) are near the only OPM jump, at pulse longitude $175^{\circ}$. In the centre of the profile, there are several kinks in the PA swing. Where the most prominent kink occurs (at pulse longitude $\sim$$180^{\circ}$), we observe a significant dip in the shape of the RM($\phi$) values. 
Around pulse longitude $182^{\circ}$ there is a steep PA gradient, coinciding with another region of high amplitude variations in RM($\phi$). The greatest $\Delta(V/I)$ variations occur in the central region, where Stokes $V$ is also changing rapidly. Considering that both RM($\phi$) and $\Delta(V/I)(\phi)$ variations  happen where the PA and Stokes $V$ vary most rapidly, in addition to the correlation between the gradient of the PA swing and apparent RM variations suggest that scattering is the cause for RM variations. 
Scatter broadening has been previously reported at lower frequencies (e.g. \citealt{mr01}).

\textbf{PSR J0907$-$5157}. 
For this pulsar, $L$ is moderately strong, peaking towards the centre of the profile, and the PA swing is relatively flat with and OPM jump at pulse longitude $130^{\circ}$. 
This pulsar was classified by \cite{nkk+09} as showing no RM($\phi$) variations. In our observation,

\begin{landscape}
\begin{table}
\centering
\caption{Results for the 98 pulsars analysed. Column 2 displays our measured value of the DM using \textsc{tempo2}. Column 3 displays the trial DM offset which minimized $\chi^2_{\text{RM}(\phi)}$. Column 4 displays the most recent published value of RM for all pulsars, taken from the ATNF catalogue \citep{mhth05}. 
The first uncertainties displayed in Columns 5 and 6 are statistical and the second are systematic.
Pulsars for which the phase-resolved RM profiles had never been investigated were marked with an asterisk (*). 
\textbf{References}: (1) \citet{nsk+15} (2) \citet{hmq99} (3) \citet{fdr15} (4) \citet{njk+08} 
(5) \citet{qmlg95} (6) \citet{jkk+07} (7) \citet{hmv+18}
(8) \citet{jhv+05} (9) \citet{hml+06} (10) \citet{hl87} (11) \citet{tml93} 
(12) \citet{cmh91} (13) \citet{rl94}.
%
}
\label{table:bigtable}
\begin{tabular}{lrrrrrrrrrrrrrl}
\hline
Pulsar name &  DM & DM$_{\text{offset}}$ & RM$_{\text{cat}}$ & RM$_{\text{profile}}$ & RM$_{\text{scatt}}$ & $\left<\text{RM}(\phi)\right>$ & $\chi^2_{\text{RM}(\phi)}$ & $\chi^2_{\text{PA}(\lambda^2)}$\\
& (cm$^{-3}$pc) & (cm$^{-3}$pc) & (rad m$^{-2}$) & (rad m$^{-2}$) & (rad m$^{-2}$) & (rad m$^{-2}$)\\
\hline & \vspace{-3mm} \\
J0034$-$0721 &  14.2 $\pm$ 0.1  &  $-0.30 $ & 10.977 $\pm$ 0.004$^1$ &        $ 8.1  \pm  0.9 \pm 1.3 $ & 63.1 $\pm$ 5.7  $\pm$ 27.4 &    $ 7.6  \pm  0.3 $ &    $ 2.1  $ &     2.7\\
J0134$-$2937  &  21.792 $\pm$ 0.003  &  0.40  & 13 $\pm$ 2$^2$ &        $ 15.9  \pm  0.3 \pm  0.3 $ & 17.6 $\pm$ 0.3 $\pm$ 0.4 &    $ 15.7  \pm  0.2 $ &    $ 1.2  $ &     0.9\\
J0152$-$1637* &  11.95 $\pm$ 0.04 &  0.30  & 6.6 $\pm$ 5.0$^3$  &        $ 9.1  \pm  3.5 \pm 1.4 $   &   8.5 $\pm$ 3.4 $\pm$ 1.5 &    $ 8.3  \pm  0.7 $ &    $ 0.6  $ &     1.5\\
J0255$-$5304* &  17.879 $\pm$ 0.009  &  0.15  & 32 $\pm$ 3$^4$  &        $ 22.1  \pm  0.9 \pm  1.6 $   & 16.7 $\pm$ 2.9 $\pm$ 9.6 &    $ 21.9  \pm  0.3 $ &    $ 13.3  $ &     1.6\\
\hline

J0401$-$7608* &  21.68 $\pm$ 0.02  &  0.05  & 19.0 $\pm$ 0.5$^5$ &        $ 24.5  \pm  0.4 \pm  0.4 $ &     26.1 $\pm$ 0.7 $\pm$ 1.0 &    $ 24.6  \pm  0.2 $ &    $ 2.7  $ &     1.4\\
J0452$-$1759* &  39.76 $\pm$ 0.02  &  0.30  & 11.1 $\pm$ 0.3$^6$ &        $ 13.6  \pm  0.1 \pm  0.4 $     & 20.8 $\pm$ 0.1 $\pm$ 1.2 &    $ 13.6  \pm  0.1 $ &    $ 54.5 $ &     27.3\\
J0536$-$7543* &  18.51 $\pm$ 0.03  &  $-0.45$  & 23.8 $\pm$ 0.9$^7$ &        $ 25.4  \pm  0.1  \pm  0.2 $    & 26.8 $\pm$ 0.2 $\pm$ 1.2  &    $ 25.4  \pm  0.1 $ &    $ 15.9  $ &     3.2\\
J0614$+$2229* &  96.88 $\pm$ 0.01 &  0.25  & 66.0 $\pm$ 0.3$^6$ &        $ 66.7  \pm  0.2 \pm  0.4  $     & 66.1 $\pm$ 0.3 $\pm$ 0.3 &    $ 66.7  \pm  0.2 $ &    $ 1.1  $ &     1.6\\
\hline

J0630$-$2834* &  35.08 $\pm$ 0.06  & $ -0.40$  & 46.53 $\pm$ 0.12$^8$ &        $ 44.8  \pm  0.1 \pm  0.3 $    & 44.1 $\pm$ 0.1 $\pm$ 0.6 &    $ 44.8  \pm  0.1 $ &    $ 1.1  $ &     10.7\\
J0729$-$1836* &  61.26 $\pm$ 0.02  &  $-0.05$  & 51 $\pm$ 4$^4$ &        $ 52.6  \pm  0.8 \pm  0.7 $     & 54.7 $\pm$ 1.1  $\pm$ 2.0 &    $ 52.5  \pm  0.3 $ &    $ 0.7  $ &     1.0\\
J0738$-$4042  &  160.785 $\pm$ 0.003  &  0.00  & 12.1 $\pm$ 0.6$^4$ &        $ 11.98  \pm  0.02 \pm  0.20 $ &  10.32 $\pm$ 0.02  $\pm$ 1.12 &    $ 11.97  \pm  0.04 $ &    $ 10.5  $ &     513.2\\
J0742$-$2822  &  73.754 $\pm$ 0.003  &  0.00  & 149.95 $\pm$ 0.05$^8$ &       $ 149.64  \pm  0.03  \pm  0.30  $   & 149.72 $\pm$ 0.03  $\pm$  0.20 &    $ 149.64  \pm  0.06 $ &    $ 1.3  $ &     55.9\\
\hline

J0745$-$5353* &  121.88 $\pm$ 0.01  &  0.40  & -71 $\pm$ 4$^4$ &        $ -75.6  \pm  0.7 \pm  0.7$ &    $ -77.4 \pm 1.0 \pm 1.5$  &    $ -75.7  \pm  0.3 $ &    $ 0.9  $ &     1.7\\
J0809$-$4753* &  228.41 $\pm$ 0.02  &  0.50  & 105 $\pm$ 5$^9$ &        $ 101.1  \pm  0.9 \pm 0.6$     & 104.3 $\pm$ 1.3 $\pm$ 1.6 &    $ 101.2  \pm  0.3 $ &    $ 0.9  $ &     1.2\\
J0820$-$1350* &  40.93 $\pm$ 0.03  &  $-0.05$  & -1.2 $\pm$ 0.4$^{10}$ &        $ -3.1  \pm  0.3 \pm 0.7 $   & $-10.6 \pm$ 1.8 $\pm$ 4.1 &    $ -3.1  \pm  0.2 $ &    $ 7.1  $ &     1.4\\
J0835$-$4510  &  67.894 $\pm$ 0.001 &  0.00  & 31.38 $\pm$ 0.01$^8$ &        $ 39.3  \pm  0.0  \pm 0.8 $ &  39.5 $\pm$ 0.0 $\pm$ 0.4 &    $ 39.3  \pm  0.0 $ &    $ 412.2  $ &     1464.3\\
\hline

J0837$-$4135* &  147.176 $\pm$ 0.003  &  0.10  & 145 $\pm$ 1$^6$ &        $ 144.6  \pm  0.1 \pm 0.8 $   & 141.9 $\pm$ 0.1 $\pm$ 2.0 &    $ 144.63  \pm  0.09 $ &    $ 91.8  $ &     183.2\\
J0907$-$5157  &  103.659 $\pm$ 0.007  &  $-0.05$  & -23.3 $\pm$ 1.0$^4$ &        $ -25.7  \pm  0.1 \pm 0.6 $  & $-26.5$ $\pm$ 0.2 $\pm$ 0.6 &    $ -25.9  \pm  0.1 $ &    $ 2.2  $ &     3.4\\
J0908$-$4913* (MP) &  180.2062 $\pm$ 0.0008  &  0.00  & 10.0 $\pm$ 1.6$^5$ &        $ 14.85  \pm  0.03 \pm 0.30 $ & 15.2 $\pm$ 0.04 $\pm$ 0.30 &    $ 14.85  \pm  0.07 $ &    $ 5.3  $ &     4.4\\
J0908$-$4913* (IP) &  180.2062 $\pm$ 0.0008  &  0.00  & 10.0 $\pm$ 1.6$^5$ &        $ 14.3  \pm  0.1  \pm $ 0.4  & 14.3 $\pm$ 0.1 $\pm$ 0.3 &    $ 14.3  \pm  0.1 $ &    $ 1.1  $ &     0.4\\
\hline

J0942$-$5552  &  180.24 $\pm$ 0.02  &  0.25  & -61.9 $\pm$ 0.2$^{11}$ &        $ -64.8  \pm  0.3 \pm 0.7 $     & $-68.4 \pm 0.4 \pm 1.2$ &    $ -64.8  \pm  0.2 $ &    $ 2.18  $ &     1.52\\
J1001$-$5507* &  130.246 $\pm$ 0.008  &  0.05  & 297 $\pm$ 18$^9$ &        $ 270.9  \pm  0.4 \pm 0.4$     &    286.7 $\pm$ 1.5 $\pm$ 2.9 &    $ 270.8  \pm  0.3 $ &    $ 11.3  $ &     1.2\\
J1043$-$6116* &  449.02 $\pm$ 0.01  &  $-0.40 $ & 257 $\pm$ 23$^9$ &        $ 189.6  \pm  2.9 \pm 0.4 $   & 189.0 $\pm$ 2.9 $\pm$ 0.6 &    $ 188.2  \pm  0.6 $ &    $ 0.6  $ &     0.9\\
J1047$-$6709* &  116.269 $\pm$ 0.007  &  $0.40$  & -79.3 $\pm$ 2.0$^4$ &        $ -79.4  \pm  0.3 \pm 0.5 $   & $-81.5 \pm 0.3 \pm 0.5$ &    $ -79.5  \pm  0.2 $ &    $ 0.5  $ &     1.3\\
\hline

J1048$-$5832* &  128.721 $\pm$ 0.006  &  $-0.10$  & -155 $\pm$ 5$^5$ &        $ -151.4  \pm  0.1 \pm 0.3 $ & $-150.7 \pm 0.1  \pm 0.3 $ &    $ -151.37  \pm  0.09 $ &    $ 0.9  $ &     5.0\\
J1056$-$6258  &  320.64 $\pm$ 0.01  &  0.45  & 4 $\pm$ 2$^{12}$ &        $ 6.5  \pm  0.1 \pm 0.8 $   & 8.1 $\pm$ 0.1 $\pm$ 0.6&    $ 6.5  \pm  0.1 $ &    $ 1.7  $ &    13.5\\
J1057$-$5226  (MP) &  29.717 $\pm$ 0.003  &  0.25  & 47.2 $\pm$ 0.8$^{11}$ &        $ 46.56  \pm  0.01  \pm 0.30 $    & 47.02 $\pm$ 0.02 $\pm$ 0.40&    $ 46.56  \pm  0.04 $ &    $ 10.9  $ &     2.3\\
J1057$-$5226* (IP) &  29.717 $\pm$ 0.003  &  0.45  & 47.2 $\pm$ 0.8$^{11}$ &        $ 45.39  \pm  0.03 \pm 0.50 $ & 45.18 $\pm$ 0.04 $\pm$ 0.60 &    $ 45.39  \pm  0.05 $ &    $ 27.3  $ &     1.1\\
\hline

J1136$-$5525* &  85.41 $\pm$ 0.02  &  0.30  & 28 $\pm$ 5$^5$ &        $ 27.3  \pm  2.0 \pm 1.3 $ & 32.9 $\pm$ 7.1 $\pm$ 7.5 &    $ 27.2  \pm  0.5 $ &    $ 1.1  $ &     0.8\\
J1146$-$6030* &  111.67 $\pm$ 0.01  &  $-0.05$  & -5 $\pm$ 4$^4$ &        $ 2.6  \pm  0.5  \pm 0.3 $   &  1.5 $\pm$ 1.0 $\pm$ 0.3 &    $ 2.7  \pm  0.3 $ &    $ 0.7  $ &     1.2\\
J1157$-$6224  &  324.32 $\pm$ 0.02  &  $-0.20$  & 508.2 $\pm$ 0.5$^{11}$ &        $ 507.2  \pm  0.3 \pm 0.3$ & 507.1 $\pm$ 0.4 $\pm$ 0.3&    $ 507.4  \pm  0.2 $ &    $ 1.1  $ &     2.9\\
J1243$-$6423  &  297.046 $\pm$ 0.002  &  0.10  & 157.8 $\pm$ 0.4$^{11}$ &        $ 161.9  \pm  0.1 \pm 0.8 $   &    167.1 $\pm$ 0.2 $\pm$ 0.8 &    $ 161.93  \pm  0.1 $ &    $ 87.3  $ &     9.7\\

\hline
\end{tabular}
\end{table}
\end{landscape}


\begin{landscape}
\begin{table}
\centering
\contcaption{}

\begin{tabular}{lrrrrrrrrrrrrl}
\hline
Pulsar name &  DM & DM$_{\text{offset}}$ & RM$_{\text{cat}}$ & RM$_{\text{profile}}$ & RM$_{\text{scatt}}$ & $<\text{RM}(\phi)>$ & $\chi^2_{\text{RM}(\phi)}$ & $\chi^2_{\text{PA}(\lambda^2)}$\\
& (cm$^{-3}$pc) & (cm$^{-3}$pc) & (rad m$^{-2}$) & (rad m$^{-2}$) & (rad m$^{-2}$) & (rad m$^{-2}$) \\
\hline & \vspace{-3mm} \\
J1306$-$6617* &  437.09 $\pm$ 0.03  &  0.45  & 396 $\pm$ 4$^4$ &        $ 393.9  \pm  0.7 \pm 0.3 $  & 390.8 $\pm$ 1.6 $\pm$ 0.4 &    $ 393.7  \pm  0.3 $ &    $ 1.6  $ &     1.9\\
J1326$-$5859  &  287.253 $\pm$ 0.006  &  $-0.50$  & -579.6 $\pm$ 0.9$^4$ &        $ -578.1  \pm  0.1 \pm 0.2 $   & $-581.4 \pm 0.1 \pm 0.3 $ &    $ -578.1  \pm  0.1 $ &    $ 124.6  $ &     13.6\\
J1326$-$6408* &   502.612  $\pm$ 0.031 &  $ -0.50 $ & 226.4 $\pm$ 3.8$^{7}$ &               $ 235.5  \pm  1.9 \pm 0.4 $ &  $ 231.7  \pm  3.2 \pm 0.8 $ & $ 235.8  \pm  0.5 $ &    $ 0.6  $ &     0.6\\
J1326$-$6700* &  209.17 $\pm$ 0.03  &  0.15  &  $-47 \pm$ 1$^9$ &  $ -51.5  \pm  0.2 \pm 0.8 $     &  $-58.3 \pm 0.3 \pm 1.5 $ &    $ -51.5  \pm  0.1 $ &    $ 1.4  $ &     1.3\\
\hline

J1327$-$6222* &  318.43 $\pm$ 0.01  &  $-0.50$  & $-306 \pm$ 8$^9$ &        $ -316.0  \pm  0.3 \pm  0.4 $    & $-$318.8 $\pm$ 0.7 $\pm$ 0.5 &    $ -316.1  \pm  0.2 $ &    $ 1.9  $ &     6.9\\
J1328$-$4357* &  41.58 $\pm$ 0.02  &  0.00  & $-22.9 \pm$ 0.9$^3$ &        $ -34.8  \pm  0.2 \pm 1.4 $   & $-31.1 \pm 0.4 \pm 1.0 $ &    $ -34.8  \pm  0.2 $ &    $ 1.0  $ &     1.4\\
J1357$-$62*   &  416.47 $\pm$ 0.03  &  $-0.45$  & $-586 \pm$ 5$^9$ &        $ -585.4  \pm  0.4 \pm 0.2 $    & $-594.0 \pm 1.3 \pm 1.0 $ &    $ -586.6  \pm  0.2 $ &    $ 2.0  $ &     1.3\\
J1359$-$6038  &  293.738 $\pm$ 0.004  &  0.50  & 33 $\pm$ 5$^2$ &        $ 38.5  \pm  0.1  \pm 0.3 $     & 36.5 $\pm$ 0.1 $\pm$ 0.3 &    $ 38.5  \pm  0.1 $ &    $ 3.4  $ &   2.7\\
\hline

J1401$-$6357* &  97.76 $\pm$ 0.01  &  $-0.35$  & 62 $\pm$ 4$^5$ &        $ 52.0  \pm  0.2 \pm 1.0 $  & 53.1 $\pm$ 0.3 $\pm$ 0.8 &    $ 52.0  \pm  0.2 $ &    $ 4.3  $ &     1.4\\
J1428$-$5530* &  82.19 $\pm$ 0.02  &  $-0.20$  & 4 $\pm$ 3$^5$ &        $ -10.8  \pm  1.4 \pm 1.5 $   & 4.4 $\pm$ 2.0 $\pm$ 4.7 &    $ -9.6  \pm  0.4 $ &    $ 2.1  $ &     3.0\\
J1430$-$6623* &  65.13 $\pm$ 0.01  &  0.10  & $-19.2 \pm$ 0.3$^8$ &        $ -22.2  \pm  0.3 \pm 0.7 $   & $-26.5 \pm 2.5 \pm 17.8 $ &    $ -22.2  \pm  0.2 $ &    $ 3.7  $ &     1.4\\
J1453$-$6413  &  71.256 $\pm$ 0.006  &  0.00  & $-18.6 \pm$ 0.2$^8$ &        $ -22.6  \pm  0.1 \pm 0.4  $  & $-41.9 \pm 0.3 \pm 1.3 $ &    $ -22.6  \pm  0.1 $ &    $ 5.2  $ &     2.3\\
\hline

J1456$-$6843* &  8.720 $\pm$ 0.007  &  $-0.15$  & $-$4.0 $\pm$ 0.3$^8$ &        $ -0.9  \pm  0.1 \pm 1.0  $ &  1.8 $\pm$ 0.1 $\pm$ 3.3 &    $ -0.9  \pm  0.1 $ &    $ 54.1  $ &     13.9\\
J1512$-$5759* &    627.535      $\pm$   0.045    &  $-0.5$  &  510.0 $\pm$ 0.7$^4$ & 		  511.7 $\pm$  1.1 $\pm$  0.3 &   500.7 $\pm$ 5.6 $\pm$  2.1 &    511.9  $\pm$  0.3  &   2.0  &     1.3\\
J1522$-$5829  &  199.83 $\pm$ 0.02  &  0.15  & $-24.2 \pm$ 2.0$^4$ &        $ -24.3  \pm  0.4 \pm 0.3 $   &   $-55.6 \pm 4.2 \pm 5.3 $ &    $ -24.2  \pm  0.2 $ &    $ 1.4  $ &     0.9\\
J1534$-$5334* &  25.39 $\pm$ 0.04  &  $-0.05$  & $-46 \pm$ 17$^9$ &        $ 21.9  \pm  0.8 \pm 1.0 $ & 17.9 $\pm$ 1.2 $\pm$ 1.7 &    $ 21.7  \pm  0.3 $ &    $ 2.0  $ &     1.2\\
\hline

J1539$-$5626* &  175.90 $\pm$ 0.01  &  $0.1$  & $-18.0 \pm$ 2.0$^4$ &        $ -14.4  \pm  0.5 \pm 0.3 $    &  $-16.7 \pm 0.7 \pm 0.8 $ &    $ -14.6  \pm  0.2 $ &    $ 1.1  $ &     1.6\\
J1544$-$5308* &  35.214 $\pm$ 0.007  &  $-0.20$  & $-29 \pm$ 7$^2$ &        $ -41.9  \pm  1.0 \pm 0.7 $   &  $-37.7 \pm 1.2 \pm 1.9 $ &    $ -41.5  \pm  0.4 $ &    $ 1.0  $ &     0.8\\
J1555$-$3134  &  73.01 $\pm$ 0.02  &  0.25  & $-49 \pm$ 6$^{10}$ &        $ -52.4  \pm  2.0 \pm 1.1 $   & $-82.3 \pm 4.9 \pm 6.9 $ &    $ -52.3  \pm  0.4 $ &    $ 1.0  $ &    1.3\\
J1557$-$4258* &  144.43 $\pm$ 0.01  &  $-0.10$  & $-41.9 \pm$ 2.0$^4$ &        $ -37.2  \pm  0.7 \pm 0.6 $ & $ -40.0 \pm 1.2 \pm 1.4 $ &    $ -37.5  \pm  0.3 $ &    $ 0.9  $ &     0.8\\
\hline

J1559$-$4438* &  55.87 $\pm$ 0.01  &  $-0.20$  & $-5.0 \pm$ 0.6$^6$ &        $ -2.9  \pm  0.1 \pm  0.4 $  &   $-7.4 \pm 0.1 \pm 0.7 $ &    $ -2.9  \pm  0.1 $ &    $ 39  $ &     2.8\\
J1602$-$5100* &  170.921 $\pm$ 0.008  &  0.10  & 71.5 $\pm$ 1.1$^{11}$ &        $ 84.8  \pm  0.5 \pm 0.8 $  & 80.1 $\pm$ 0.9 $\pm$ 0.5 &    $ 84.7  \pm  0.3 $ &    $ 4.1  $ &     1.6\\
J1604$-$4909* &  140.730 $\pm$ 0.007  &  0.10  & 34 $\pm$ 1$^6$ &        $ 13.4  \pm  1.0 \pm 1.1 $  & 30.8 $\pm$ 3.1 $\pm$ 3.7 &    $ 13.8  \pm  0.4 $ &    $ 3.9  $ &     0.9\\
J1605$-$5257* &  35.03 $\pm$ 0.04  &  0.20  & 1.0 $\pm$ 2.0$^4$ &        $ 3.7  \pm  0.2 \pm  0.3 $   & 4.8 $\pm$ 0.5 $\pm$ 0.9 &    $ 3.76  \pm  0.14 $ &    $ 1.8  $ &     1.8\\
\hline

J1633$-$4453* &   474.022     $\pm$  0.025     & $-0.45$  &  159.0 $\pm$ 0.6$^4$ &         $ 157.4  \pm  1.8 \pm 1.0 $ &  $ 153.7  \pm  1.7 \pm 1.1 $ & $ 156.5  \pm  0.5 $ & $ 1.0  $ &     2.0\\
J1633$-$5015* &  399.04 $\pm$ 0.01  &  $-0.35$  & 406.1 $\pm$ 2.0$^4$ &        $ 406.8  \pm  0.3 \pm 0.3 $ & 407.4 $\pm$ 0.4 $\pm$ 0.3  &    $ 406.7  \pm  0.2 $ &    $ 1.8  $ &     5.7\\
J1644$-$4559 & 478.6673 $\pm$ 0.0071 &  $-0.5$ & $-626.9$ $\pm$ 0.8$^{7}$ & $-623.3$ $\pm$ 0.1 $\pm$ 0.2 & $-621.4$ $\pm$ 0.1 $\pm$ 0.2 & $-623.4$ $\pm$ 0.1 &  404      &  2200
 \\
J1646$-$6831* &  42.18 $\pm$ 0.07  &  $-0.50$  & 105 $\pm$ 3$^5$ &        $ 100.1  \pm  0.4 \pm  0.3  $ & 99.5 $\pm$ 0.5 $\pm$ 0.4 &    $ 100.0  \pm  0.2 $ &    $ 1.3  $ &     1.4\\
\hline

J1651$-$4246* &  481.74 $\pm$ 0.04  &  0.50  & $-167.4  \pm$ 1.1$^7$ &        $ -166.9  \pm  0.1 \pm  2.5  $ & $-170.0 \pm 0.3 \pm 9.2 $ &    $ -166.9  \pm  0.1 $ &    $ 1.4  $ &     1.1\\
J1651$-$5222* &  178.84 $\pm$ 0.03  &  $-0.35$  & $-38 \pm$ 5$^2$ &        $ -47.8  \pm  1.1 \pm 0.5 $ & $-41.1 \pm 2.7 \pm 2.1 $ &    $ -48.6  \pm  0.4 $ &    $ 3.8  $ &     1.1\\
J1653$-$3838* &  206.83 $\pm$ 0.01  &  $-0.15$  & $-82 \pm$ 3$^4$ &        $ -81.2  \pm  0.7 \pm  0.7  $ & $-82.3 \pm 0.7 \pm 0.8 $ &    $ -81.1  \pm  0.4 $ &    $ 0.8  $ &     1.9\\
J1701$-$3726* &  301.1 $\pm$ 0.2  &  $-0.50$  & $-605.9 \pm$ 2.0$^4$ &        $ -607.4  \pm  0.6 \pm  0.3 $   & $ -632.4 \pm 2.7 \pm 0.7 $ &    $ -607.6  \pm  0.3 $ &    $ 5.0 $ &     1.8\\
\hline

J1703$-$3241* &  110.01 $\pm$ 0.03  &  0.20  & $-21.7 \pm$ 0.5$^{10}$ &        $ -22.6  \pm  0.1 \pm  0.7  $ & $-22.4 \pm 0.5 \pm 1.6 $  &    $ -22.6  \pm  0.2 $ &    $ 2.7  $ &     1.2\\
J1705$-$1906  (MP) &  22.906 $\pm$ 0.007  &   0.10  & $-19.2 \pm$ 1.0$^4$ &        $ -21.1 \pm  0.2 \pm  1.5 $ & $-$19.9 $\pm$ 0.2 $\pm$ 0.5  &      $20.8 \pm 0.2$ &    $2.7  $ &     2.6\\
J1705$-$1906* (IP) &  22.906 $\pm$ 0.007  &  0.45  & $-19.2 \pm$ 1.0$^4$ &        $ -20.8  \pm  0.4 \pm  0.3 $  &  $-21.1 \pm 0.3 \pm 0.3 $ &    $ -20.8  \pm  0.3 $ &    $ 1.3 $ &     0.8\\
J1705$-$3423* &  146.34 $\pm$ 0.01 &  0.25  & $-44 \pm$ 8$^9$ &        $ -43.9  \pm  1.1 \pm  1.8 $ & $-44.1 \pm 1.4  \pm 2.0 $ &    $ -44.2  \pm  0.3 $ &    $ 0.8  $ &     1.0\\

\hline
\end{tabular}
\end{table}
\end{landscape}

\begin{landscape}
\begin{table}
\centering
\contcaption{}

\begin{tabular}{lrrrrrrrrrrrrl}
\hline
Pulsar name &  DM & DM$_{\text{offset}}$ & RM$_{\text{cat}}$ & RM$_{\text{profile}}$ & RM$_{\text{scatt}}$ & $<\text{RM}(\phi)>$ & $\chi^2_{\text{RM}(\phi)}$ & $\chi^2_{\text{PA}(\lambda^2)}$\\
& (cm$^{-3}$pc) & (cm$^{-3}$pc) & (rad m$^{-2}$) & (rad m$^{-2}$) & (rad m$^{-2}$) & (rad m$^{-2}$)\\
\hline & \vspace{-3mm} \\
J1709$-$1640* &  24.95 $\pm$ 0.01  &  0.40  & $-1.3 \pm$ 0.3$^{10}$ &        $ -6.3  \pm  0.4 \pm  1.5 $  & $ -1.5 \pm 0.4 \pm 2.0 $ &    $ -5.9  \pm  0.2 $ &    $ 4.4  $ &     4.0\\
J1709$-$4429* &  75.645 $\pm$ 0.005  &  $-0.30$  & 0.70 $\pm$ 0.07$^8$ &        $ -1.7  \pm  0.1 \pm  0.4 $  & $-1.3 \pm 0.1 \pm 0.3 $ &    $ -2.0  \pm  0.1 $ &    $ 0.7  $ &     3.2\\
J1717$-$3425* &   585.20       $\pm$   0.14    &  $-0.05$  &  $-191$ $\pm$ 14$^9$ &       $ -192.1  \pm  2.0 \pm 0.7 $ & $ -192.7  \pm  2.5 \pm 0.9 $ & $ -192.4  \pm  0.6 $ & $ 1.3  $ &   1.9\\
J1721$-$3532* &   496.046       $\pm$  0.016     & $0.40$  & $159$ $\pm$ 4$^4$ &       $ 169.3  \pm  0.9 \pm 0.7 $ & $ 170.1  \pm  1.0 \pm 1.2 $ & $ 169.5  \pm  0.4 $ & $ 1.2  $ &     2.9\\
\hline

J1722$-$3207* &  126.11 $\pm$ 0.02  &  0.00  & 70.4 $\pm$  0.5$^3$ &        $ 67.8  \pm  0.5 \pm 0.4 $ & 63.9 $\pm$ 1.3 $\pm$ 2.3 &    $ 67.6  \pm  0.23 $ &    $ 3.0  $ &     0.9\\
J1722$-$3712* (MP) &  99.47 $\pm$ 0.01  &  $-0.50$  & 104 $\pm$ 3$^2$ &        $ 108.2  \pm  0.5 \pm 0.5 $   & 110.6 $\pm$ 0.7 $\pm$ 0.6 &    $ 108.2  \pm  0.3 $ &    $ 0.9  $ &     1.3\\
J1722$-$3712* (IP) &  99.47 $\pm$ 0.01  &  $-0.25$  & 104 $\pm$ 3$^2$ &        $ 111.1  \pm  2.4 \pm 0.3 $ & 111.5 $\pm$ 2.2 $\pm$ 0.3 &    $ 111.4  \pm  0.7 $ &    $ 0.4  $ &     1.3\\
J1731$-$4744* &  122.87 $\pm$ 0.01  &  $-0.05$  & $-429.1 \pm$ 0.5$^{11}$ &        $ -445.9  \pm  0.1 \pm 0.3 $ & $-445.5 \pm 0.2 \pm 0.3 $ &    $ -445.9  \pm  0.1 $ &    $ 10.5  $ &     14.3\\
\hline

J1739$-$2903* (MP) &  138.58 $\pm$ 0.01  &  0.05  & $-236 \pm$ 18$^9$ &        $ -301.0  \pm  4.8 \pm 0.7 $ & $-301.6 \pm$ 4.7 $\pm$ 0.8 &    $ -300.5  \pm  0.7 $ &    $ 0.8  $ &     1.5\\
J1739$-$2903* (IP) &  138.58 $\pm$ 0.01  &  0.45  & $-236 \pm$ 18$^9$ &        $ -303.3  \pm  1.7 \pm 1.8 $ & $-298.9 \pm 2.2 \pm 0.5 $ &    $ -303.4  \pm  0.5 $ &    $ 0.8  $ &     1.0\\
J1740$-$3015* &  151.82 $\pm$ 0.01  &  0.50  & $-168.0 \pm$ 0.7$^4$ &        $ -158.4  \pm  0.1 \pm 0.3 $ & $-156.9 \pm 0.1 \pm 0.2 $ &    $ -158.4  \pm  0.1 $ &    $ 2.7  $ &     9.5\\
J1741$-$3927* &  158.52 $\pm$ 0.02  &  $-0.50$  & 204 $\pm$ 6$^9$ &        $ 211.2  \pm  0.7 \pm 0.4 $ & 213.5 $\pm$ 2.4 $\pm$ 1.6  &    $ 210.8  \pm  0.3 $ &    $ 1.7  $ &     1.2\\
\hline

J1745$-$3040  &  88.074 $\pm$ 0.008 &  $-0.20$  & 101 $\pm$ 7$^{10}$ &        $ 96.0  \pm  0.2 \pm 0.4  $ & 94.7 $\pm$ 0.2 $\pm$ 0.7 &    $ 96.0  \pm  0.1 $ &    $ 2.5  $ &     1.8\\
J1751$-$4657* &  20.61 $\pm$ 0.01  &  $-0.10$  & 19 $\pm$ 1$^2$ &        $ 17.1  \pm  0.3 \pm 1.0 $  & 27.7 $\pm$ 0.9  $\pm$ 2.8 &    $ 17.3  \pm  0.3 $ &    $ 7.5  $ &     1.7\\
J1752$-$2806* & 50.302 $\pm$ 0.003 & 0.00 & 96 $\pm$ 2$^{10}$ & 87.1  $\pm$ 0.3  $\pm$ 2.8 & 78.4 $\pm$ 0.6  $\pm$ 7.1 & 86.6 $\pm$ 0.2 & 72   &  6.0 \\
J1807$-$0847  &  112.344 $\pm$ 0.003  &  0.25  & 166 $\pm$ 9$^{10}$ &        $ 163.3  \pm  0.2 \pm  0.3 $ & 168.7 $\pm$ 0.8 $\pm$ 2.2 &    $ 163.3  \pm  0.1 $ &    $ 2.0  $ &     4.5\\
\hline

J1817$-$3618* &  94.29 $\pm$ 0.02  &  0.45  & 66 $\pm$  4$^5$ &        $ 66.0  \pm  1.1 \pm 0.8 $ & 60.6 $\pm$ 2.6 $\pm$ 3.4 &    $ 65.5  \pm  0.4 $ &    $ 2.6  $ &     0.9\\
J1820$-$0427* &  84.48 $\pm$ 0.02  &  $-0.35 $ & 69.2 $\pm$ 0.2$^8$ &        $ 62.5  \pm  0.2 \pm 1.2 $ & 59.8 $\pm$ 0.3 $\pm$ 1.6 &    $ 62.4  \pm  0.2 $ &    $ 7.3  $ &     1.4\\
J1822$-$2256* &  120.9 $\pm$ 0.1  &  $-0.40$  & 124 $\pm$ 3$^4$ &        $ 134.0  \pm  0.8 \pm 0.6  $ & 133.5 $\pm$ 0.9 $\pm$ 0.7 &    $ 133.6  \pm  0.3 $ &    $ 0.7  $ &     2.8\\
J1823$-$3106* &  50.275 $\pm$ 0.006  &  0.50  & 95 $\pm$ 5$^8$ &        $ 89.2  \pm  0.2 \pm 0.5 $ & 91.9 $\pm$ 0.2 $\pm$ 0.7 &    $ 89.1  \pm  0.2 $ &    $ 1.5  $ &     1.5\\
\hline

J1824$-$1945* &  224.31 $\pm$ 0.02  &  $-0.25$  & $-302.2 \pm$ 0.7$^3$ &        $ -297.1  \pm  0.3 \pm 0.7 $   & $-296.6 \pm 0.9 \pm 1.1 $ &    $ -297.1  \pm  0.2 $ &    $ 9.5  $ &     1.0\\
J1825$-$0935* (MP) &  19.62 $\pm$ 0.03  &  $-0.50$  & 65.2 $\pm$ 0.2$^{10}$ &        $ 68.3  \pm  0.7 \pm 1.2 $ & 69.1 $\pm$ 0.8 $\pm$ 1.3 &    $ 68.4  \pm  0.3 $ &    $ 1.3  $ &     2.0\\
J1825$-$0935* (IP) &  19.62 $\pm$ 0.03  &  0.35  & 65.2 $\pm$ 0.2$^{10}$ &        $ 51.1  \pm  10.6 \pm 2.0 $ & 59.1 $\pm$ 9.7 $\pm$ 3.0 &    $ 58.5  \pm  1.18 $ &    $ 0.8  $ &     3.9\\
J1829$-$1751* &  216.79 $\pm$ 0.01  &  0.00  & 304.7 $\pm$ 0.4$^3$ &        $ 303.1  \pm  0.2 \pm  0.2 $ & 300.1 $\pm$ 1.0 $\pm$ 1.3 &    $ 303.1  \pm  0.2 $ &    $ 1.0  $ &     0.8\\
\hline

J1830$-$1059* &  159.71 $\pm$ 0.02  &  0.30  & 47 $\pm$ 5$^{13}$ &        $ 44.6  \pm  0.5 \pm  0.4  $ & 45.9 $\pm$ 0.6 $\pm$ 0.4 &    $ 44.4  \pm  0.3 $ &    $ 1.0  $ &     1.0\\
J1832$-$0827* &  301.01 $\pm$ 0.02  &  $-0.45 $ & 39 $\pm$ 7$^{13}$ &        $ 17.8  \pm  0.8 \pm  0.4 $ & 16.6 $\pm$ 1.7 $\pm$ 1.4 &    $ 17.7  \pm  0.4 $ &    $ 0.7  $ &     1.0\\
J1845$-$0743* &  280.99 $\pm$ 0.01  &  $-0.20 $ & 448.4 $\pm$ 1.8$^7$ &        $ 449.5  \pm  0.8 \pm 0.3 $ & 436.5 $\pm$ 8.0 $\pm$ 2.8  &    $ 449.4  \pm  0.3 $ &    $ 0.7  $ &     1.0\\
J1847$-$0402* &  141.42 $\pm$ 0.03  &  $-0.15 $ & 117 $\pm$ 8$^{10}$ &        $ 106.9  \pm  0.9  \pm  0.3 $ & 113.9 $\pm$ 1.6 $\pm$ 1.7 &    $ 106.9   \pm  0.3 $ &    $ 1.0  $ &     1.2\\
\hline

J1848$-$0123* &  159.88 $\pm$ 0.02  &  0.25  & 580 $\pm$ 30$^{11}$ &        $ 513.6  \pm  0.5 \pm  0.2 $ & 521.0 $\pm$ 1.3 $\pm$ 0.9 &    $ 513.9  \pm  0.2 $ &    $ 3.5  $ &     2.0\\
J1852$-$0635* &  173.14 $\pm$ 0.03  &  0.45  & 414.5 $\pm$ 0.7$^{7}$  &        $ 413.6  \pm  0.3 \pm  0.3 $   & 414.9 $\pm$ 0.4 $\pm$ 0.3  &    $ 413.6  \pm  0.1 $ &    $ 0.9  $ &     2.0\\
J1900$-$2600*&  38.22 $\pm$ 0.03  &  $-0.50 $ & $-9.3 \pm$ 0.2$^7$ &        $ -5.9  \pm  0.2 \pm  0.3 $  & $-1.3 \pm 0.3 \pm 0.6 $ &    $ -5.9  \pm  0.1 $ &    $ 5.8  $ &     0.9\\
J1913$-$0440* &  89.49 $\pm$ 0.02  &  $-0.20$  & 3.98 $\pm$ 0.05$^1$ &        $ 4.6  \pm  0.5 \pm  1.1 $ & 0.0 $\pm$ 1.5 $\pm$ 4.6 &    $ 4.7  \pm  0.3 $ &    $ 5.6  $ &     1.0\\
\hline

J1941$-$2602* &  50.12 $\pm$ 0.01  &  $0.00 $ & $-33.5 \pm$ 0.8$^8$ &        $ -33.9  \pm  0.3  \pm 0.3 $   & $-34.0 \pm 0.3 \pm 0.3 $  &    $ -33.7  \pm  0.2 $ &    $ 0.7  $ &     2.0\\
J2048$-$1616  &  11.86 $\pm$ 0.01  &  $-0.50$  & $-10.0 \pm$ 0.3$^6$ &        $ -10.5  \pm  0.1 \pm  0.4 $ & $-11.0 \pm$ 0.1 $\pm$ 0.7 &    $ -10.5  \pm  0.1 $ &    $ 12.7  $ &     22.8\\
J2330$-$2005* &  8.81 $\pm$ 0.09  &  0.25  & 9.2 $\pm$ 0.8$^3$ &        $ 8.7  \pm  1.3 \pm  0.6 $  & 9.0 $\pm$ 1.8 $\pm$ 2.1  &    $ 7.6  \pm  0.5 $ &    $ 0.8  $ &     1.0\\
J2346$-$0609* &  22.73 $\pm$ 0.03  &  0.50  & $-5 \pm$ 1$^2$ &        $ -6.3  \pm  0.6 \pm  0.3  $ &  $-12.1 \pm  3.6 \pm  1.5 $ &    $ -6.6  \pm  0.2 $ &    $ 1.0  $ &     1.1\\
\hline
\end{tabular}
\end{table}
\end{landscape}

\noindent the RM($\phi$) curve remains generally consistent with $\left<\text{RM}(\phi)\right>$ in the second component, however at earlier pulse longitudes there are significant variations coinciding with variations in $\Delta(V/I)$.
Stokes $V$ is smooth across the whole profile, hence if scattering affected the pulsar, variations in $\Delta(V/I)$ would not be confined to the second component. This is therefore suggestive of magnetospheric effects as a primary cause for the apparent RM variations. 

\textbf{PSR J0908$-$4913*}. This pulsar has a MP and an IP, both being completely linearly polarized. The PA swing is steep, without any OPM jumps. \citet{kj08} determined the geometry of this pulsar at two frequencies, 1.4 GHz and 8.4 GHz, and concluded that this pulsar is an orthogonal rotator and the geometry is independent of frequency, if the effects of interstellar scattering were considered.
The RM($\phi$) curve is flat for most pulse longitudes. The only observed apparent variations are in the MP, towards the trailing edge, where there is a significant upward deviation ($\sim 5$ rad m$^{-2}$), starting with pulse longitude $90^{\circ}$, coinciding with where the PA swing is the steepest. 
The IP appears to be similar to the MP (highly polarized, steep PA), but significant apparent RM($\phi$) variations are not observed.
For the MP,  $\Delta(V/I)$ variations occur at the same pulse longitudes as the RM($\phi$) variations. Stokes $V$ is low and not varying rapidly across the profile, hence scattering would not necessarily be able to create such variations in $\Delta(V/I)(\phi)$, and if it would, it should start earlier. It is possible that magnetospheric effects are the cause of the apparent RM variations.

\textbf{PSR J0942$-$5552}. The profile of this pulsar has three components: a strong central one and two weaker outriders. 
$L$ is moderately high and the PA swing is relatively flat, broken by one OPM jump at pulse longitude 172$^{\circ}$. Around longitude 187$^{\circ}$, a dip can be seen in the PA swing. 
\cite{nkk+09} observed a varying RM($\phi$) curve, by as much as 20 rad m$^{-2}$  in the trailing component, as well as a change in the leading component, but not the dip because of the lack of S/N.
We observe similar trends in the leading component, near the OPM jump, of the order $\sim15$ rad m$^{-2}$ and towards the centre of the profile in a shape of a downward gradient of few rad m$^{-2}$, however we do not observe any significant apparent RM($\phi$) variations in the trailing component.
The statistical uncertainties on RM($\phi$) are relatively large, hence the significance of the variations is moderate.
There are no significant variations observed in the $\Delta(V/I)$. If scattering was responsible for the apparent RM variations, we expect $\Delta(V/I)(\phi)$ variations towards the centre of the profile, where Stokes $V$ is changing. Given the large observed uncertainties on $\Delta(V/I)(\phi)$, this is difficult to verify.
As \citet{mr01} measured a low level of scatter broadening at lower frequencies, it is possible that the observed RM variations were caused by low level scattering.

\textbf{PSR J1001$-$5507*}. This pulsar has a three component profile, with a strong central component and two weaker components. 
$L$ is very low as is Stokes $V$, which displays the common sign reversal towards the centre of the profile.
The PA swing has a complex shape.
The highest apparent RM($\phi$) variations coincide with the first OPM jump and with where the PA gradient is steep. 
The values of RM$_{\text{profile}}$ and RM$_{\text{scatt}}$ suggest that interstellar scattering could affect the RM measurements. Scatter broadening has been previously reported at lower frequencies, as this pulsar is located in the direction of the Gum nebula (e.g. \citealt{mr01}).
We see marginally significant variations in $\Delta(V/I)$ towards the trailing half of the profile, where Stokes $V$ is changing, as expected if scattering was  responsible for the apparent RM variations. Since this coincides with where the RM($\phi$) curve is changing most, it is difficult to distinguish between scattering and a magnetospheric effect as the dominant effect (see Section~\ref{sect:method}).

\textbf{PSR J1057$-$5226}. 
This pulsar has a completely polarized four component MP, with a smooth PA swing. There is also a three component  IP, with a similar flux density to the MP at several observing frequencies \citep{ww09}. The first component of the IP is highly polarized, however $L$ decreases significantly towards the trailing edge. The shape of the PA swing of the IP is peculiar, as towards the centre of the profile there is a sharp gradient change in the PA sweep, accompanied by a drop in $L$, which is hard to fit with the RVM model (e.g. \citealt{rwj15a}). 
 \cite{nkk+09} only analysed the MP and did not find RM($\phi$) variations. From our observations with better S/N, both the MP and IP show significant deviations in RM($\phi$). The amplitude of the observed variations in the MP are $\sim 3$ rad m$^{-2}$ and appear as a shallow gradient in RM($\phi$).
The observed variations in the IP are much larger ($\sim 30$ rad m$^{-2}$) and coincide with the peculiar change of gradient of the PA swing. 
An extreme scattering event with a duration of $\sim 3$ years, was reported in the direction of this pulsar \citep{kcw+18}. Hence, it is very likely that during the time of our observations, the radiation was affected by a low amount of scattering.
For the MP, there are significant $\Delta(V/I)$ variations around longitudes $110^{\circ}$ and possibly $90^{\circ}$, where Stokes $V$ is changing, which is what we expect if scattering was the effect responsible for the apparent RM variations. 
Since this is where the largest RM($\phi$) variations occur, we cannot dismiss magnetospheric effects as a possible cause (see Section~\ref{sect:method}). 
For the IP, there are $\Delta(V/I)(\phi)$ and RM($\phi$) variations only up to pulse longitude $270^{\circ}$, where Stokes $V$ is changing most rapidly.
Evidence points towards low level of interstellar scattering as the reason for the observed RM variations, however we cannot dismiss magnetospheric effects.

\textbf{PSR J1243$-$6423}. 
For this pulsar, $L$ is low ($\sim20\%$), with significant depolarization observed starting from pulse longitude $180^{\circ}$. 
The PA swing is complex, with flat regions and a very steep region in the centre of the profile.  
The observed RM($\phi$) variations have a similar shape to that reported by \cite{nkk+09}. One discrepancy is the amplitude of variations: in our observation it is $\sim 20$ rad m$^{-2}$, while in \cite{nkk+09} it is $\sim 60$ rad m$^{-2}$. 
The largest variations in RM($\phi$) occur at a similar pulse phase where there is a kink in the PA swing.
At higher frequencies, an OPM is seen at the location of this kink \citep{kj06}.
The values of RM$_{\text{profile}}$ and RM$_{\text{scatt}}$ indicate that the pulsar might be affected by scattering. The pulsar appears to be located behind an HII region, however a scattering deficit is reported by \citet{cws+16}.
If scattering affected the pulsar, we would expect to see significant variations in $\Delta(V/I)(\phi)$ around pulse longitudes $179^{\circ}$, where Stokes $V$ is changing most rapidly, as observed. 
However, despite Stokes $V$ being less variable, the largest  RM($\phi$) variations occur where the greatest $\Delta(V/I)(\phi)$ changes occur. This indicates that the observed RM variations are caused by a mixture of scattering and magnetospheric effects.

\textbf{PSR J1326$-$5859}. The profile of this pulsar has three components: one bright central one and two weak outriders. 
Using multi-frequency observations, \citet{lkk15} estimated the scattering timescale for this pulsar at 1 GHz to be 9.47 ms. 
RM$_{\text{profile}}$ and RM$_{\text{scatt}}$ are inconsistent with each other, indicating that this pulsar may indeed be significantly affected by scattering.
The pulsar has a moderately high $L$ and a complex PA swing. There is an OPM jump towards the leading edge of the profile, around pulse longitude $173^{\circ}$, in a region where the PA swing is flat and there is a drop in the amount of $L$. In this region, there are large RM($\phi$) variations, similar to the ones presented in \citet{nkk+09}.
$\Delta(V/I)(\phi)$ varies where Stokes $V$ is most changing as function of pulse longitude. This starts before the largest RM variations. So far, this is consistent with scattering. 
However, the largest RM($\phi$) variations do not coincide with where the PA is the steepest, suggesting that they could have been caused by a mixture of scattering and magnetospheric effects.

\textbf{PSR J1357$-$62*}. 
For this pulsar, the PA swing is generally flat, with the exception of longitude 175$^{\circ}$, where a very steep swing can be seen. In this region, two OPM jumps occur where $L$ is low. 
The first OPM is in the leading part of the profile, while the second one coincides with the bridge between the second and third profile components.   
The RM($\phi$) curve is flat, except where the PA swing is steep and where the second OPM jump occurs.
The peak-to-peak amplitude of these apparent variations is $\sim 60$ rad m$^{-2}$.
We also see variations in $\Delta(V/I)$ where Stokes $V$ is changing the most,  coinciding also with the largest amplitude variations in RM($\phi$). 
Comparing the values of RM$_{\text{profile}}$ and RM$_{\text{scatt}}$, we see that they are inconsistent with each other, therefore all indicators suggests that scattering is likely the reason for the apparent RM variations, however we cannot dismiss magnetospheric effects (see Section~\ref{sect:method}).

\textbf{PSR J1359$-$6038}. The pulsar has a single component profile and a high degree of $L$. 
It was classified by \cite{nkk+09} as having small variations in RM($\phi$) towards the trailing edge of the profile, with an amplitude of $\sim 40$ rad m$^{-2}$. 
In our observation, we see a similar behaviour, however the overall amplitude of the apparent variations is around $\sim 10$ rad m$^{-2}$.  
RM$_{\text{profile}}$ and RM$_{\text{scatt}}$ are inconsistent, indicating that this pulsar could be affected by scattering. \citet{lkk15} estimated a scattering timescale of 1 ms at 1 GHz.
If scattering was responsible for the observed RM variations, $\Delta(V/I)(\phi)$ variations should occur where Stokes $V$ is changing most rapidly, around pulse longitudes $180^{\circ}$ and $185^{\circ}$. $\Delta(V/I)(\phi)$ variations are only observed around pulse longitude $185^{\circ}$, where the greatest RM($\phi$) variations occur. This points towards magnetospheric effects playing a role in producing the observed RM variations.

\textbf{PSR J1401$-$6357*}. The profile of this pulsar consists of one weak leading component and several other blended components. $L$ is very weak in the leading part of the profile.  After pulse longitude 178$^{\circ}$, where an OPM jump occurs in the PA swing, the degree of $L$ is relatively high. 
Moderately significant variations in the RM($\phi$) curve coincide with the OPM jump. 
The degree of Stokes $V$ is very low and we do not see significant $\Delta(V/I)(\phi)$ variations, preventing us saying more about the origin of the apparent RM($\phi$) variations.

\textbf{PSR J1428$-$5530*}.
Both the degree of $L$ and Stokes $V$ of this pulsar are low ($\sim10\%$). The PA swing is relatively flat with several kinks. Where $L$ drops to zero, apparent variations can be seen in RM($\phi$) ($\sim 50$ rad m$^{-2}$). The statistical uncertainties on RM($\phi$) are large, hence this is only moderately significant. There are no significant $\Delta(V/I)(\phi)$ variations to help comment on the origin of the apparent RM($\phi$) variations.

\textbf{PSR J1453$-$6413}. This pulsar has a profile with one strong component, one weak pre-cursor and an extended tail. $L$ is moderately high, with a drop to zero coinciding with the OPM jump in the PA swing at pulse longitude $177^{\circ}$. 
In the extended tail of this pulsar, $L$ is weak and the PA swing becomes very steep around longitude $188^{\circ}$. 
The observed RM($\phi$) variations have a similar shape to the ones presented in \citet{nkk+09}. 
The largest RM($\phi$) variations are where the OPM jump is and where the PA swing is the steepest. 
The values of RM$_{\text{profile}}$ and RM$_{\text{scatt}}$ are inconsistent with each other, indicating that this pulsar may well be affected by scattering.
Hence, the largest $\Delta(V/I)(\phi)$ variations should be in the central region of the profile, where Stokes $V$ displays several steep sign reversals. This is the case. Interestingly, at pulse longitude $185^{\circ}$, where there is a kink in the PA swing, there is a peak in the RM($\phi$) curve and a significant dip in the $\Delta(V/I)(\phi)$ curve. At this pulse longitude, Stokes $V$ is relatively smooth, indicating at least some of the observed RM variations are caused by magnetospheric effects.

\textbf{PSR J1456$-$6843*}. 
For this pulsar, $L$ is low, with one drop to nearly zero at pulse longitude $182^{\circ}$, coincident with a dip in the PA swing. 
The shape of the PA sweep is complex, with one OPM jump around pulse longitude $158^{\circ}$. 
The RM($\phi$) variations display one of the most complex shapes in our sample, with the largest variations $(\sim 20$ rad m$^{-2}$) near where the PA gradient is the steepest.
There are $\Delta(V/I)(\phi)$ variations across the entire profile, however they do not coincide with longitudes where Stokes $V$ is changing the most. The largest $\Delta(V/I)(\phi)$ variations occur around longitude $168^{\circ}$, coincident with significant RM($\phi$) variations, but Stokes $V$ is relatively constant. This suggests that the effect responsible is of magnetospheric origin.

\textbf{J1512$-$5759*}. 
This pulsar has a single peaked profile with a long tail. $L$ is low ($\sim10\%$), and it vanishes at higher frequencies \citep{kjm05}. At pulse longitude $172^{\circ}$, $L$ drops to zero, coincident with an OPM jump in the PA swing. At pulse longitude $179^{\circ}$, the PA swing is steep and some depolarization can be observed. Here, deviations of $\sim$80 rad m$^{-2}$ in RM($\phi$) occur.
More deviations occur towards the trailing edge, coincident with a wiggle in the PA swing. 
The $\Delta(V/I)$ variations are seen where Stokes $V$ is changing the most, indicating that the possible cause for the observed RM($\phi$) variations is scattering.

\textbf{PSR J1534$-$5334*}. The profile of this pulsar has three components: a strong leading one, which peaks at the same pulse longitude as Stokes $V$; and two weaker and wider trailing components.  
$L$ is low, hence the statistical uncertainty on RM($\phi$) is high, especially in the trailing part of the profile. 
Nevertheless, there is a region between pulse longitudes $178^{\circ}$ and $185^{\circ}$ where there are significant RM($\phi$) variations with an amplitude of $\sim 10$ rad m$^{-2}$, coincident with the wiggle in the PA swing. 
Where Stokes $V$ is changing sharply, there are moderately significant $\Delta(V/I)(\phi)$  variations, indicating that  scattering could well be the cause for the apparent RM variations.

\textbf{PSR J1559$-$4438*}. The profile of this pulsar consists of a strong central component and weaker pre- and post-cursors. 
$L$ is moderately high with two drops to zero, which coincide with the OPM jumps in the PA swing. 
In the centre of the profile, the PA swing shows a dip close to where Stokes $V$ changes sign.
RM($\phi$) varies significantly where the PA gradient is steep, with the largest deviations coincident with the dip in the PA swing.  
Significant  $\Delta(V/I)(\phi)$ variations occur at longitudes where Stokes $V$ is changing the most, as expected for scattering. 
The measured values of RM$_{\text{profile}}$ and RM$_{\text{scatt}}$ indicate that our observation might be affected by interstellar scattering. \citet{jkm+08} found that at low frequencies, scatter broadening can be seen in the profile of this pulsar, hence all indicators are suggestive of scattering being the primary cause for the observed RM($\phi$) variations.

\textbf{PSR J1604$-$4909*}. The profile of this pulsar consists of multiple components. $L$ is generally weak, with the exception of the central region of the profile. Here, the PA swing is steep with several kinks. 
Across most of the profile, the statistical uncertainty on RM($\phi$) points is high. 
Nevertheless, significant apparent RM($\phi$) variations ($\sim 15$ rad m$^{-2}$) occur where the PA gradient is steep. 
There are significant  $\Delta(V/I)(\phi)$ variations in the central region of the profile, where Stokes $V$ is changing  rapidly, as expected if scattering affected the pulsar. This is not where the RM($\phi$) curve is changing the most, pointing to scattering as the main cause.
RM$_{\text{profile}}$ and RM$_{\text{scatt}}$ are inconsistent ($10\sigma$ deviation), indicating that the RM measurements could be indeed affected by scattering. \citet{kjm17} estimated that the scattering timescale at 1GHz is 0.02 ms, which is small. RM$_{\text{scatt}}$ is consistent with the similarly derived value by \citet{jkk+07}.

\textbf{PSR J1644$-$4559}.
The profile of this pulsar has an scatter extended tail and a very weak pre-cursor. $L$ is weak and the PA swing is relatively shallow with several bumps and an OPM transition at pulse longitude $172^{\circ}$. The RM($\phi$) curve displays variations ($\sim 20$ rad m$^{-2}$) across the whole profile and has a similar shape to what \cite{nkk+09} presented, with the largest variations coincident with the bump in the PA swing at longitude $\sim186^{\circ}$.  The pulsar is highly scattered at lower frequencies \citep[e.g][]{rjt+09}.
RM$_{\text{profile}}$ and RM$_{\text{scatt}}$ are inconsistent, indicating that the pulsar is indeed affected by scattering. 
Where Stokes $V$ changes most rapidly, $\Delta(V/I)(\phi)$ variations are seen, as expected from scattering. Since RM($\phi$) and $\Delta(V/I)(\phi)$ variations coincide, magnetospheric effects cannot be entirely dismissed, but it is clear scattering contributes significantly to the observed RM($\phi$).

\textbf{PSR J1651$-$5222*}. This pulsar has a profile consisting of several components blended into one feature. $L$ is low ($\sim20\%$) and the PA swing is relatively flat up to pulse longitude $176^{\circ}$. 
The RM($\phi$) curve in this part of the profile has a U-shaped structure and is higher compared to $\left<\text{RM}(\phi)\right>$. After longitude $176^{\circ}$, the PA swing is steeper and there are some RM($\phi$) variations. The highest amplitude variations ($\sim 40$ rad m$^{-2}$) occur around the notch in the PA curve where the gradient changes. 
\citet{kjm17} measured scatter broadening at lower frequencies, however the scattering should be small above 600 MHz.
There are some moderately significant $\Delta(V/I)(\phi)$ variations between longitudes $177^{\circ}$ and $180^{\circ}$, in a region where Stokes $V$ is changing most rapidly. However, this region is also where RM($\phi$) is varying and we cannot indicate which effect was responsible for the RM variations (see Section~\ref{sect:method}).

\textbf{PSR J1701$-$3726*}. This pulsar has a complex profile. 
$L$ is moderately high and Stokes $V$ has a comparable magnitude with regions where it exceeds $L$.
The PA swing is steep in the central region of the profile and displays several kinks. 
The largest RM($\phi$) variations ($\sim 30$ rad m$^{-2}$) occur in the shape of a dip  where the PA swing is the steepest.
RM$_{\text{profile}}$ and RM$_{\text{scatt}}$ are inconsistent, indicating that scattering could be responsible.
Significant $\Delta(V/I)(\phi)$ variations appear towards the centre of the profile, with the largest variations around longitude $\sim184^{\circ}$, where Stokes $V$ is changing rapidly. The steepest variations in  RM($\phi$) occur at an earlier longitude, suggesting that scattering is  the main cause for the observed RM variations.

\textbf{PSR J1703$-$3241*}. 
For this pulsar, $L$ is moderately high and the PA has a smooth and steep swing, without any OPM jumps. There are two wiggles in the PA swing towards the trailing edge of the profile, and this region is where significant RM($\phi$) variations ($\sim 5$ rad m$^{-2}$) occur. 
\citet{kjm17} estimated a scattering timescale of 0.13 ms at 1GHz, which is low.
If the observed RM variations were caused by scattering, $\Delta(V/I)(\phi)$ variations would be expected at most longitudes, as Stokes $V$ is steeply changing across the entire profile. However, $\Delta(V/I)$ only varies towards the trailing half of the profile, where RM($\phi$) is changing the most. This suggests that the observed variations are caused by a magnetospheric origin effect.

\textbf{PSR J1722$-$3207*}. The profile of this pulsar has two components: a strong leading one and a weaker and wider trailing one. $L$ is relatively low, peaking in the leading edge of the profile. The PA swing and RM($\phi$) curve remain flat until pulse longitude $181^{\circ}$. After this longitude, the PA gradient becomes steep and a dip in the shape of the apparent RM($\phi$) variations can be seen, followed by an upward deviation. 
The amplitude of the overall variations is $\sim 40$ rad m$^{-2}$.
The $\Delta(V/I)(\phi)$ variations do not coincide with where the largest RM($\phi$) variations occur. The only significant deviations can be seen around longitude $186^{\circ}$, where Stokes $V$ is changing the most, indicating that scattering is the likely cause for the apparent RM variations. 
\citet{kjm17} estimated a scattering timescale of 0.3 ms at 1 GHz.

\textbf{PSR J1731$-$4744*}. This pulsar has a complex profile, with the central component being the weakest. 
$L$ is generally low ($\sim20\%$) and the PA swing has a complex shape: it is flat until pulse longitude $182^{\circ}$, followed by a steep region and several kinks and a dip around longitude $187^{\circ}$.
The RM($\phi$) curve shows significant variations across the whole profile, with the highest apparent variations ($\sim 20$ rad m$^{-2}$) occurring coincident with the dip in the PA curve.
There are no significant $\Delta(V/I)(\phi)$ variations across the profile, since Stokes $V$ is very low, hence the origin of the RM variations cannot be determined.

\textbf{PSR J1745$-$3040}. This pulsar has a three component profile with a moderately high $L$. 
The PA swing is relatively flat and it is broken by two OPM jumps, at pulse longitudes $168^{\circ}$ and $187^{\circ}$, and has a
bump towards the centre of the profile. 
\cite{nkk+09} classified this pulsar as showing high RM($\phi$) variations, with a downwards gradient change in the central region of the profile of the order $\sim 20$ rad m$^{-2}$, and no apparent variations in the leading component of the pulse. 
In our observation, the RM($\phi$) curve is very different. In the leading part of the profile, the statistical uncertainties on the RM values are high, consistent with $\left<\text{RM}(\phi)\right>$. Towards the centre of the profile, we see two bumps in the RM($\phi$) curve, right before and after the bump in the PA swing. There is no obvious gradient. This suggests the RM($\phi$) curve is potentially time variable.
\citet{kjm17} observed scatter broadening at lower frequencies for this pulsar and estimated a timescale of 0.06 ms at 1 GHz, which is relatively low. 
The largest $\Delta(V/I)(\phi)$ variations occur at longitude $\sim173^{\circ}$, which is not where $V$ changes most rapidly, suggesting a magnetospheric origin for the RM($\phi$) variations. 
If magnetospheric effects are the cause of the time variability, one might expect the profiles to change as well (a similar argument applies to changes in scattering). This is not obvious from the observations. It should be noted that a change in frequency dependence (causing RM($\phi$) variations) does not imply a noticeable change in frequency average (profile shape).

\textbf{PSR J1751$-$4657*}. This pulsar has a double component profile, with a stronger leading component.
$L$ is low ($\sim 20$$\%$) and Stokes $V$ is most intense in the leading peak of the profile.
The PA swing is relatively steep across the entire profile with a kink, which coincides with the peak in Stokes $V$, and the minimum in $L$. This is also where the highest variations in RM($\phi$) occur $(\sim 20$ rad m$^{-2}$).
At all other longitudes, the RM($\phi$) curve remains flat. 
The large systematic uncertainties indicate that the RM($\phi$) variations are only moderately significant.
If the pulsar was affected by scattering, $\Delta(V/I)(\phi)$ variations are expected to occur in the centre of the profile, where Stokes $V$ is changing, especially around longitude $\sim178^{\circ}$, where the most rapid changes happen. Moderately significant $\Delta(V/I)(\phi)$ variations are observed between pulse longitudes $180^{\circ}$ and $183^{\circ}$. It is unclear as to the origin of the RM variations.

\textbf{PSR J1752$-$2806*}. This pulsar has very low $L$ and a PA swing which is relatively shallow, but with an OPM jump at pulse longitude $178^{\circ}$ and several steep kinks in the central part of the profile. Significant RM($\phi$) variations can be seen before the OPM jump. In the centre of the profile, the systematic uncertainties on RM($\phi$) are higher making these variations only moderately significant ($\sim 80$ rad m$^{-2}$). 
The largest $\Delta(V/I)(\phi)$ variations coincide with where Stokes $V$ changes most rapidly, but do not occur where the largest RM($\phi$) variations are seen. This suggests that the RM variations are caused scattering.

\textbf{PSR J1807$-$0847}. The profile of this pulsar has several components, with the central one the strongest. $L$ is low ($\sim20\%$) and there are two longitudes where it is minimal, coincident with the OPM jumps in the PA curve. The PA swing is shallow except for several kinks under the central component. 
This pulsar was classified by \cite{nkk+09} as a low varying RM($\phi$) pulsar with similar looking RM($\phi$) curve to ours. Most apparent variations can be seen where the PA swing displays wiggles starting at longitude $\sim178^{\circ}$.
The statistical uncertainties on RM($\phi$) are large, hence these apparent variations are only moderately significant.
\citet{kjm17} estimated a scattering timescale of 0.3 ms at 1 GHz, which is small.  
Large $\Delta(V/I)(\phi)$ variations occur across the whole profile, and they do not all coincide with Stokes $V$ changing steeply. This indicates that the RM variations are unlikely to be entirely because of scattering.

\textbf{PSR J1817$-$3618*}. The pulsar has two components, as well as a long tail. $L$ is relatively low, with a drop to zero at the OPM jump in the PA swing. The PA gradient becomes steep after the OPM jump, and this is where an upward deviation of the RM($\phi$) curve occurs ($\sim 10$ rad m$^{-2}$).
If the pulsar was affected by scattering, the largest $\Delta(V/I)(\phi)$ variations should be in the centre of the profile, where Stokes $V$ is changing most. No significant variations are detected, so without
detailed modelling we cannot comment further on the origin of the RM variations.

\textbf{PSR J1820$-$0427*}. The profile consists of multiple blended components, with the central one having the highest amplitude. 
As discussed by \citet{jkk+07}, at lower frequencies an OPM jump can be seen in the PA swing towards the leading edge of the profile around pulse longitude $178^{\circ}$, however at our observing frequency the jump is less than $90^{\circ}$ and $L$ does not completely disappear.
At this longitude we see the highest variations in RM($\phi$) $(20$ rad m$^{-2}$), in the shape of a dip.
This coincides with the largest $\Delta(V/I)(\phi)$ variations, and where Stokes $V$ is changing the most. For this pulsar, everything is consistent with scattering being the cause. However, since the rapid Stokes $V$ and PA changes coincide, magnetospheric effects cannot be ruled out (see Section~\ref{sect:method}).

\textbf{PSR J1824$-$1945*}. 
The profile of this pulsar has a strong central component and a weaker leading component. The PA swing is relatively flat and it is broken by two OPM jumps at pulse longitudes $177^{\circ}$ and $180^{\circ}$, which is where
the largest RM($\phi$) variations occur, with the largest variations at the second OPM jump $(\sim 35$ rad m$^{-2}$).
At the pulse longitudes where significant $\Delta(V/I)(\phi)$ variations occur, Stokes $V$ changes steeply. This suggests that scattering could be the cause for the observed variations.
Several authors (e.g. \citealt{lmg+04, wse07}) indeed reported finding scatter broadening at lower frequencies for this pulsar.

\textbf{PSR J1848$-$0123*}. The profile of this pulsar consists of multiple components. $L$ is very low, and rapid steep changes are observed in the shape of the PA swing. 
RM$_{\text{profile}}$ and RM$_{\text{scatt}}$ are inconsistent, indicating that this pulsar could have been affected by interstellar scattering. 
\citet{lkk15} did not detect any scatter broadening at higher frequencies, hence could not predict a scattering timescale at 1 GHz.
The statistical uncertainties on the RM($\phi$) values are high, especially towards the trailing part of the profile.
The RM($\phi$) curve is flat, with the exception of some significant deviations around the OPM jumps and near a very steep part of the PA swing ($\sim 40$ rad m$^{-2}$). 
At a pulse longitude of $179^{\circ}$ the largest $\Delta(V/I)(\phi)$ variations occur, coincident with the largest changes in Stokes $V$. Since this coincides with where the RM($\phi$) variations are observed, 
we cannot distinguish the effect responsible for the RM variations (see Section~\ref{sect:method}).

\textbf{PSR J1900$-$2600*}. 
This pulsar has a multi-component profile with moderately high $L$ and Stokes $V$. 
The PA swing is steep and it is broken by a jump around pulse longitude $171^{\circ}$. This jump is less than $90^{\circ}$ and $L$ does not completely disappear. After the jump, the slope of the PA swing changes sign, which is different from a typical OPM jump where the slope is conserved.
The largest variations in RM($\phi$) occur at the quasi-OPM jump in the shape of a peak in RM($\phi$) of $\sim 30$ rad m$^{-2}$.
At a similar longitude the largest $\Delta(V/I)(\phi)$ variations occur, and the Stokes $V$ is steeply changing. Hence, we cannot distinguish which effect was responsible for the apparent RM variations (see Section~\ref{sect:method}).
RM$_{\text{profile}}$ and RM$_{\text{scatt}}$ are inconsistent, indicating that this pulsar could be affected by interstellar scattering. This would confirm findings from other authors (e.g. \citealt{phs+16,ymw17}), who found scattering at lower frequencies. 
Our measurements are consistent with scattering being the dominant cause of RM variations, but magnetospheric effects are also possible.

\textbf{PSR J1913$-$0440*}. 
This pulsar has a profile consisting of three overlapping components.
$L$ is low ($\sim$15$\%$) at all observed frequencies \citep{jkm+08}.  
The PA swing is steep, with several kinks and one OPM jump situated at a pulse longitude $176^{\circ}$. 
The RM($\phi$) curve is generally flat, except a $\sim 30$ rad m$^{-2}$ deviation where the PA swing is the steepest ($\sim181^{\circ}$).
The largest $\Delta(V/I)(\phi)$ variations occur at an earlier ($\sim178^{\circ}$) and later ($\sim182^{\circ}$) longitude, where Stokes $V$ is changing most rapidly. This suggest that scattering is the reason for the apparent RM variations.
Several authors (e.g. \citealt{lrk+15, nsk+15}) observed this pulsar to be scattered at low frequencies. Hence it is possible that the observed RM variations were caused by low level scattering.

\textbf{PSR J2048$-$1616}. 
The profile of this pulsar has three components, with the trailing one being the strongest. 
The PA swing has an `S'-shape and can be fit with the RVM model at multiple frequencies \citep{jkk+07}.
The shape of the apparent variations in RM($\phi$) is similar to those of the observations from \cite{nkk+09}. The RM($\phi$) curve  has a dip at pulse longitude $176^{\circ}$ of around $\sim 5$ rad m$^{-2}$, coincident with the steepest region of the PA swing.
We see the largest $\Delta(V/I)(\phi)$ variations where Stokes $V$ changes most rapidly. This indicates that the possible dominant cause for the observed RM variations is interstellar scattering.

\subsection{Pulsars without significant RM variations}
\label{sect:noRM}

The remaining pulsars, for which $\chi^2_{\text{RM}(\phi)} < 2$, were classified as not showing any significant RM variations. A list is given in Table \ref{table:noRM}.  
One of these pulsars, PSR J1056$-$6258 (see Fig.~A.7), was classified by \cite{nkk+09} as having high RM($\phi$) variations of $\sim 100$ rad m$^{-2}$, especially away from the centre of the profile. In our observation, the RM($\phi$) curve is relatively flat across the whole profile. There appears to be a small upwards gradient of the entire curve, however given the size of the systematic uncertainty, this effect is not significant. 
PSR J0134$-$2937 (Fig.~A.1) has a shallow PA swing with two OPM jumps and shows no apparent RM($\phi$) variations. However, around pulse longitude $175^{\circ}$, there are significant $\Delta(V/I)(\phi)$ variations where Stokes $V$ is changing most rapidly. This indicates that the pulsar may be affected by interstellar scattering, but not at a level which is enough to produce RM($\phi$) variations, which is not surprising considering the flat PA swing.

\begin{table}

\caption{A list of pulsars which show no significant phase-resolved RM variations.  Pulsars for which the phase-resolved RM profiles have not previously been investigated are marked with an asterisk (*).}
\label{table:noRM}
\begin{tabular}{|l|l|l|l|}
\hline 
J0134$-$2937    &   J0152$-$1637*    &	J0614$+$2229*	&   J0630$-$2834*  \\
J0729$-$1836*	&	J0742$-$2822     &	J0745$-$5353*	&	J0809$-$4753* \\
J0835$-$4510	&	J1043$-$6116* 	 &	J1047$-$6709*   &	J1048$-$5832* \\
J1056$-$6258	&   J1136$-$5525*    &	J1146$-$6030*	&	J1157$-$6224 \\
J1306$-$6617*	&	J1326$-$6408*    &	J1326$-$6700*	&	J1327$-$6222* \\ 
J1328$-$4357*	&	J1430$-$6623*    &	J1522$-$5829*	&	J1539$-$5626* \\
J1544$-$5308*	&	J1555$-$3134*    &	J1557$-$4258*   &	J1602$-$5100* \\
J1605$-$5257* 	&	J1633$-$4453*    &	J1633$-$5015*   &	J1646$-$6831* \\
J1651$-$4246*	&	J1653$-$3838*    &	J1705$-$1906*	&	J1705$-$3423* \\ 
J1709$-$1640*	&	J1709$-$4429*    &	J1717$-$3425*	&	J1721$-$3532* \\
J1722$-$3712*   &	J1739$-$2903*	 &	J1740$-$3015*	&	J1741$-$3927* \\
J1822$-$2256*   &	J1823$-$3106*	 &	J1825$-$0935*	&	J1829$-$1751* \\
J1830$-$1059*	&	J1832$-$0827* 	 &	J1845$-$0743*	&	J1847$-$0402* \\
J1852$-$0635*   &	J1941$-$2602* 	 &	J2330$-$2005*	&	J2346$-$0609* \\

\hline
\end{tabular}
\end{table}

The Vela pulsar, PSR J0835$-$4510 (Fig.~A.4), is very bright, has high $L$ and a steep `S'-shaped PA swing without OPM jumps.
\cite{nkk+09} observed RM($\phi$) variations of $\sim 13$ rad m$^{-2}$ in their 2004 observation and $\sim 6$ rad m$^{-2}$ in their 2006 observation. In addition, the authors did not find any non-Faraday behaviour affecting the PA.
Although we see statistically significant RM($\phi$) variations, the systematic effects are large and completely dominate the results. 
There are deviations from the expected $\lambda^2$ dependence of the PA as a function of frequency at all pulse longitudes, but again these are introduced by the large systematic errors. 
In the centre of the profile the deviations are so large, that the values of $\chi^2_{\text{PA}\left(\lambda^2,\phi\right)}$ greatly exceed few hundreds, and are not displayed in Fig.~A.4.
Ultimately, the systematic effects prevent us from seeing any significant RM($\phi$) variations.

There are a few other pulsars which do not show RM($\phi$) variations, however significant $\Delta(V/I)(\phi)$ variations occur coincident with the most rapid changes in Stokes $V$, indicating that it is possible that low level scattering is affecting them. These pulsars are PSRs J0630$-$2834, J1157$-$6224, J1602$-$5100, J1605$-$5257 and J2330$-$2005. 
In addition, for PSR J1741$-$3927, we also see tentative evidence for RM($\phi$) variations where the PA swing is steep. This strongly suggests that our ability to detect RM($\phi$) is S/N limited.

\section{Discussion}
\label{sect:discussion}

From our sample of 98 pulsars, 78 pulsars had their RM($\phi$) curves determined for the first time.
Of the 98 pulsars, 42 showed significant RM($\phi$) variations. This is a similar fraction (9/19) from the much smaller sample of \cite{nkk+09}. Currently, for the majority of our sample, the statistical errors dominate over the systematic errors. We provided evidence that with increased S/N more examples of pulsars with RM($\phi$) variations would be detected. 
\cite{nkk+09} concluded that interstellar scattering is the dominant cause of the RM($\phi$) variations they observed.
We re-examine this conclusion below. 

From the results, it is clear that scattering alone is not enough to explain the apparent RM variations.
We have identified some clear examples of pulsars for which the RM($\phi$) and $\Delta(V/I)(\phi)$ curves were not consistent with scattering as the cause of the observed apparent RM variations.  
In the case of the interpulse pulsar, PSR J0908$-$4913 (Fig~A.5), for the MP, we observe that the greatest apparent RM variations occur towards the trailing part of pulse where the PA gradient is the steepest. This is consistent with a picture where scattering is the cause for the observed variations. However, for the IP, in the region where the PA swing is steeper than in the MP, the RM($\phi$) curve is flat. This is inconsistent with scattering.
We also expect that if the linear polarization and Stokes $V$ are affected by scattering, the largest $\Delta(V/I)(\phi)$ variations should occur where Stokes $V$ is changing the most. For the MP, we instead see the largest $\Delta(V/I)(\phi)$ variations coincident with the greatest change in RM($\phi$) rather than Stokes $V$, which varies more at earlier pulse longitudes. This points towards magnetospheric effects as a cause for the apparent RM variations of this pulsar. 

Another good example is PSR J1703$-$3241 (Fig~A.17). For this pulsar, Stokes $V$ is changing throughout the profile and is sharply varying in the central region. If scattering was the dominant cause for the apparent RM variations, $\Delta(V/I)(\phi)$ variations should occur at almost all pulse longitudes equally. However, large $\Delta(V/I)(\phi)$ variations only occur coincident with the greatest gradient in the RM($\phi$) curve. Using similar arguments we also identified magnetospheric effects as a cause for apparent RM variations for PSRs J0738$-$4042, J0820$-$1350,  J0907$-$5157, J1243$-$6423, J1326$-$5859, J1359$-$6038, J1453$-$6413,  J1456$-$6843, J1745$-$3040 and J1807$-$0847 (see Section~\ref{sect:highRM} for more details), so far a total of 12 out of 42 pulsars with RM($\phi$) variations. The variations in 12 pulsars are caused by scattering, with the results for the final 18 pulsars being ambiguous (see Table~\ref{table:disc}).

\cite{nkk+09} considered two intrinsic effects of magnetospheric origin which could cause RM($\phi$) variations. The first is the superposition of two quasi-orthogonal OPMs with different spectral indices.
The authors assumed completely linearly polarized OPMs with a spectral index of $\sim-0.5$, as derived by \citet{sse+06}. Using simulations, \cite{nkk+09} concluded that in order to create apparent RM variations of the observed amplitudes, the fractional linear polarization should remain under $10\%$, which occurs only rarely in either their sample or ours.
Hence, it was deemed unlikely that this effect could produce apparent RM variations on a large scale.

The second intrinsic effect considered was generalized Faraday rotation. Based on theory discussed in \citet{km98}, both linear and circular polarization are expected to be frequency dependent. 
Hence, one can take a correlation between the RM($\phi$) and $\Delta(V/I)(\phi)$ curves as evidence for generalized Faraday rotation. 
\cite{nkk+09} did not observe, based on a limited sample of five pulsars, any obvious correlations and concluded that this effect is not able to produce the observed RM variations on a large scale. 
However, in our sample these correlations appear to be observed for many of our pulsars (based on inspecting the curves, see Section~\ref{sect:highRM}), hence the lack of correlation appears as an exception from the general rule. One of these pulsars is J1243$-$6423, for which \cite{nkk+09} did not find a correlation, yet in our data clearly reveals a correlation between the region of greatest RM($\phi$) change and greatest $\Delta(V/I)(\phi)$ change, but it occurs at a pulse longitude where \cite{nkk+09} did not have significant data points (see Section~\ref{sect:highRM} and Fig.~A.8). 
Although these correlations are common, we note that this does not imply generalized Faraday rotation must be operating in the magnetosphere of pulsars. 
This is because also the presence of scattering will introduce a frequency dependence in both linear and circular polarization, and as a consequence, for both generalized Faraday rotation and scattering one expects a correlation between the RM($\phi$) and $\Delta(V/I)(\phi)$ curves.

Although there is now good evidence that scattering is not the only effect  causing apparent RM variations, we note that without detailed simulations for each pulsar, one cannot be sure on a case by case basis which effect is dominant.  However, additional evidence for the importance of magnetospheric effects can be found by using results drawn from the sample as a whole.

\cite{nkk+09} supported the idea that the RM($\phi$) variations we observe are caused by interstellar scattering, given that they observed the largest  RM variations coincident with the steepest gradients of the PA swing. \citet{dgm+15} also noted this coincidence in their millisecond pulsar sample.
We tried to verify whether a similar trends exists in our larger sample. By inspection, we classified a PA swing as steep if we identified a region of the swing which had a gradient larger than $\sim4$ deg/deg \footnote{Scattering will flatten the observed PA curve \cite[e.g.][]{kar09}. However, since most pulsars from our sample are not heavily scattered, this effect will be small.}. 
This classification can be found in Table~\ref{table:disc}.
The trend was confirmed, considering that out of the 42 pulsars which were classified as showing phase-resolved RM variations, 34 of these have a steep PA curve and only 8 of the pulsars display a shallow PA curve. In contrast, for the 56 pulsars classified as not showing apparent RM($\phi$) variations, 33 had a steep PA swing and 23 had a shallow PA swing. 
However, it is not correct that this implies that scattering is the primary cause for RM variations. If effects intrinsic to the pulsar
magnetosphere distort the PA as function of both frequency and pulse
longitude, then we would also expect the greatest variations in
RM to occur near the steepest part of the PA swing.

\begin{figure}
\begin{center}
\includegraphics[width=9cm,angle=0]{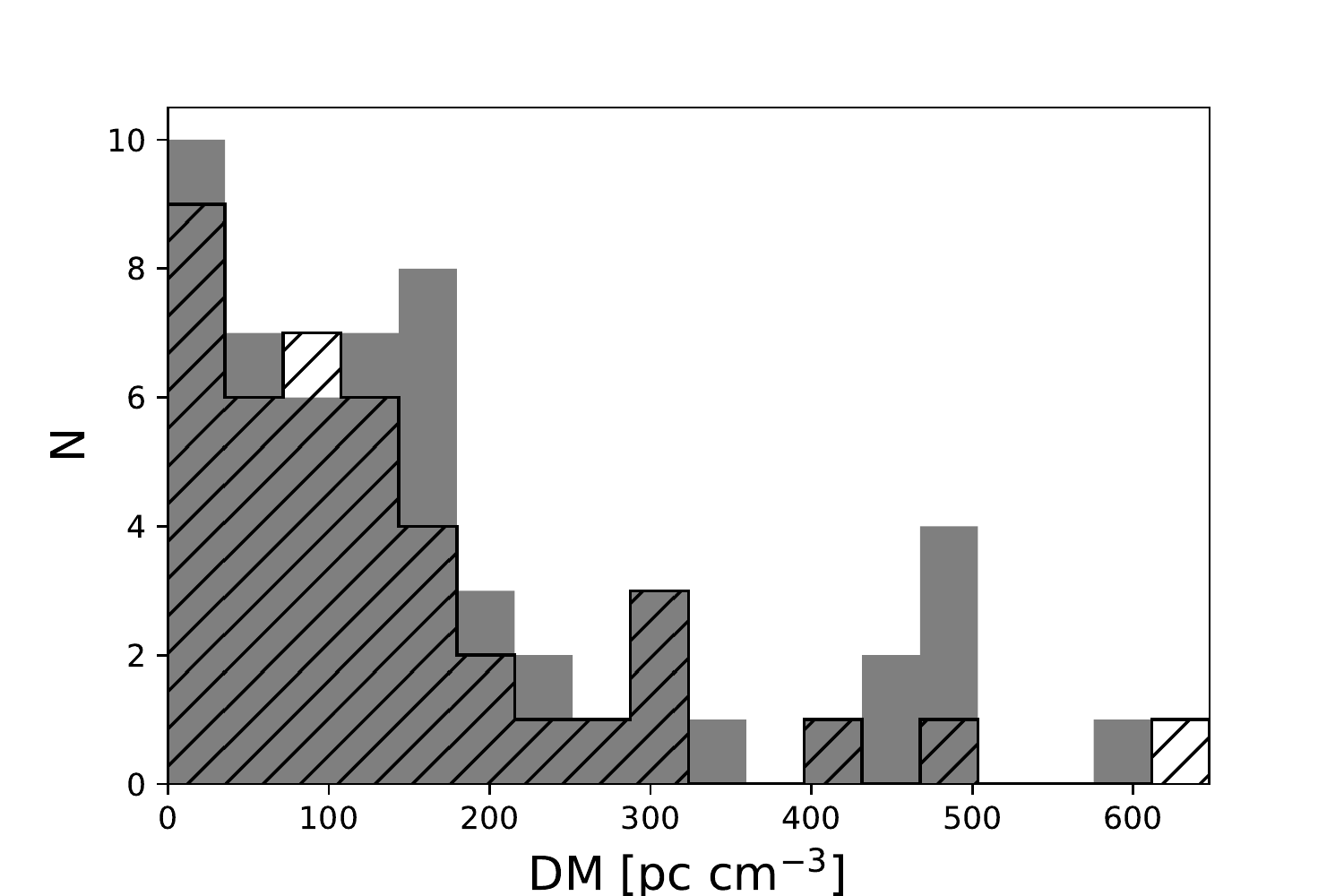} 
\end{center}
\caption{DM distributions. The pulsars with apparent RM variations are represented with the clear hatched distribution, while the pulsars without apparent RM variations are represented in grey.}
\label{fig:DM_RMvar}
\end{figure}

If scattering was the dominant cause for most cases of RM($\phi$) variations, we expect that the pulsars which show apparent RM variations to be high-DM pulsars. This is because pulsars with higher DM are more affected by interstellar scattering (e.g. \citealt{bcc+04,gkk+17}).   
\cite{nkk+09} observed a weak correlation, since the two strongest RM varying pulsars from their sample were also the highest DM pulsars. 
In Fig. \ref{fig:DM_RMvar}, a DM histogram is displayed for pulsars showing (42 pulsars) and not showing RM variations (56 pulsars). 
A Kolmogorov$-$Smirnov (KS) test shows the DM distributions are very similar,  pointing to magnetospheric effects being important.

\begin{figure*}
\begin{center}
\includegraphics[width=8.8cm,angle=0]{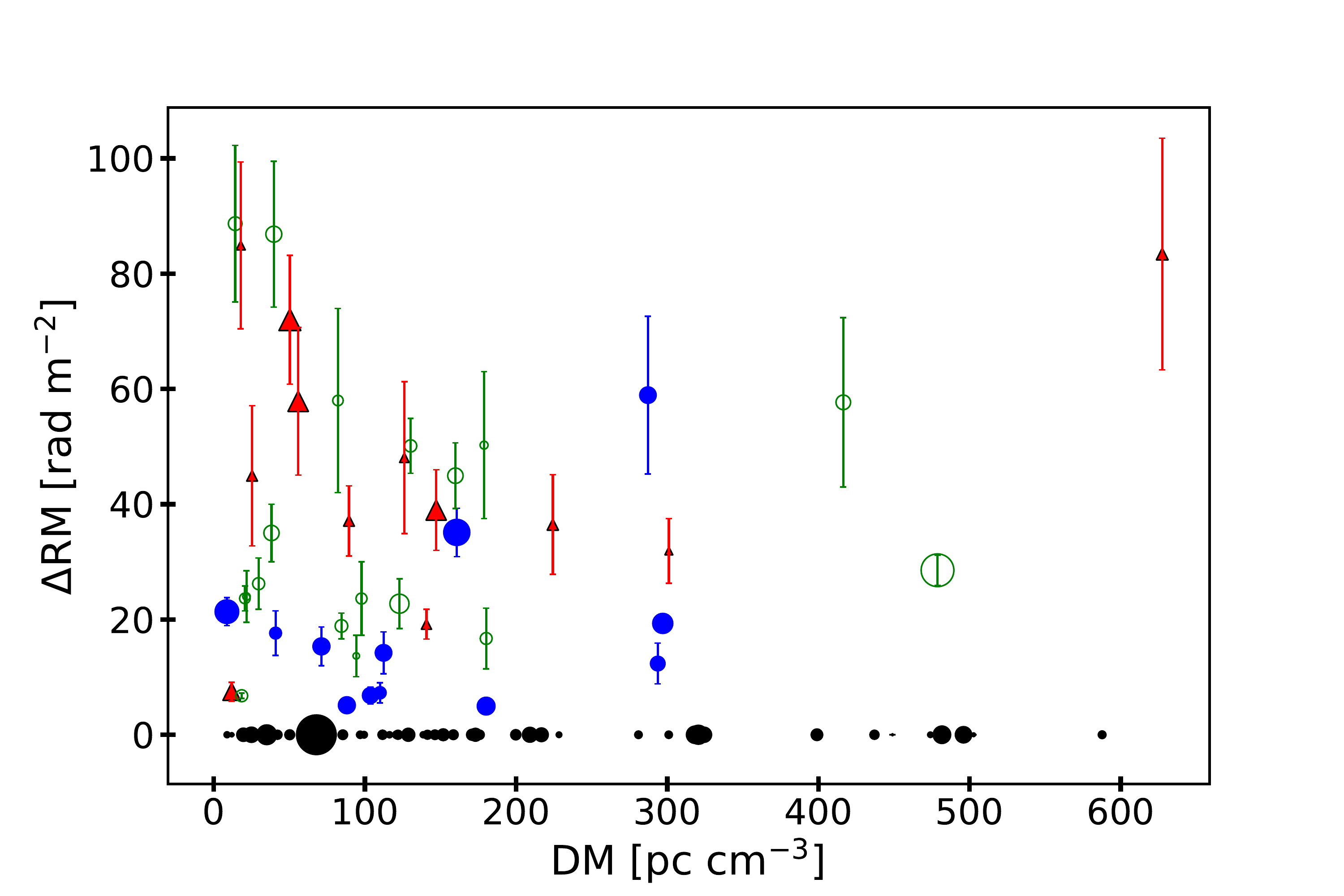} 
\includegraphics[width=8.8cm,angle=0]{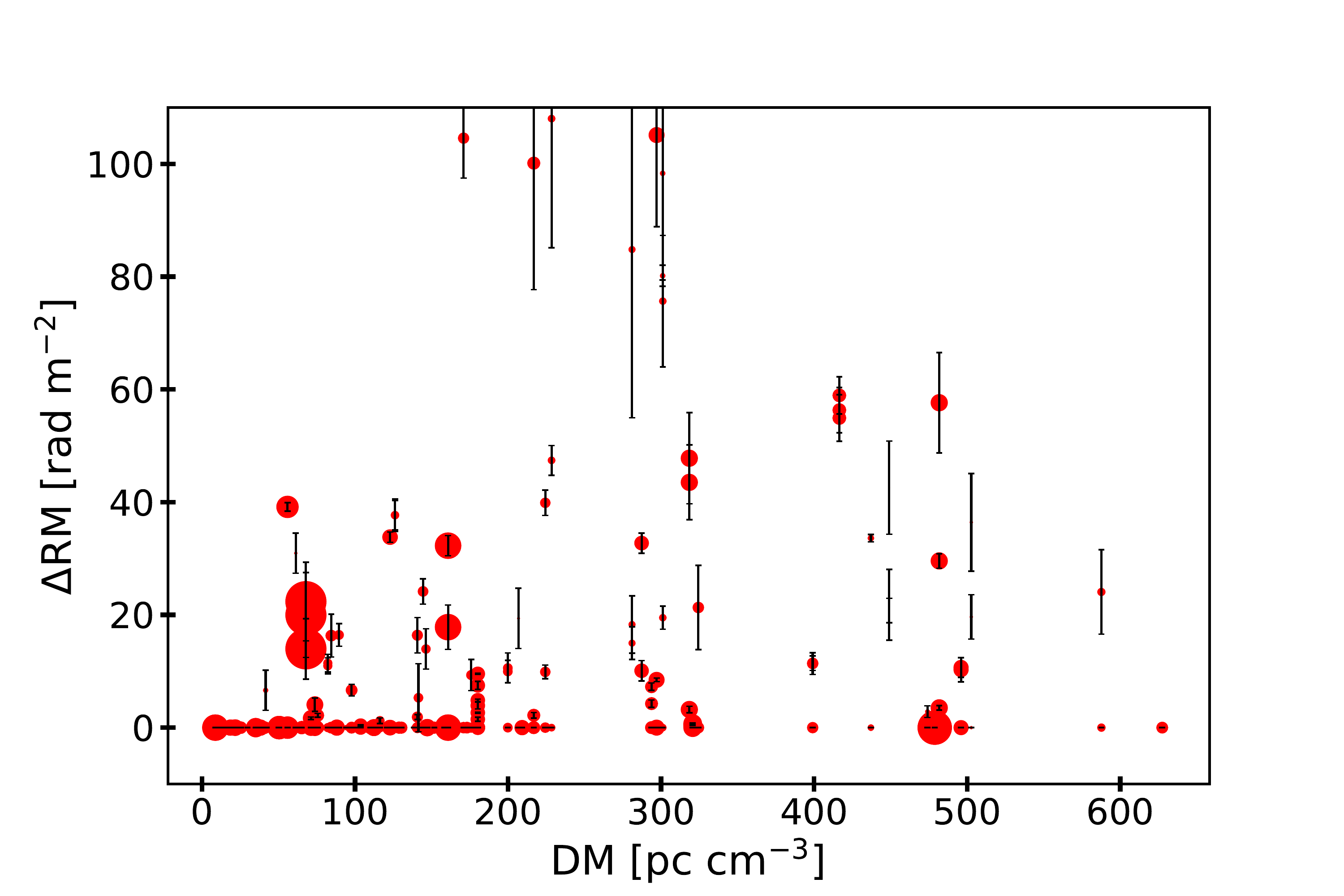} 
\end{center}
\caption{The amplitude of phase-resolved apparent RM variations, $\Delta$RM, as a function of their DM. The radius of the symbols  is proportional to the $\log_{10}(S_{1400})$ of each pulsar. The triangles (red in the online version) and filled circles (blue in the online version) are pulsars for which scattering or magnetospheric effects are responsible for the RM variations.  For the pulsars indicated by the open circles (green in the online version) it could not be distinguished which effect was responsible. \textit{Left}: The measured $\Delta$RM values for the observed 98 pulsars.  \textit{Right}: The expected $\Delta$RM values for the pulsars based on simulations including only the effects of scattering.}
\label{fig:deltaRM}
\end{figure*}

Similar to Fig. 9 in \cite{nkk+09}, the amount of observed RM($\phi$) variations of the 42 pulsars as a function of the DM of the pulsar is shown in the left-hand panel of Fig.~\ref{fig:deltaRM}. The peak-to-peak amplitude of the apparent RM variations, $\Delta$RM, was obtained by only considering RM($\phi$) values which were situated more than $3\sigma$ away from $\left<\text{RM}(\phi)\right>$. 
For the 56 pulsars with no significantly varying RM($\phi$) curves, 
$\Delta$RM was set to zero.
The uncertainties on the $\Delta$RM values were obtained by combining the statistical uncertainties of the two RM($\phi$) values in quadrature. For all pulsars with non-zero measurements of $\Delta$RM, the statistical uncertainties dominated over the expected systematics. 
The S/N of each pulsar was also represented in the plot, by displaying a symbol with a size proportional to the $\log_{10}(S_{1400})$, where $S_{1400}$ is the mean flux density at 1.4 GHz taken from \citet{jk18} in Fig.~\ref{fig:deltaRM}.

From the left-hand side plot in Fig.~\ref{fig:deltaRM}, there is no correlation between the DM of a pulsar and $\Delta$RM. There are several low-DM pulsars which display comparable magnitude apparent RM variations to the high-DM pulsars. 
Regardless of including the pulsars with lower flux densities, for which the detection of phase-resolved apparent RM variations is more difficult, we do not observe a correlation.
Pulsars for which the variations were caused by magnetospheric effects (blue circles in Fig.~\ref{fig:deltaRM}) appear to have in general smaller RM($\phi$) variations
compared to the ones for which scattering was identified as the dominant effect. 
This suggests that magnetospheric effects are most evident in cases of pulsars with low levels of scattering. If both effects are present, scattering potentially masks the other effect.

\begin{figure}
\begin{center}
\includegraphics[width=8.7cm,angle=0]{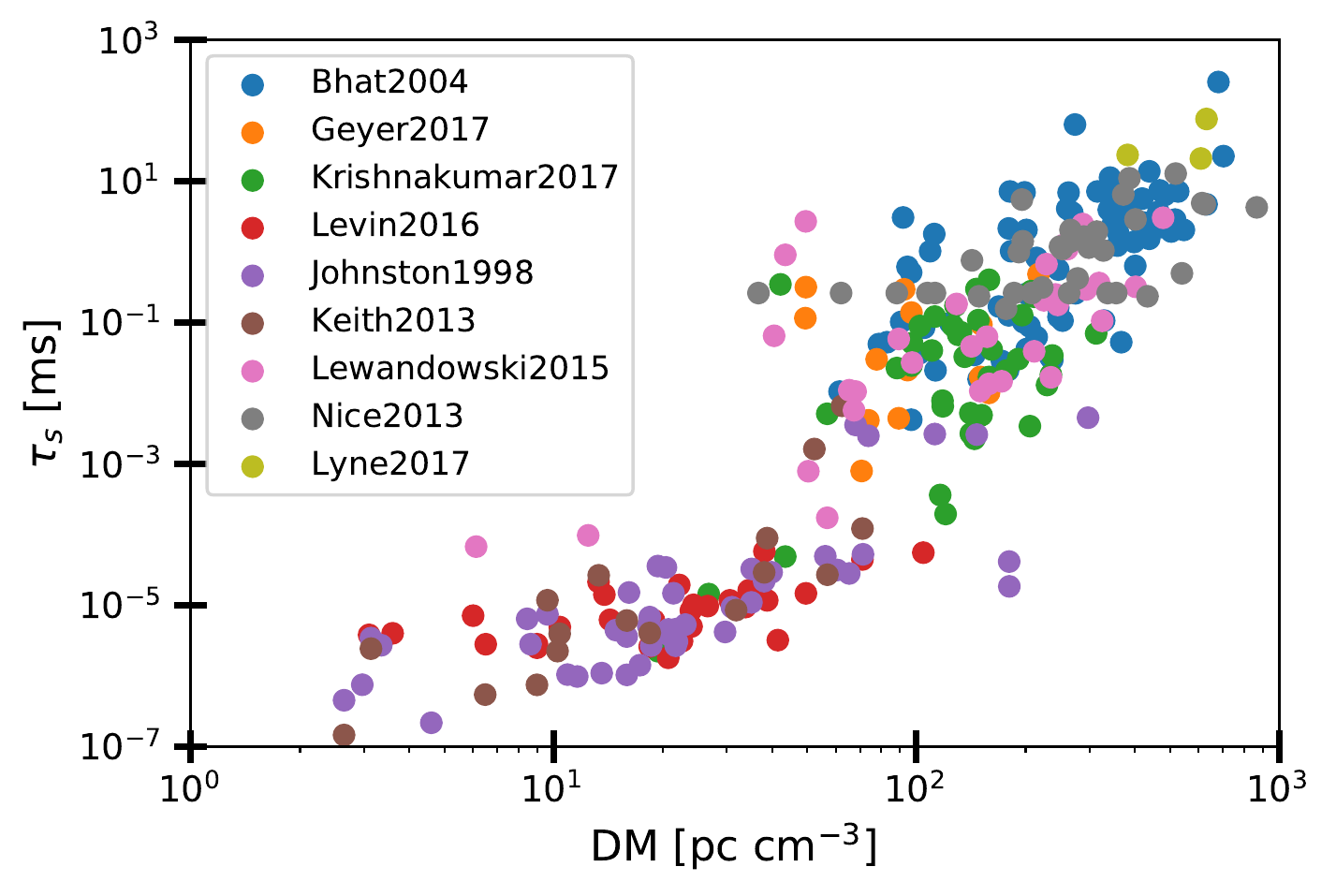} 
\end{center}
\caption{Measured scattering timescales, $\tau_s$, scaled to 1.4 GHz, as a function of DM. In the online version, the blue circles are measurements from \citet{bcc+04}, orange circles from \citet{gkk+17}, green circles from \citet{kjm17}, pink circles from \citet{lkk15}, grey circles from \citet{nab+13} and the light green circles from \citet{lsb+17}. Most of the points with a DM $< 200 $ pc cm$^{-3}$ were derived from measured values of scintillation bandwidths. The purple points were taken from \citet{jnk98}, the red points from \citet{lmj+16} and the brown circles from \citet{kcs+13}.}
\label{fig:bhatPlot}
\end{figure}

The absence of a correlation between DM and the magnitude of the detected apparent RM variations is surprising, given that scattering should theoretically have an effect.
We therefore simulate what level of correlation could be expected assuming scattering is the main reason for the apparent RM variations.
We used a distribution between measured scattering timescales, $\tau_{\text{scatt}}$, and DM. Measurements from \citet{bcc+04} were supplemented by more recent measurements, in order to quantify the range of expected scattering timescales for a certain DM. 
The obtained distribution of $\tau_{\text{scatt}}$ versus DM, scaled at a frequency of 1.4 GHz, is shown in Fig.~\ref{fig:bhatPlot}.
Most of the points with a DM $< 200 $ pc cm$^{-3}$ were derived from measured values of scintillation bandwidths, $\nu_{d}$. These correspond to scattering timescales of $C/2\pi\nu_{d}$, with $C$ being a constant with values between 0.6 and 1.5 \citep{lr99}. These values depend on various assumptions about the spectrum or geometry of the electron density variations. We assume $C=1$.

Before simulating the expected effects of interstellar scattering on our pulsars, the profile in each frequency channel was replaced by the frequency-averaged profile. This ensured that all RM($\phi$) variations were eliminated, while the shape of the average PA swing remained unaffected. 
We randomly selected three possible scattering timescales for each DM of every pulsar using the distribution from Fig.~\ref{fig:bhatPlot}. Scattering was added, as described in the Method section, Section~\ref{sect:method}.
Each of the three different scattered versions of each pulsar was analysed as previously described, and the plot of $\Delta$RM versus DM is displayed as the right-hand panel of Fig.~\ref{fig:deltaRM}. 

It is clear from this panel that the expected RM($\phi$) variations are too low compared to what is observed for low, DM $< 200 $ pc cm$^{-3}$ pulsars.
Furthermore, we were also not able to reproduce the observed amplitude of RM variations for the highest-DM pulsar of our sample (PSR J1512$-$5759 with DM $= 627.47 $ pc cm$^{-3}$). This was because the expected $\tau_{\text{scatt}}$ for this pulsar was large enough to completely flatten the PA swing and not produce any apparent RM variations, suggesting that this pulsar has an unusually low scatter timescale for its DM. 
The fact that scattering alone cannot reproduce the amount of apparent RM variations found at the low DM end of the distribution supports the conclusion that additional magnetospheric effects are important, especially for pulsars that have low levels of scattering.

\begin{figure}
\begin{center}
\includegraphics[width=9cm,angle=0]{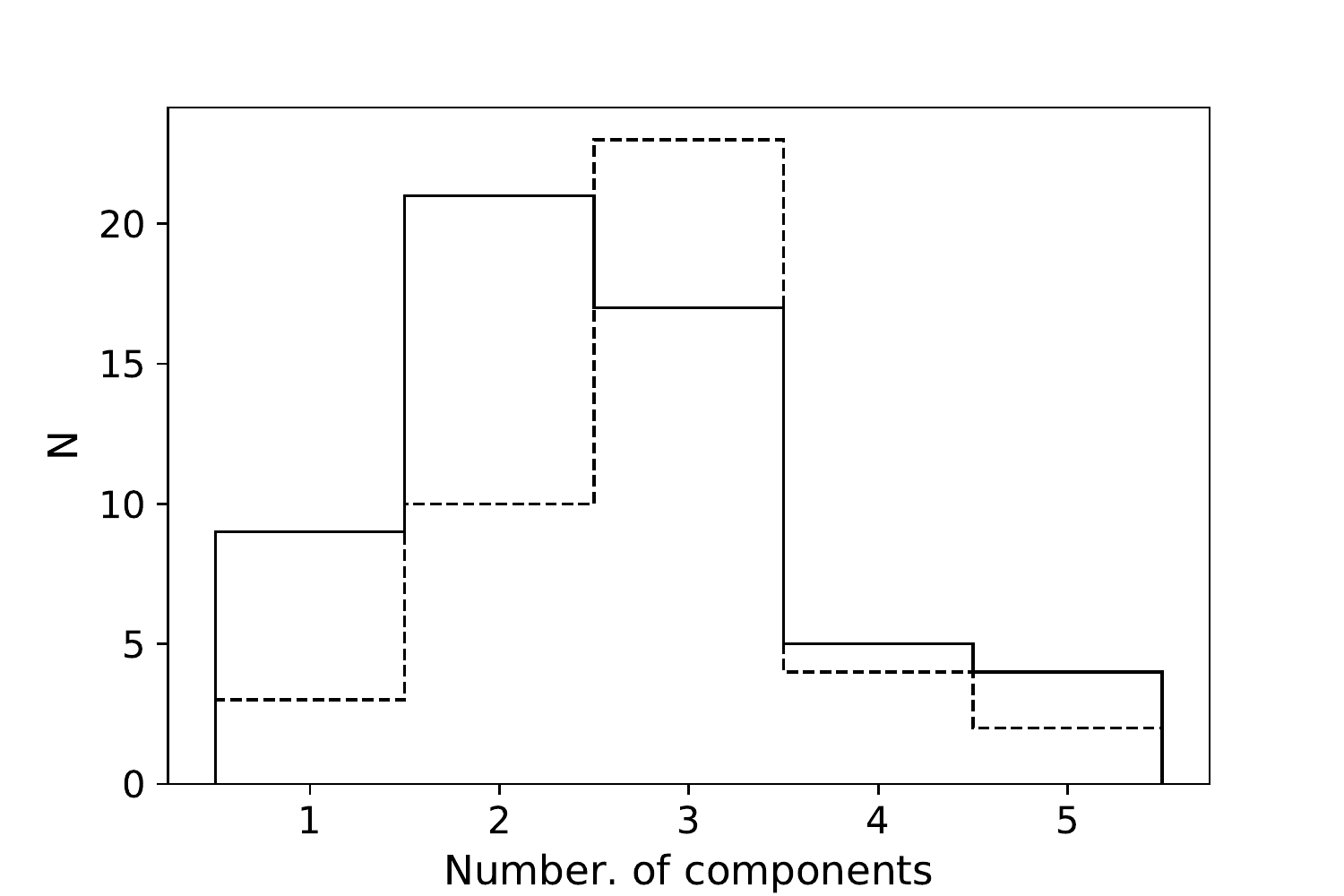} 
\end{center}
\caption{A histogram of the number of profile components for pulsars with apparent RM variations (dashed line) and without apparent RM variations (solid black line).}
\label{fig:multi}
\end{figure}

\begin{figure}
\begin{center}
\includegraphics[width=9cm,angle=0]{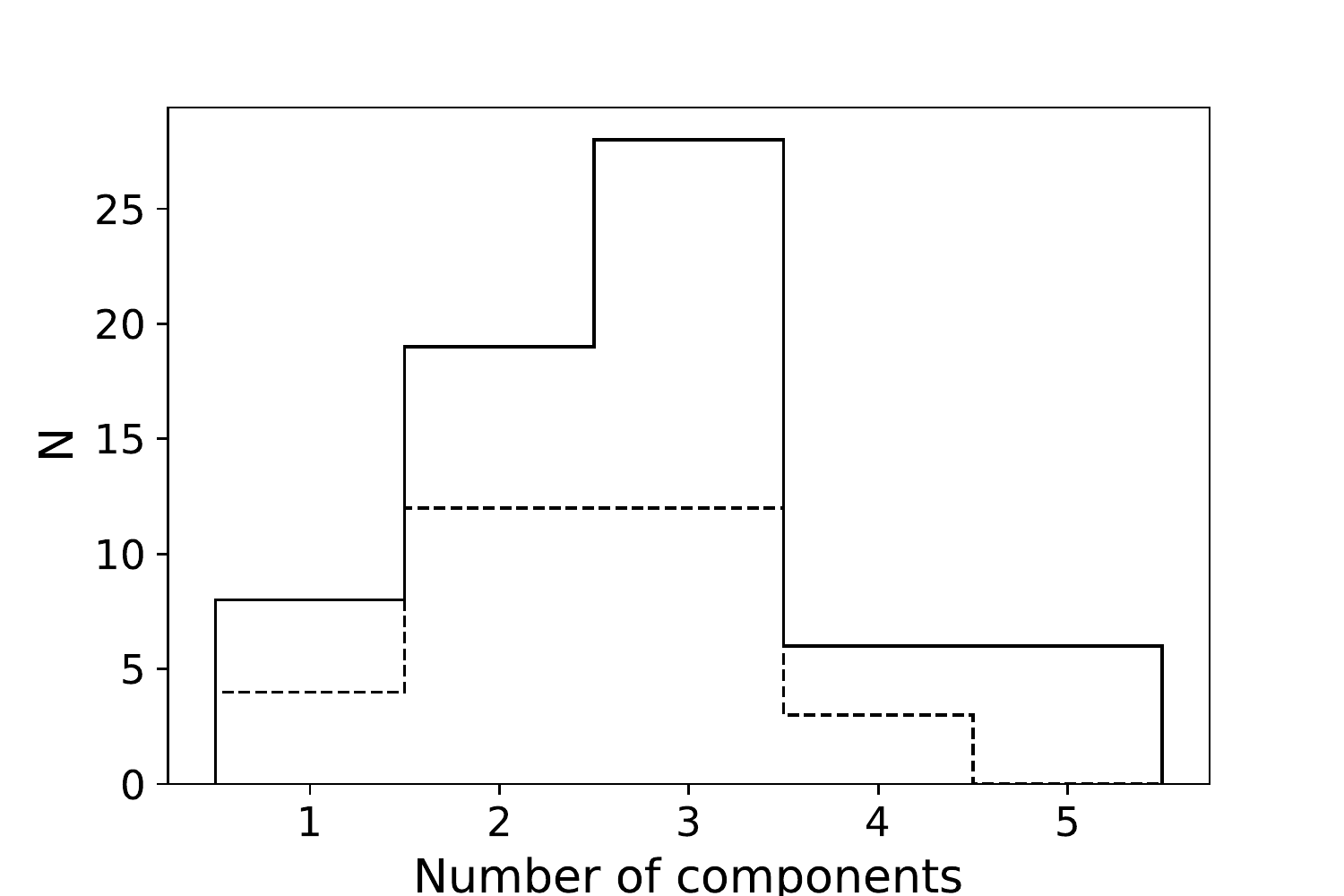} 
\end{center}
\caption{A histogram of the number of profile components for pulsars with a shallow PA swing (dashed line) and pulsars with a steep PA swing (solid black line).
} 
\label{fig:multi_steep}
\end{figure}

The magnetospheric effects that can cause the PA to be distorted in a frequency dependent way might well be most prominent in pulsars which show evidence for other complex emission properties. Motivated by this we made a distribution (Fig.~\ref{fig:multi}) of the number of profile components of pulsar which show, or not show, significant apparent RM variations (Table~\ref{table:disc} in the Appendix). We counted the number of profile components by visual inspection.
It is clear from Fig.~\ref{fig:multi} that pulsars which show
RM($\phi$) variations tend to have more complex profiles. Pulsars with single-peaked or double-peaked profiles tend to not display RM variations, 
suggesting that profile shape complexity and apparent RM variations are related. For scattering to cause apparent RM variations only requires a complex or steep PA-swing. So unless complex profiles have complex PA-swings, scattering alone cannot explain Fig.~\ref{fig:multi}. Fig.~\ref{fig:multi_steep} shows a histogram of the classification of a PA swing as either steep or shallow as a function of the number of profile components. There is no evidence for such a correlation, indicating interstellar scattering is not enough to explain the observed apparent RM variations.

\begin{figure}
\begin{center}
\includegraphics[width=0.45\textwidth,angle=0]{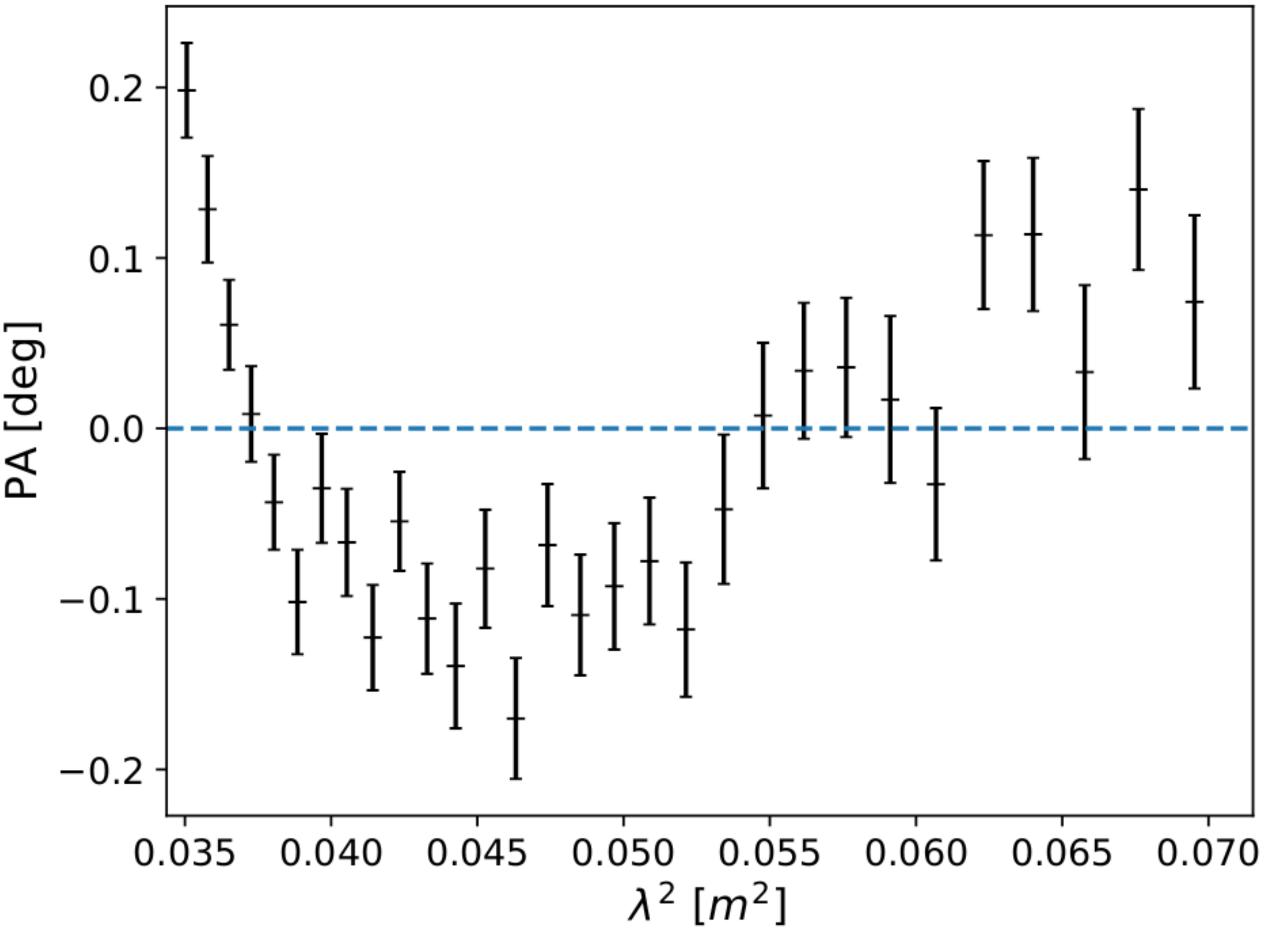} 
\end{center}
\caption{PA residuals after removing a $\lambda^2$ dependence according to equation~(\ref{eg:basicRM}), for pulse longitude 150$^{\circ}$ and scattering timescale of 8 ms in the simulations displayed in Fig.~\ref{fig:example_sim}.} 
\label{fig:palambda2}
\end{figure}

In all cases where we observed RM($\phi$) variations, there are significant deviations from the Faraday law expected $\lambda^2$ dependence of the PA as a function of frequency (as shown in the panels where $\chi^2_{\text{PA}\left(\lambda^2,\phi\right)}$ is displayed). This implies that another frequency and pulse longitude dependent effect is present, 
Furthermore, as expected, we also observe large deviations from a $\lambda^2$ dependence for pulsars with large systematic uncertainties on their RM($\phi$) values (e.g. J0742$-$2822 and J0835$-$4510).
These findings are different from \cite{nkk+09}, who did not find any evidence for non-Faraday components, at least for the two pulse longitudes they investigated for PSRs J0835$-$4510 and J1243$-$6423. Since our PA uncertainties are an order of magnitude smaller compared to the measurements presented by \cite{nkk+09}, it is maybe not surprising a different conclusion is reached. 
We observe similar deviations in the lowest panels of Fig.~A.1 $-$ Fig.~A.26, in the online supplementary material, for $\chi^2_{\text{V/I}\left(\nu,\phi\right)}$ in regions where $\Delta(V/I)(\phi)$ is deviating from zero, indicating that both linear and circular polarization are often both affected. These findings are as expected, considering that for both scattering (see Fig.~\ref{fig:palambda2} for an example) and magnetospheric effects, there is no reason to expect a $\lambda^2$ relation to describe their induced frequency dependence of the PA.

\section{Summary and Conclusions}
\label{sect:summary}

A large sample of pulsars was analysed for phase-resolved apparent RM variations, RM($\phi$). We used a basic version of the Rotation Measure Synthesis Technique (publicly available as part of the PSRSALSA software package) and quantified both statistical uncertainties, using bootstrapping, as well as systematic uncertainties. 
The way we computed the statistical uncertainties avoided making a-priori assumptions about the underlying signal.
The dominant source of systematic uncertainties was identified to be due to the impurities of the receiver. The RM($\phi$) curves of 78 pulsars which have never been published before are presented in this analysis and 42 pulsars out of our sample of 98 showed significant phase-resolved RM variations.
Given sufficient S/N, this fraction will increase, but will ultimately be limited by the systematic uncertainties.

It has been suggested in the literature, that if the largest changes in circular polarization across the frequency band, $\Delta(V/I)(\phi)$, coincide with RM($\phi$) variations, this can be taken as evidence for generalized Faraday rotation in the pulsar magnetosphere. However, we have argued that scattering is also capable of producing such changes, complicating distinguishing between magnetospheric effects and scattering as the main reason for apparent RM($\phi$) variations.
However, we were able in some individual cases to make this distinction.
This is different to \cite{nkk+09}, who concluded based on a smaller sample that all apparent RM variations are consistent with interstellar scattering. 
We identified magnetospheric effects as a cause for apparent RM($\phi$) variations for PSRs J0738$-$4042, J0820$-$1350,  J0907$-$5157, J0908$-$4913, J1243$-$6423, J1326$-$5859, J1359$-$6038, J1453$-$6413,  J1456$-$6843, J1703$-$3241, J1745$-$3040 and J1807$-$0847, a total of 12 out of 42 pulsars with RM($\phi$) variations. The variations in 12 pulsars were consistent with scattering, while the results for the final 18 pulsars were ambiguous.

If scattering, which is most prominent in high-DM pulsars, was the dominant cause for phase-resolved apparent RM variations, then the amplitude of apparent RM variations should be correlated with DM. This is not observed. 
A simulation revealed that if scattering was the dominant cause for apparent RM($\phi$) variations, then these variations should be significantly lower for pulsars with DM $\leq200$ pc cm$^{-3}$.
This confirms the conclusion that magnetospheric effects are important, especially for pulsars that have low levels of scattering. 

We have pointed out that significant deviations from the expected $\lambda^2$ dependence of the PA as a function of frequency are expected both for interstellar scattering and magnetospheric effects.  Although  \cite{nkk+09} concluded that scattering was the main cause for the observed RM($\phi$) variations, they did not find any evidence for non-Faraday components.
We found clear evidence for these expected deviations in our higher S/N data.

We found a correlation between complexity of profiles and whether the pulsars showed RM variations or not, which cannot be explained if scattering is the only cause of the apparent RM variations, considering that in that case only the shape of the PA-swing is important.
Pulsars with multiple profile components potentially have more complex emission properties and hence the magnetospheric effects could cause the PA to be distorted with frequency. It has been confirmed that RM($\phi$) variations are associated with complex and steep dependencies of PA with pulse longitude. We found that in general both linear and circular polarization as a function of frequency are affected by the effect responsible for the RM($\phi$) variations.

All this evidence strongly points towards magnetospheric effects, such as generalized Faraday rotation, as a reason for apparent RM($\phi$) variations in many pulsars.
In most cases, the amplitude of these variations is relatively small, with the exception of a few pulsars. 
PSR J1512$-$5759, for which RM($\phi$) variations are $\sim80 \,\, \,\text{rad m}^{-2}$, has one of the largest amplitude variations from the sample.
Even then the impact on the phase-averaged RM is only modest. Therefore, apparent RM($\phi$) variations will not have a great impact on, for example, the strength of the Galactic  magnetic field as derived from RM, especially when compared with the other uncertainties involved. 
Our results imply that the frequency dependence of PA gives a way to probe what is happening to the radiation in pulsar magnetospheres.

\section*{Acknowledgements}

The Parkes radio telescope is part of the Australia Telescope which is funded by the Commonwealth of Australia for operation as a National Facility managed by CSIRO. Pulsar research at Jodrell Bank Centre for Astrophysics and Jodrell Bank Observatory is supported by a consolidated grant from the UK Science and Technology Facilities Council (STFC). 




\bibliographystyle{mnras}
\bibliography{poln.bib} 




\appendix

\section{Additional Table}

In this section, we display an additional table. Table~\ref{table:disc} contains various classifications for the pulsars, discussed in more detail in Section~\ref{sect:discussion}.

\begin{table*}

\caption{Table summarising the observational results. The classification of whether a pulsar showed significant phase-resolved RM variations or not is displayed in columns 2 and 7, as discussed in Section~\ref{sect:results}. In columns 3 and 8, the number of identified profile components is displayed. The classification of whether the PA swing of a pulsar was steep or shallow is shown in columns 4 and 9. For the 42 pulsars with significant RM($\phi$) variations, columns 5 and 10 display whether the observed variations can be explained by scattering alone, or whether magnetospheric effects play an important role. For the pulsars marked with an asterisk (*), we concluded that it is likely that scattering plays a role in the observed RM($\phi$) variations, besides magnetospheric effects.}
\label{table:disc}
\begin{tabular}{lrrrrlrrrr}
PSR name & RM var & No$_{c}$  & PA swing & Scatt. enough? & PSR name & RM var  & No$_{c}$  & PA swing & Scatt. enough? \\
\hline & \vspace{-3mm} \\

J0034$-$0721 &  Yes & 1 & Steep  & Ambiguous &  J1557$-$4258 &  No &  3 & Steep & \\
J0134$-$2937 &  No & 2 & Shallow &  & J1559$-$4438 &  Yes &  3 & Steep & Yes  \\
J0152$-$1637 &  No & 2 & Shallow &  & J1602$-$5100 &  No &  5 & Steep &\\
J0255$-$5304 &  Yes & 2 & Steep  & Yes & J1604$-$4909 &  Yes &  4 & Steep & Yes \\
J0401$-$7608 &  Yes & 3 & Steep  & Ambiguous &  J1605$-$5257 &  No &  5 & Steep &\\
J0452$-$1759 &  Yes & 3 & Steep  & Ambiguous & J1633$-$4453 &  No &  2 & Shallow &\\
J0536$-$7543 &  Yes & 3 & Steep  & Ambiguous & J1633$-$5015 &  No &  3 & Shallow &\\
J0614+2229 &  No &  1 & Steep & & J1644$$-$$4559 &  Yes & 3 & Shallow & Ambiguous\\
J0630$-$2834 &  No &  2 & Steep & &  J1646$-$6831 &  No &  4 & Steep &\\
J0729$-$1836 &  No &  2 & Steep & & J1651$-$4246 &  No &  2 & Steep &\\
J0738$-$4042 &  Yes &  3 & Steep & No & J1651$-$5222 &  Yes &  3 & Steep & Ambiguous \\
J0742$-$2822 &  No &  4 & Steep & & J1653$-$3838 &  No &  2 & Shallow &\\
J0745$-$5353 &  No &  1 & Steep & &  J1701$-$3726 &  Yes &  3 & Steep & Yes \\
J0809$-$4753 &  No &  3 & Steep & & J1703$-$3241 &  Yes &  2 & Steep & No \\
J0820$-$1350 &  Yes &  2 & Steep & No & J1705$-$1906 &  No &  5 & Steep &\\
J0835$-$4510 &  No &  3 & Steep & &    J1705$-$3423 &  No &  1 & Shallow &\\
J0837$-$4135 &  Yes &  3 & Steep & Yes &    J1709$-$1640 &  No &  1 & Steep &\\
J0907$-$5157 &  Yes &  3 & Shallow & No &  J1709$-$4429 &  No &  1 & Steep &\\
J0908$-$4913 &  Yes &  2 & Steep & No &   J1717$-$3425 &  No &  1 & Shallow &\\
J0942$-$5552 &  Yes &  3 & Shallow & Ambiguous & J1721$-$3532 &  No &  1 & Steep  &\\
J1001$-$5507 &  Yes &  3 & Steep & Ambiguous & J1722$-$3207 &  Yes &  2 & Steep & Yes \\
J1043$-$6116 &  No &  3 & Shallow & & J1722$-$3712 &  No &  2 & Steep &\\
J1047$-$6709 &  No &  2 & Shallow & & J1731$-$4744 &  Yes &  4 & Steep & Ambiguous \\
J1048$-$5832 &  No &  3 & Steep & & J1739$-$2903 &  No &  2 & Shallow &\\
J1056$-$6258 &  No &  3 & Steep & & J1740$-$3015 &  No &  2 & Steep &\\
J1057$-$5226 &  Yes &  5 & Steep & Ambiguous & J1741$-$3927 &  No &  3 & Steep &\\
J1136$-$5525 &  No &  4 & Shallow & & J1745$-$3040 &  Yes &  3 & Shallow & No \\
J1146$-$6030 &  No &  4 & Steep & & J1751$-$4657 &  Yes &  2 & Steep &  Ambiguous\\
J1157$-$6224 &  No &  3 & Shallow & & J1752$-$2806 &  Yes & 2 & Steep & Yes \\
J1243$-$6423* &  Yes &  2 & Steep & No & J1807$-$0847 &  Yes &  4 & Shallow & No\\
J1306$-$6617 &  No &  2 & Steep & & J1817$-$3618 &  Yes &  3 & Steep  & Ambiguous\\
J1326$-$5859* &  Yes &  3 & Steep & No & J1820$-$0427 &  Yes &  3 & Steep  & Ambiguous\\
J1326$-$6408 &  No &  3 & Shallow & & J1822$-$2256 &  No &  1 & Shallow &\\
J1326$-$6700 &  No &  3 & Steep & & J1823$-$3106 &  No &  1 & Steep &\\
J1327$-$6222 &  No &  2 & Steep & & J1824$-$1945 &  Yes &  2 & Shallow & Yes \\
J1328$-$4357 &  No &  2 & Steep & & J1825$-$0935 &  No &  2 & Shallow &\\
J1357$-$62 &    Yes &  3 & Steep & Ambiguous & J1829$-$1751 &  No &  3 & Shallow &\\
J1359$-$6038 &  Yes &  1 & Shallow & No & J1830$-$1059 &  No &  2 & Shallow &\\
J1401$-$6357 &  Yes &  3 & Steep & Ambiguous & J1832$-$0827 &  No &  2 & Steep &\\
J1428$-$5530 &  Yes &  2 & Shallow & Ambiguous & J1845$-$0743 &  No &  3 & Shallow &\\
J1430$-$6623 &  No &  3 & Steep & & J1847$-$0402 &  No &  5 & Steep &\\
J1453$-$6413 &  Yes &  3 & Steep & No & J1848$-$0123 &  Yes &  3 & Steep & Ambiguous\\
J1456$-$6843 &  Yes &  5 & Steep & No & J1852$-$0635 &  No &  3 & Steep &\\
J1512$-$5759 &  Yes &  1 & Steep & Yes & J1900$-$2600 &  Yes &  4 & Steep & Ambiguous\\
J1522$-$5829 &  No &  3 & Shallow & & J1913$-$0440 &  Yes &  3 & Steep & Yes\\
J1534$-$5334 &  Yes &  3 & Steep & Yes & J1941$-$2602 &  No &  2 & Shallow  &\\
J1539$-$5626 &  No &  2 & Shallow & & J2048$-$1616 &  Yes &  3 & Steep & Yes\\
J1544$-$5308 &  No &  3 & Shallow & & J2330$-$2005 &  No &  2 & Steep &\\
J1555$-$3134 &  No &  4 & Shallow & & J2346$-$0609 &  No &  2 & Steep &\\

\hline
\end{tabular}
\end{table*}


\bsp	
\label{lastpage}
\end{document}


\begin{titlepage}
    \begin{center}
        \vspace*{1cm}
        
        \Huge
        \textbf{Supplementary material}

    \end{center}
\end{titlepage}

\clearpage

\appendix
\counterwithin{figure}{section}

\section{Phase-resolved RM Plots}

\begin{figure}
\begin{center}
\begin{tabular}{cc}
\includegraphics[width=0.45\textwidth,angle=0]{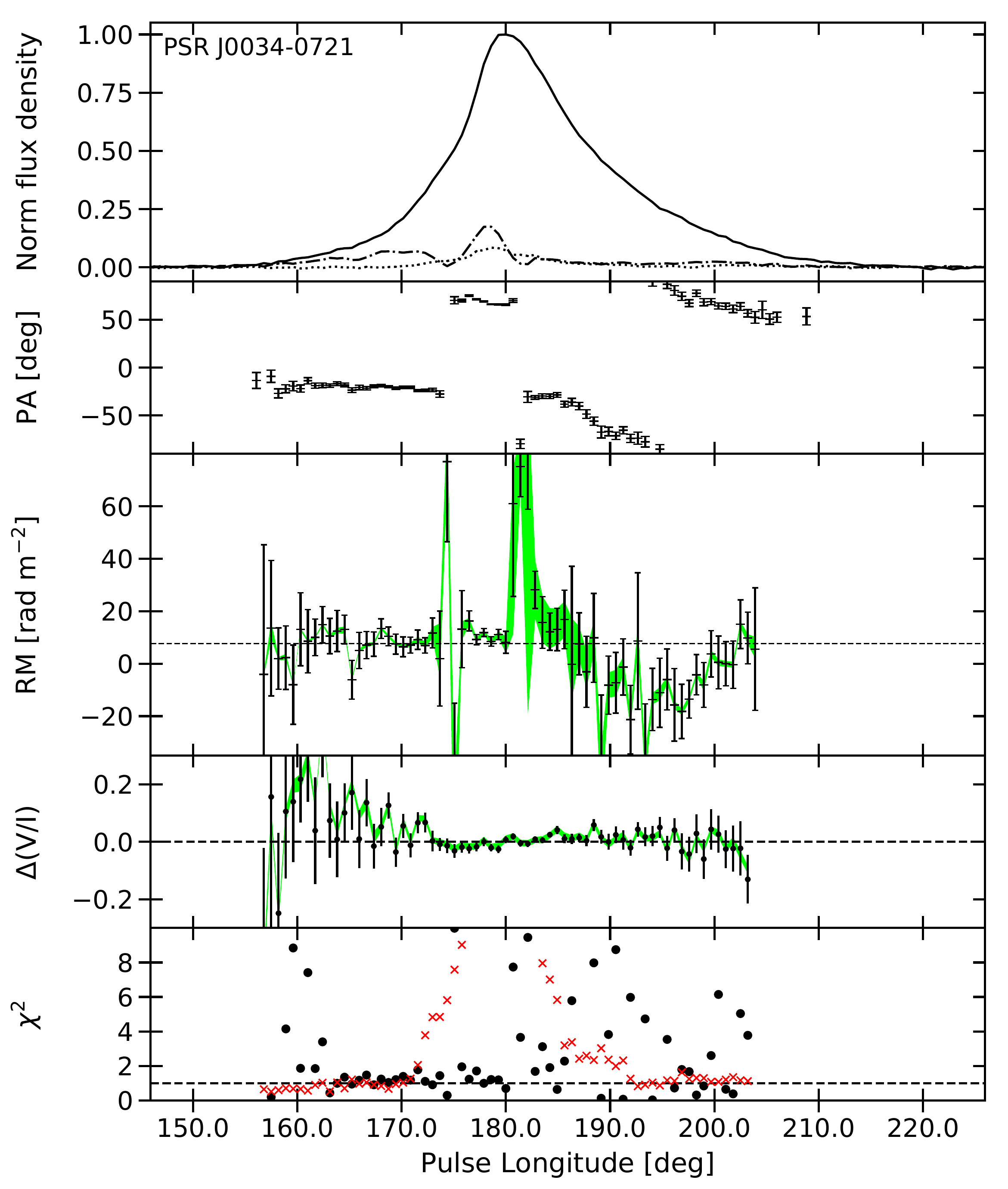} &
\includegraphics[width=0.45\textwidth,angle=0]{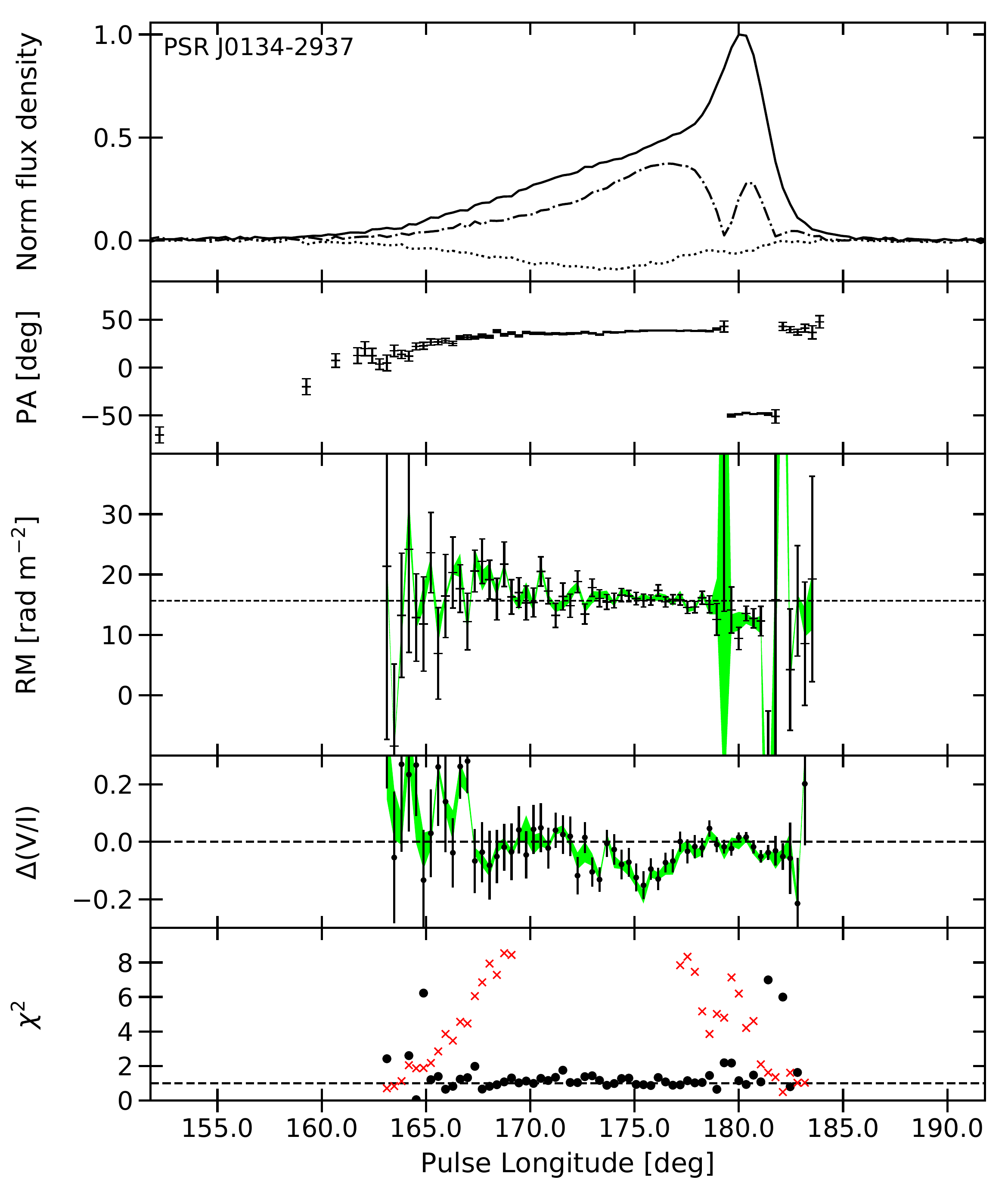} \\
\includegraphics[width=0.45\textwidth,angle=0]{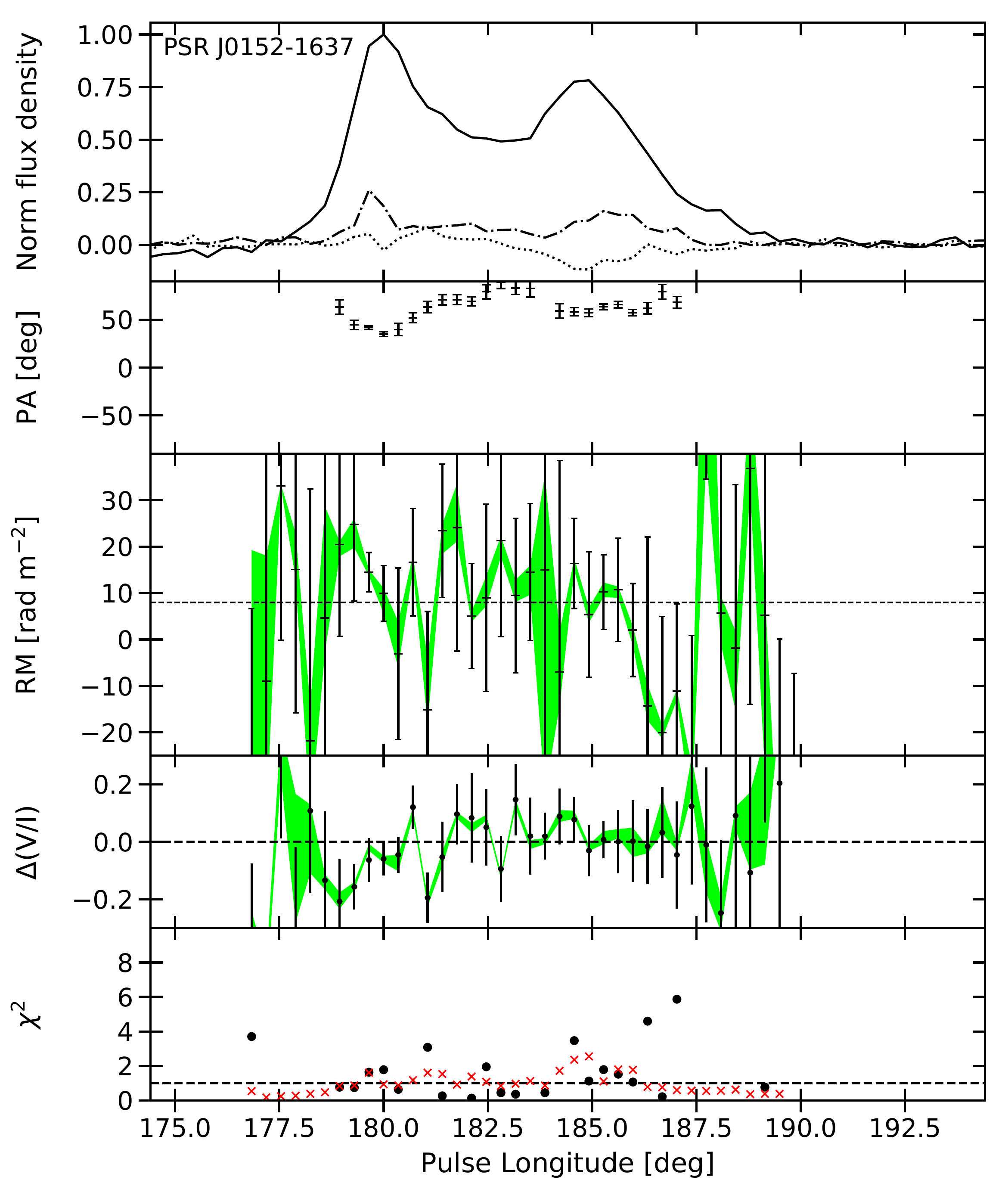} &
\includegraphics[width=0.45\textwidth,angle=0]{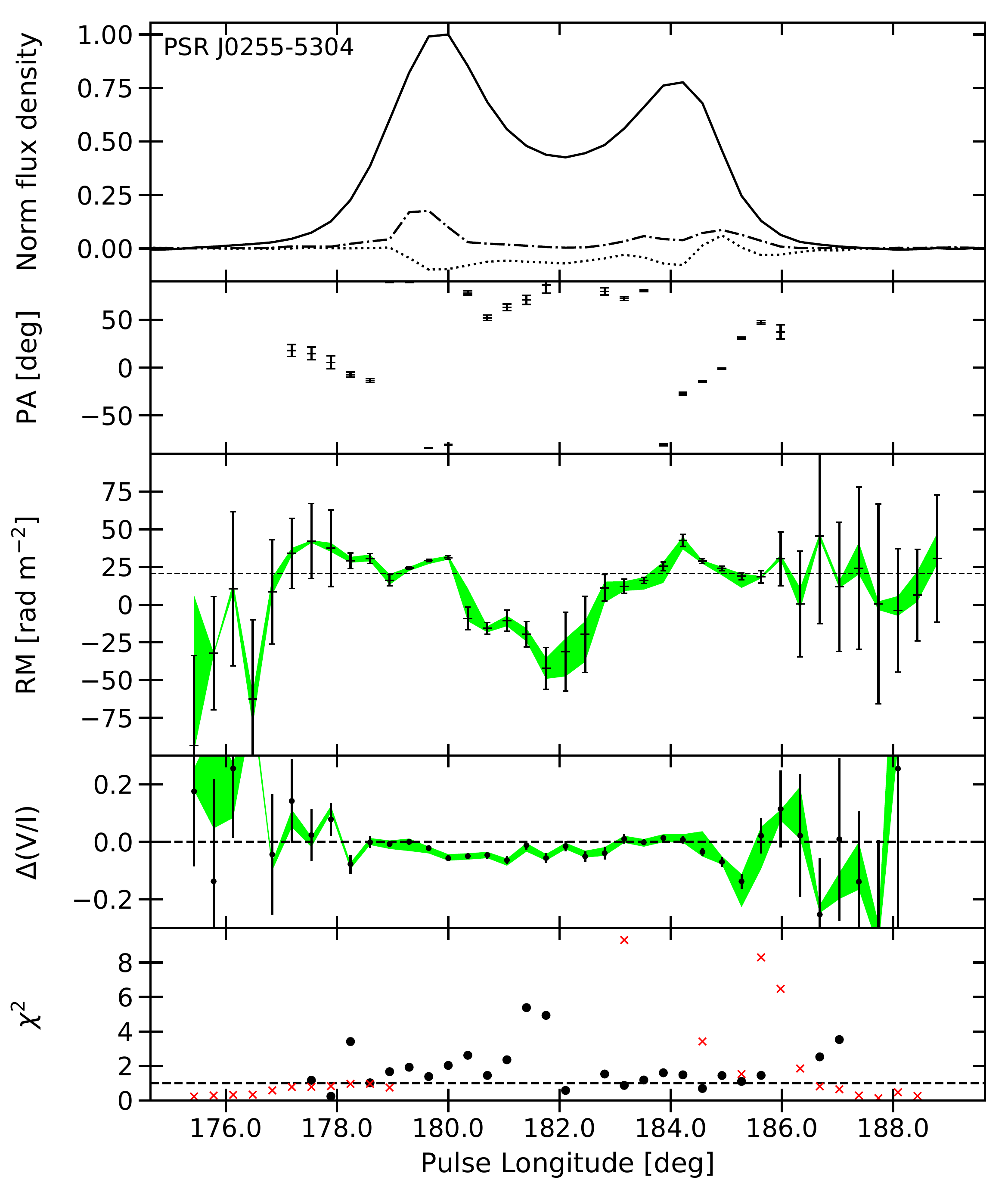} \\
\end{tabular}
\end{center}
\caption{Phase-resolved RM plots for PSR J0034$-$0721, PSR J0134$-$2937, PSR J0152$-$1637 and PSR J0255$-$5304. For more details on what is displayed in the individual panels, see Fig. 1.}
\label{fig:allPSR1}
\end{figure}

\begin{figure}
\begin{center}
\begin{tabular}{cc}
\includegraphics[width=9cm, height=11cm,angle=0]{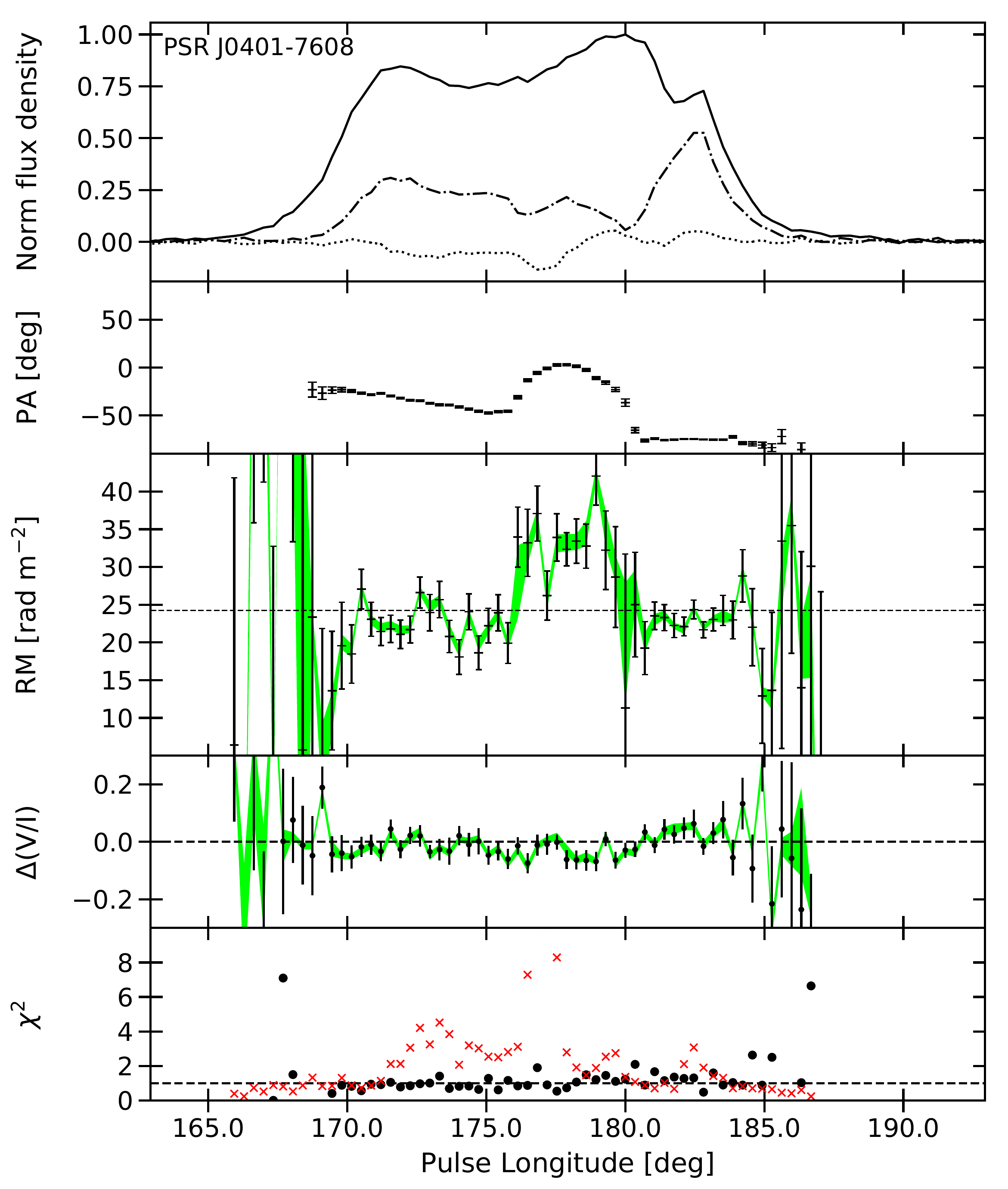} &
\includegraphics[width=9cm, height=11cm,angle=0]{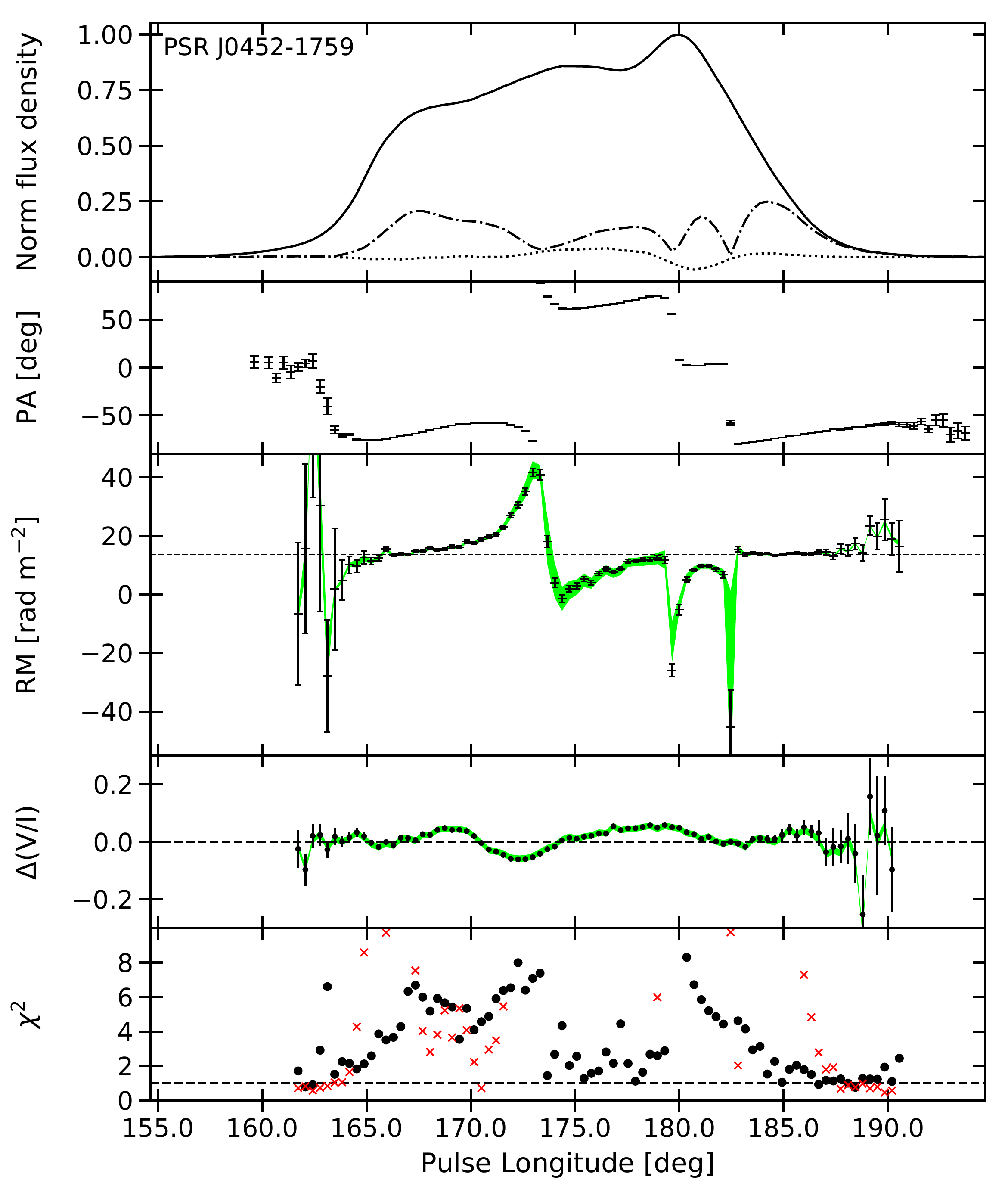} \\
\includegraphics[width=9cm, height=11cm,angle=0]{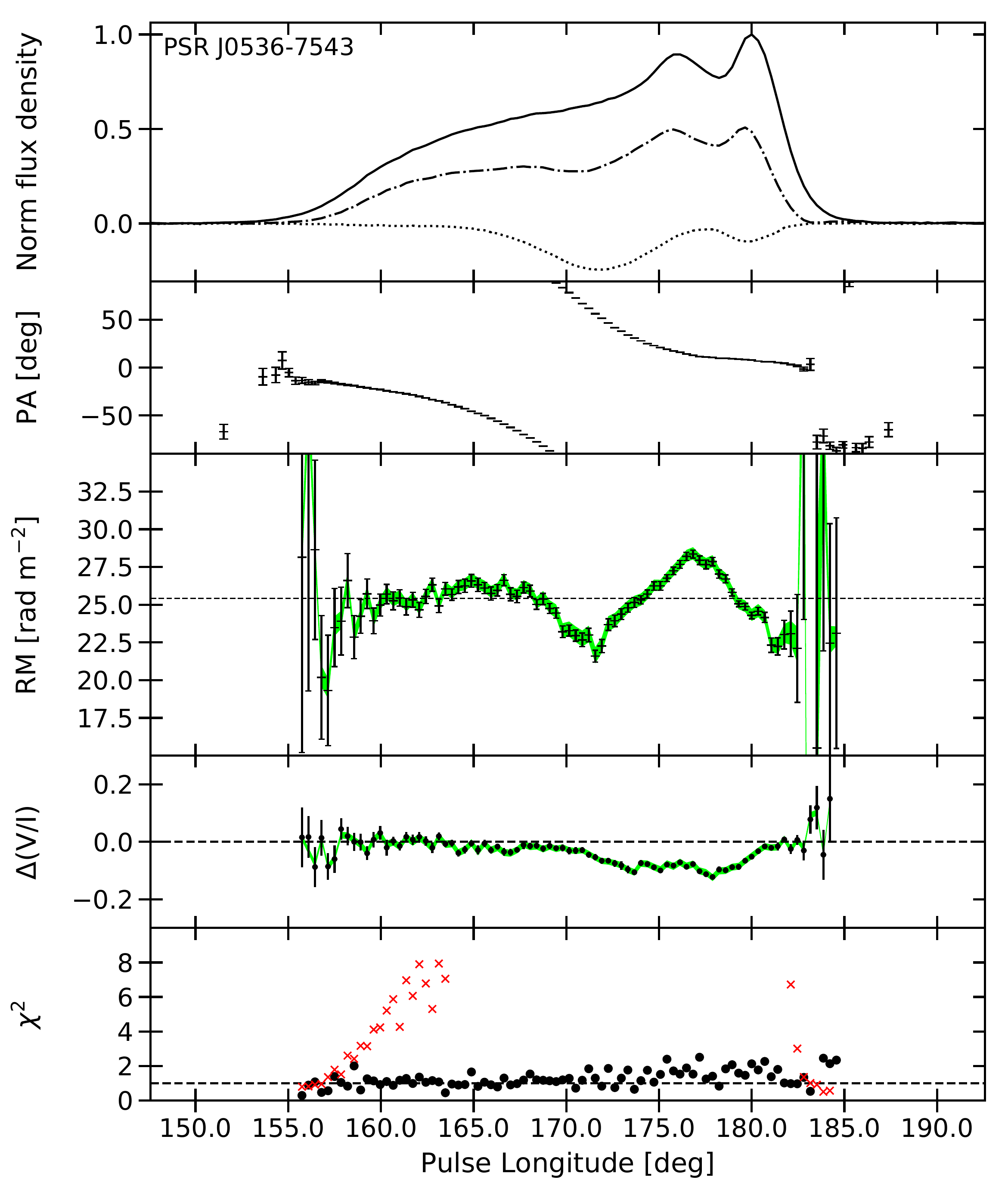} &
\includegraphics[width=9cm, height=11cm,angle=0]{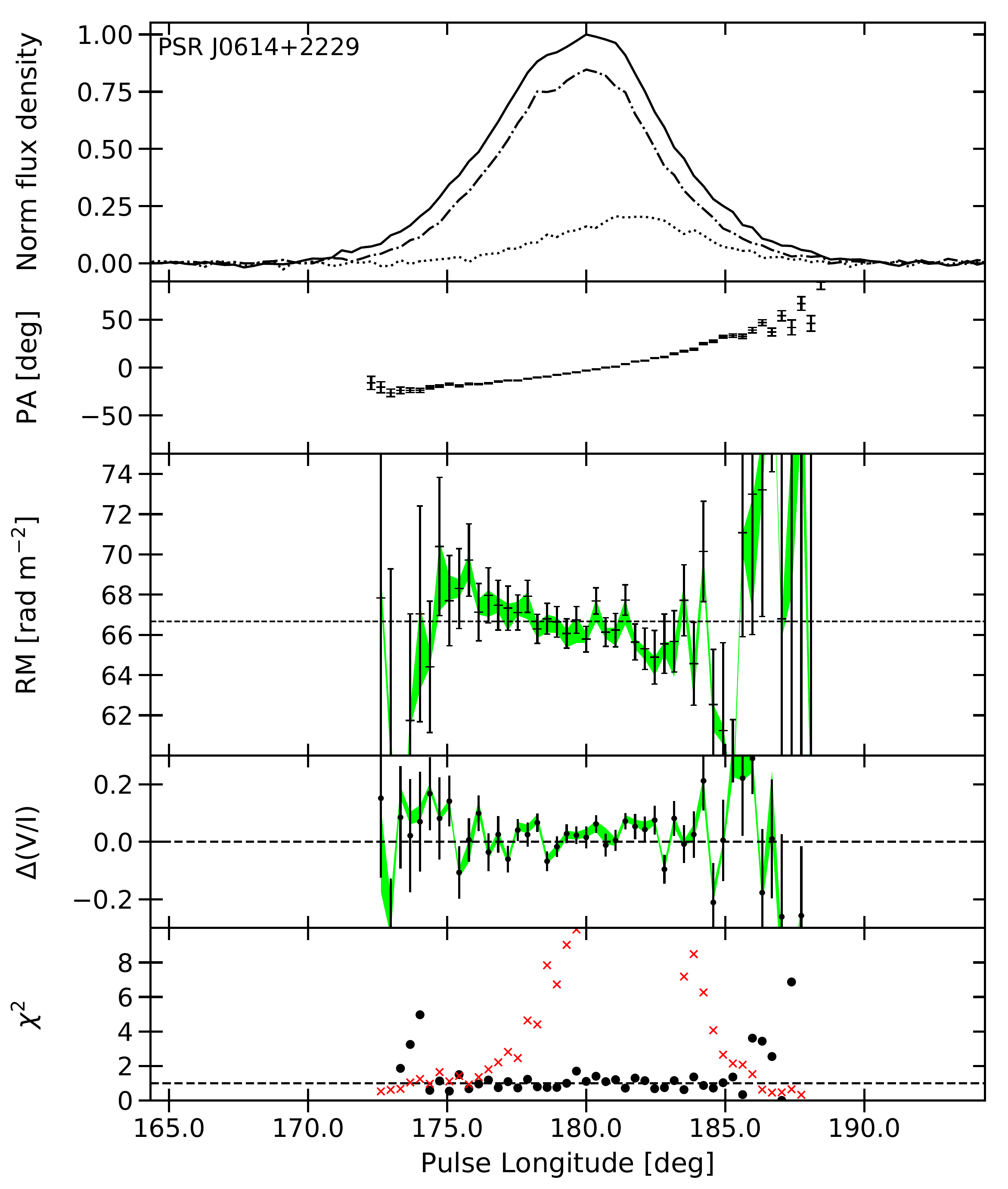} \\
\end{tabular}
\end{center}
\caption{Phase-resolved RM plots for PSR J0401$-$7608, PSR J0452$-$1759, PSR J0536$-$7543 and PSR J0614+2229. For more details on what is displayed in the individual panels, see Fig. 1.}
\label{fig:allPSR2}
\end{figure}

\begin{figure}
\begin{center}
\begin{tabular}{cc}
\includegraphics[width=0.45\textwidth,angle=0]{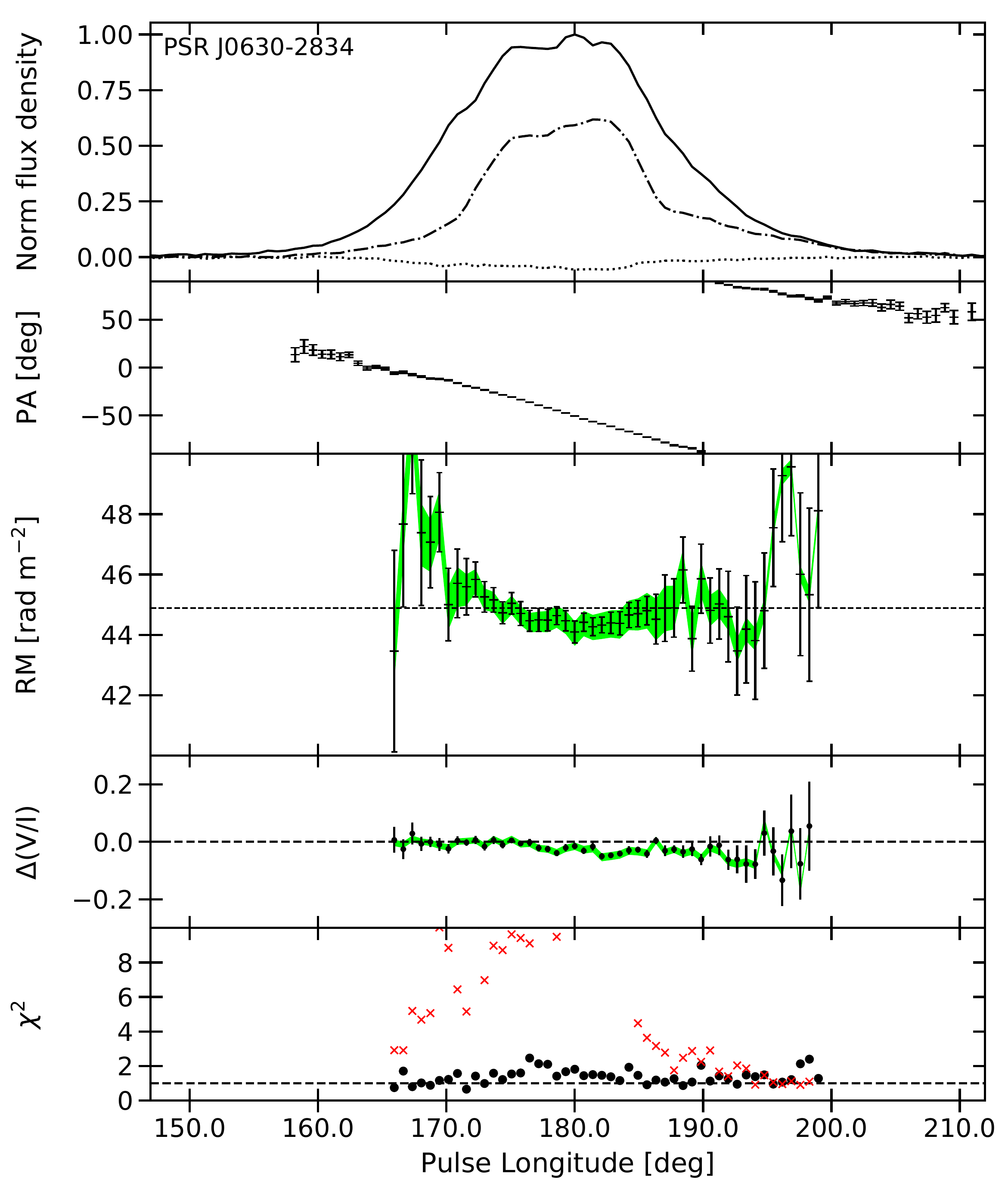} &
\includegraphics[width=0.45\textwidth,angle=0]{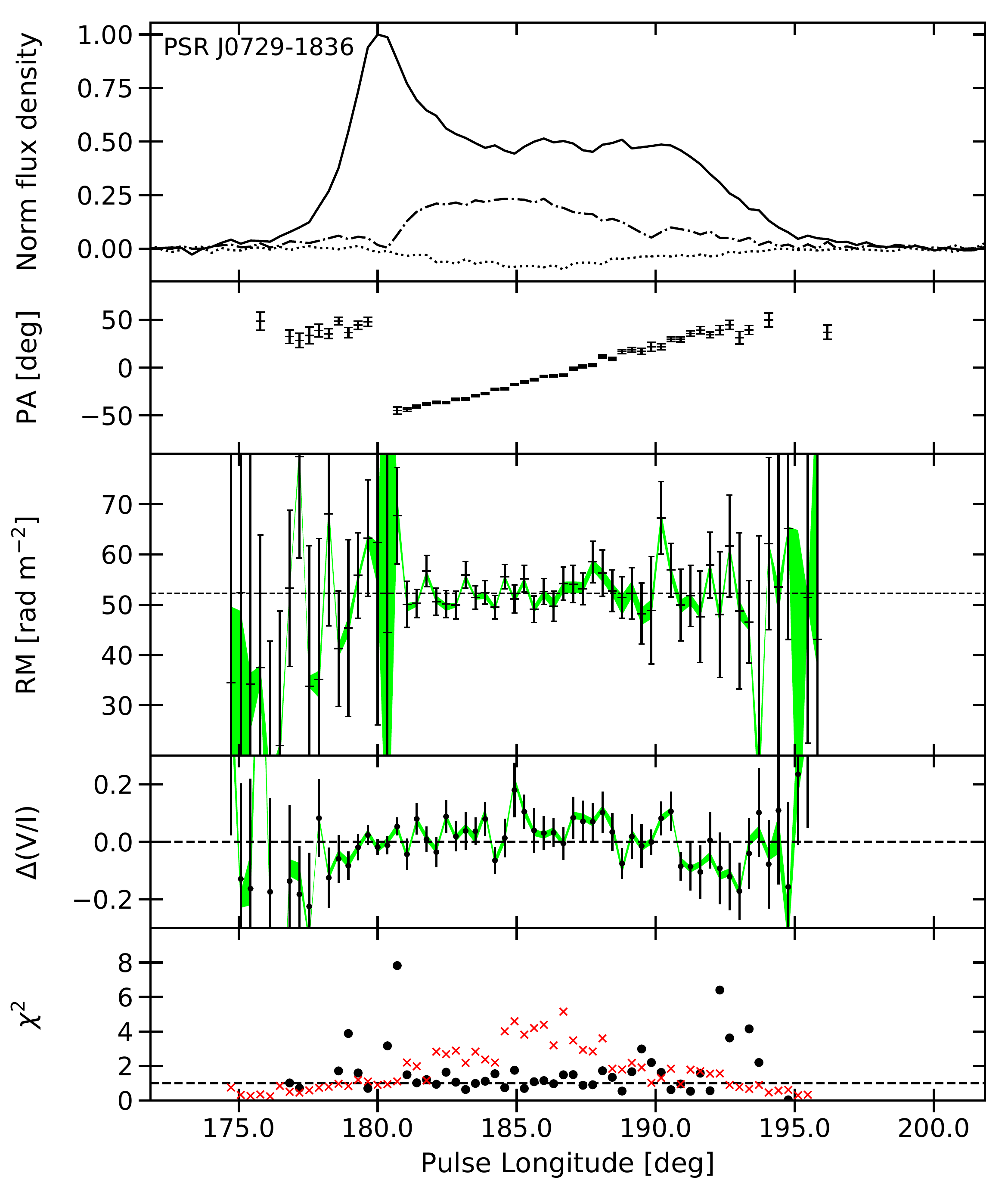} \\
\includegraphics[width=0.45\textwidth,angle=0]{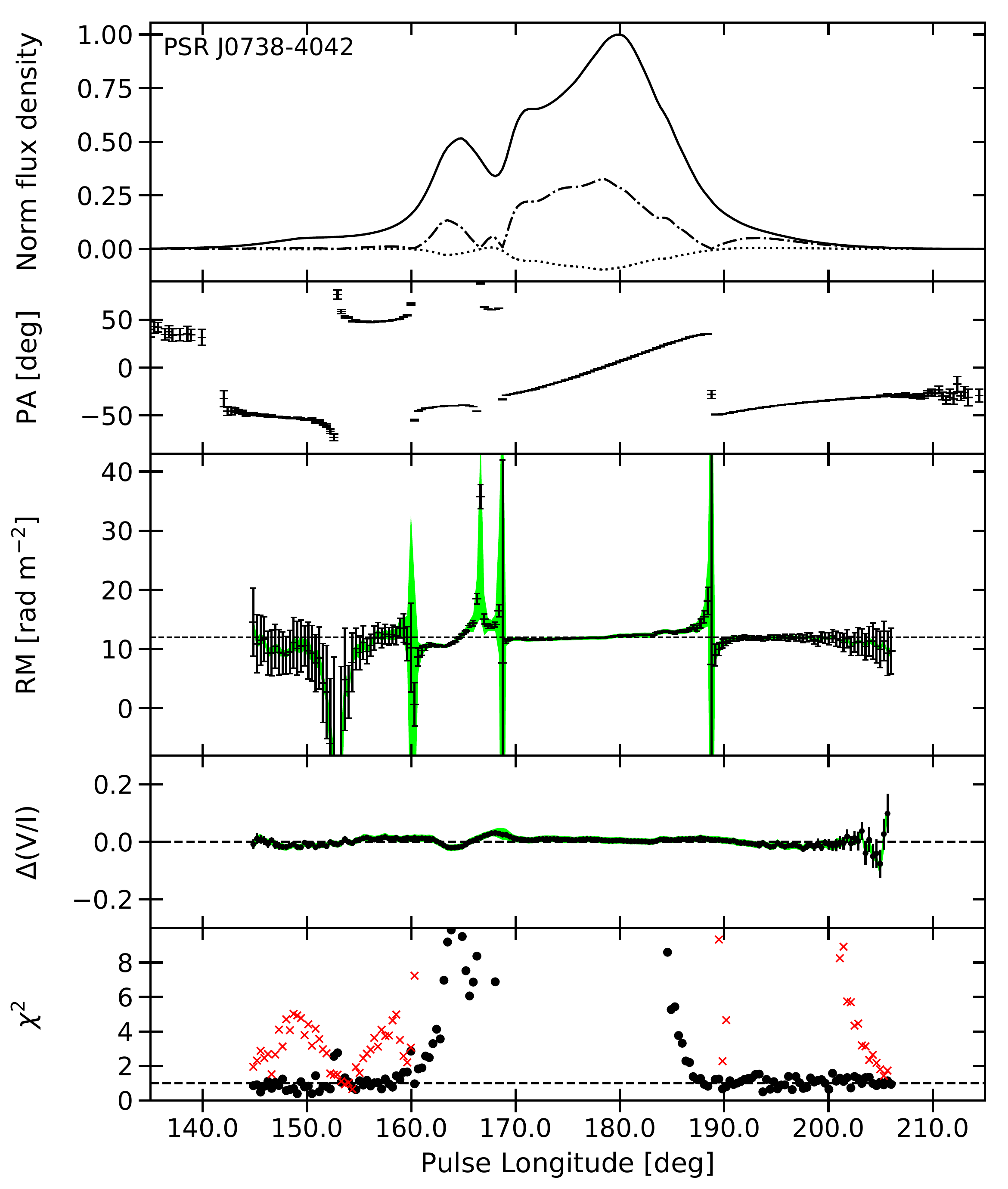} &
\includegraphics[width=0.45\textwidth,angle=0]{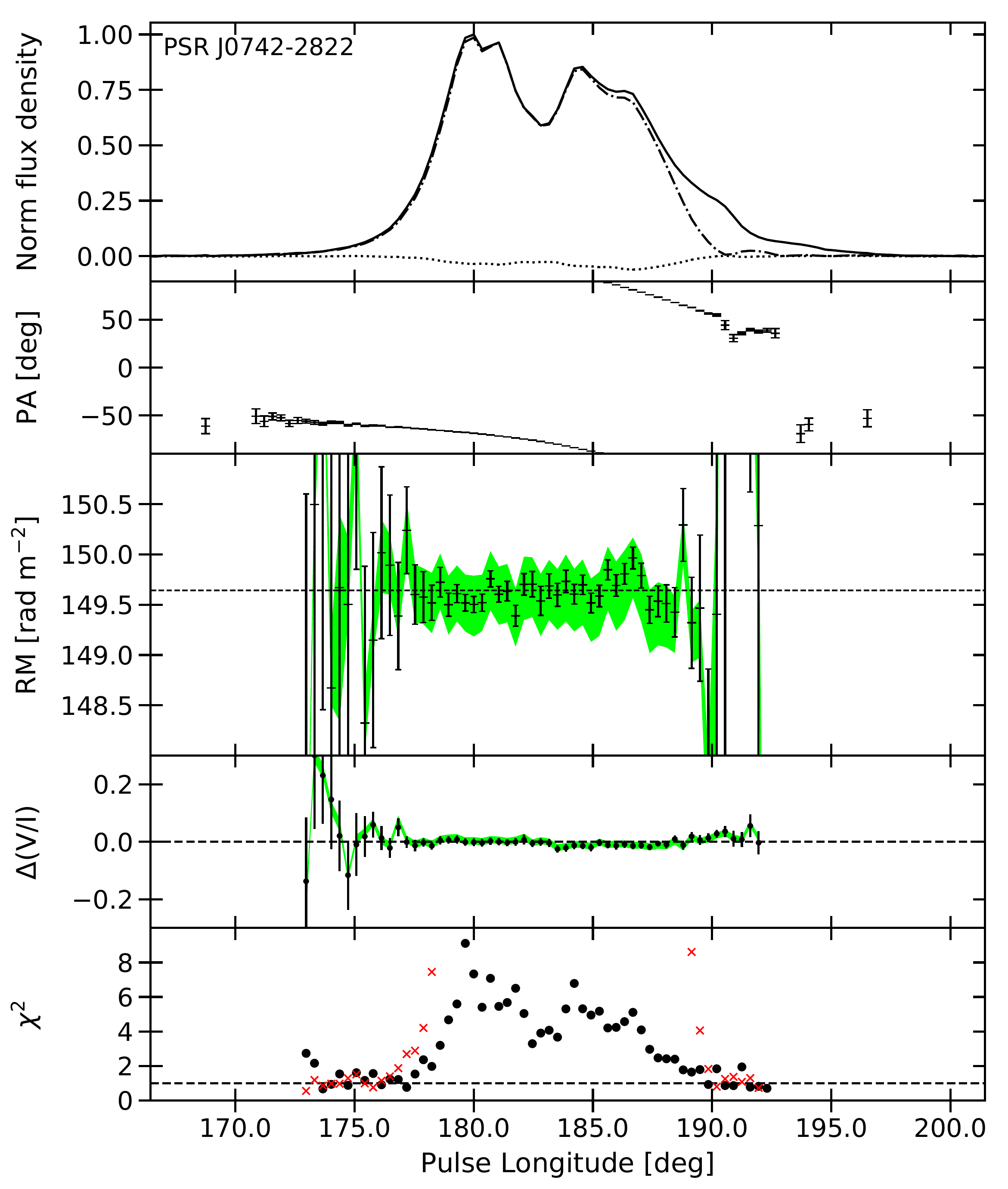} \\
\end{tabular}
\end{center}
\caption{Phase-resolved RM plots for PSR J0630$-$2834, PSR J0729$-$1836, PSR J0738$-$4042 and PSR J0742$-$2822. For more details on what is displayed in the individual panels, see Fig. 1.}
\label{fig:allPSR3}
\end{figure}

\begin{figure}
\begin{center}
\begin{tabular}{cc}
\includegraphics[width=0.45\textwidth,angle=0]{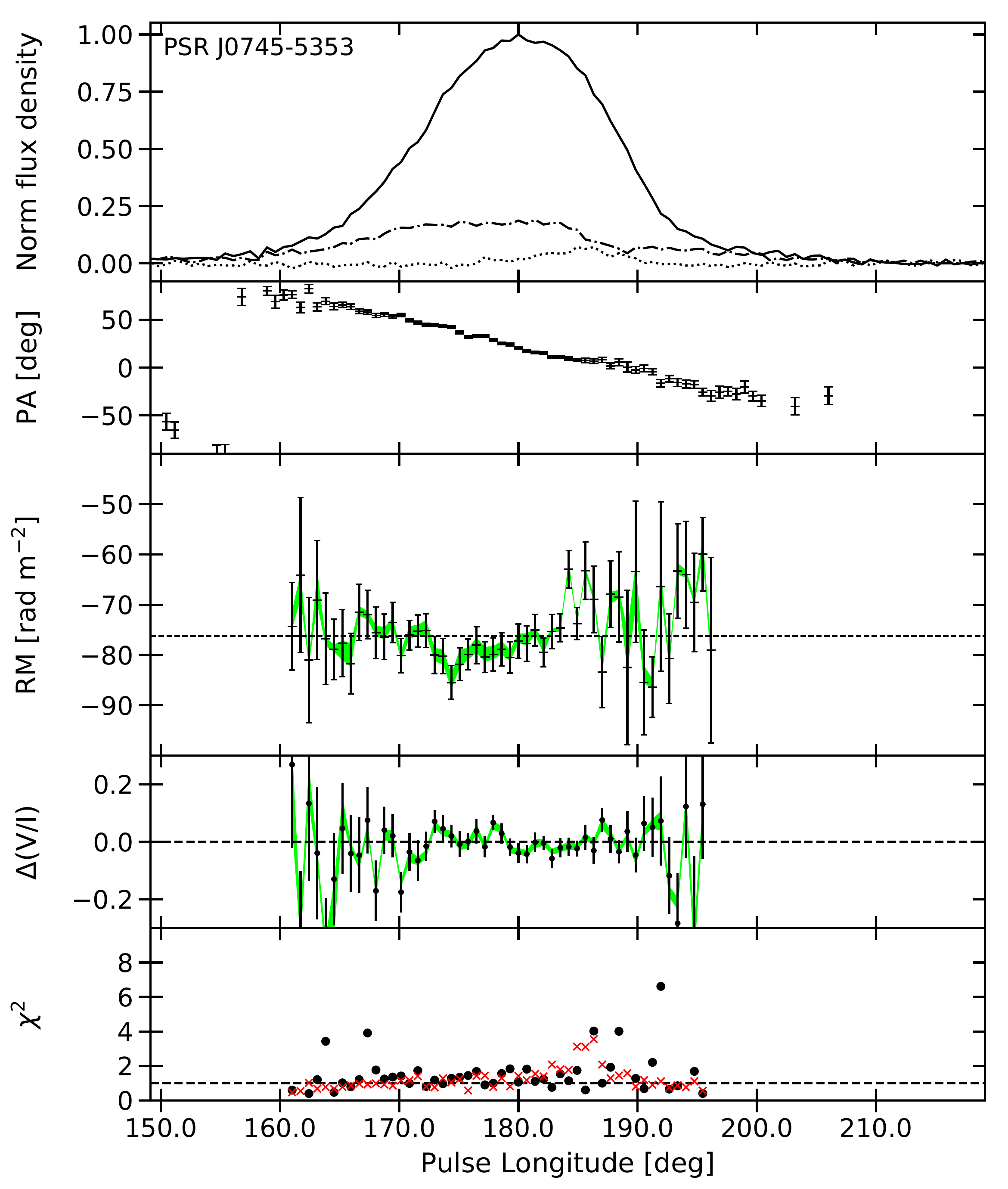} &
\includegraphics[width=0.45\textwidth,angle=0]{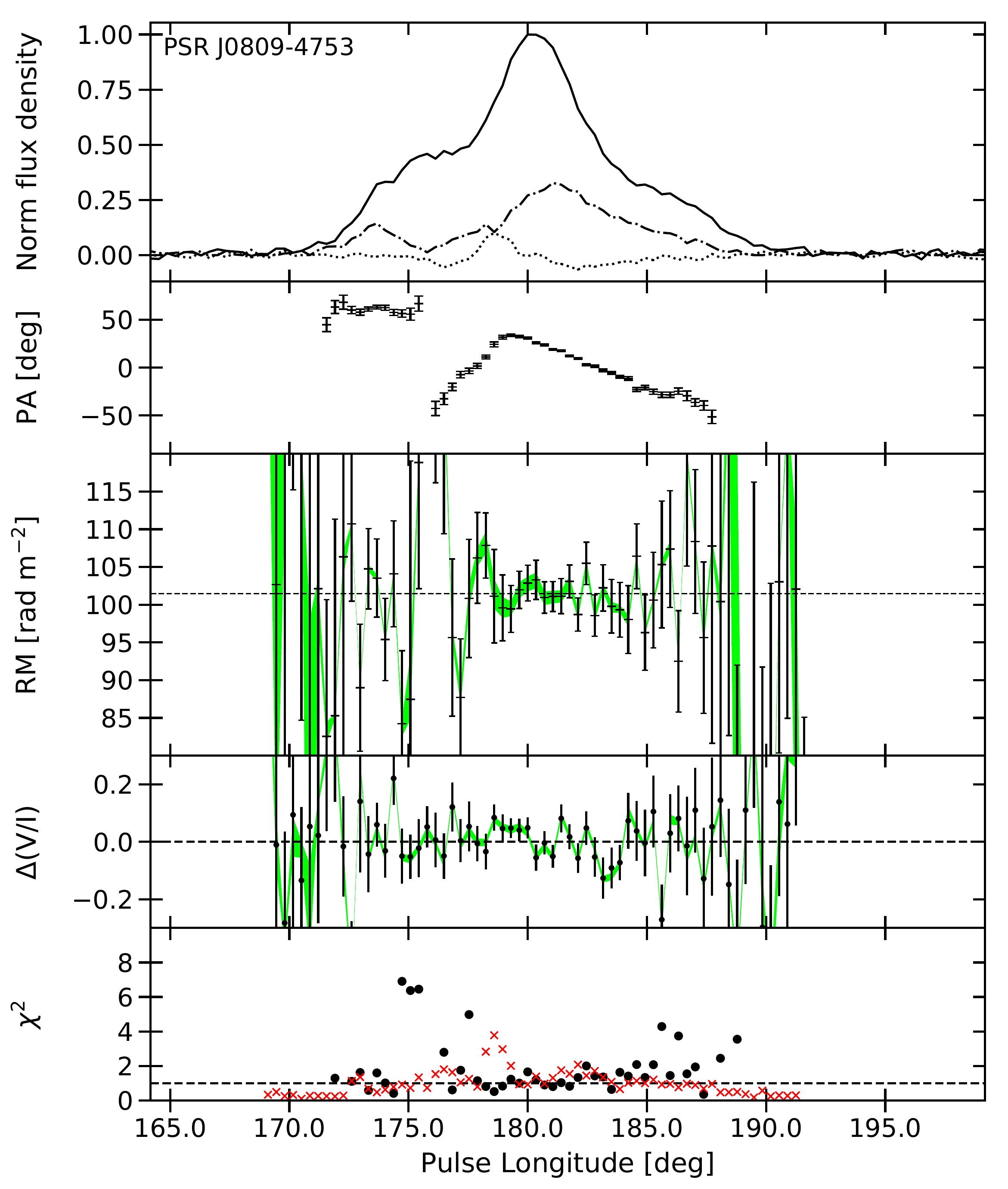} \\
\includegraphics[width=0.45\textwidth,angle=0]{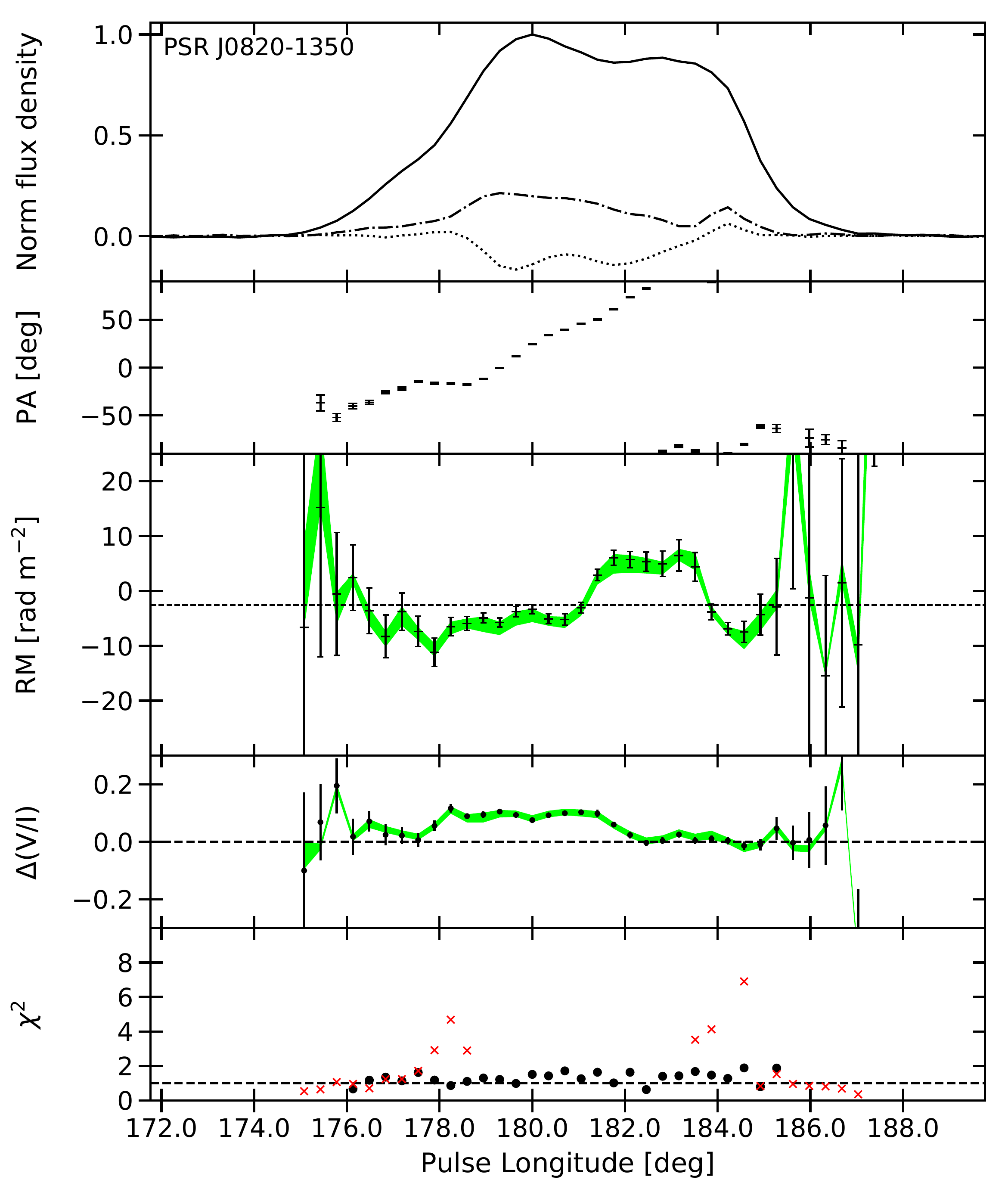} &
\includegraphics[width=0.45\textwidth,angle=0]{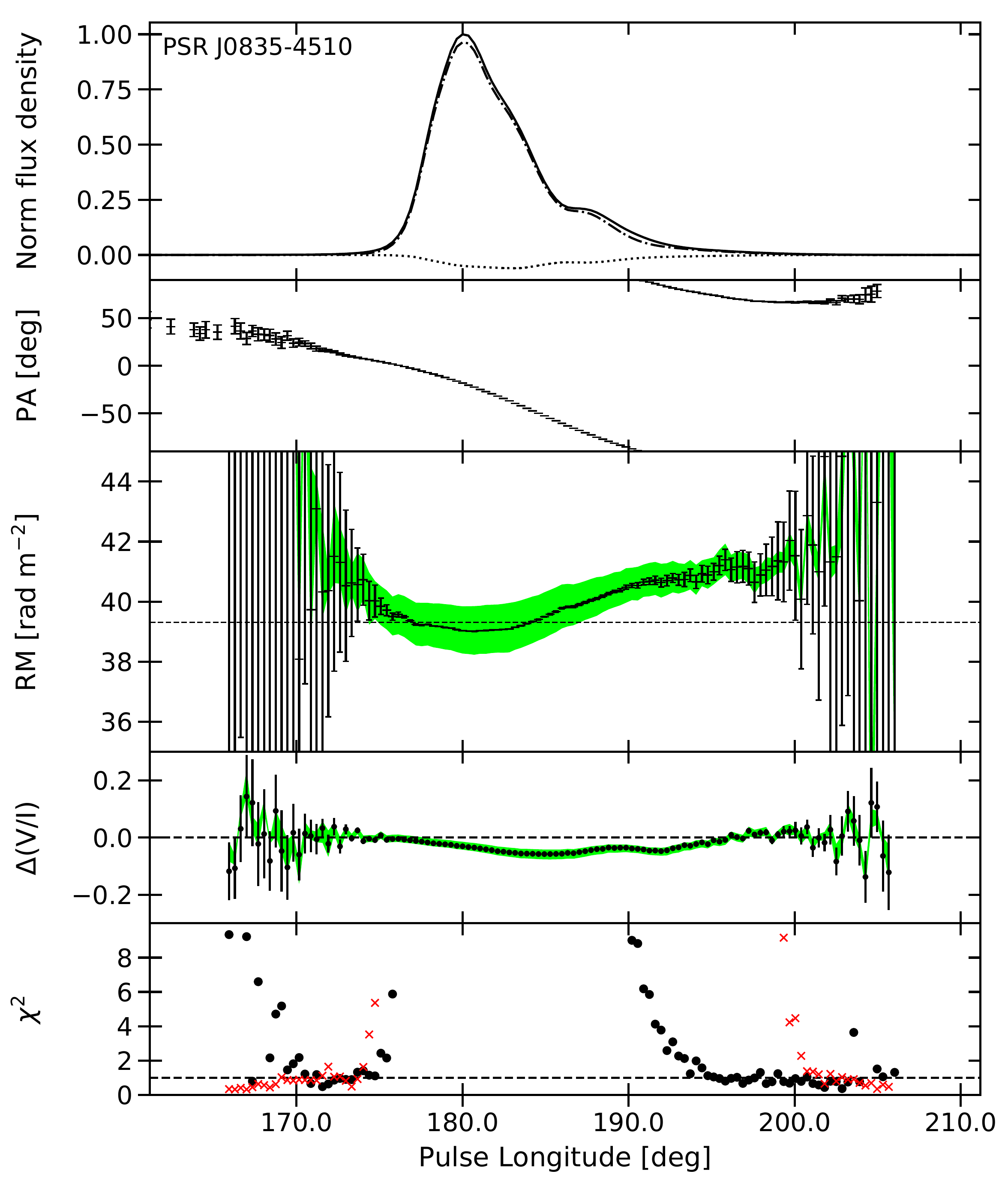} \\
\end{tabular}
\end{center}
\caption{Phase-resolved RM plots for PSR J0745$-$5353, PSR J0809$-$4753, PSR J0820$-$1350 and PSR J0835$-$4510. For more details on what is displayed in the individual panels, see Fig. 1.}
\label{fig:allPSR4}
\end{figure}

\begin{figure}
\begin{center}
\begin{tabular}{cc}
\includegraphics[width=0.45\textwidth,angle=0]{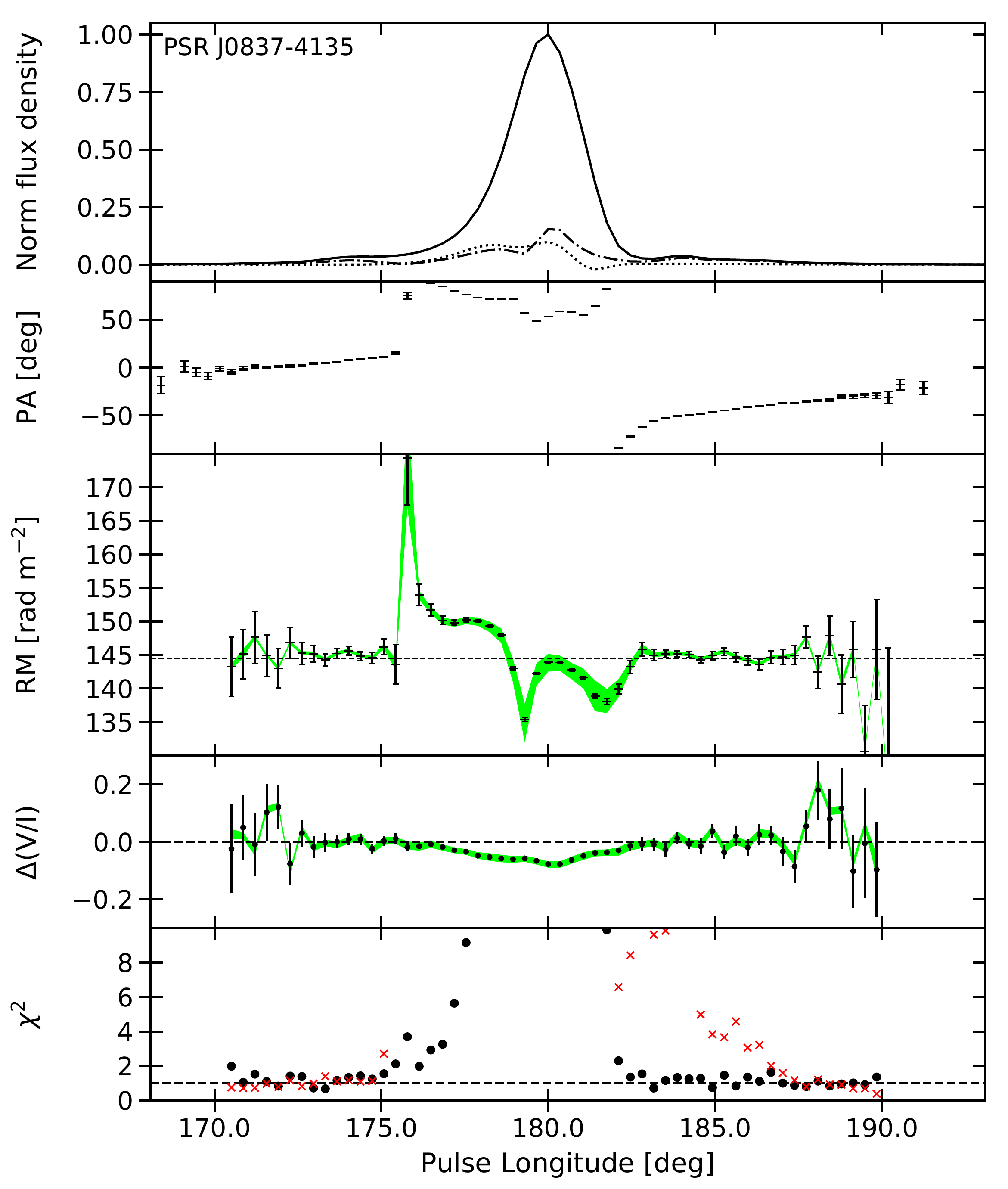} &
\includegraphics[width=0.45\textwidth,angle=0]{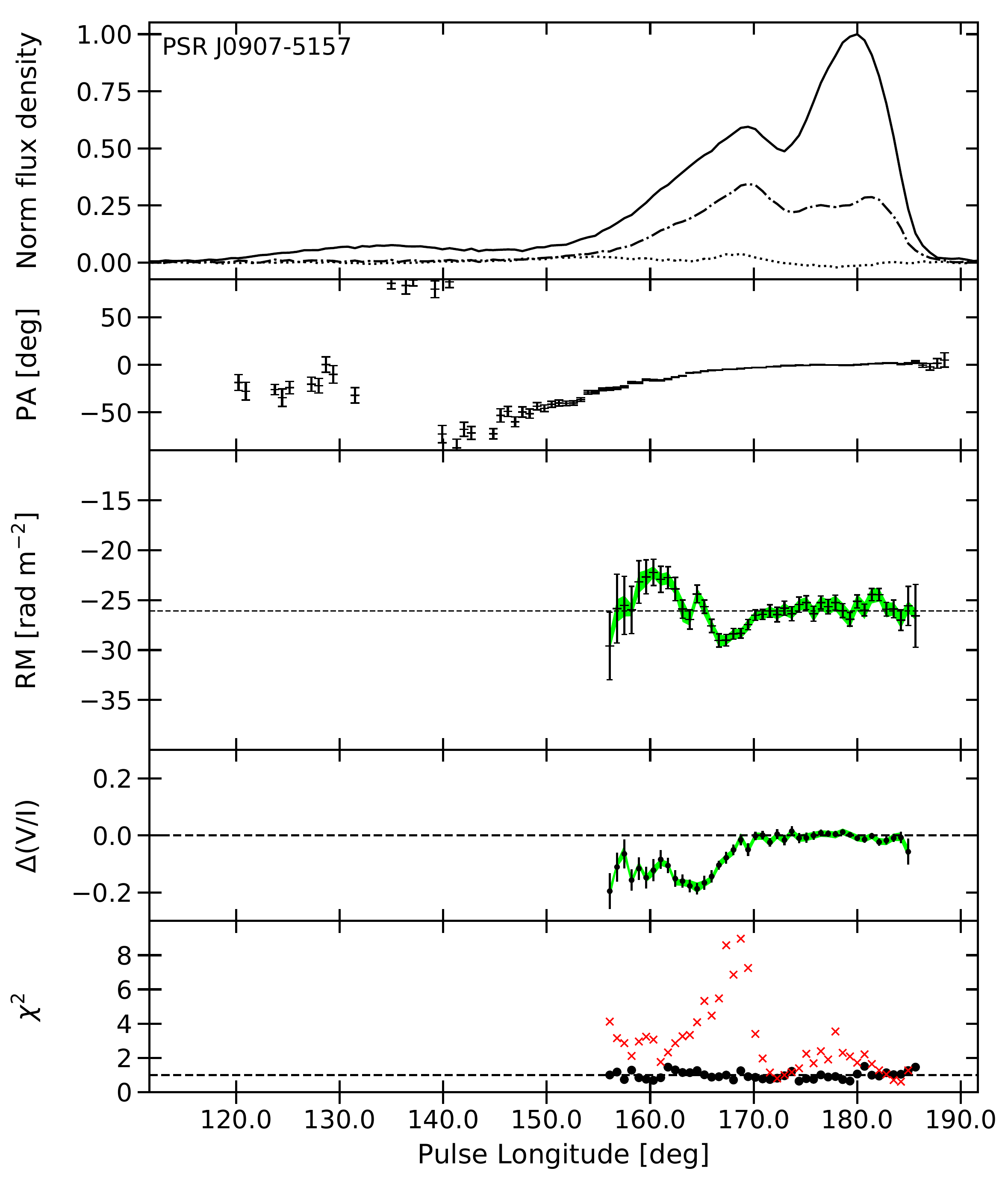} \\
\includegraphics[width=0.45\textwidth,angle=0]{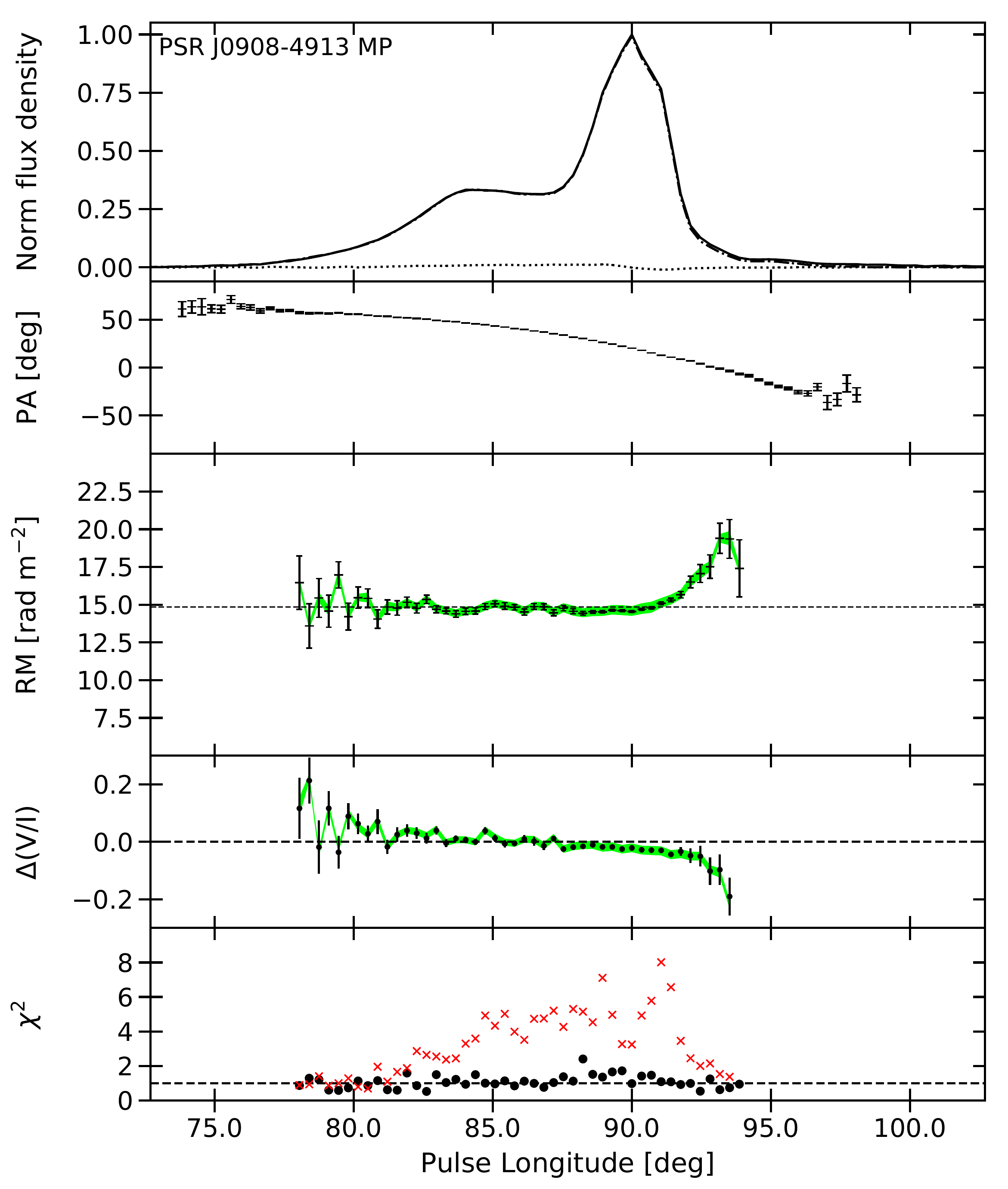} &
\includegraphics[width=0.45\textwidth,angle=0]{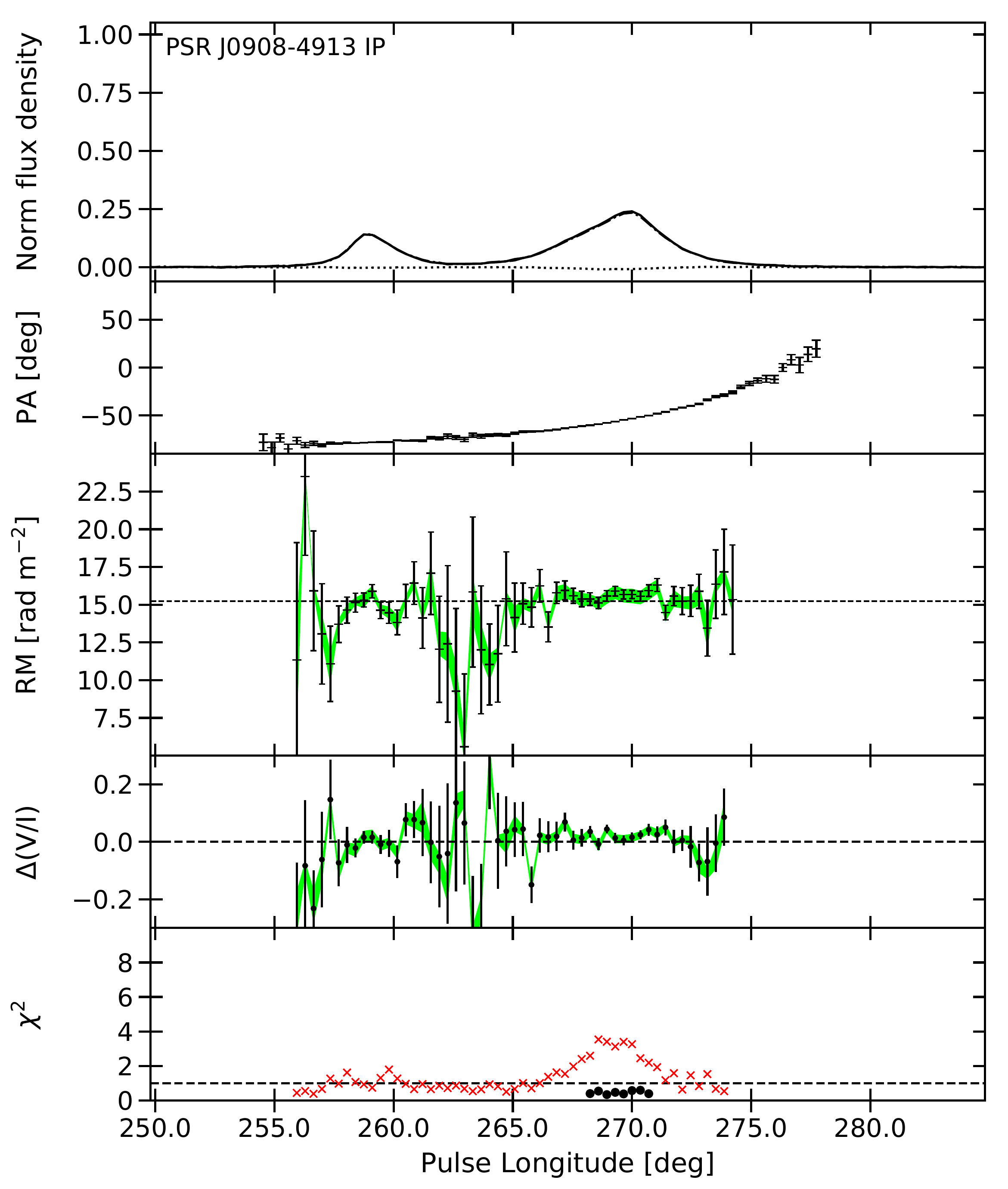} \\
\end{tabular}
\end{center}
\caption{Phase-resolved RM plots for PSR J0837$-$4135, PSR J0907$-$5157, PSR J0908$-$4913 (MP) and PSR J0908$-$4913 (IP). For more details on what is displayed in the individual panels, see Fig. 1.}
\label{fig:allPSR5}
\end{figure}

\begin{figure}
\begin{center}
\begin{tabular}{cc}
\includegraphics[width=0.45\textwidth,angle=0]{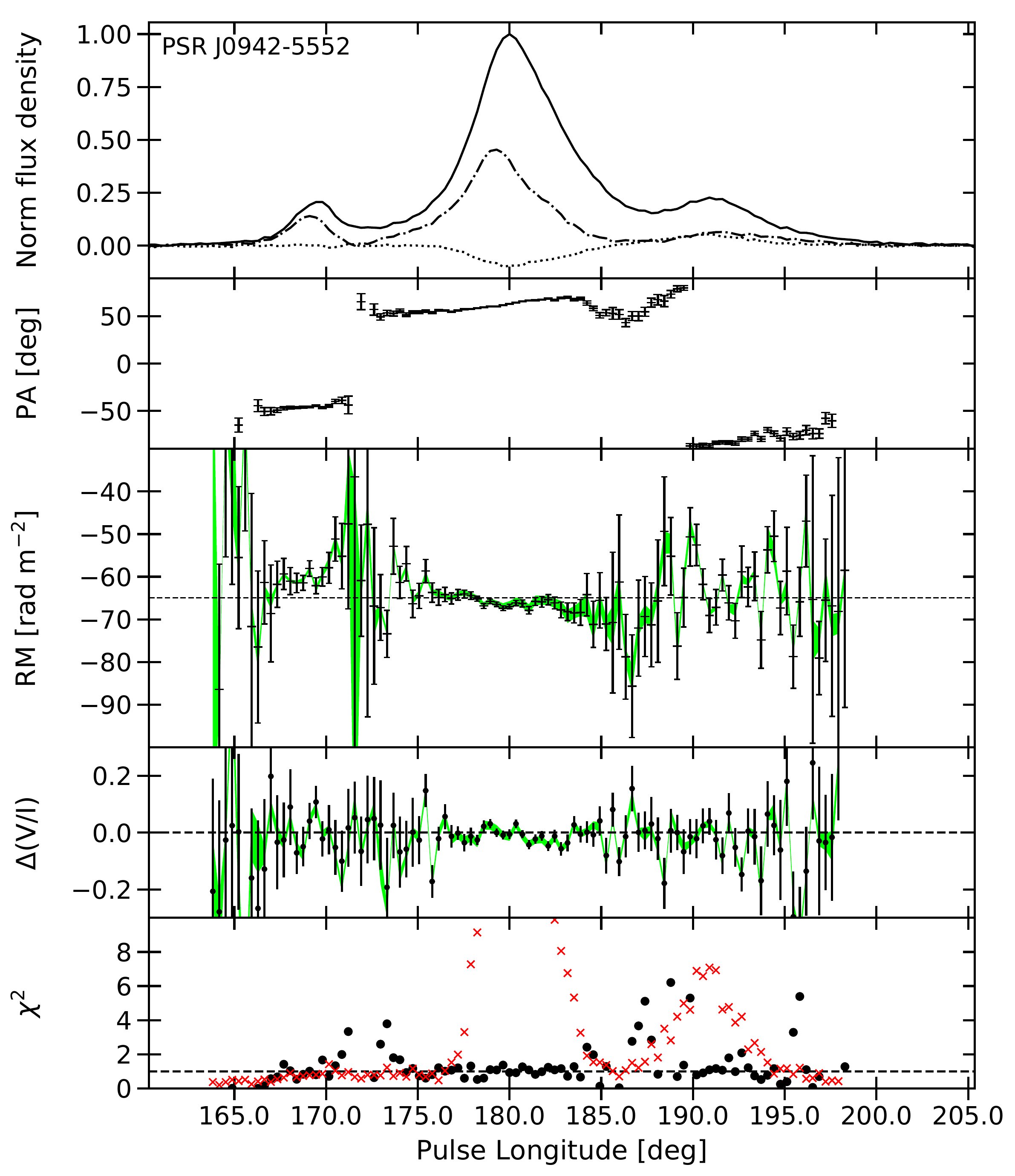} &
\includegraphics[width=0.45\textwidth,angle=0]{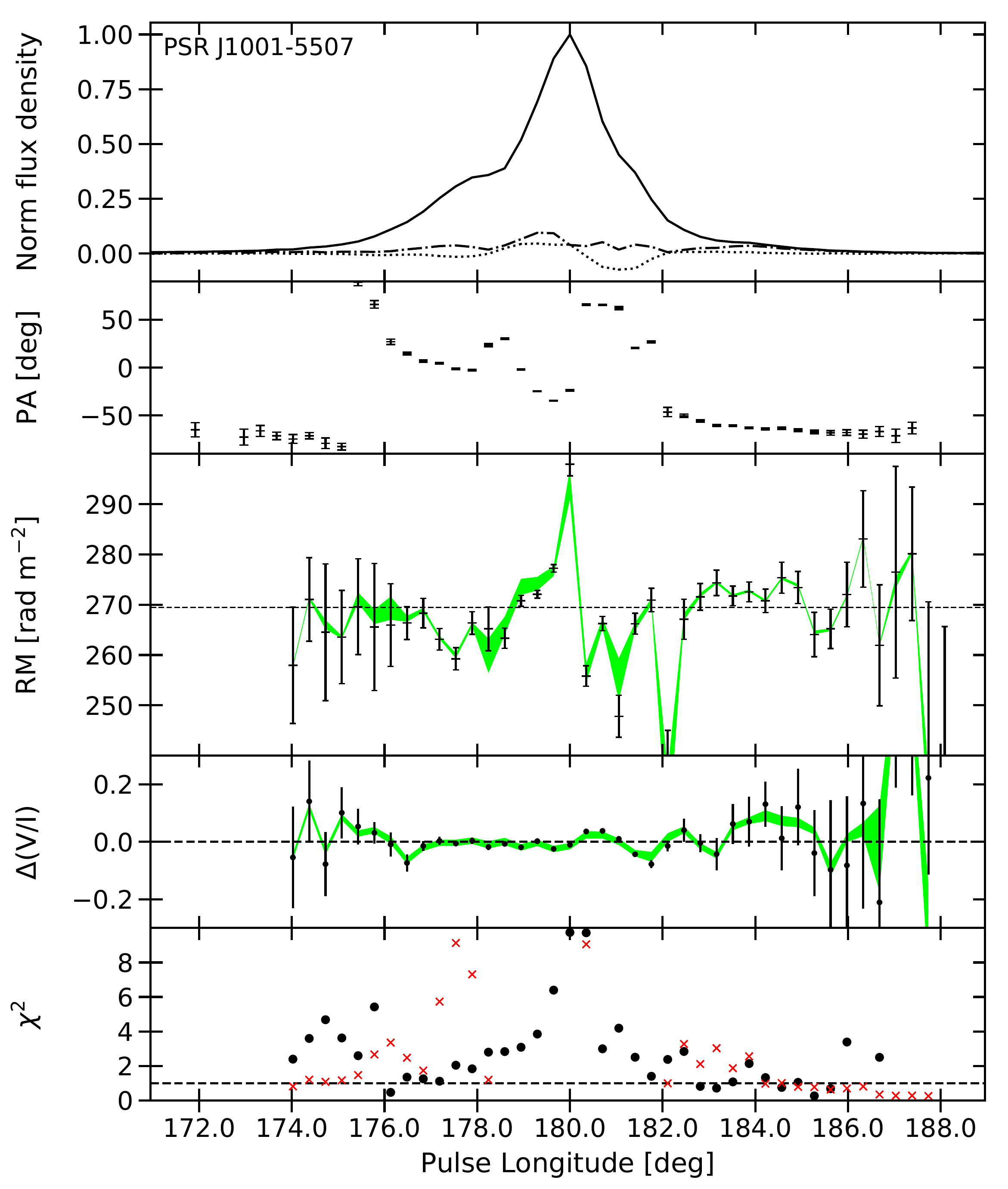} \\
\includegraphics[width=0.45\textwidth,angle=0]{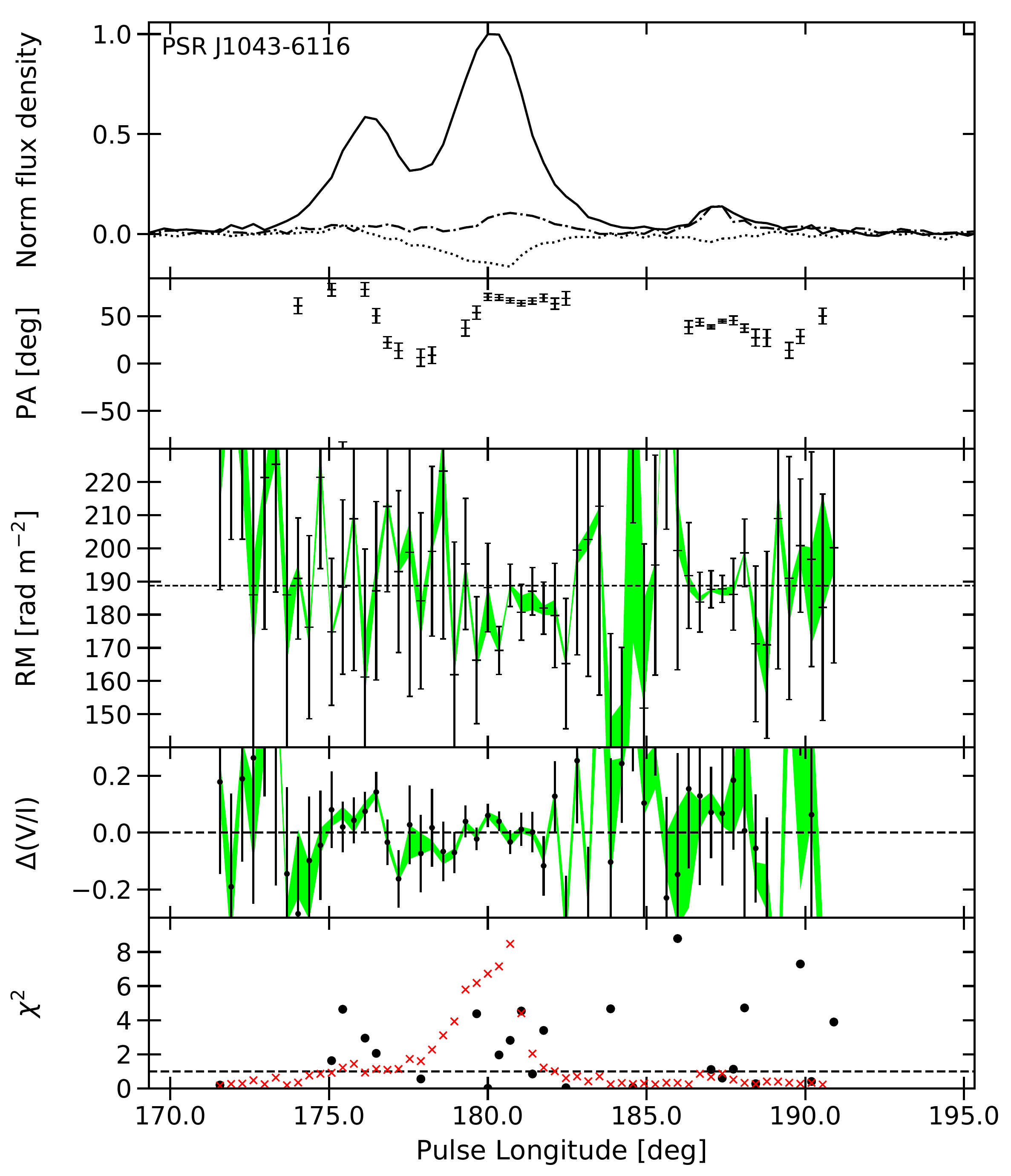} &
\includegraphics[width=0.45\textwidth,angle=0]{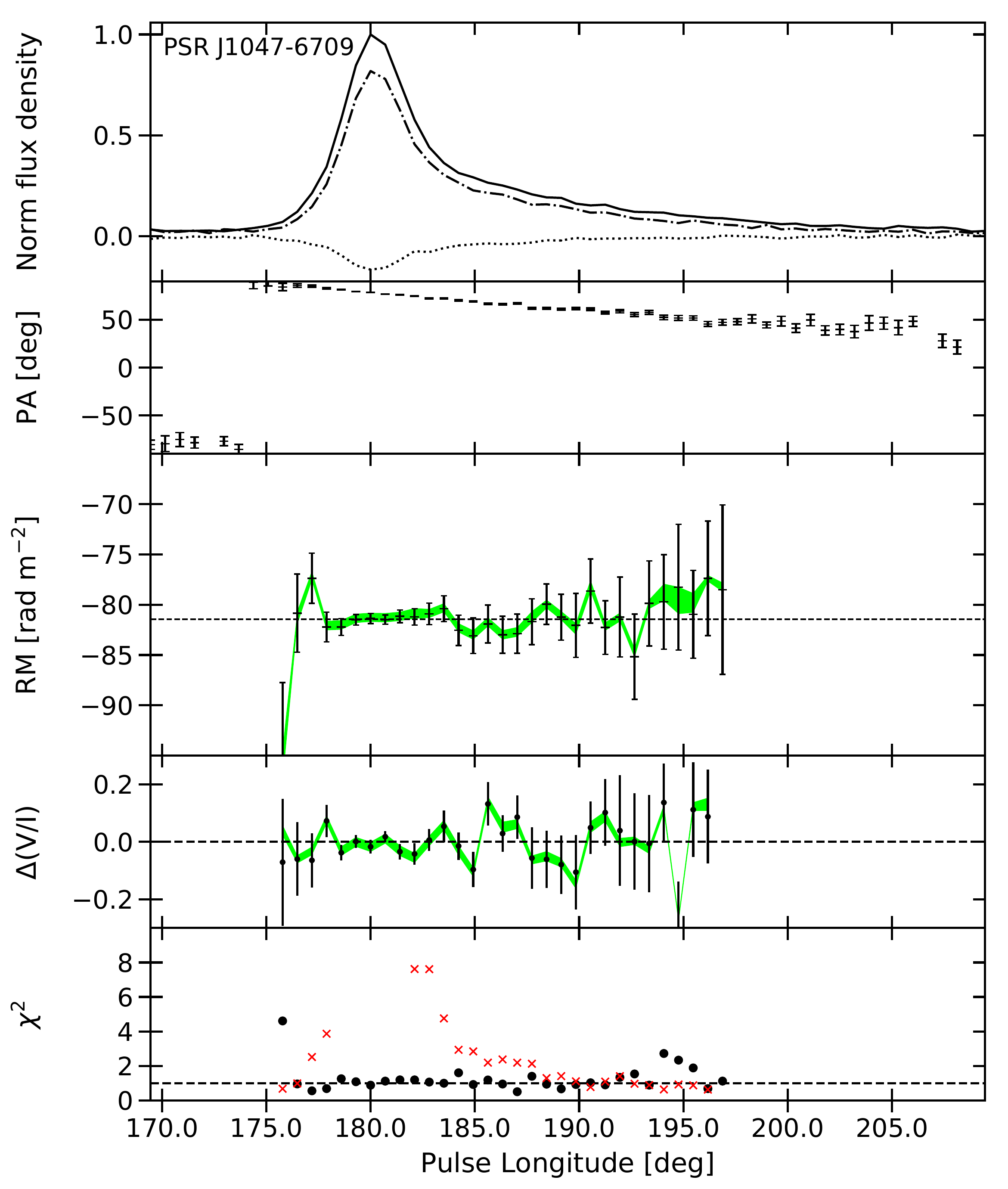} \\
\end{tabular}
\end{center}
\caption{Phase-resolved RM plots for PSR J0942$-$5552, PSR J1001$-$5507, PSR J1043$-$6116 and PSR J1047$-$6709. For more details on what is displayed in the individual panels, see Fig. 1.}
\label{fig:allPSR6}
\end{figure}

\begin{figure}
\begin{center}
\begin{tabular}{cc}
\includegraphics[width=0.45\textwidth,angle=0]{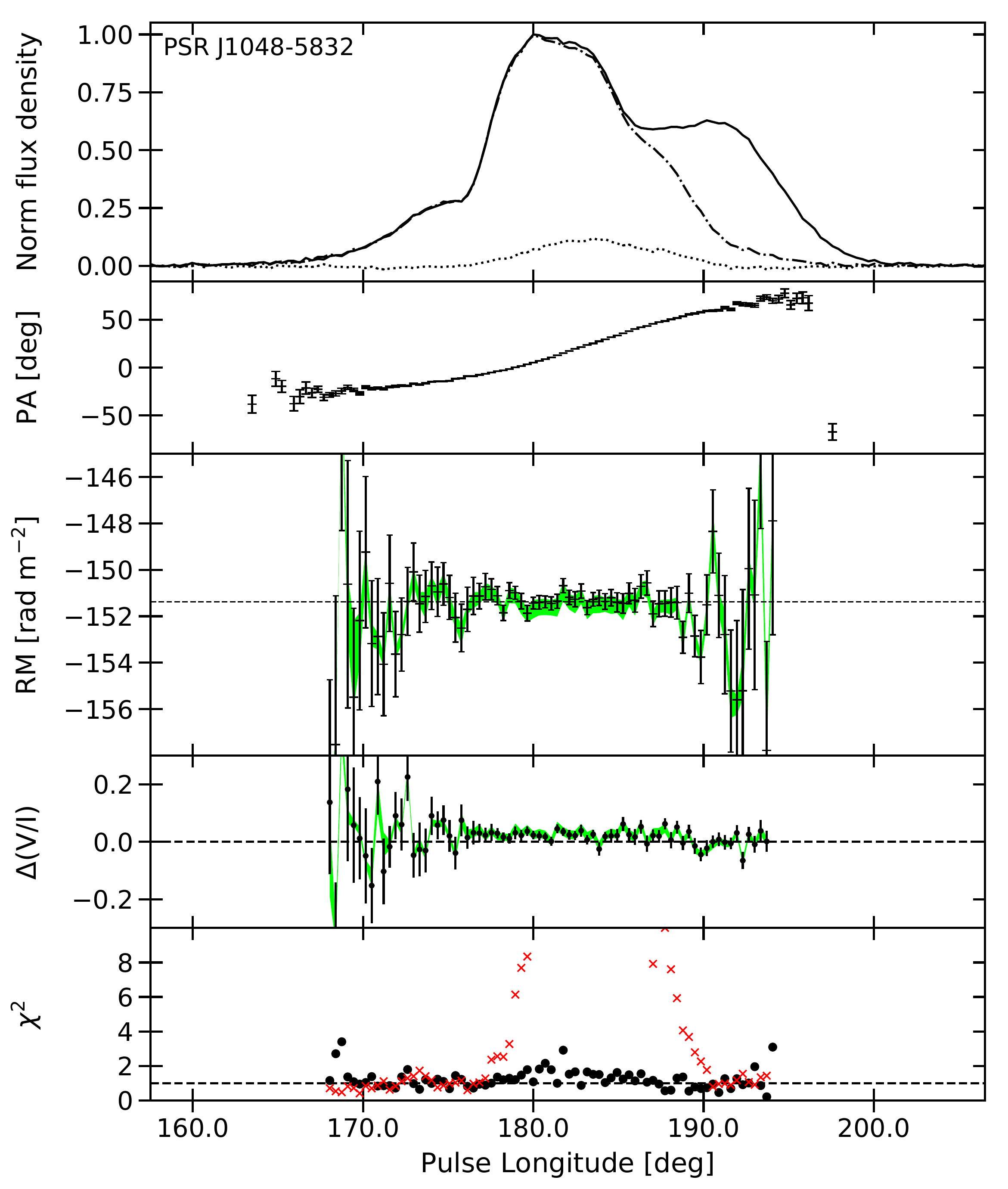} &
\includegraphics[width=0.45\textwidth,angle=0]{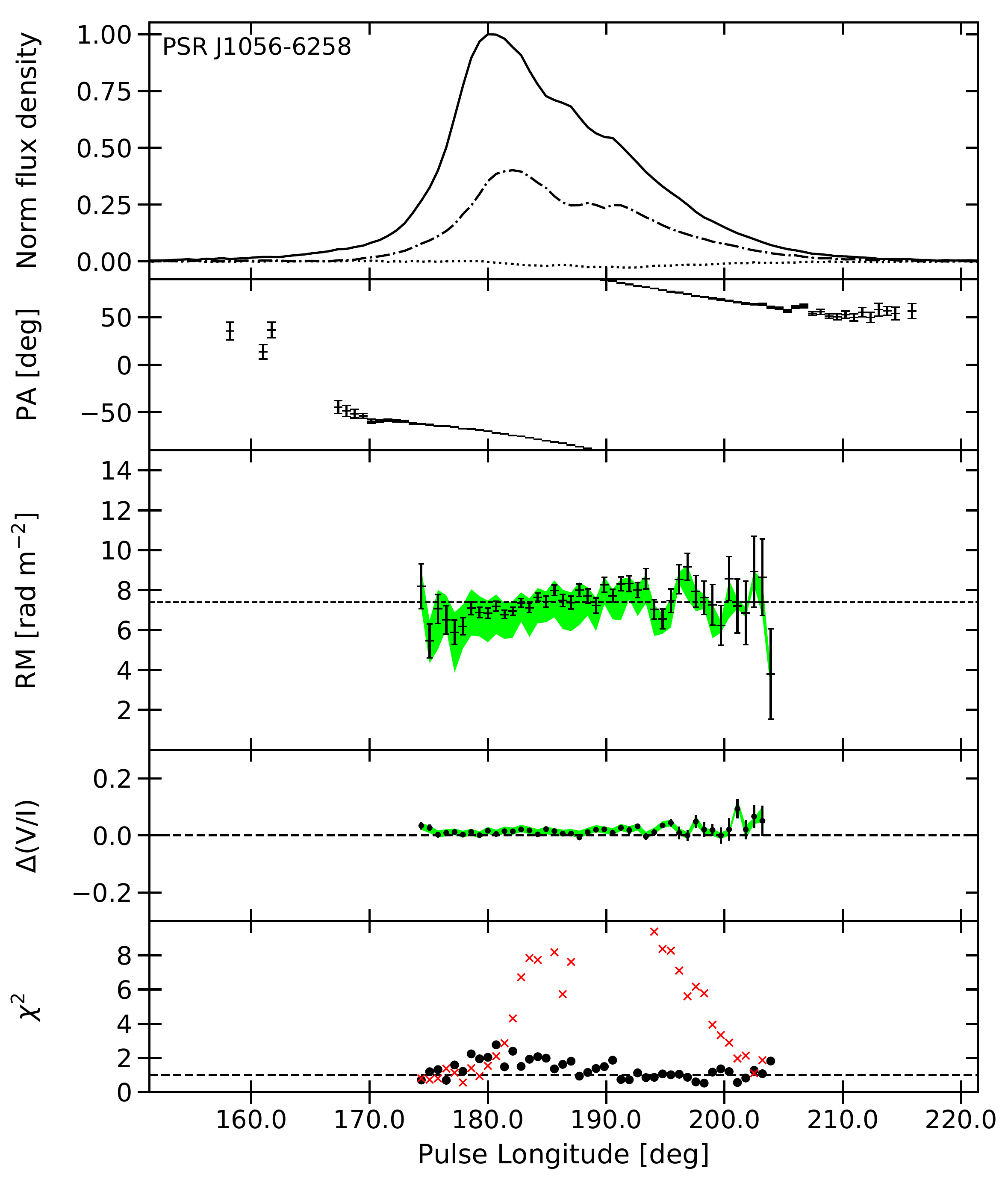} \\
\includegraphics[width=0.45\textwidth,angle=0]{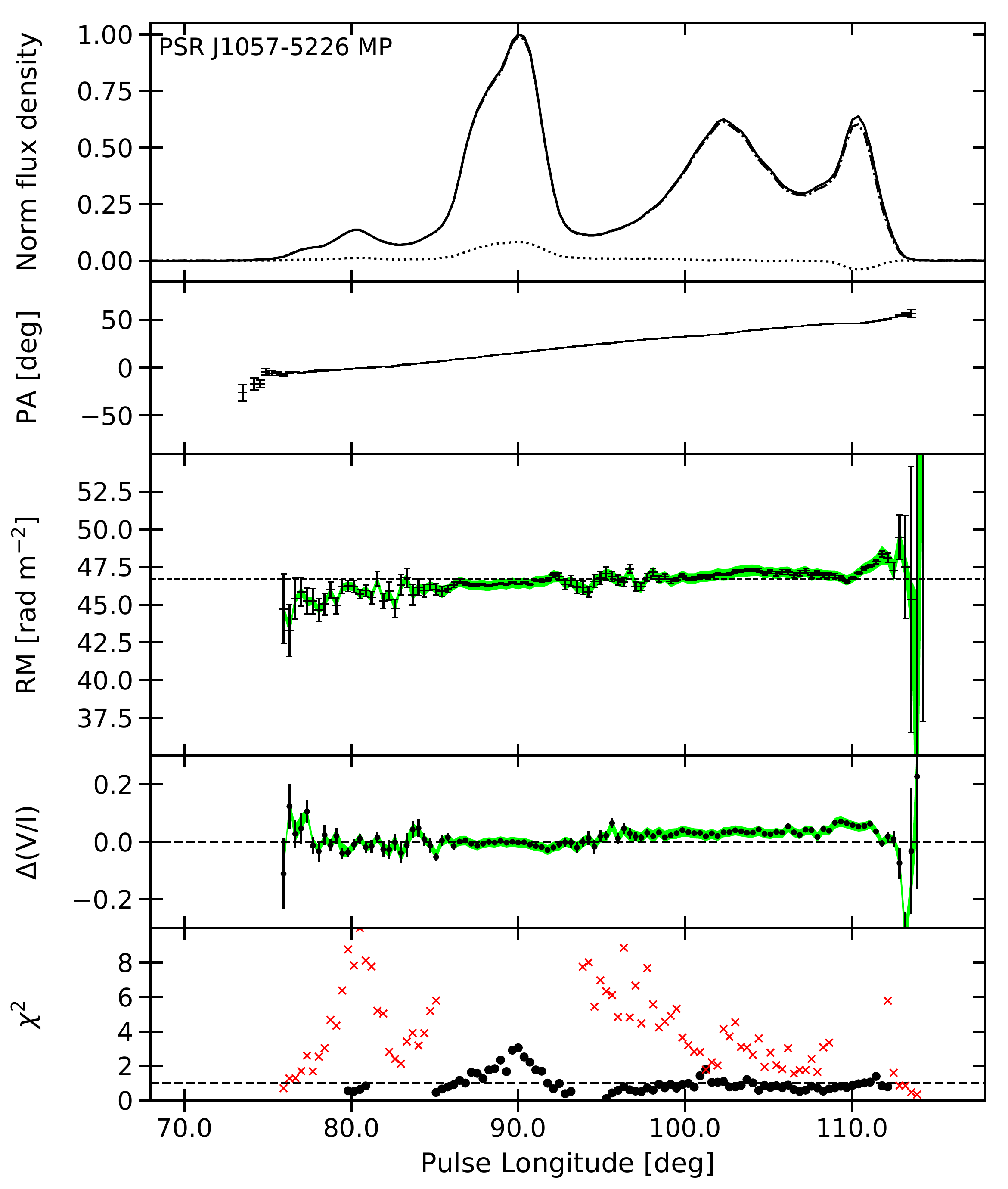} &
\includegraphics[width=0.45\textwidth,angle=0]{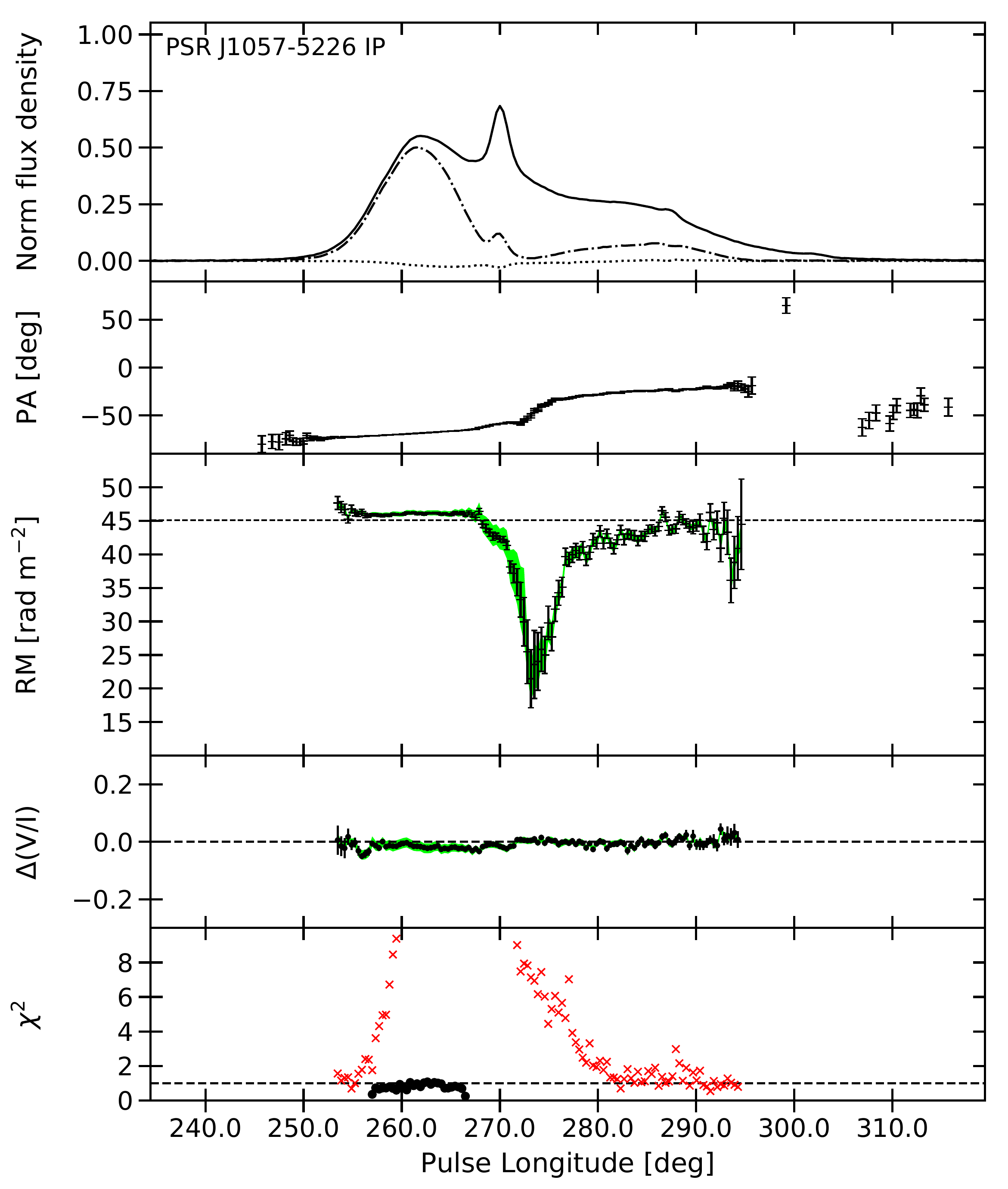} \\
\end{tabular}
\end{center}
\caption{Phase-resolved RM plots for PSR J1048$-$5832, PSR J1056$-$6258, PSR J1057$-$5226 (MP) and PSR J1057$-$5226 (IP). For more details on what is displayed in the individual panels, see Fig. 1.}
\label{fig:allPSR7}
\end{figure}

\begin{figure}
\begin{center}
\begin{tabular}{cc}
\includegraphics[width=0.45\textwidth,angle=0]{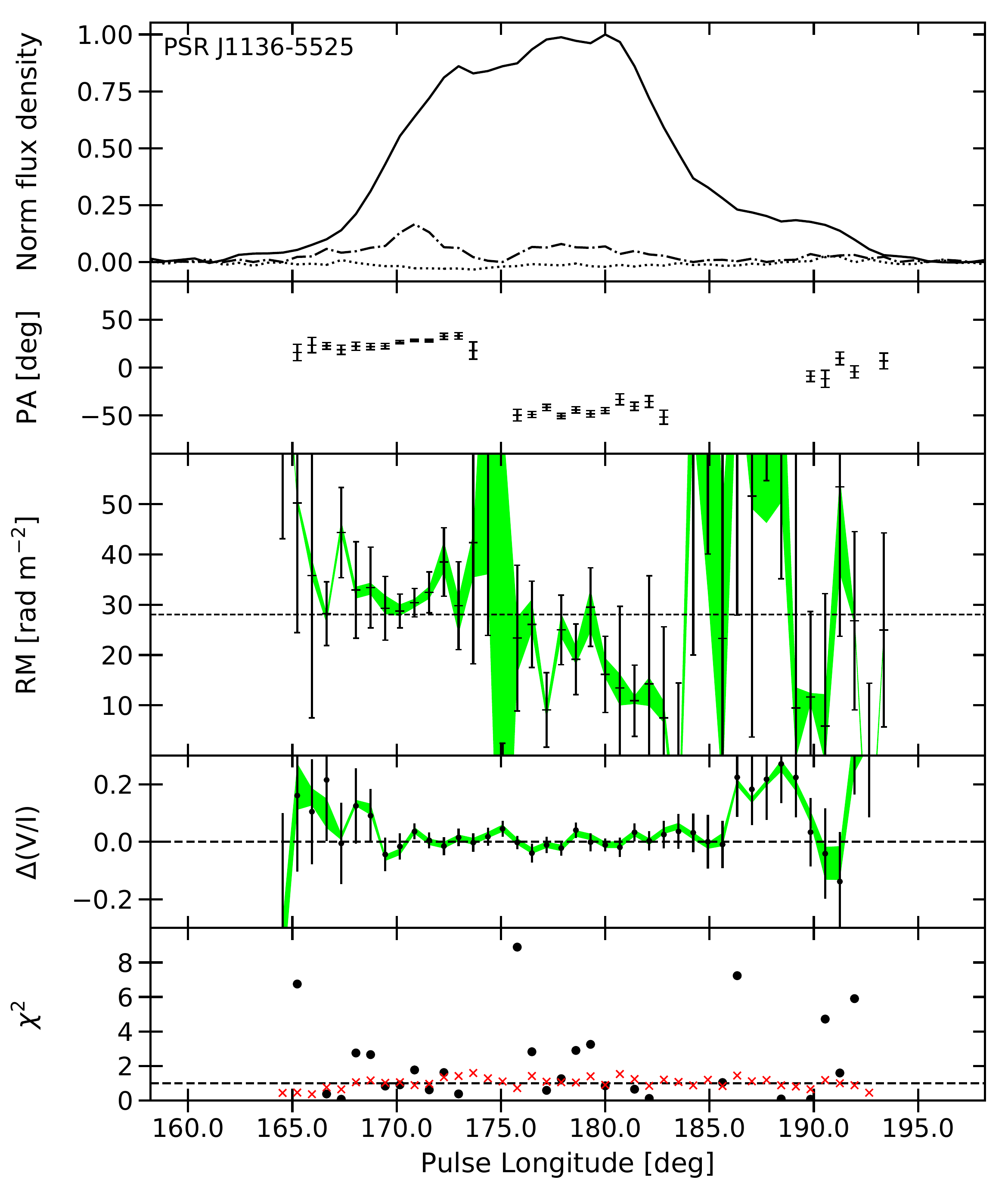} &
\includegraphics[width=0.45\textwidth,angle=0]{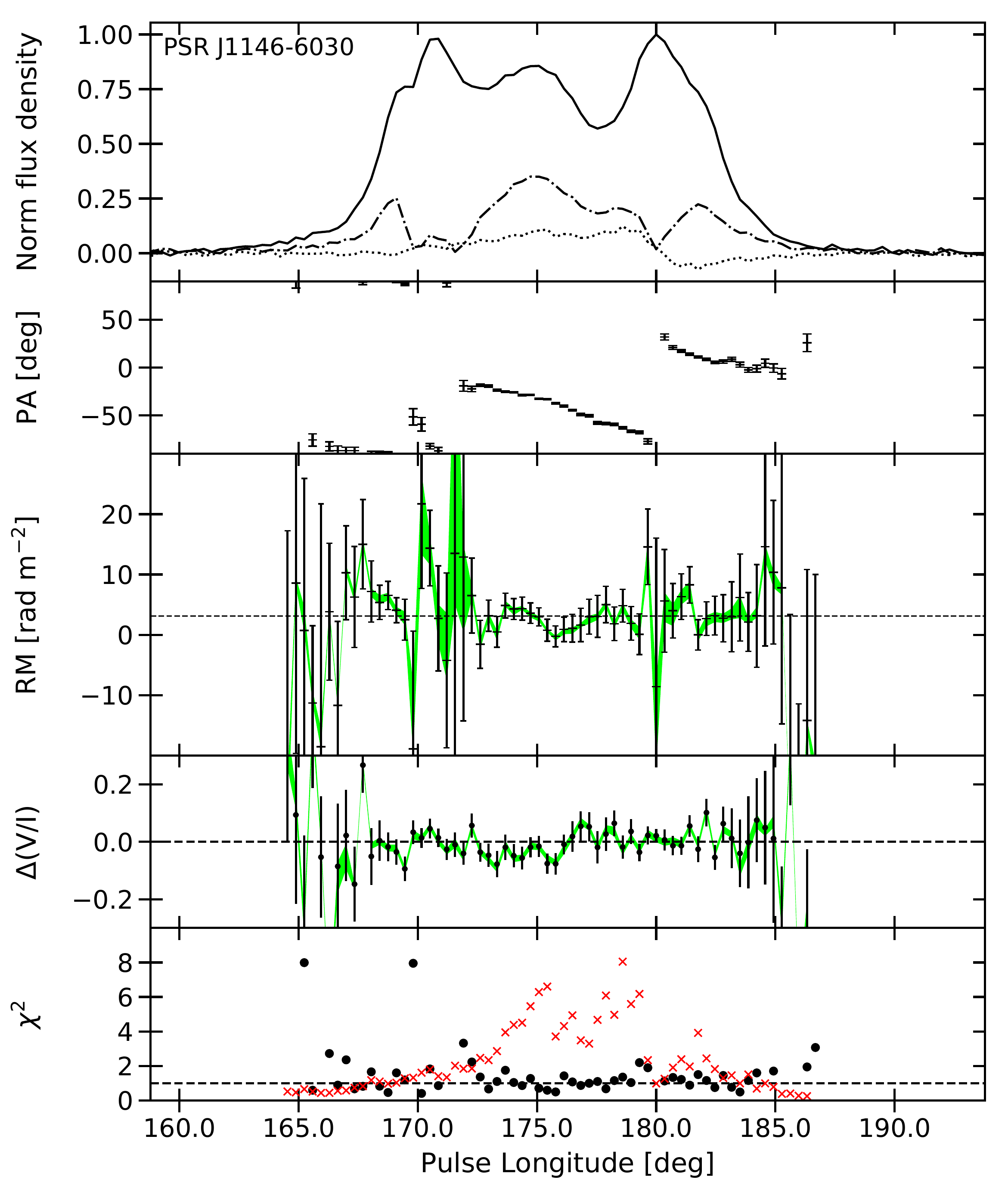} \\
\includegraphics[width=0.45\textwidth,angle=0]{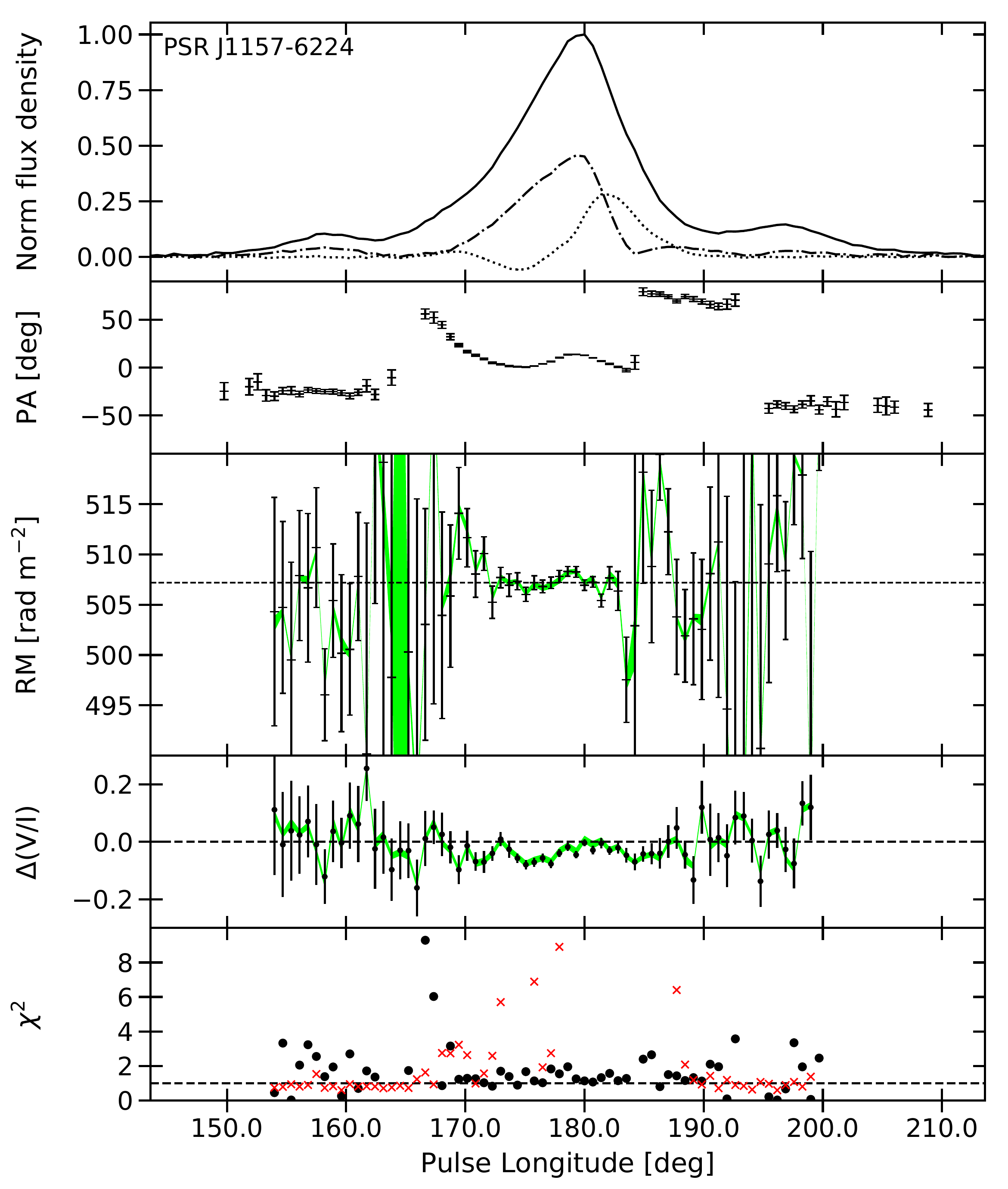} &
\includegraphics[width=0.45\textwidth,angle=0]{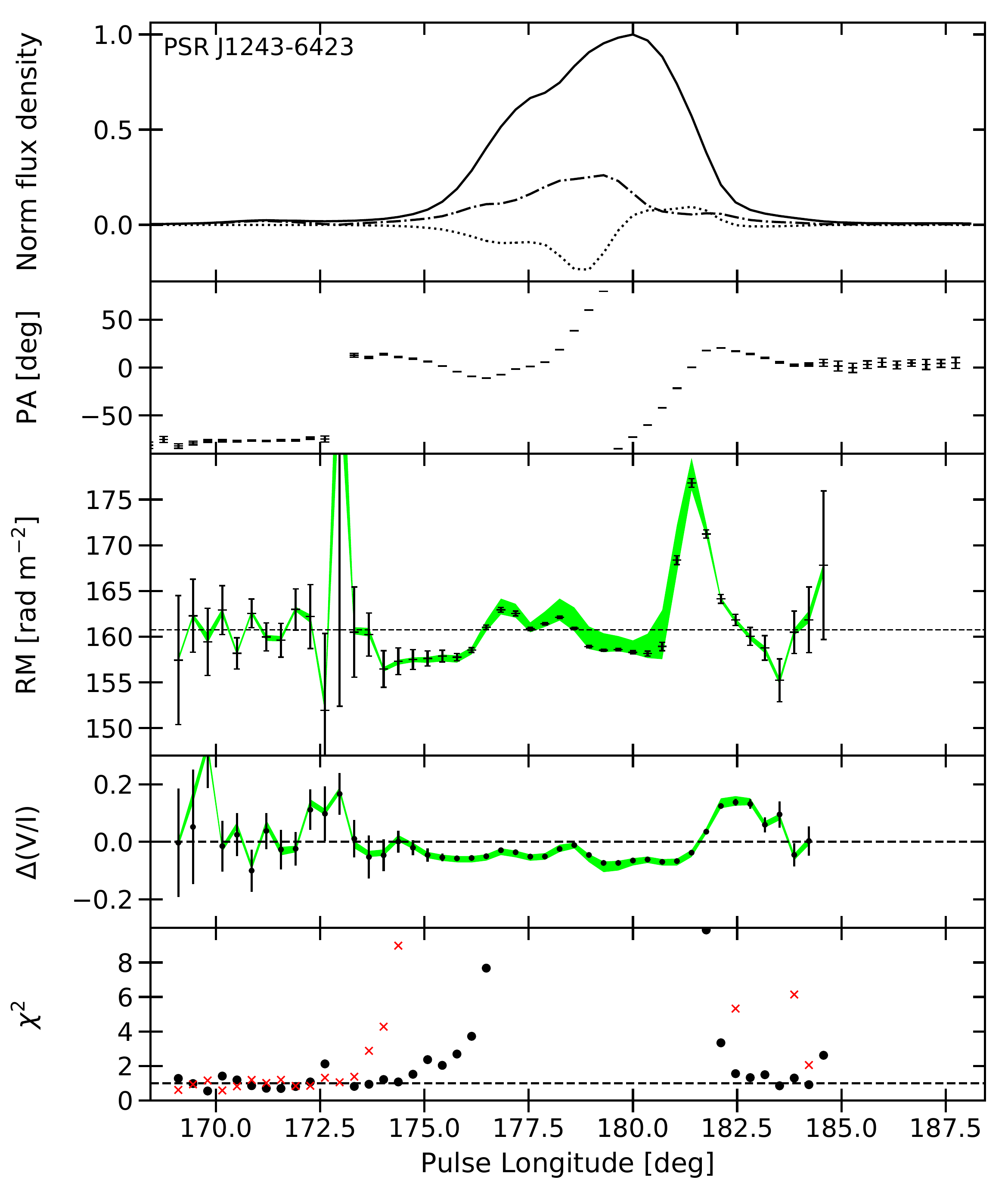} \\
\end{tabular}
\end{center}
\caption{Phase-resolved RM plots for PSR J1136$-$5525, PSR J1146$-$6030, PSR J1157$-$6224 and PSR J1243$-$6423. For more details on what is displayed in the individual panels, see Fig. 1.}
\label{fig:allPSR8}
\end{figure}

\begin{figure}
\begin{center}
\begin{tabular}{cc}
\includegraphics[width=0.45\textwidth,angle=0]{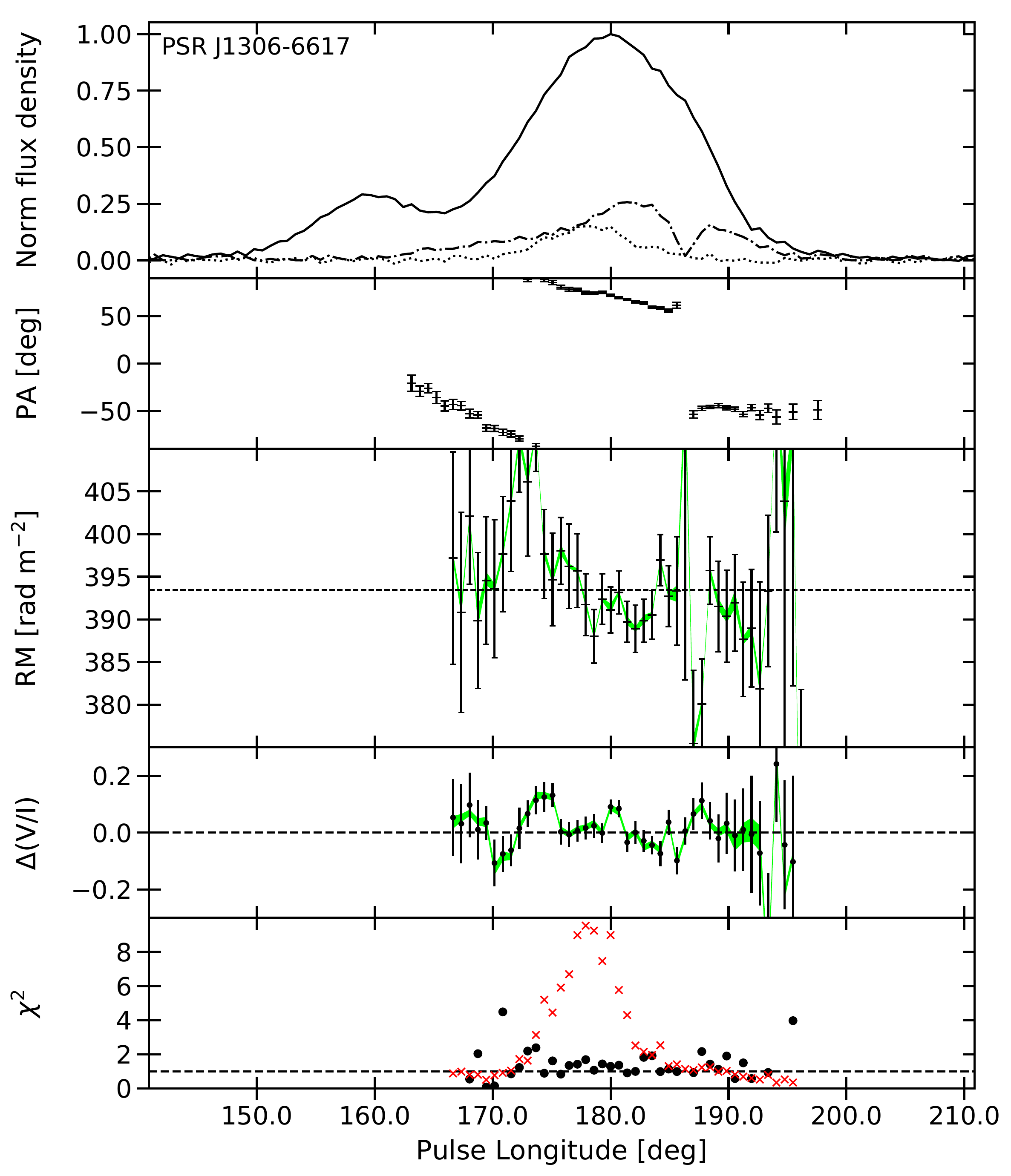} &
\includegraphics[width=0.45\textwidth,angle=0]{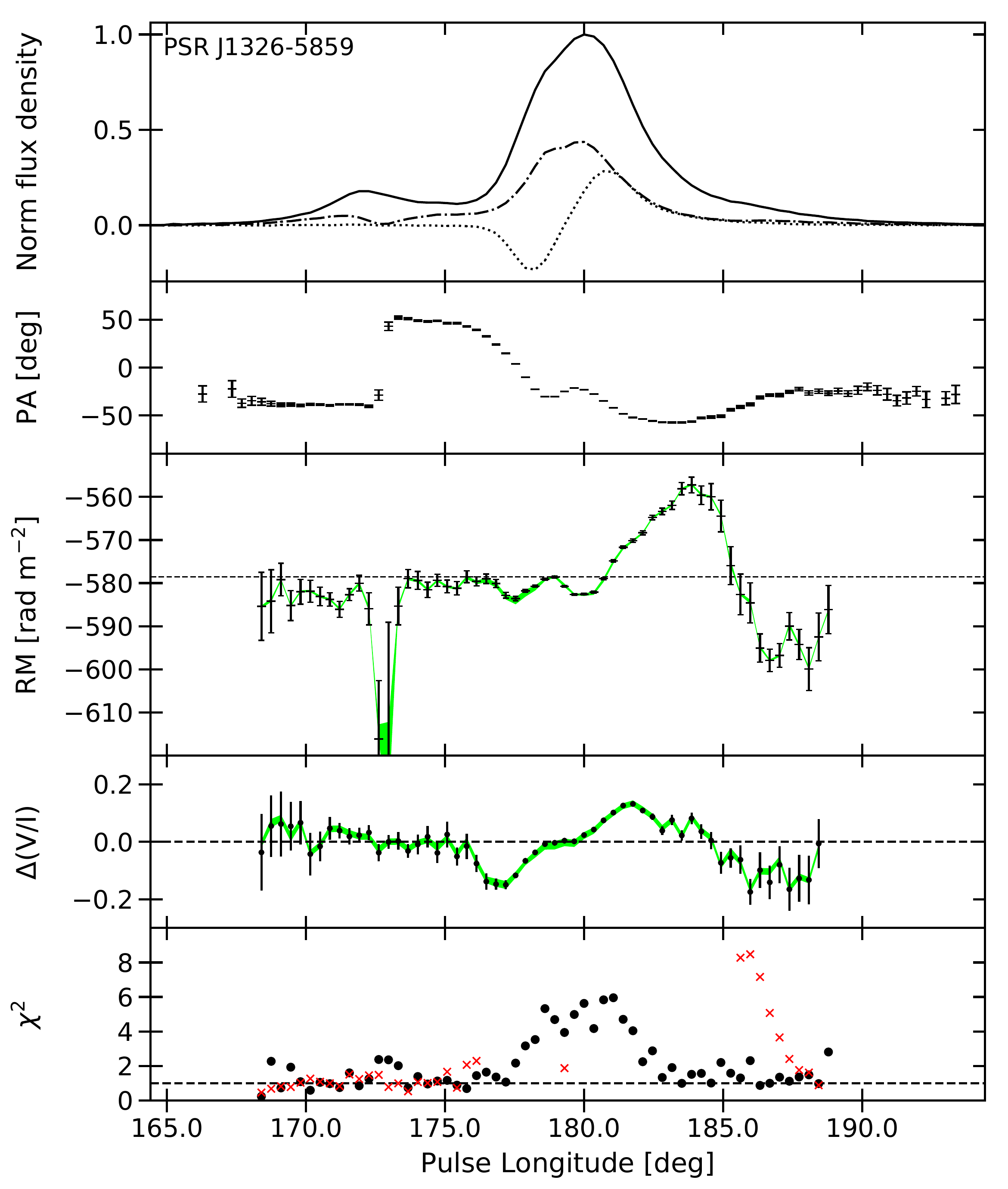} \\
\includegraphics[width=0.45\textwidth,angle=0]{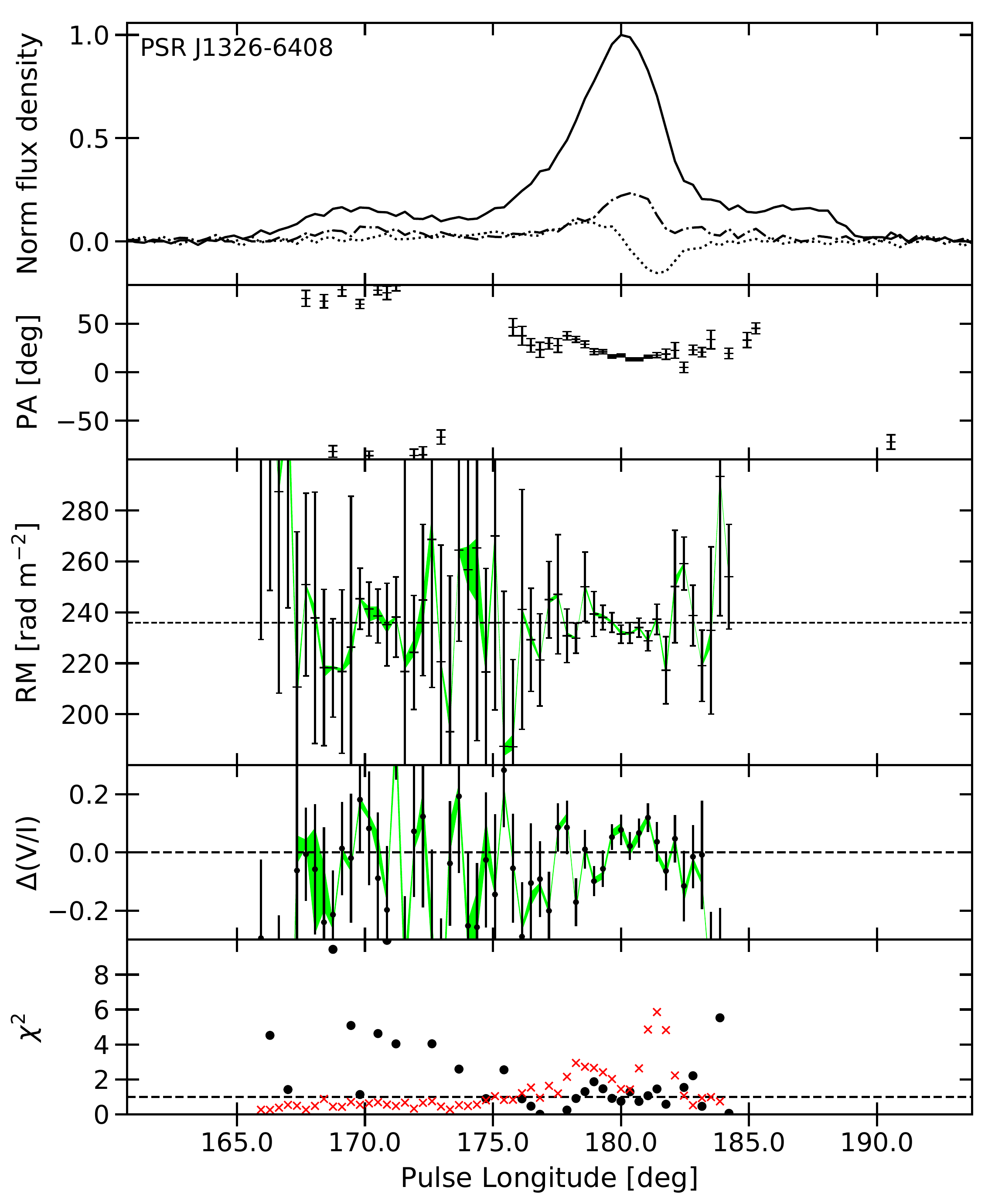} &
\includegraphics[width=0.45\textwidth,angle=0]{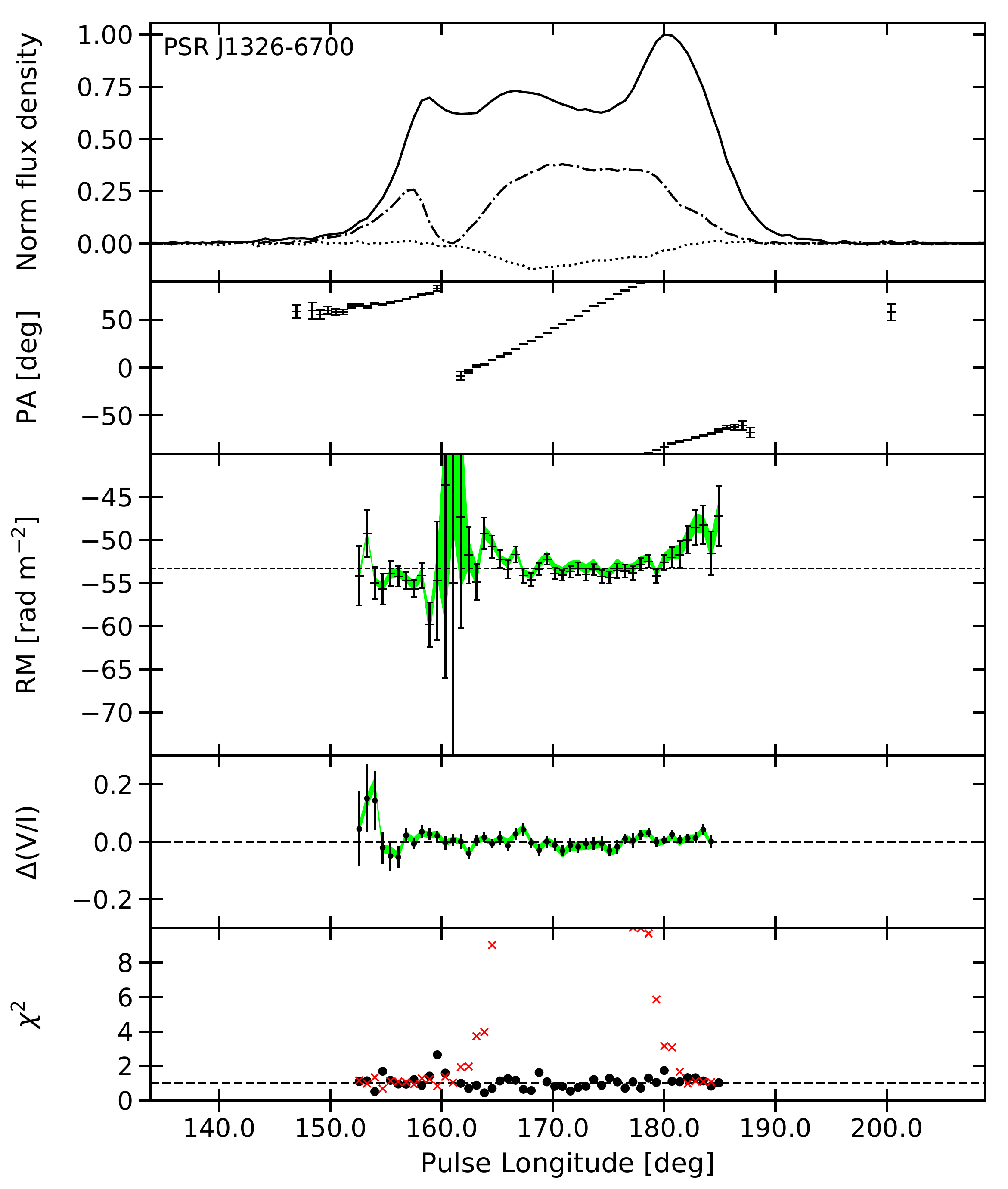} \\
\end{tabular}
\end{center}
\caption{Phase-resolved RM plots for PSR J1306$-$6617, PSR J1326$-$5859, PSR J1326$-$6408 and PSR J1326$-$6700. For more details on what is displayed in the individual panels, see Fig. 1.}
\label{fig:allPSR9}
\end{figure}

\begin{figure}
\begin{center}
\begin{tabular}{cc}
\includegraphics[width=0.45\textwidth,angle=0]{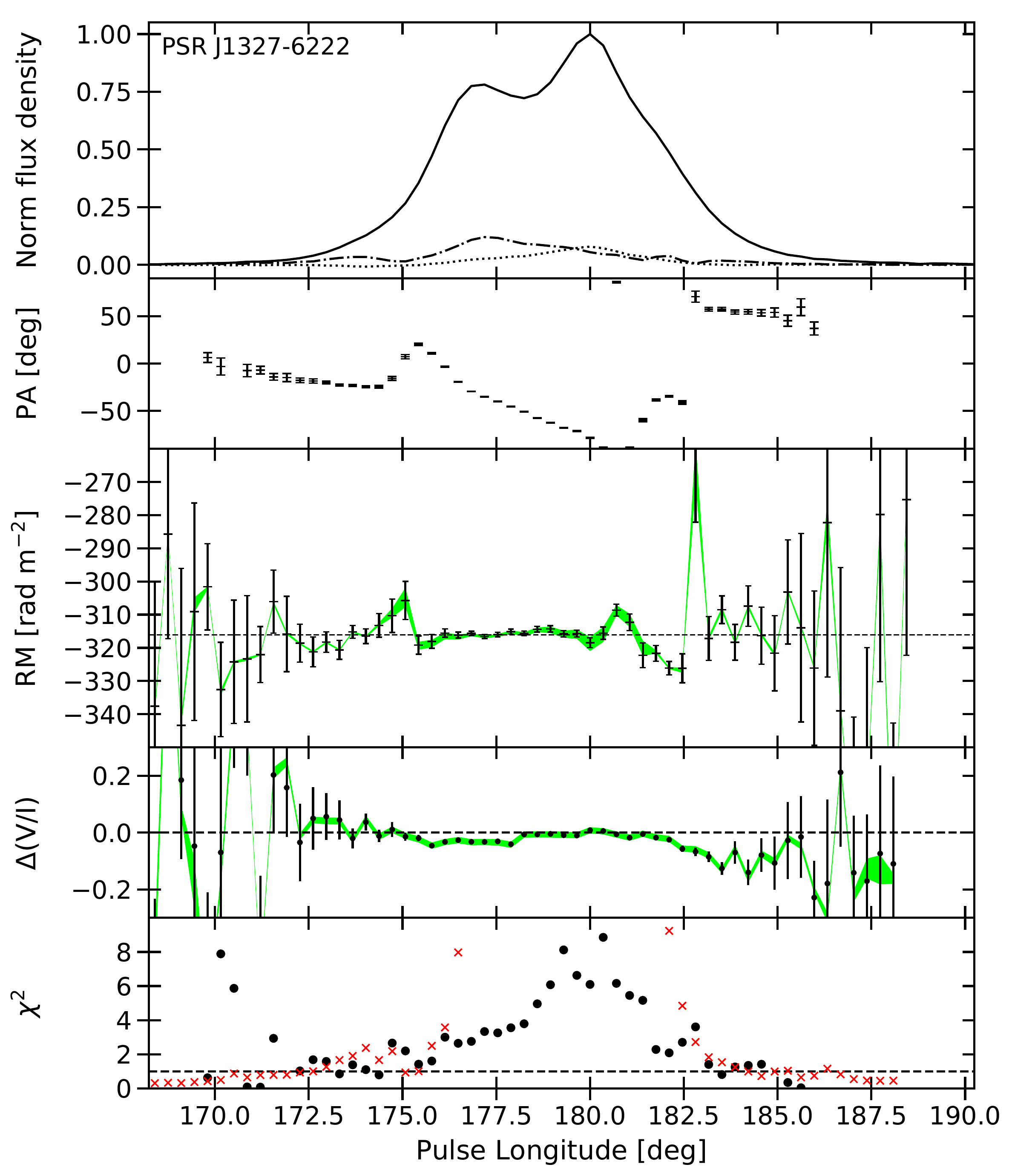} &
\includegraphics[width=0.45\textwidth,angle=0]{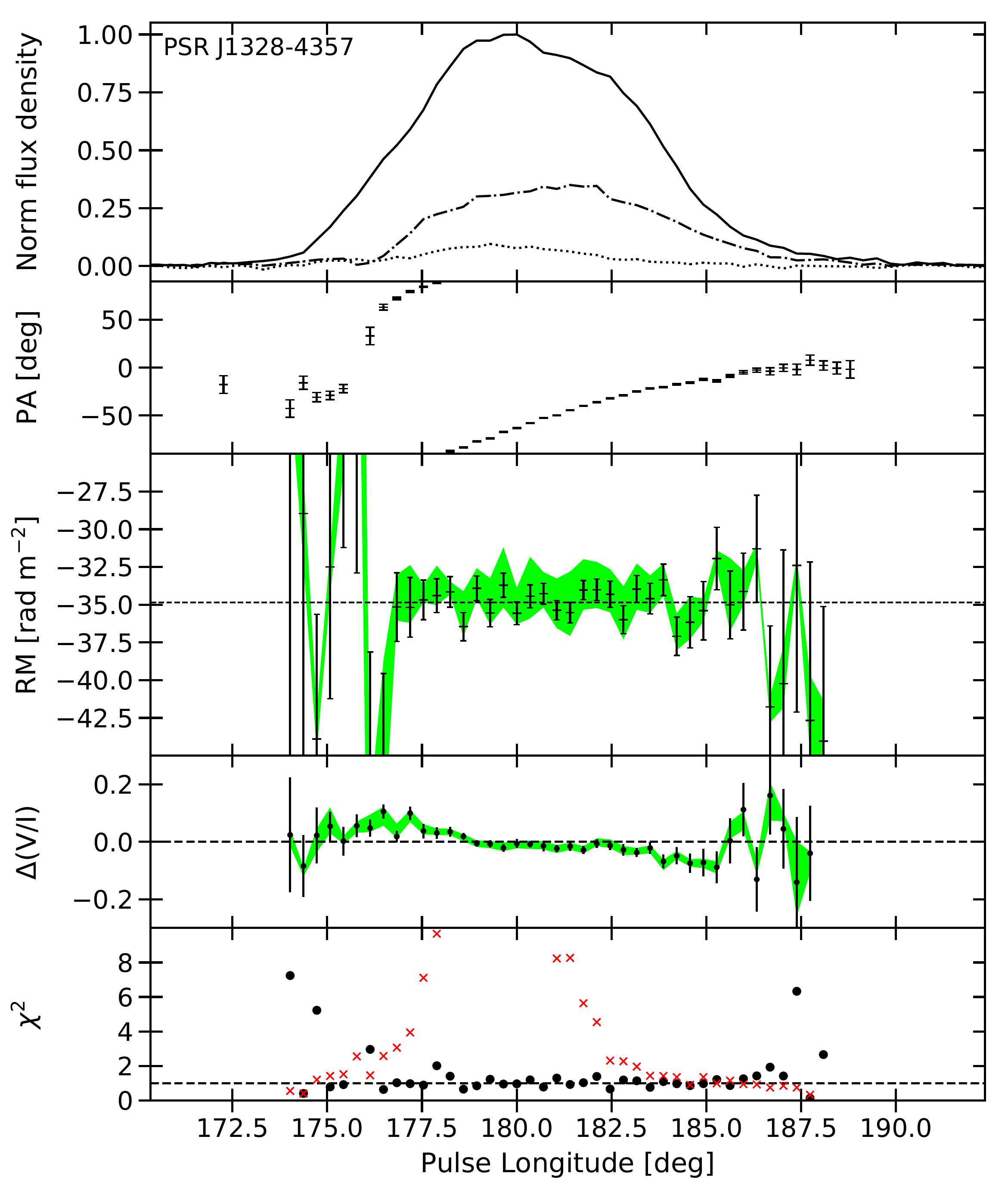} \\
\includegraphics[width=0.45\textwidth,angle=0]{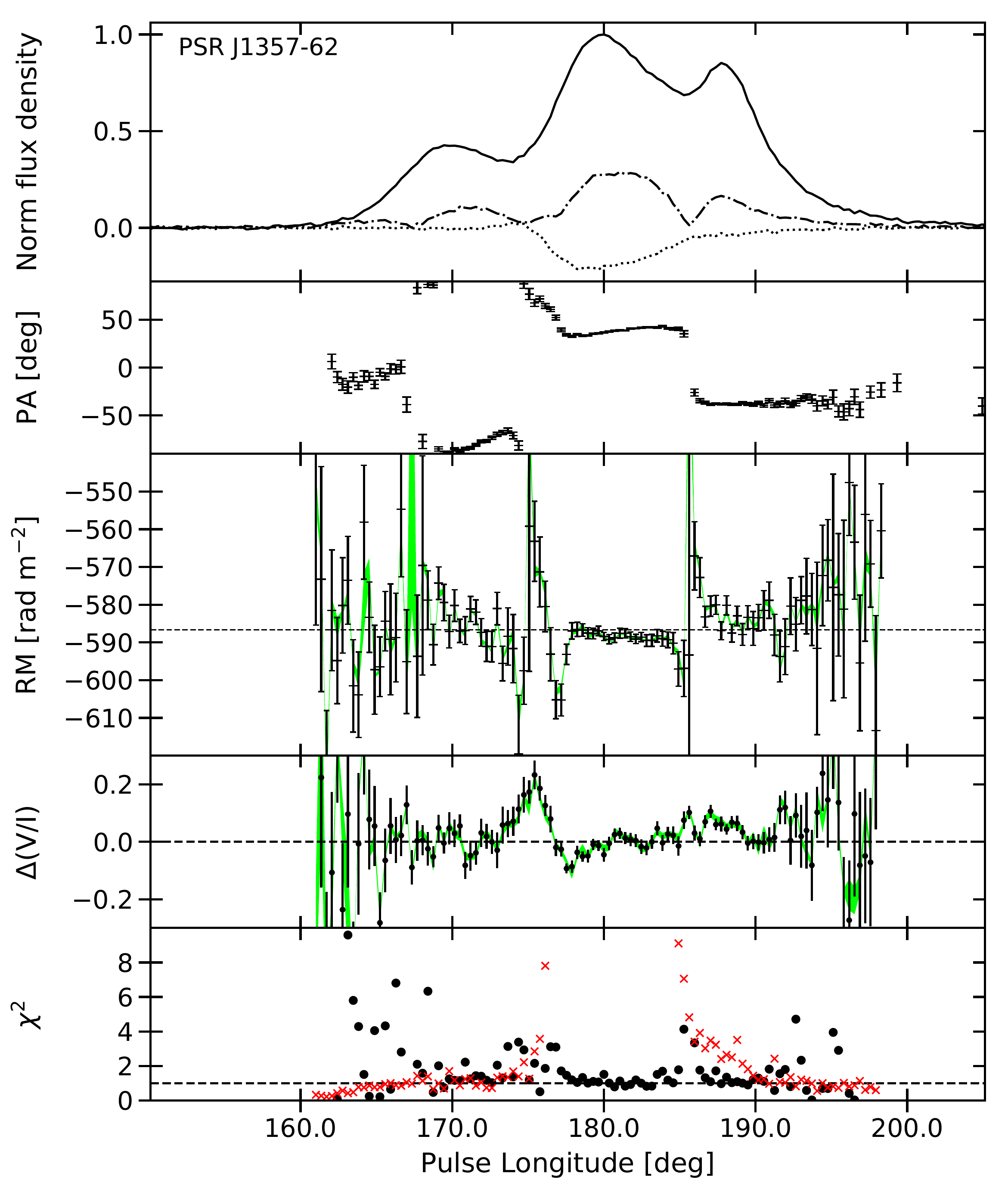} &
\includegraphics[width=0.45\textwidth,angle=0]{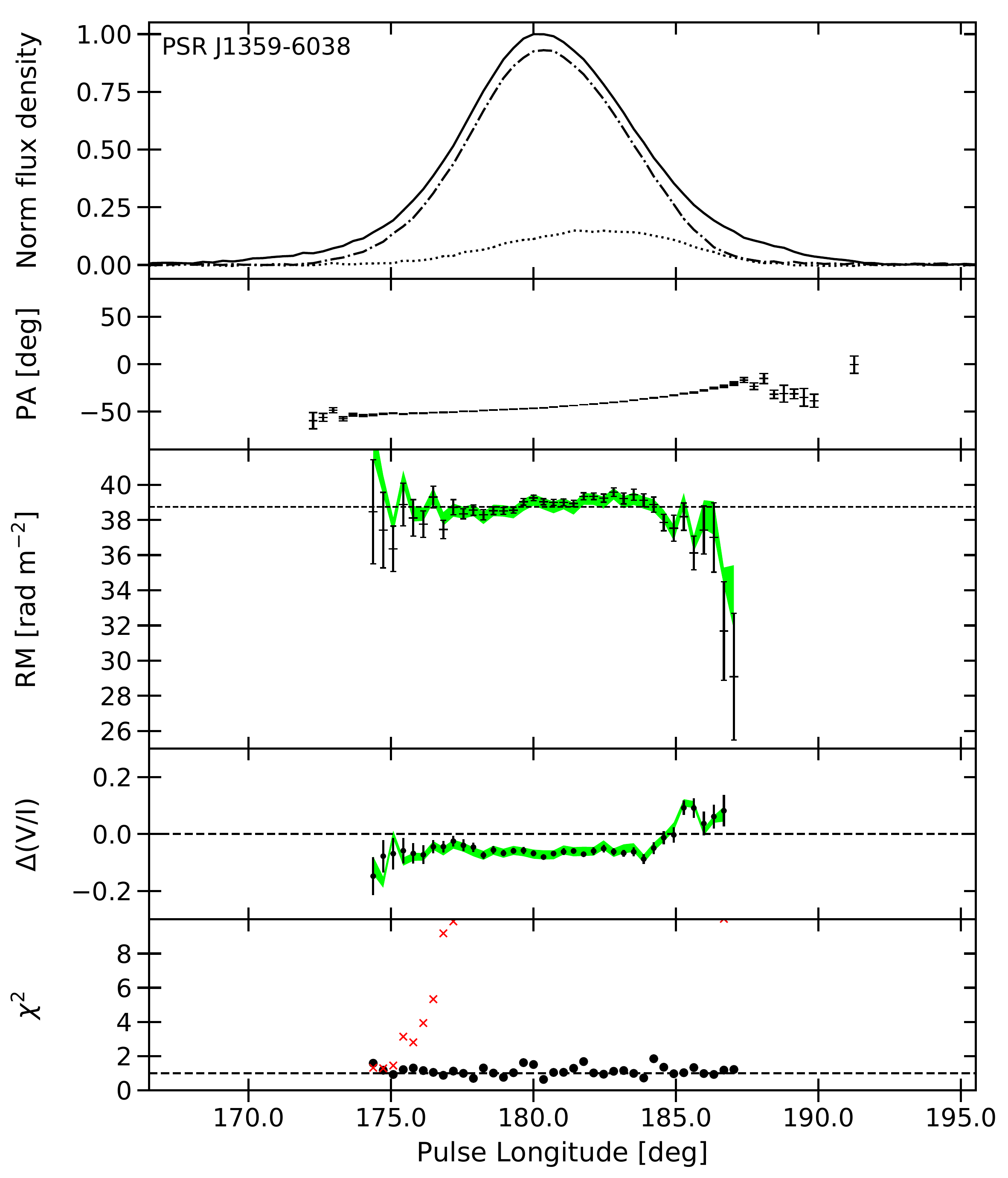} \\
\end{tabular}
\end{center}
\caption{Phase-resolved RM plots for PSR J1327$-$6222, PSR J1328$-$4357, PSR J1357$-$62 and PSR J1359$-$6038. For more details on what is displayed in the individual panels, see Fig. 1.}
\label{fig:allPSR10}
\end{figure}

\begin{figure}
\begin{center}
\begin{tabular}{cc}
\includegraphics[width=0.45\textwidth,angle=0]{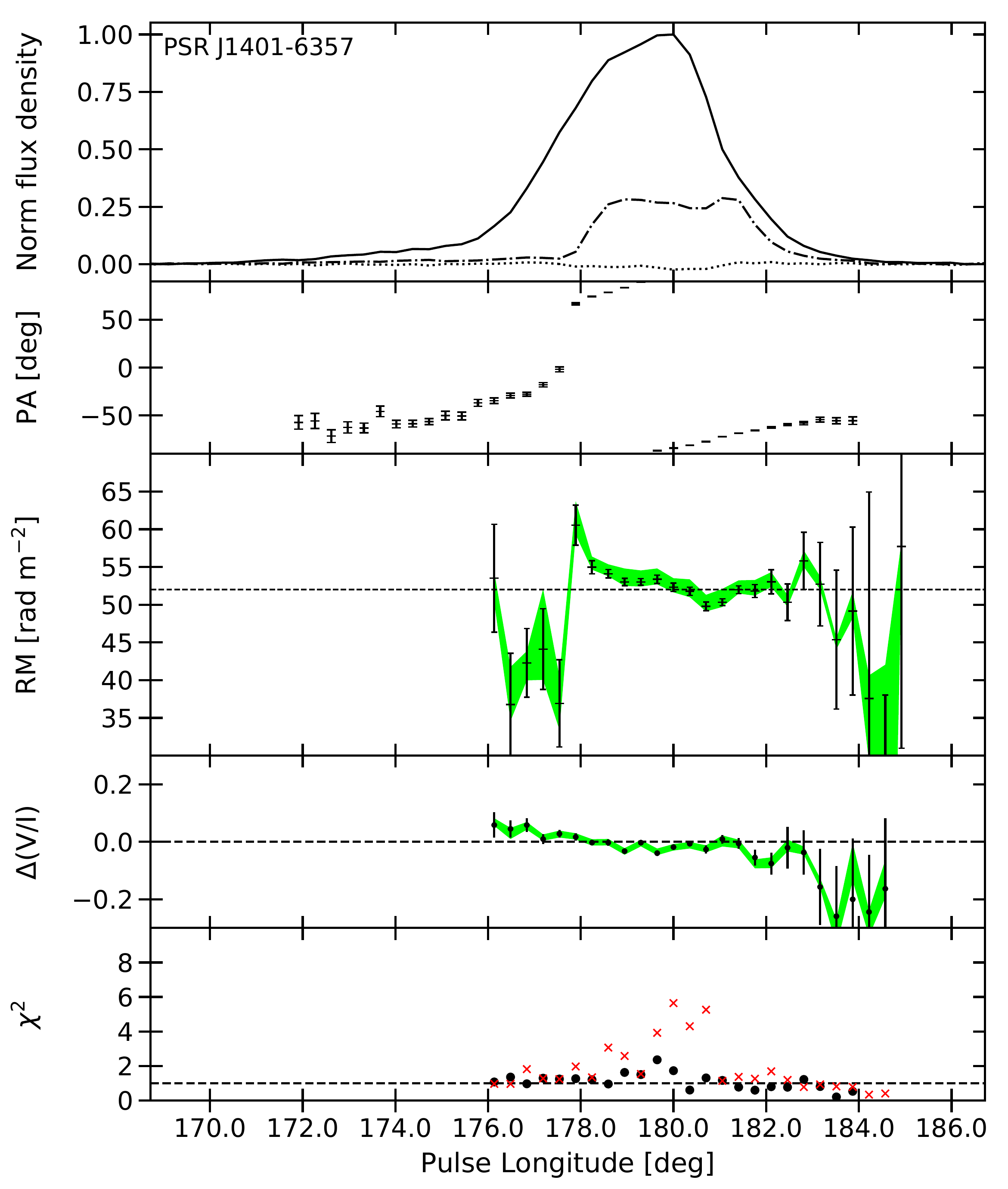} &
\includegraphics[width=0.45\textwidth,angle=0]{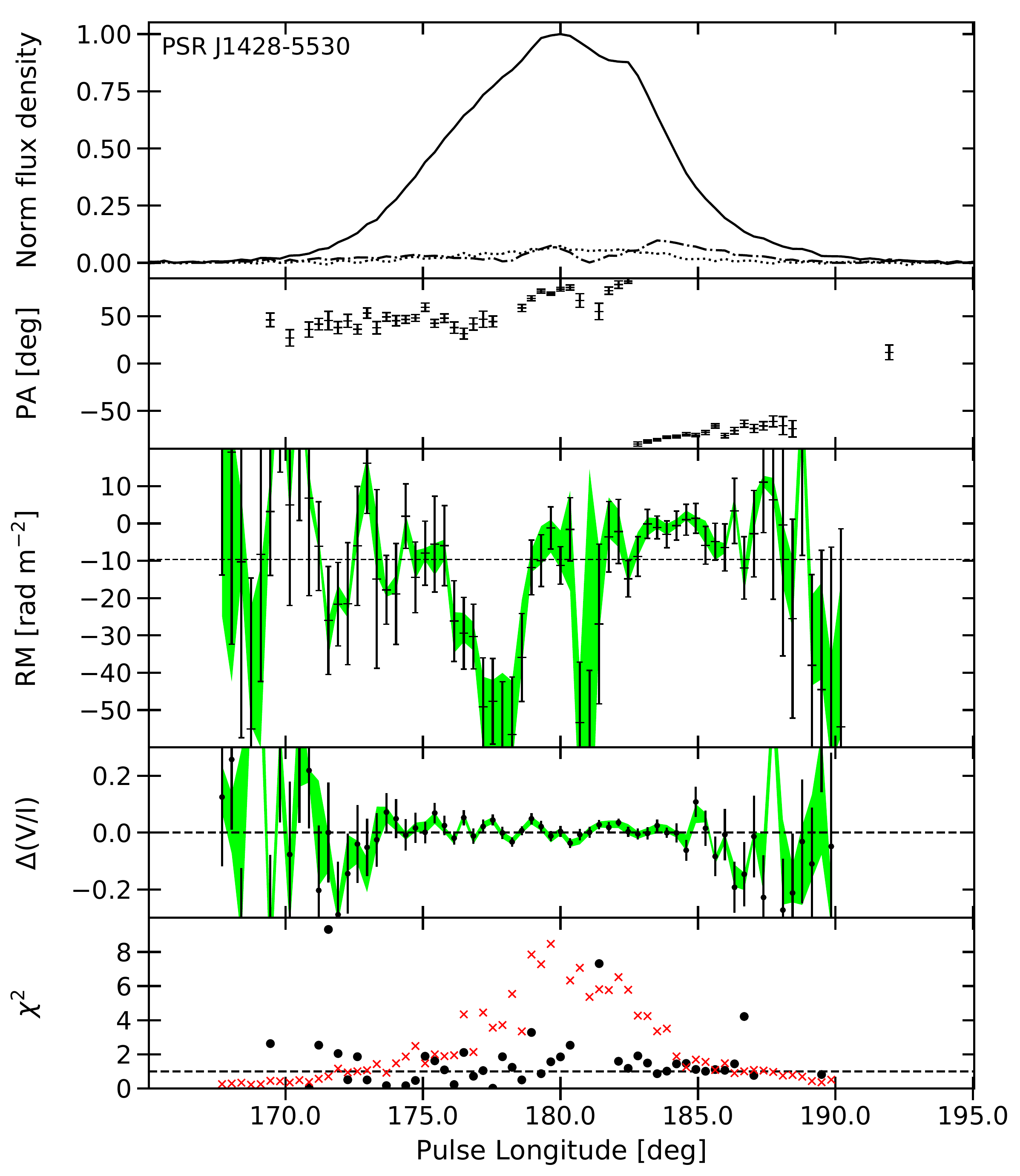} \\
\includegraphics[width=0.45\textwidth,angle=0]{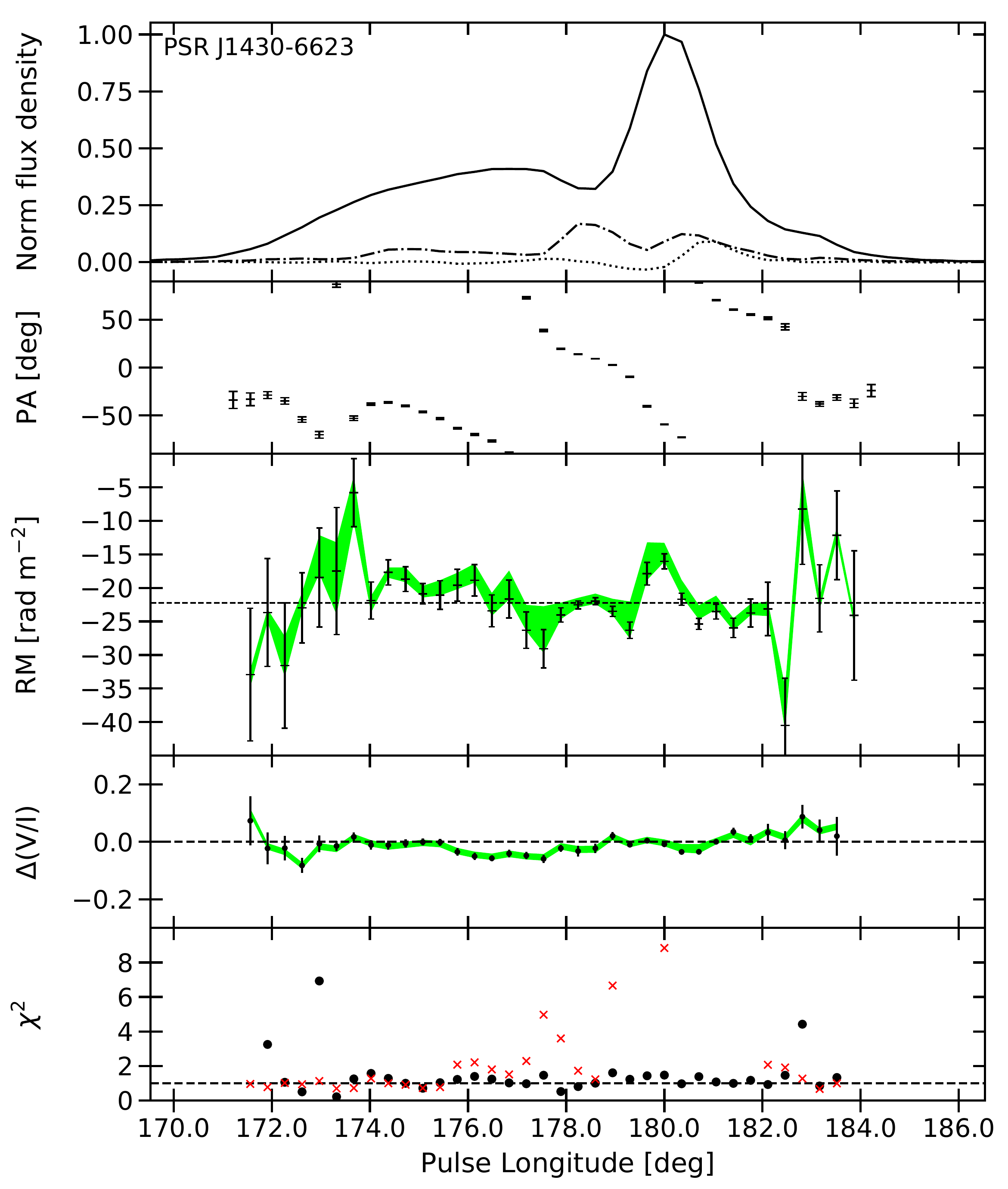} &
\includegraphics[width=0.45\textwidth,angle=0]{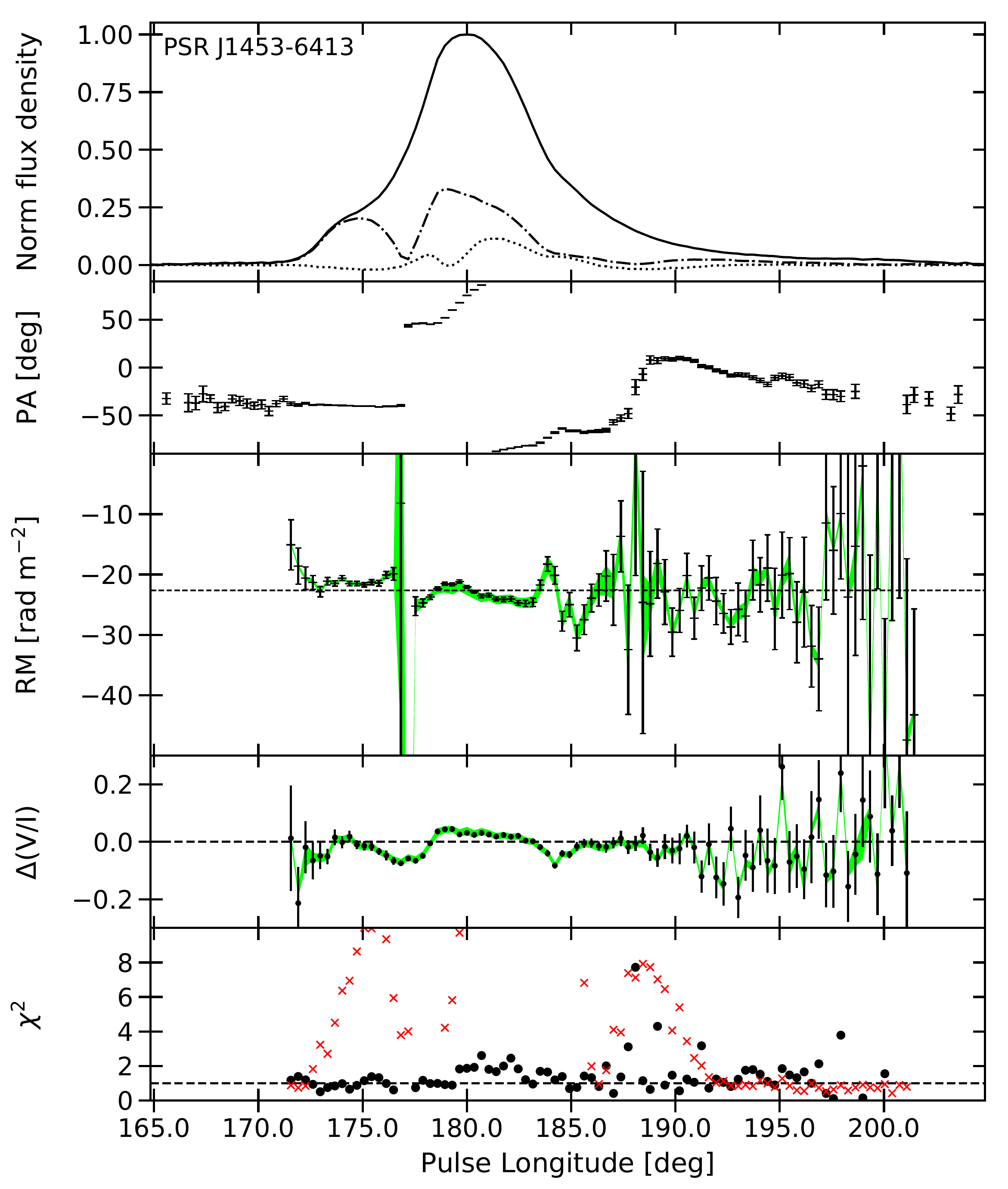} \\
\end{tabular}
\end{center}
\caption{Phase-resolved RM plots for PSR J1401$-$6357, PSR J1428$-$5530, PSR J1430$-$6623 and PSR J1453$-$6413. For more details on what is displayed in the individual panels, see Fig. 1.}
\label{fig:allPSR11}
\end{figure}

\begin{figure}
\begin{center}
\begin{tabular}{cc}
\includegraphics[width=0.45\textwidth,angle=0]{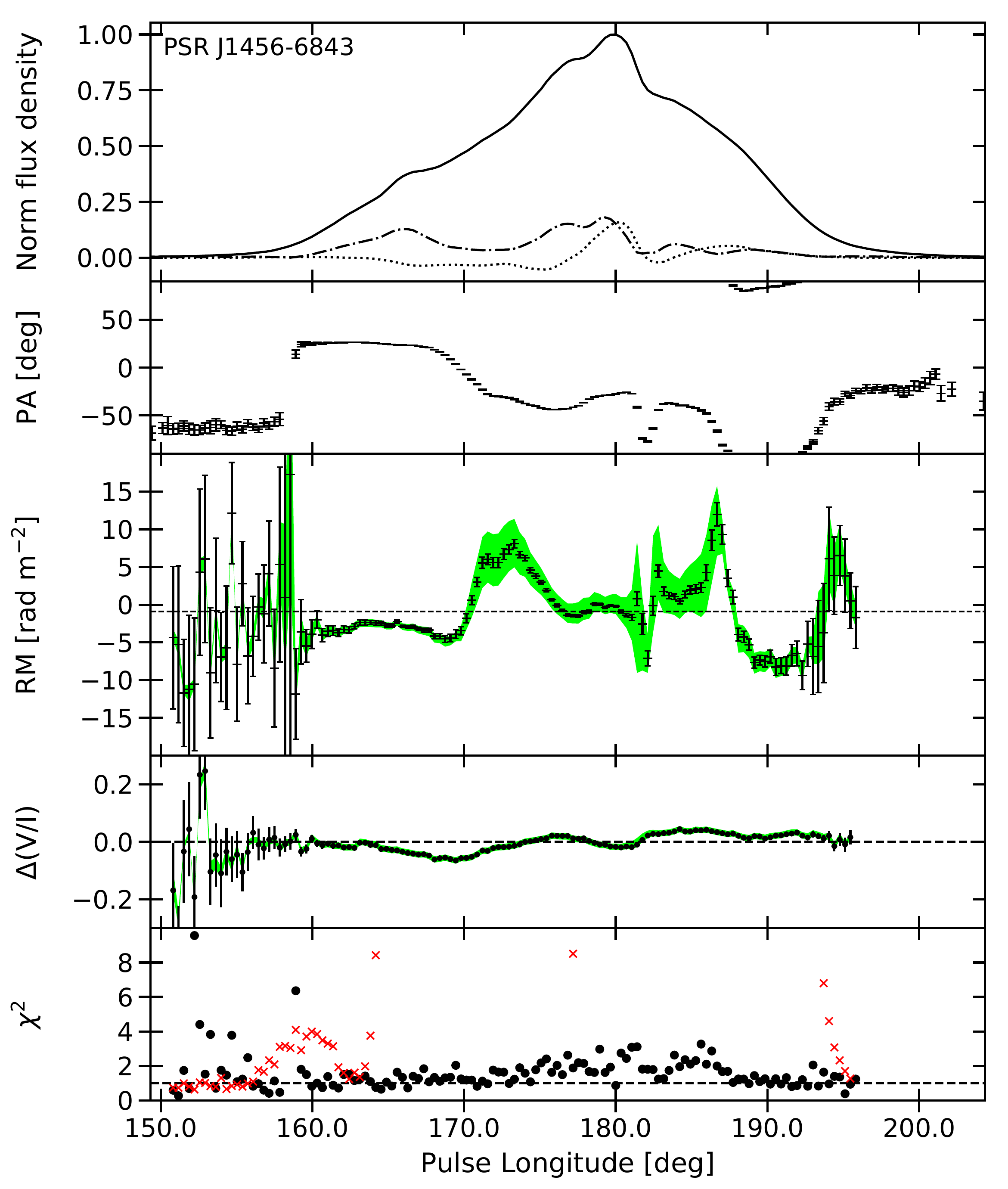} &
\includegraphics[width=0.45\textwidth,angle=0]{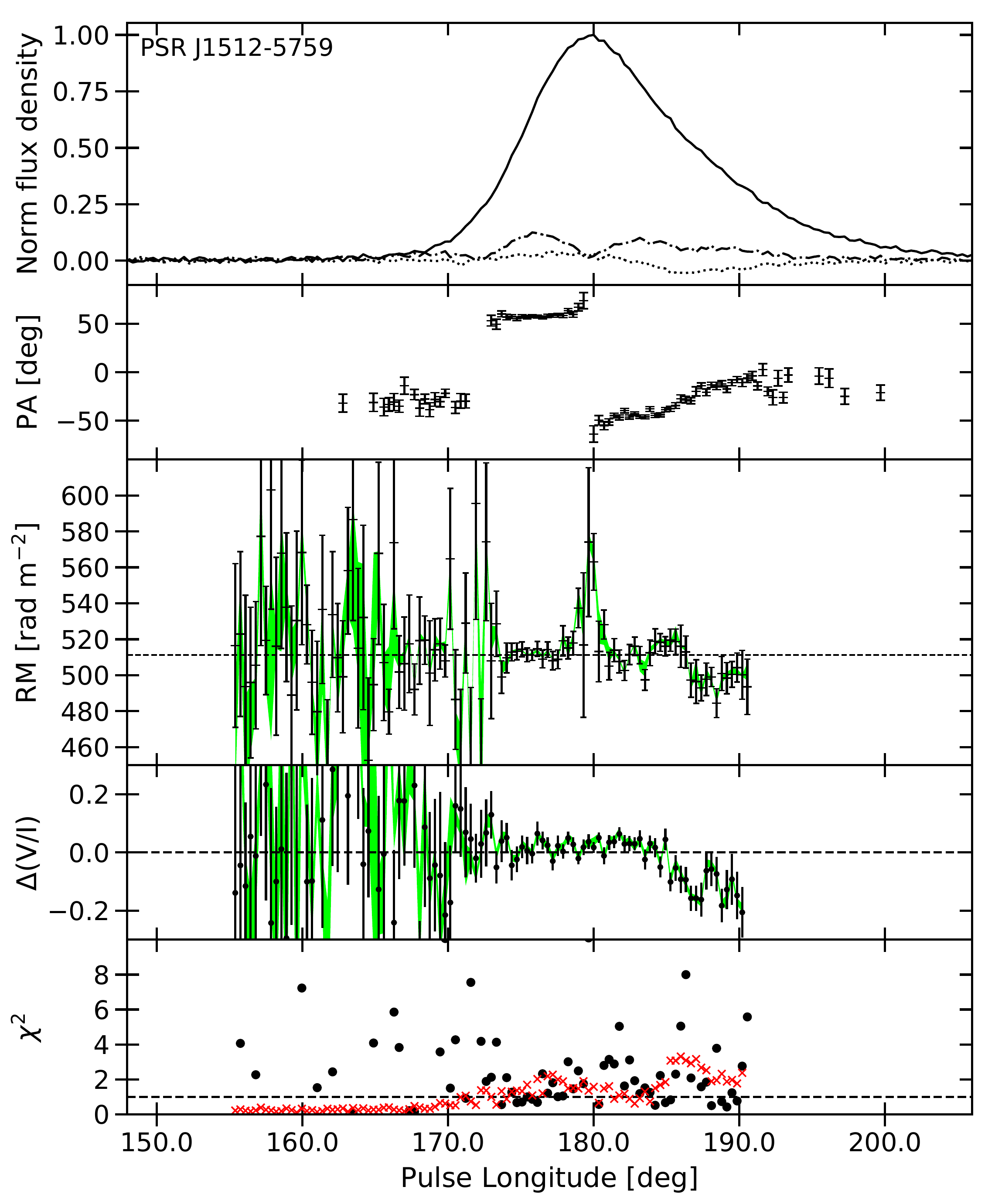} \\
\includegraphics[width=0.45\textwidth,angle=0]{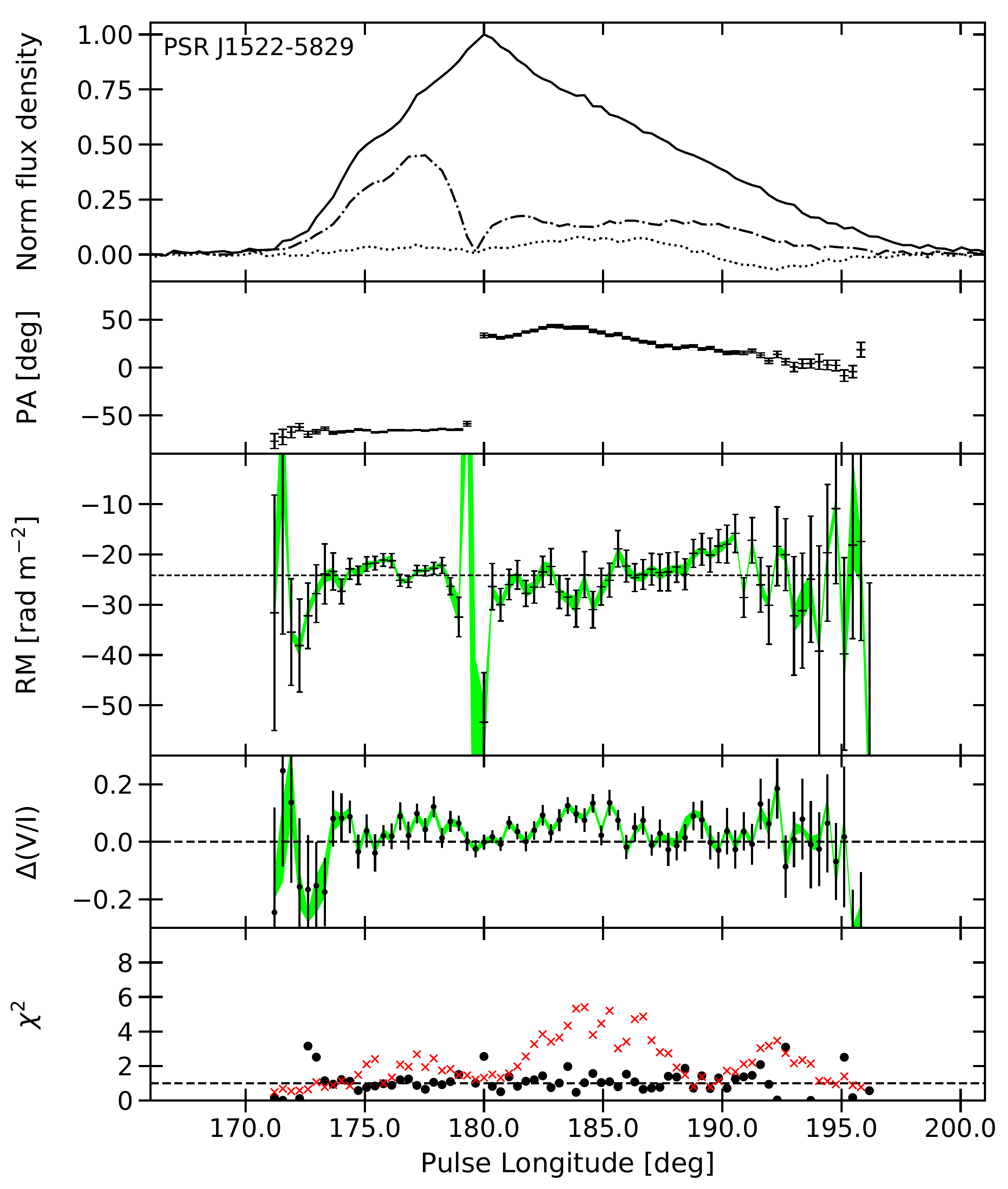} &
\includegraphics[width=0.45\textwidth,angle=0]{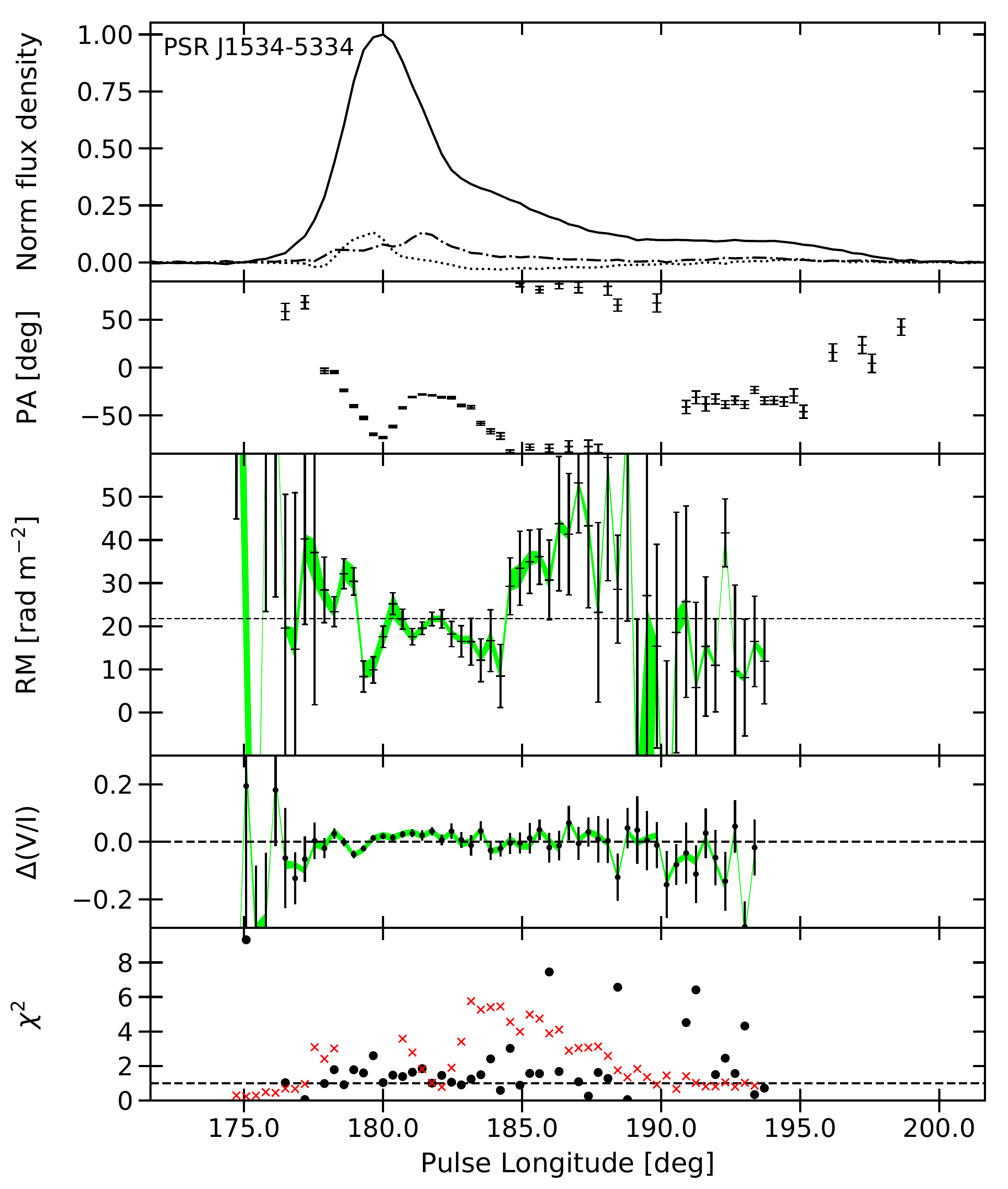} \\
\end{tabular}
\end{center}
\caption{Phase-resolved RM plots for PSR J1456$-$6843, PSR J1512$-$5759, PSR J1522$-$5829 and PSR J1534$-$5334. For more details on what is displayed in the individual panels, see Fig. 1.}
\label{fig:allPSR12}
\end{figure}

\begin{figure}
\begin{center}
\begin{tabular}{cc}
\includegraphics[width=0.45\textwidth,angle=0]{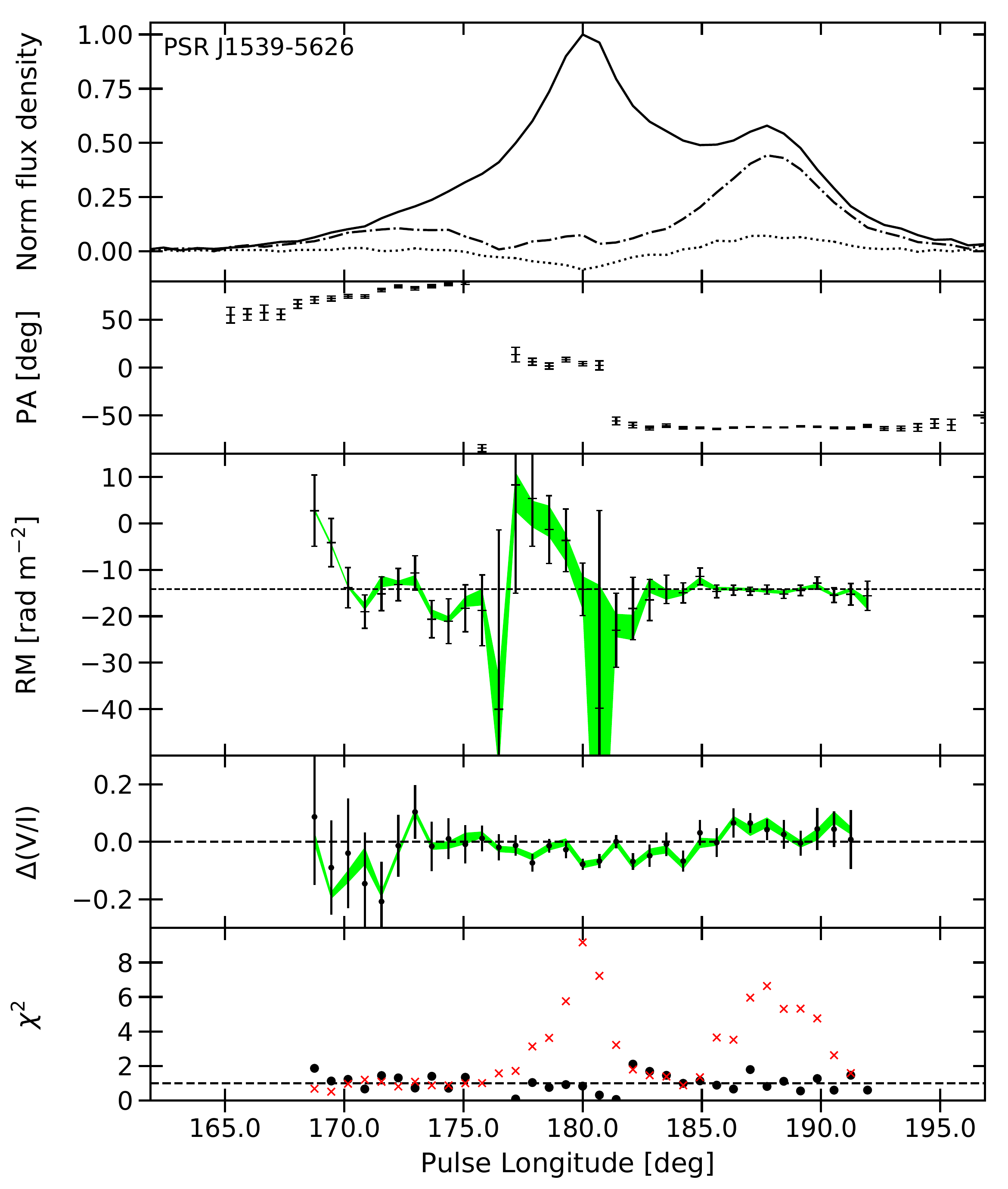} &
\includegraphics[width=0.45\textwidth,angle=0]{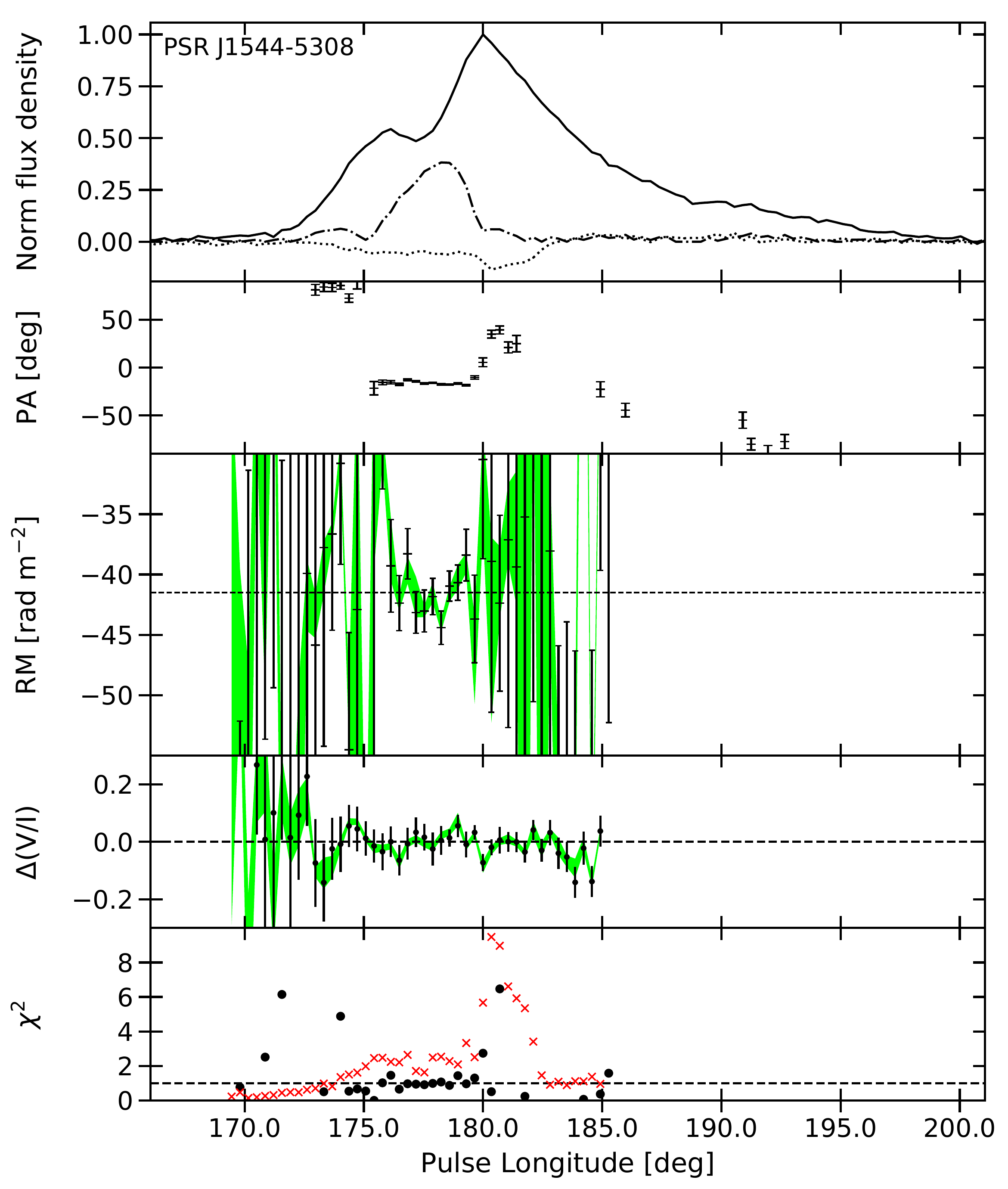} \\
\includegraphics[width=0.45\textwidth,angle=0]{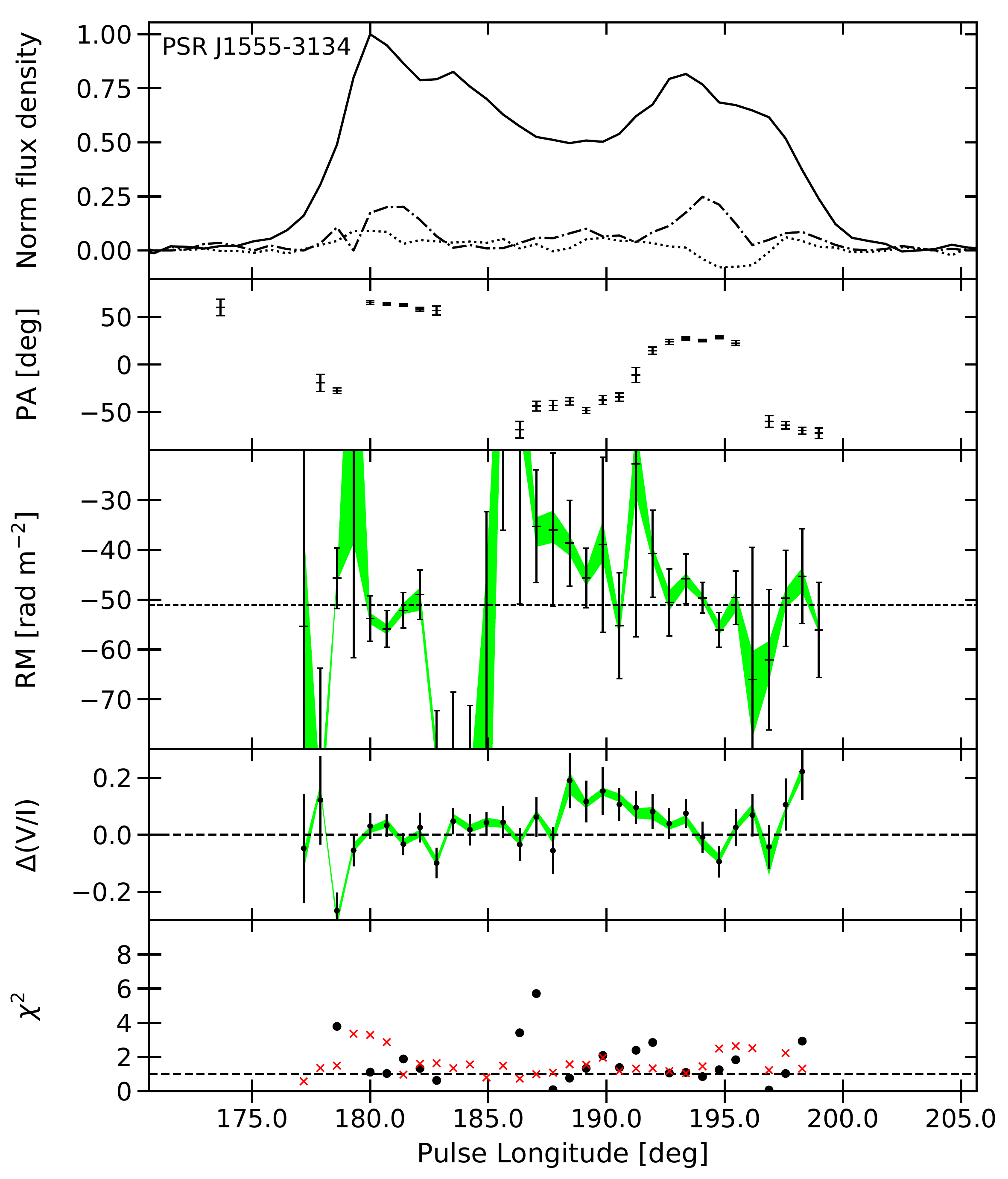} &
\includegraphics[width=0.45\textwidth,angle=0]{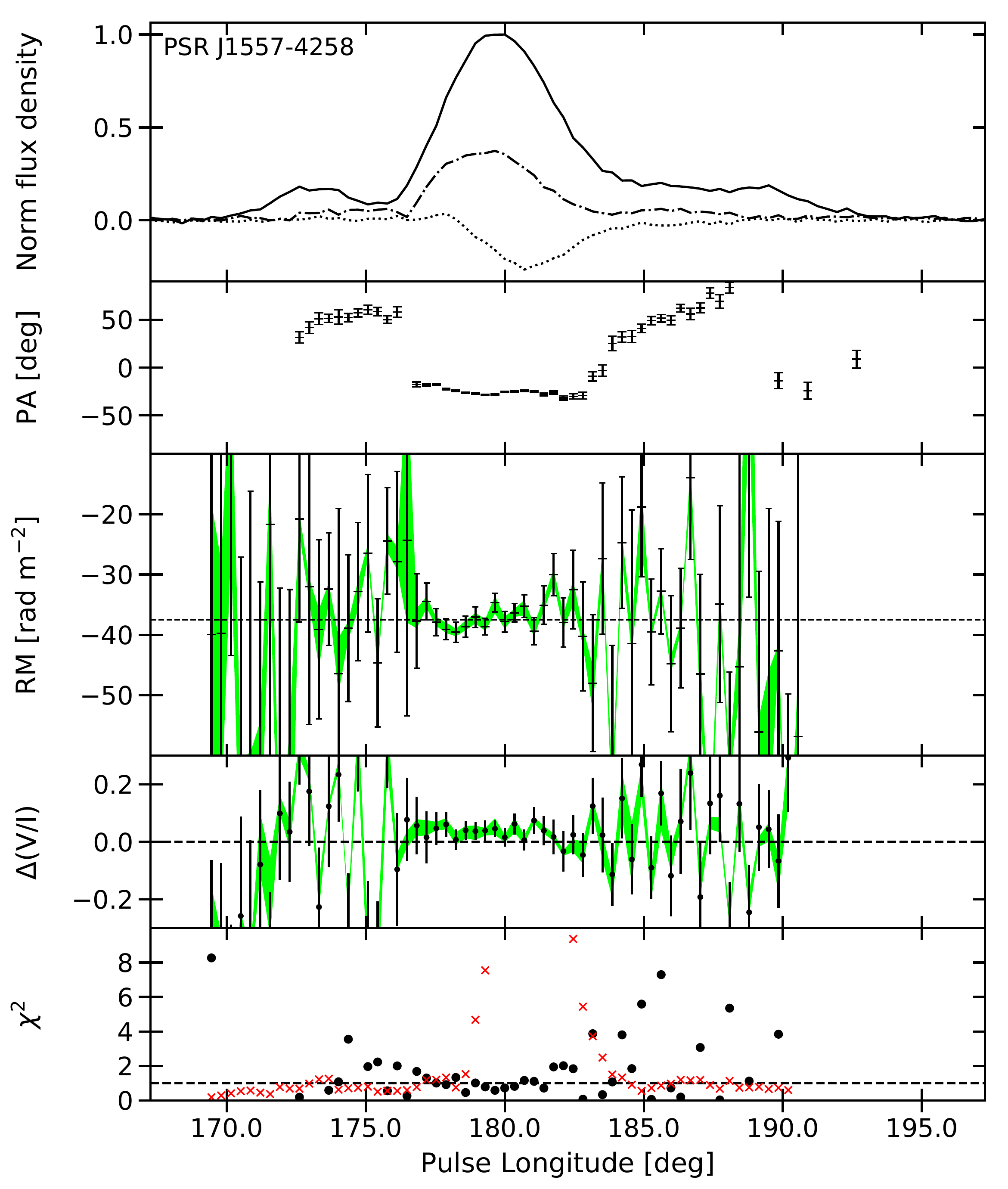} \\
\end{tabular}
\end{center}
\caption{Phase-resolved RM plots for PSR J1539$-$5626, PSR J1544$-$5308, PSR J1555$-$3134 and PSR J1557$-$4258. For more details on what is displayed in the individual panels, see Fig. 1.}
\label{fig:allPSR13}
\end{figure}

\begin{figure}
\begin{center}
\begin{tabular}{cc}
\includegraphics[width=0.45\textwidth,angle=0]{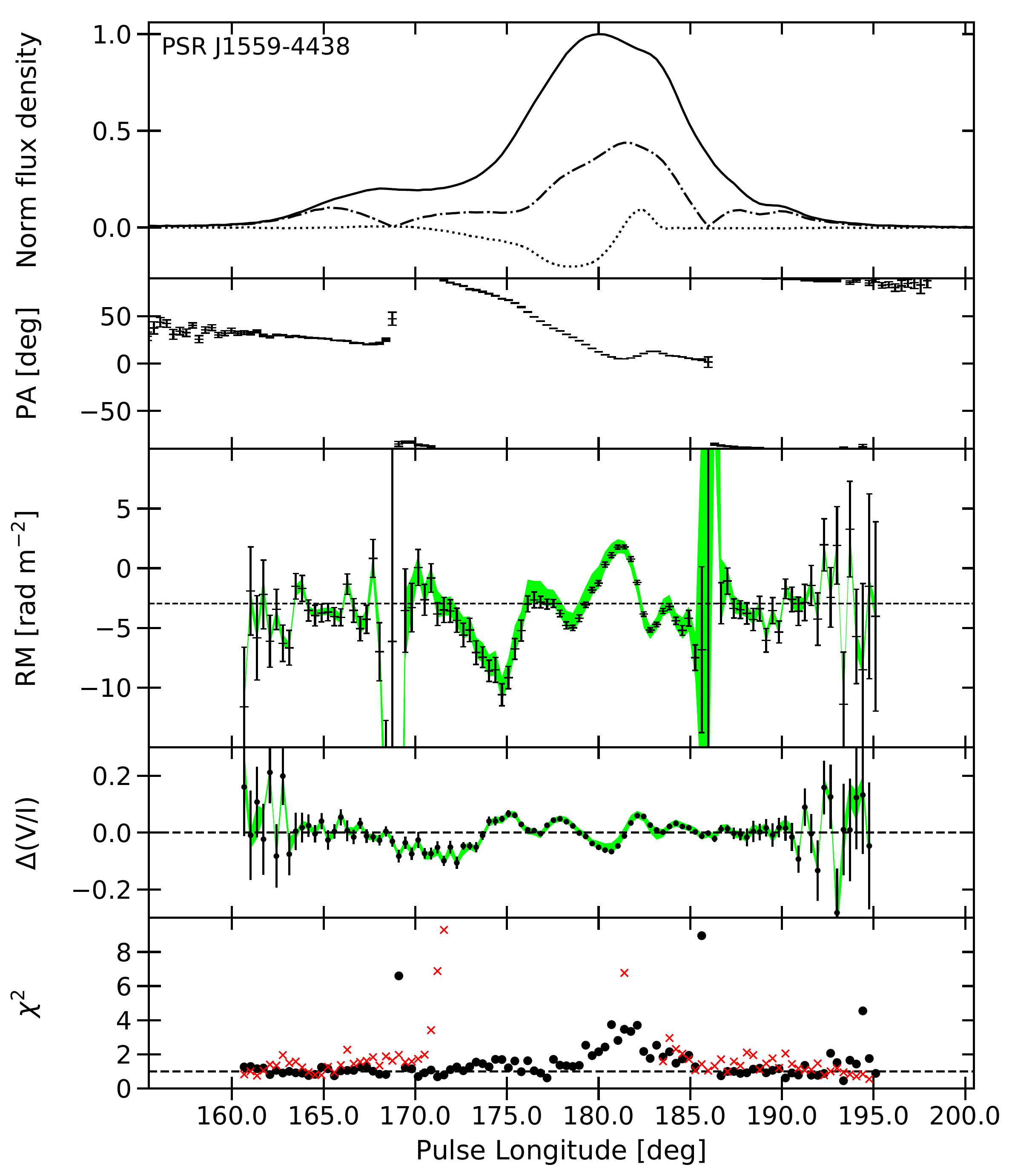} &
\includegraphics[width=0.45\textwidth,angle=0]{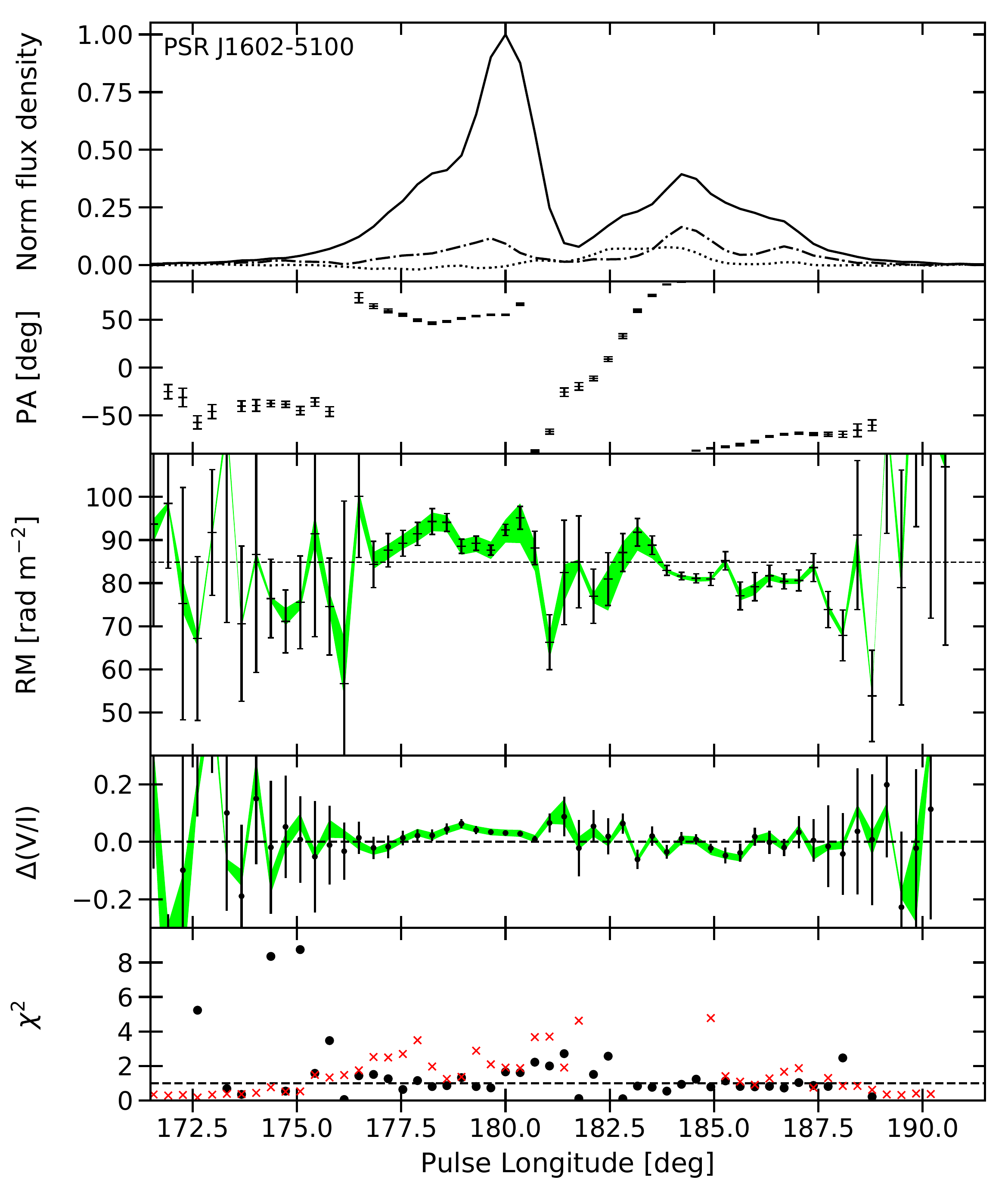} \\
\includegraphics[width=0.45\textwidth,angle=0]{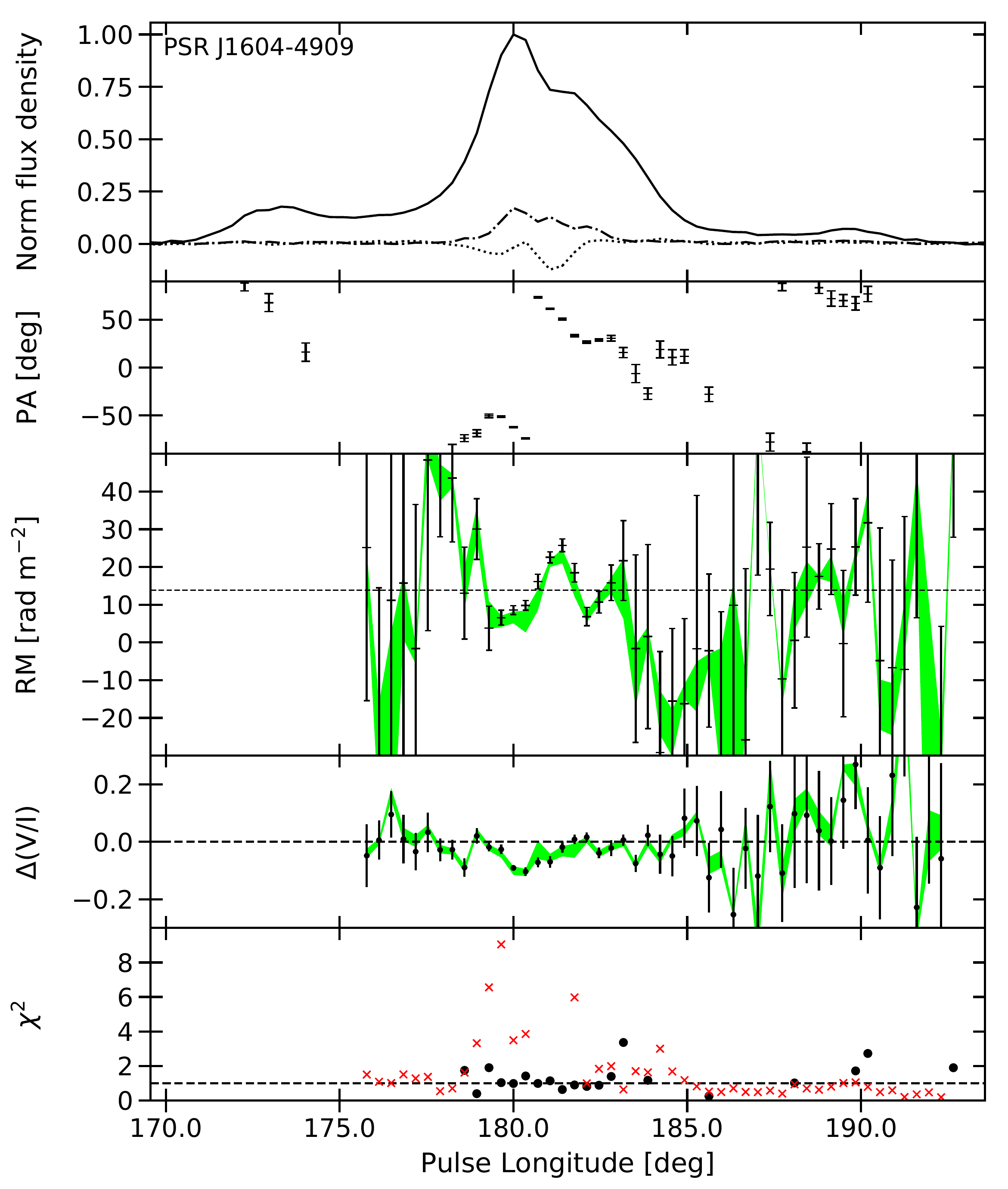} &
\includegraphics[width=0.45\textwidth,angle=0]{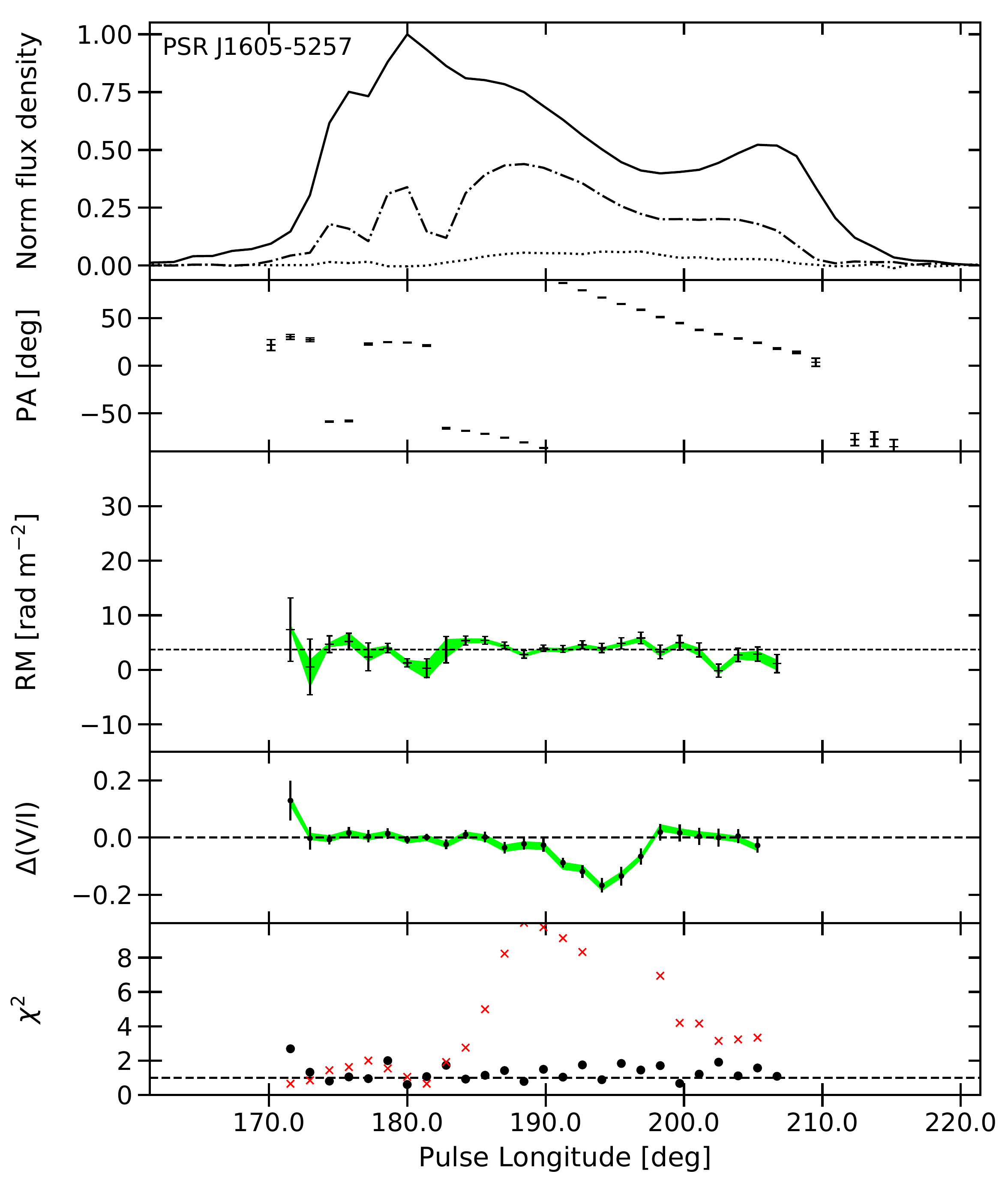} \\
\end{tabular}
\end{center}
\caption{Phase-resolved RM plots for PSR J1559$-$4438, PSR J1602$-$5100, PSR J1604$-$4909 and PSR J1605$-$5257. For more details on what is displayed in the individual panels, see Fig. 1.}
\label{fig:allPSR14}
\end{figure}

\begin{figure}
\begin{center}
\begin{tabular}{cc}
\includegraphics[width=0.45\textwidth,angle=0]{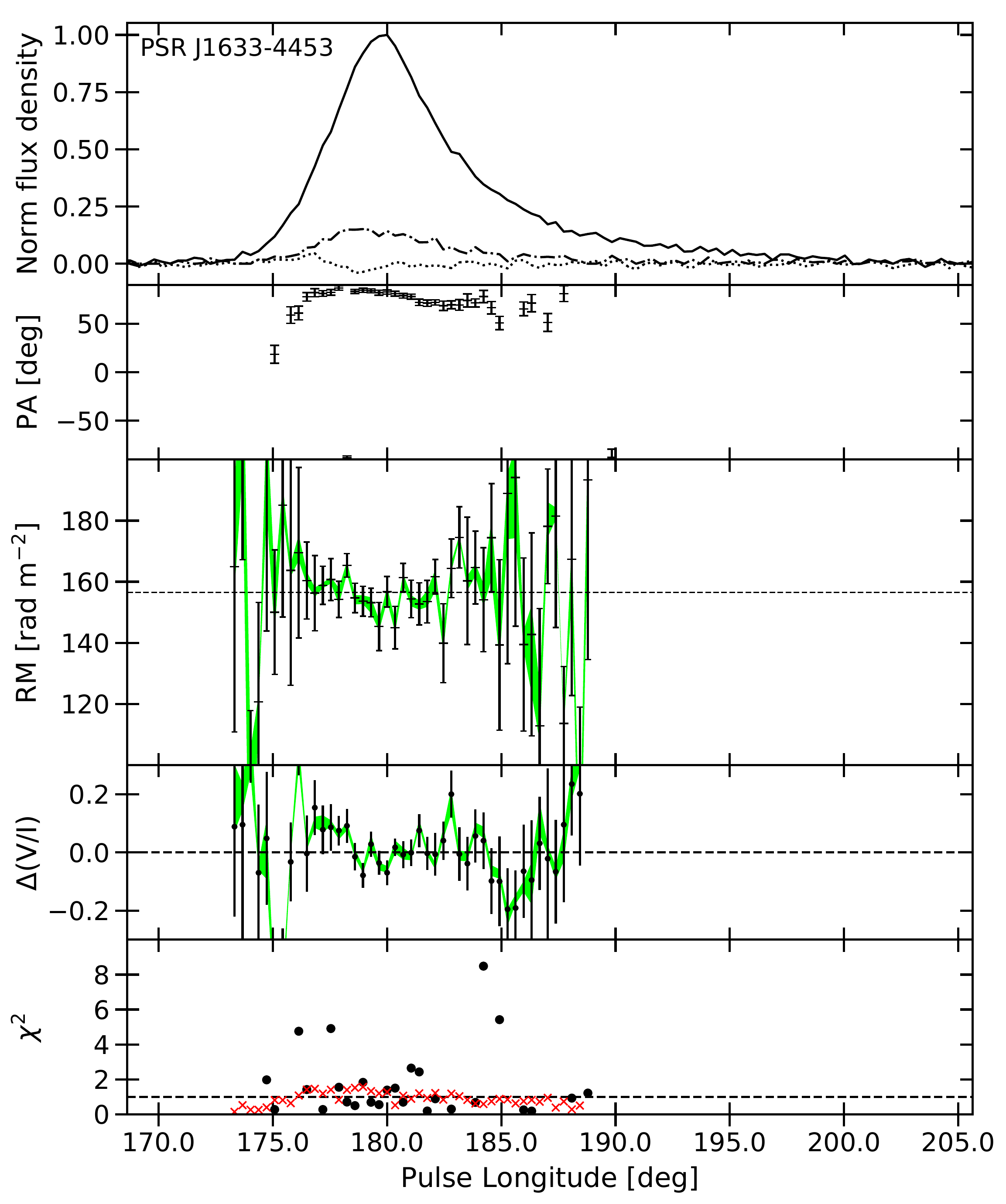} &
\includegraphics[width=0.45\textwidth,angle=0]{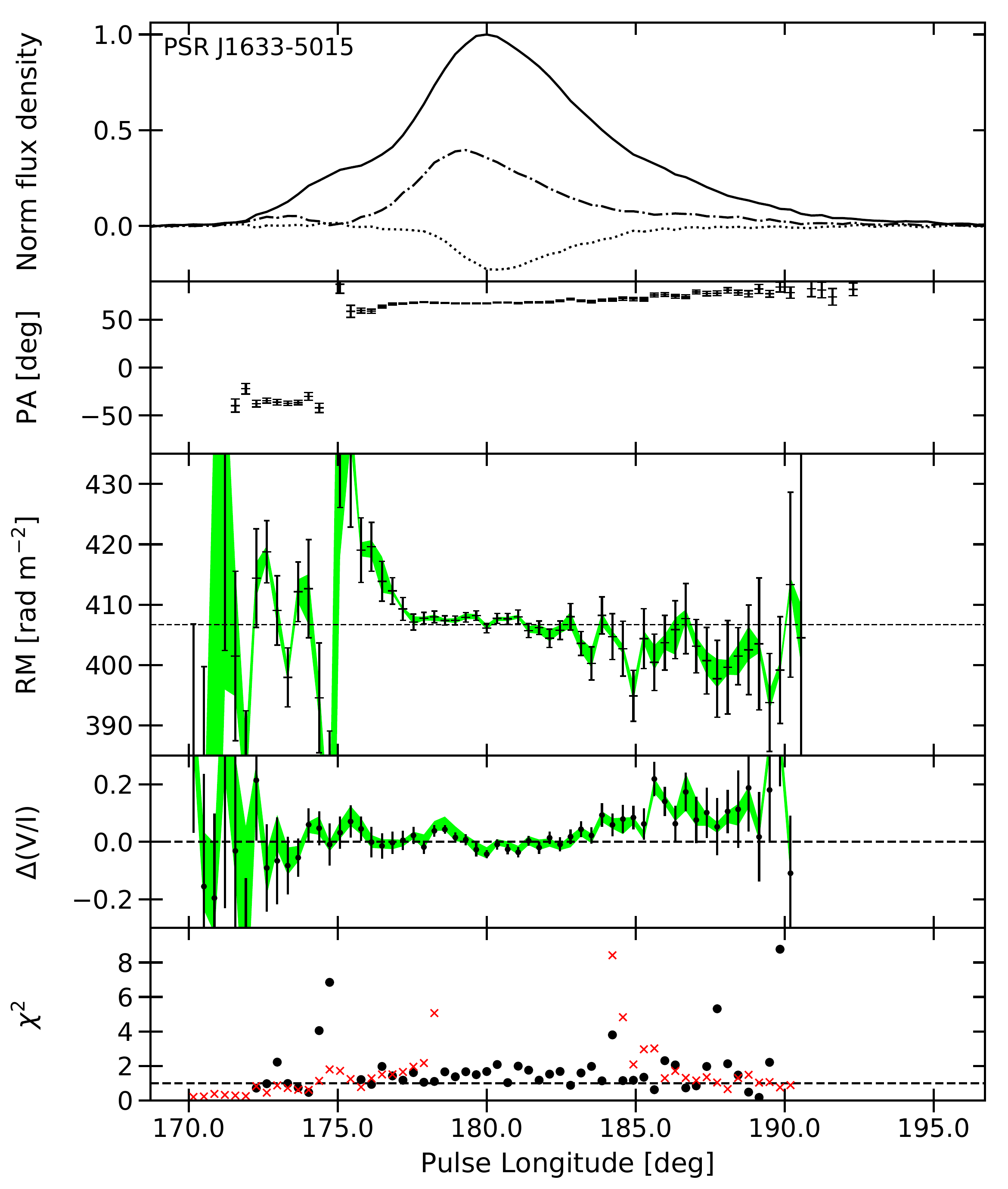} \\
\includegraphics[width=0.45\textwidth,angle=0]{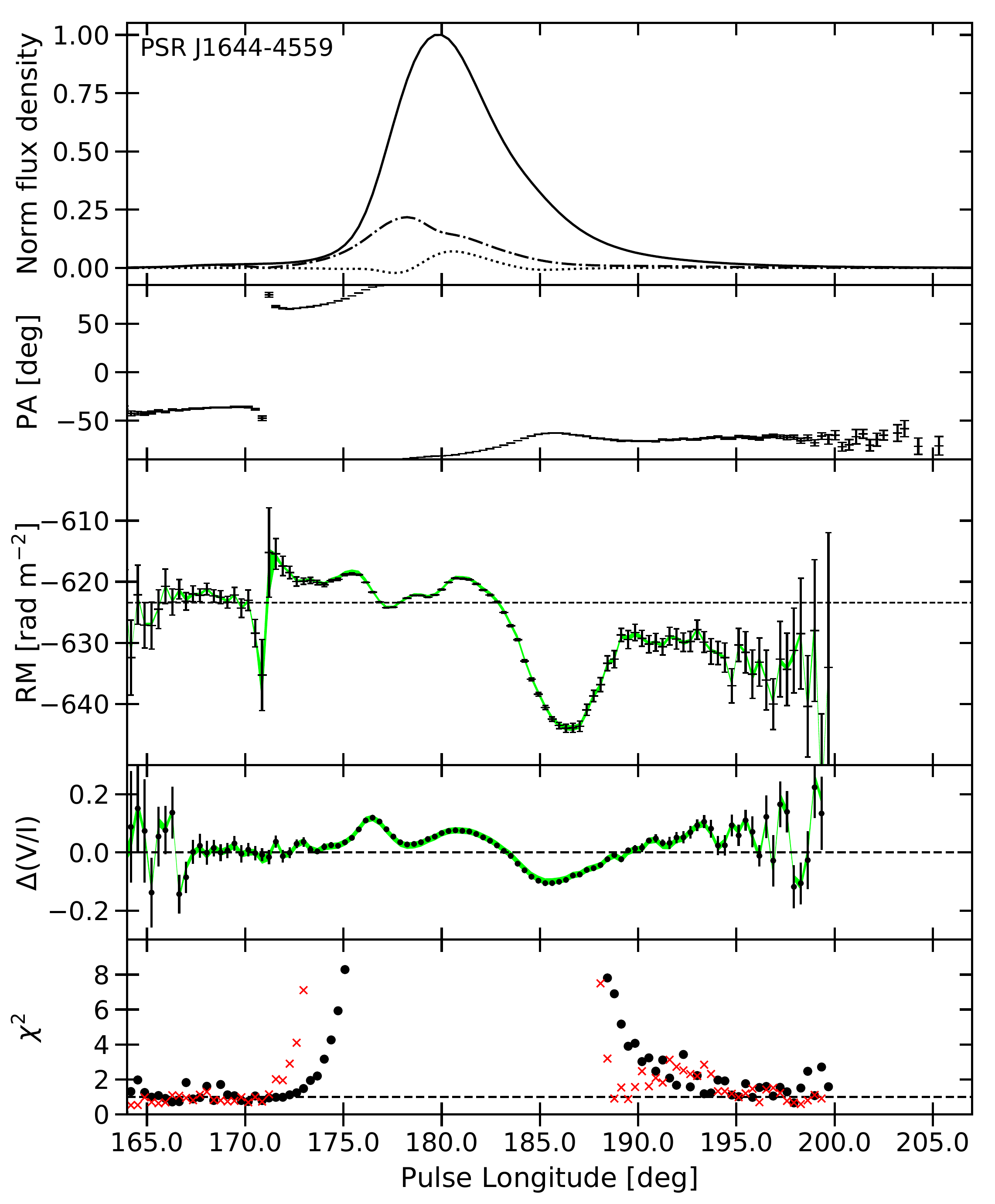} &
\includegraphics[width=0.45\textwidth,angle=0]{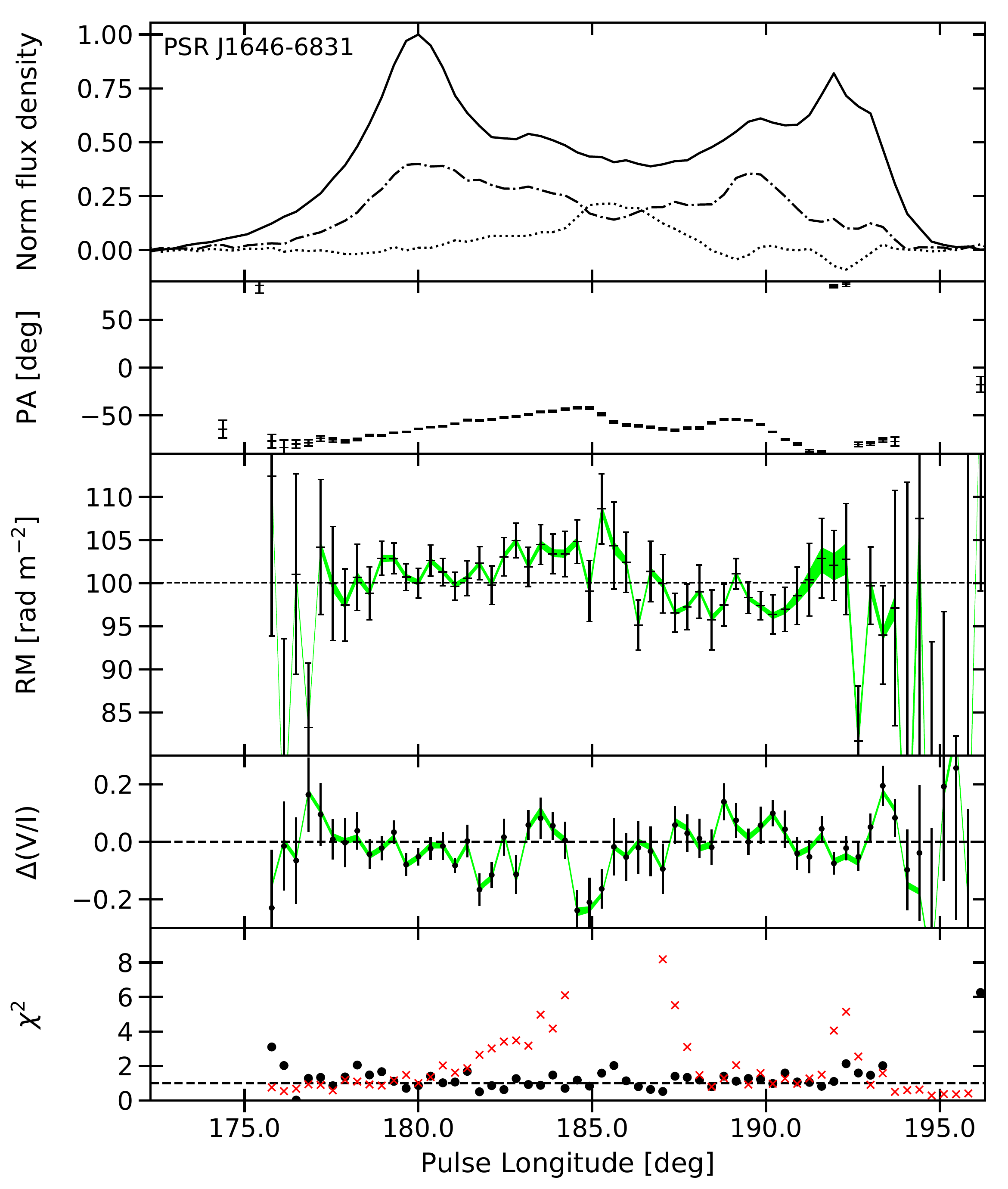} \\
\end{tabular}
\end{center}
\caption{Phase-resolved RM plots for PSR J1633$-$4453, PSR J1633$-$5015, PSR J1644$-$4559 and PSR J1646$-$6831. For more details on what is displayed in the individual panels, see Fig. 1.}
\label{fig:allPSR15}
\end{figure}

\begin{figure}
\begin{center}
\begin{tabular}{cc}
\includegraphics[width=0.45\textwidth,angle=0]{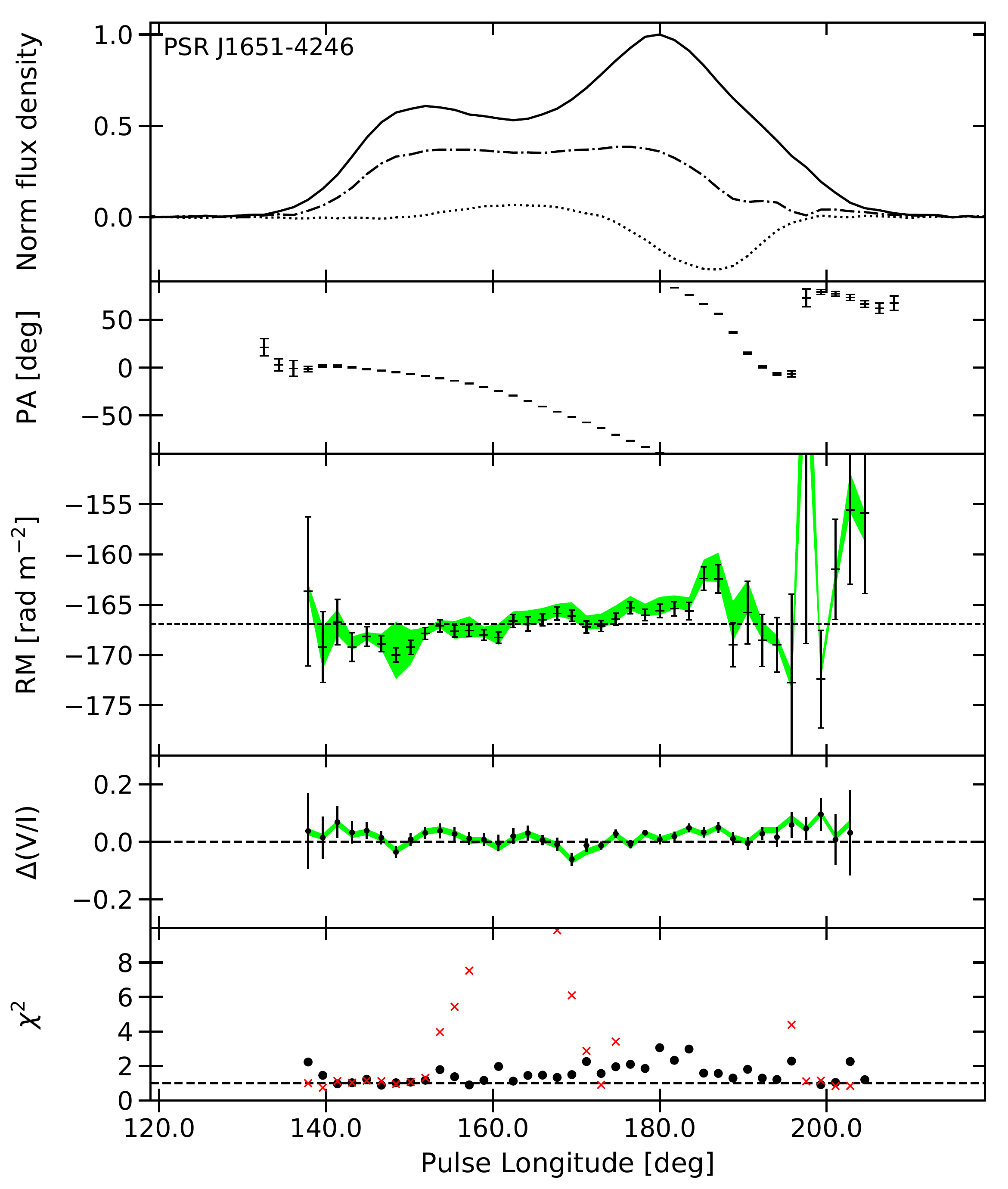} &
\includegraphics[width=0.45\textwidth,angle=0]{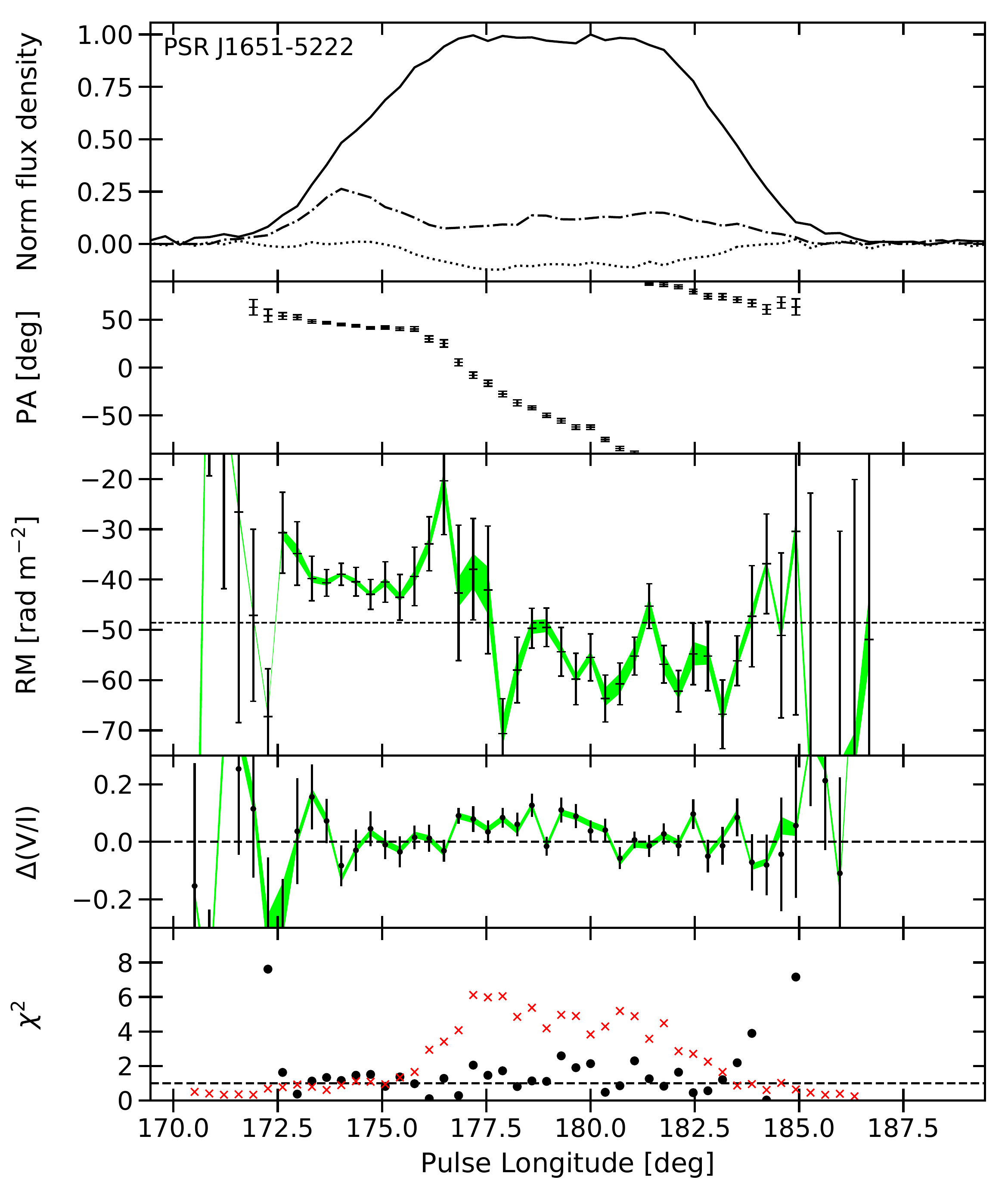} \\
\includegraphics[width=0.45\textwidth,angle=0]{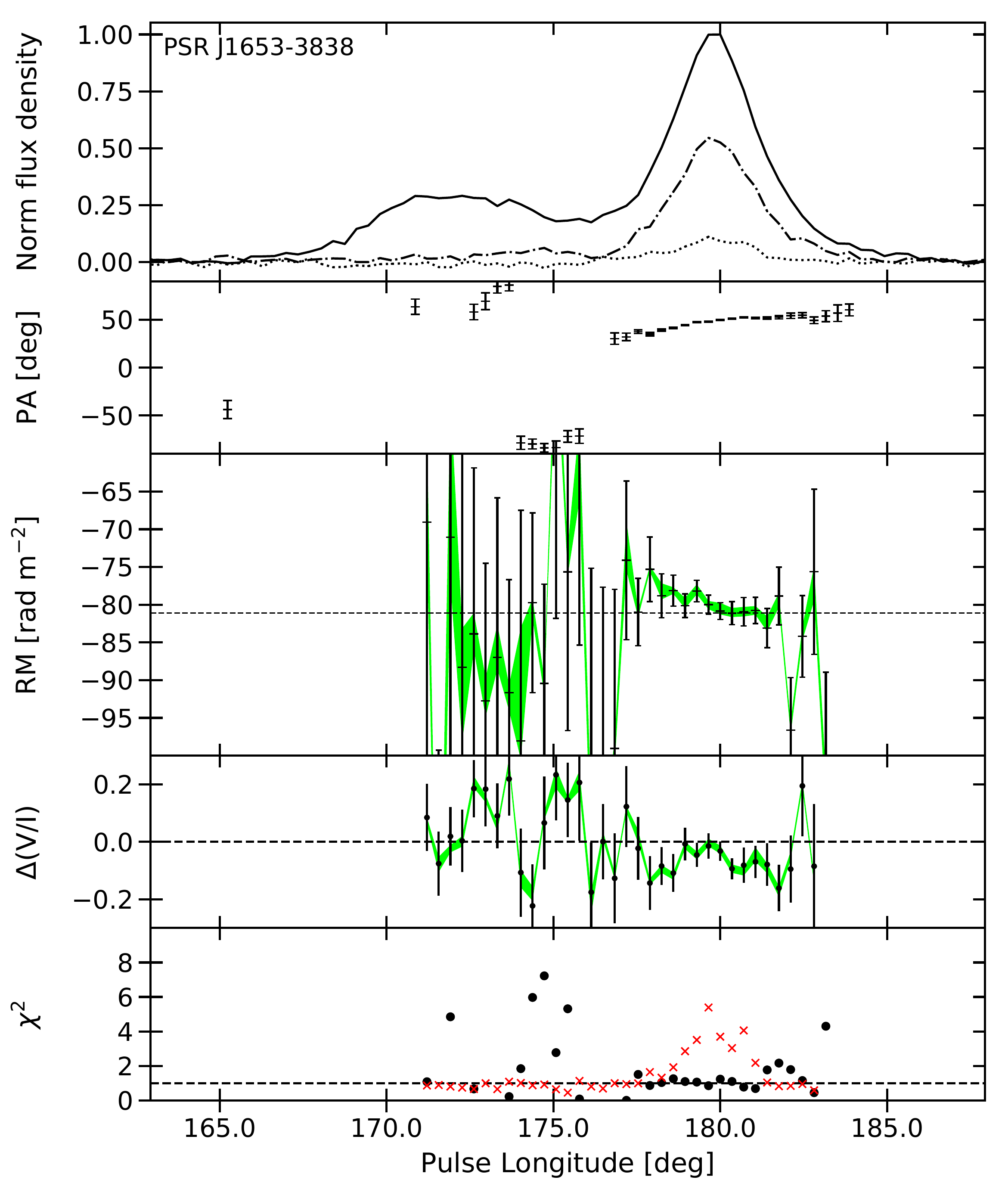} &
\includegraphics[width=0.45\textwidth,angle=0]{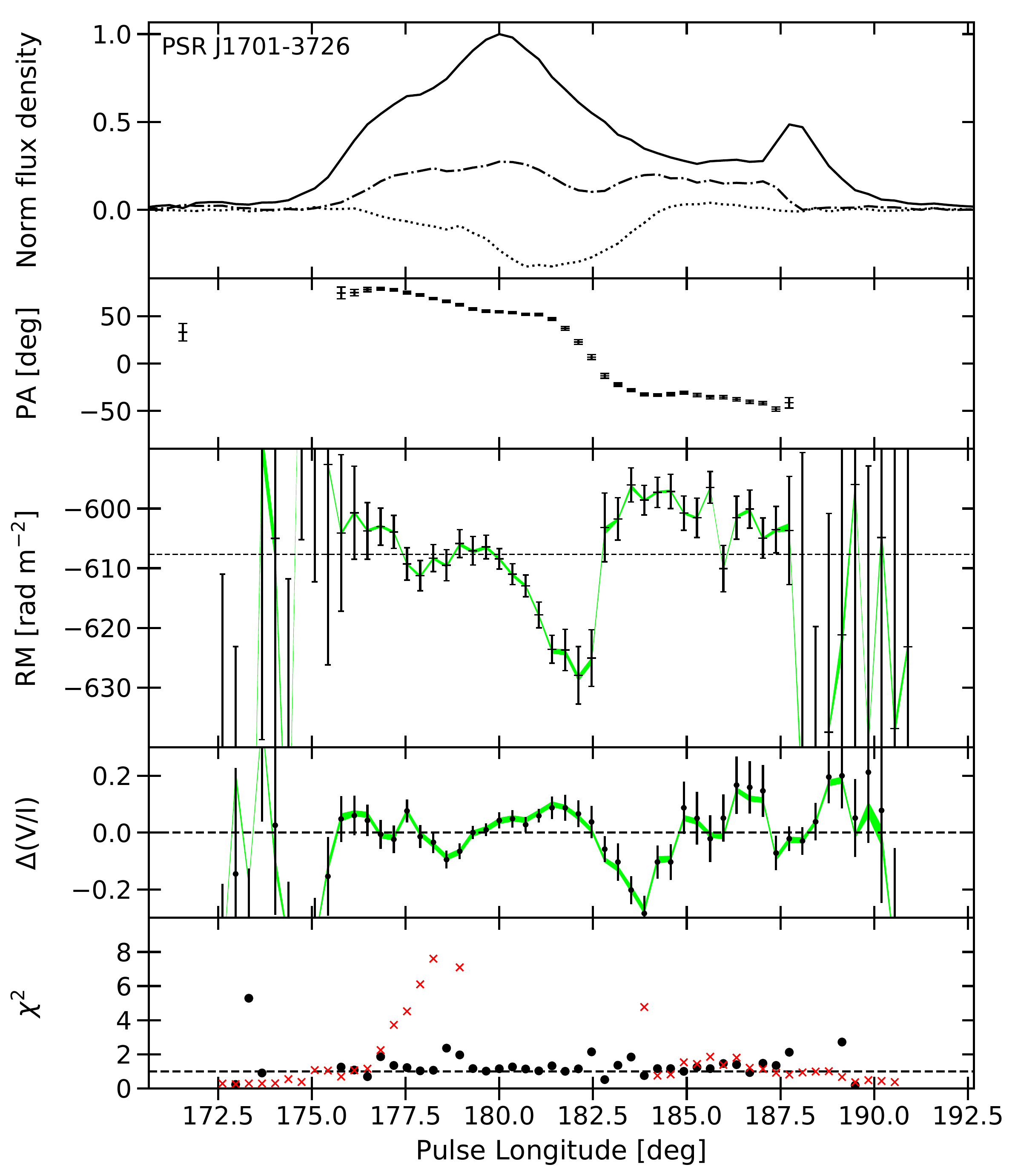} \\
\end{tabular}
\end{center}
\caption{Phase-resolved RM plots for PSR J1651$-$4246, PSR J1651$-$5222, PSR J1653$-$3838 and PSR J1701$-$3726. For more details on what is displayed in the individual panels, see Fig. 1.}
\label{fig:allPSR16}
\end{figure}

\begin{figure}
\begin{center}
\begin{tabular}{cc}
\includegraphics[width=0.45\textwidth,angle=0]{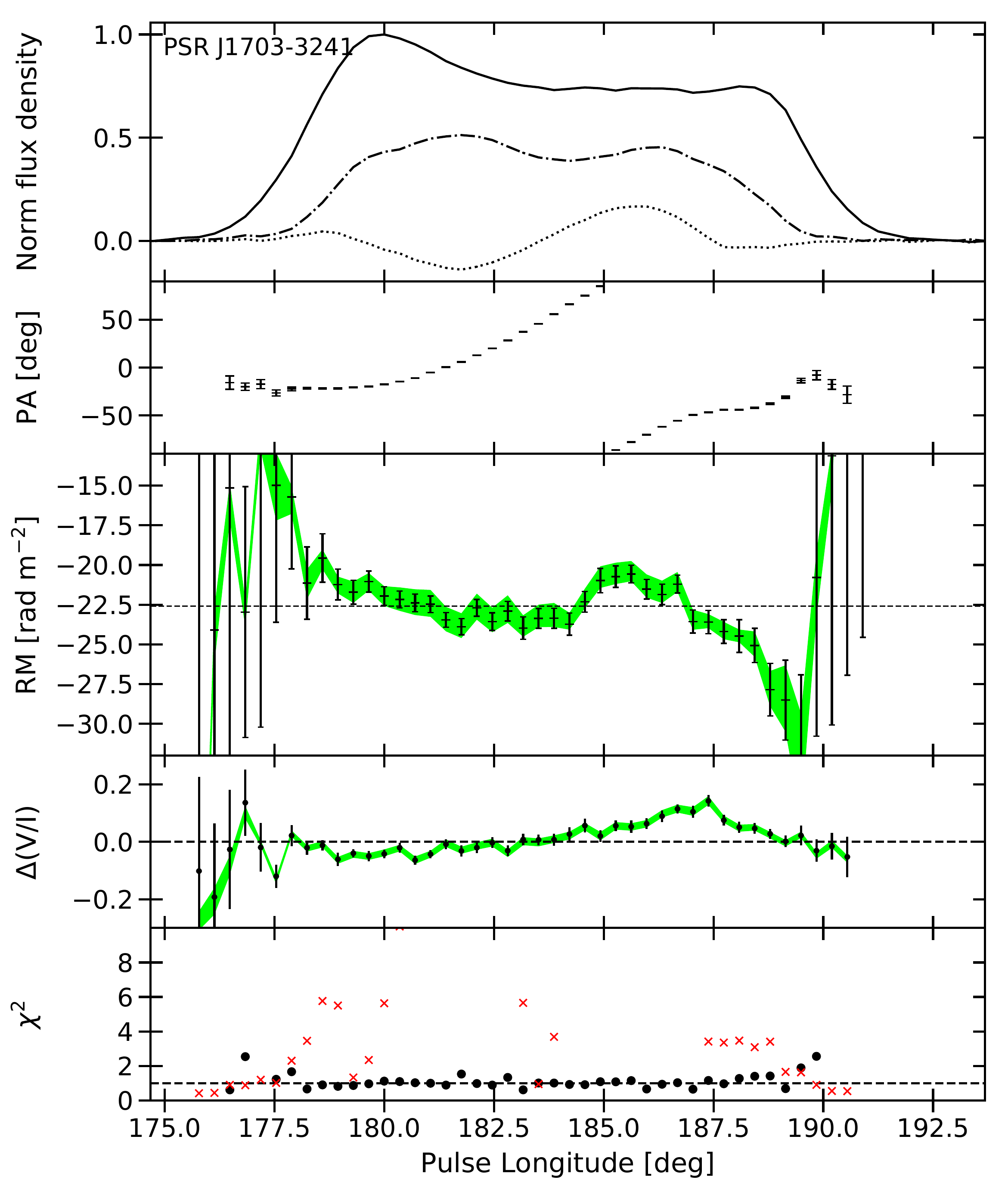} &
\includegraphics[width=0.45\textwidth,angle=0]{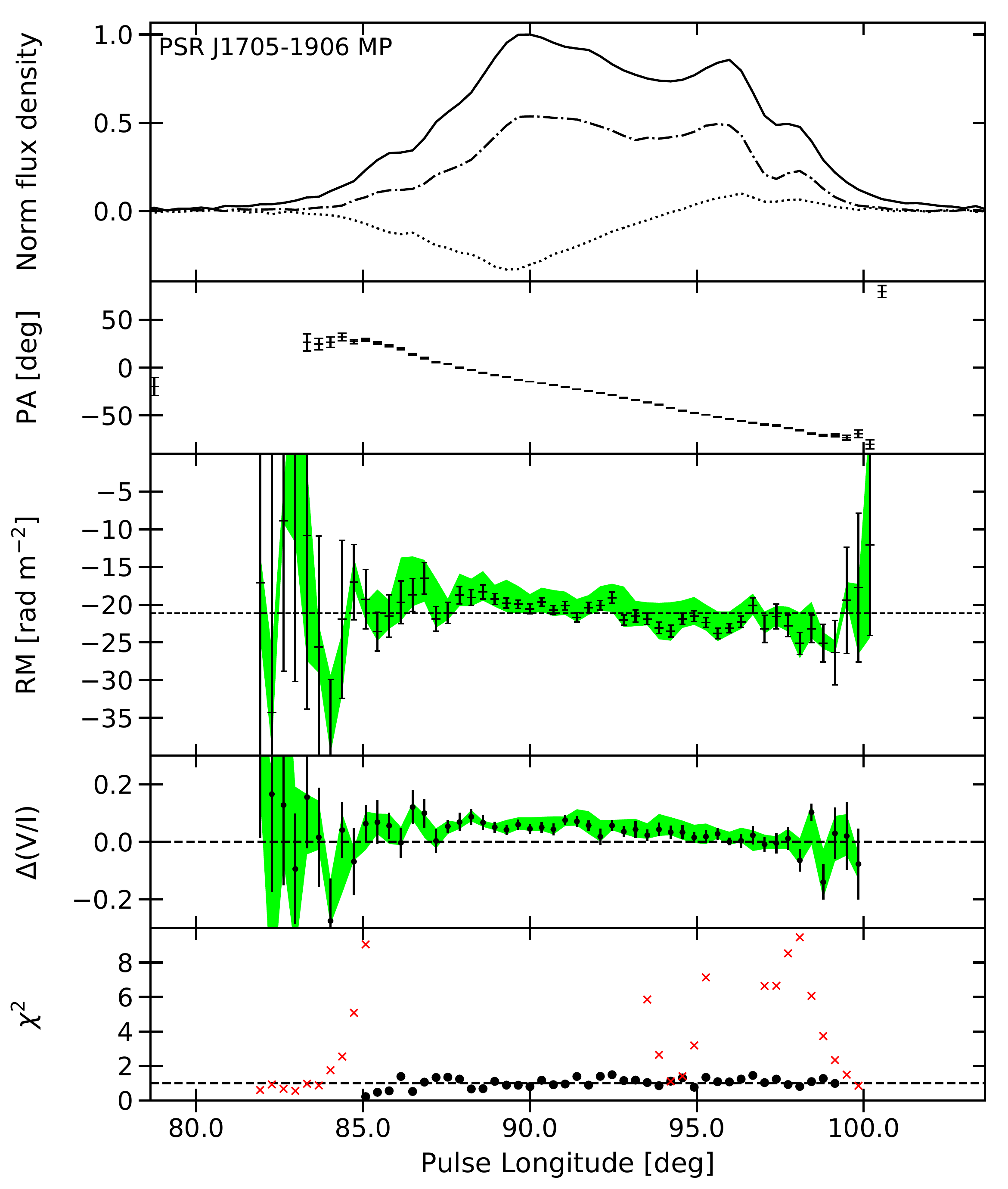} \\
\includegraphics[width=0.45\textwidth,angle=0]{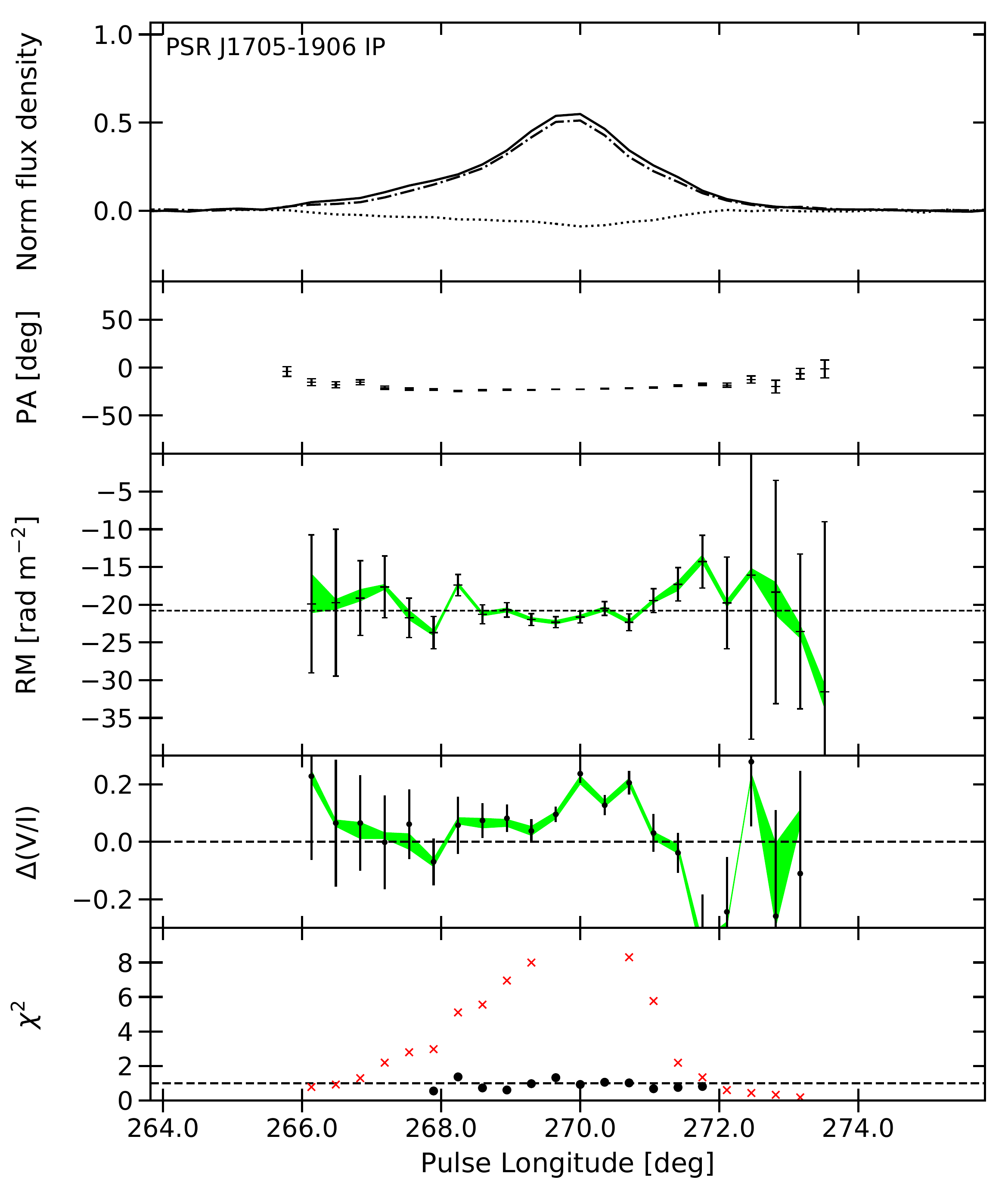} &
\includegraphics[width=0.45\textwidth,angle=0]{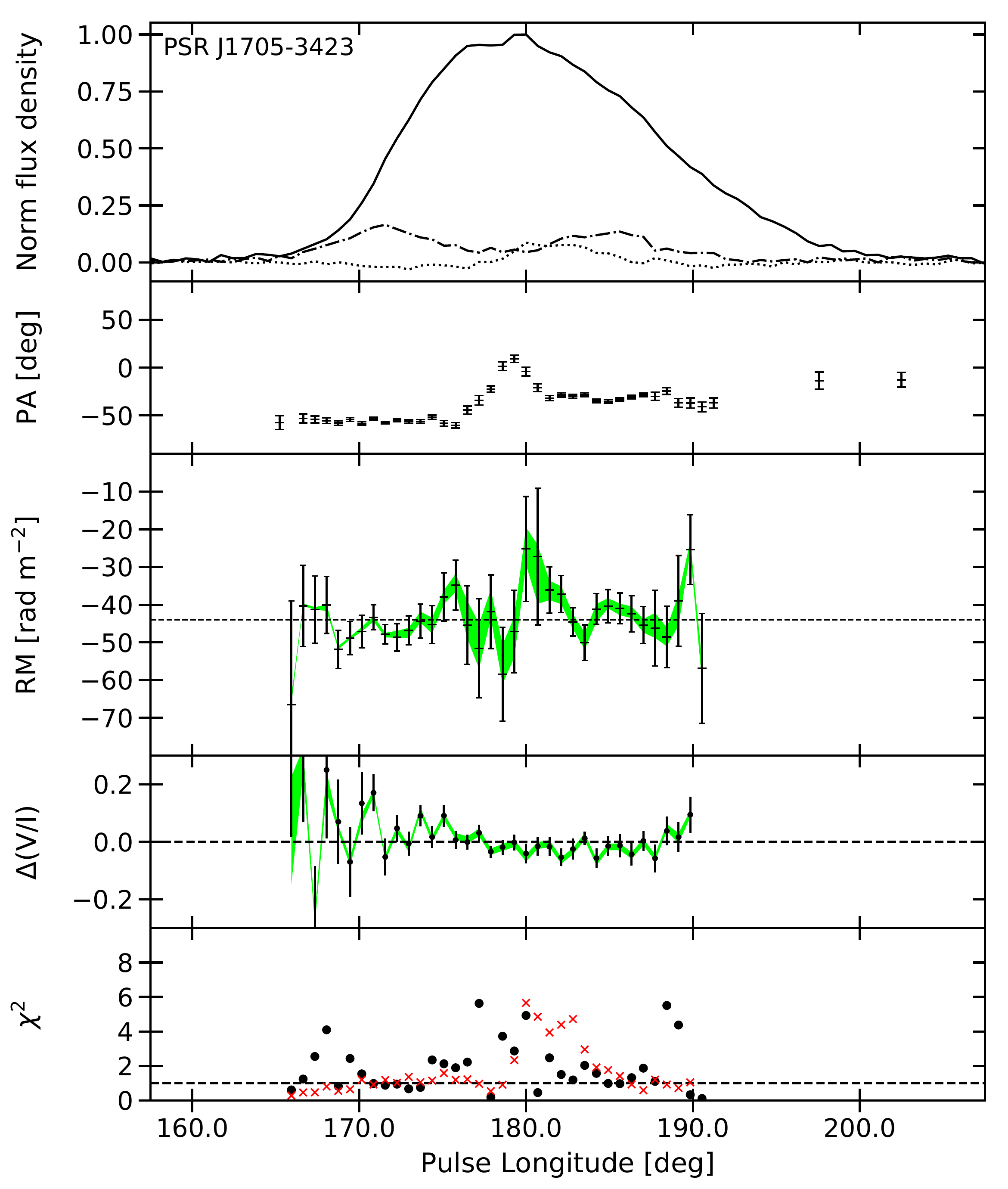} \\
\end{tabular}
\end{center}
\caption{Phase-resolved RM plots for PSR J1703$-$3241, PSR J1705$-$1906 (MP), PSR J1705$-$1906 (IP) and PSR J1705$-$3423. For more details on what is displayed in the individual panels, see Fig. 1.}
\label{fig:allPSR17}
\end{figure}

\begin{figure}
\begin{center}
\begin{tabular}{cc}
\includegraphics[width=0.45\textwidth,angle=0]{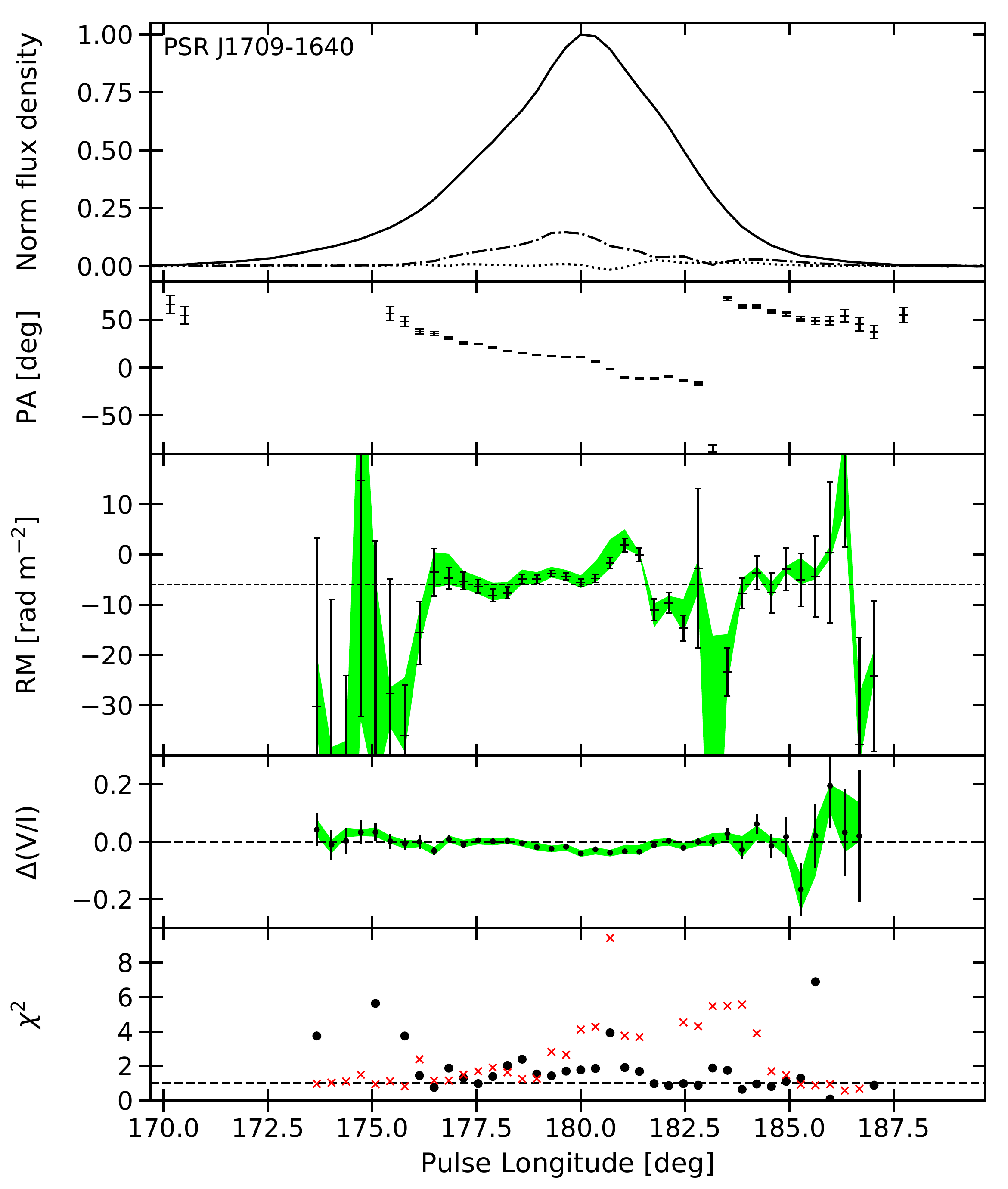} &
\includegraphics[width=0.45\textwidth,angle=0]{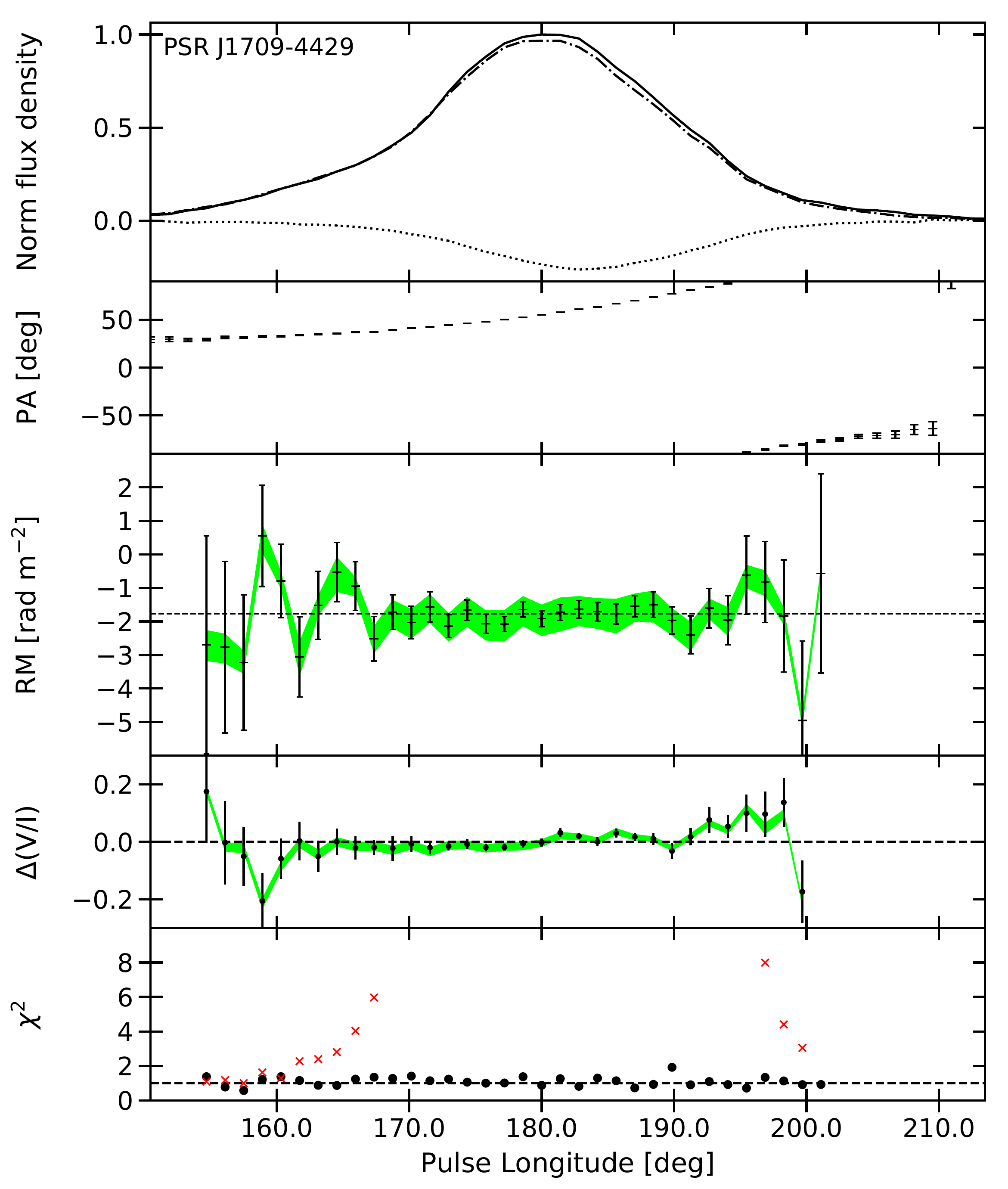} \\
\includegraphics[width=0.45\textwidth,angle=0]{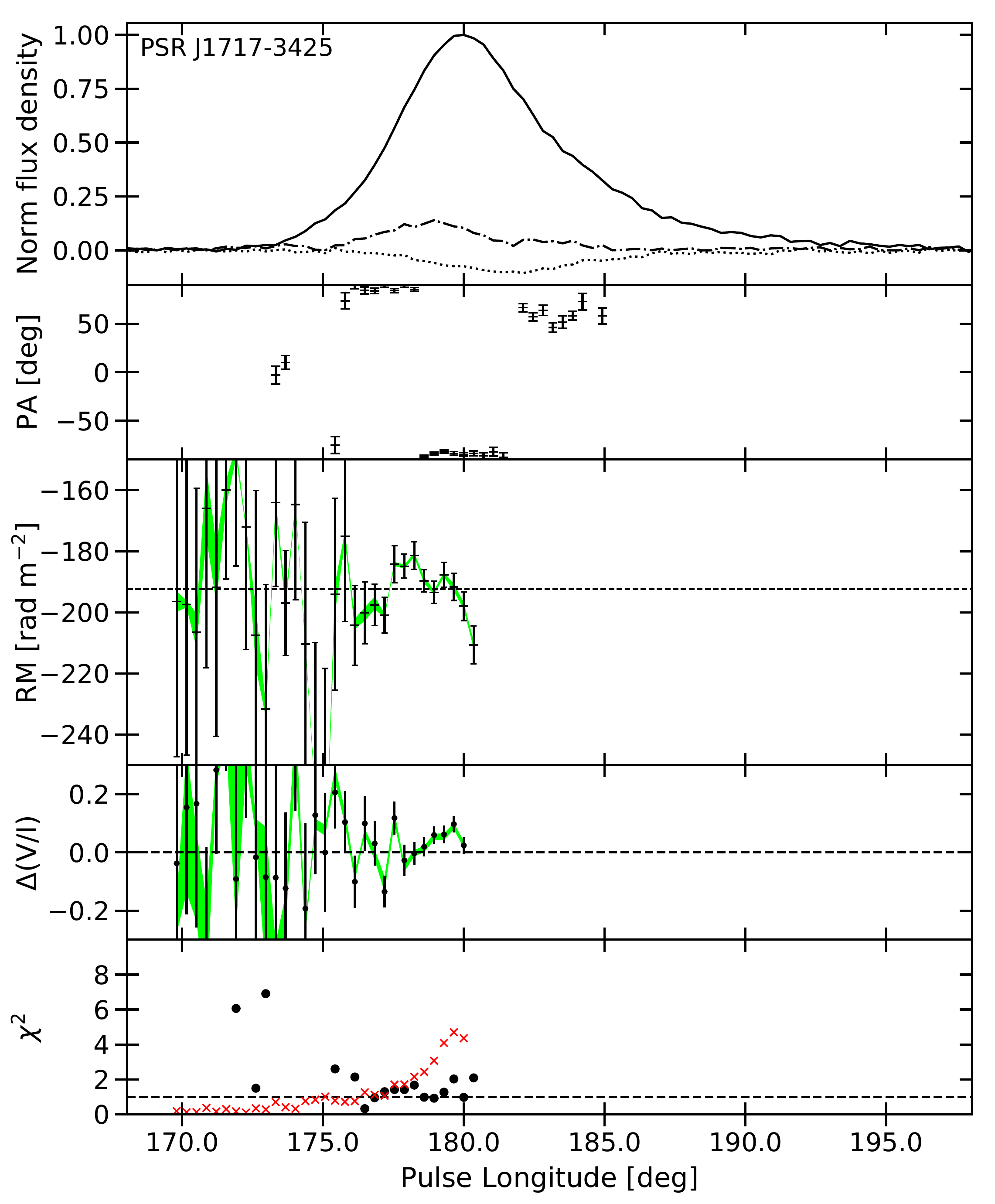} &
\includegraphics[width=0.45\textwidth,angle=0]{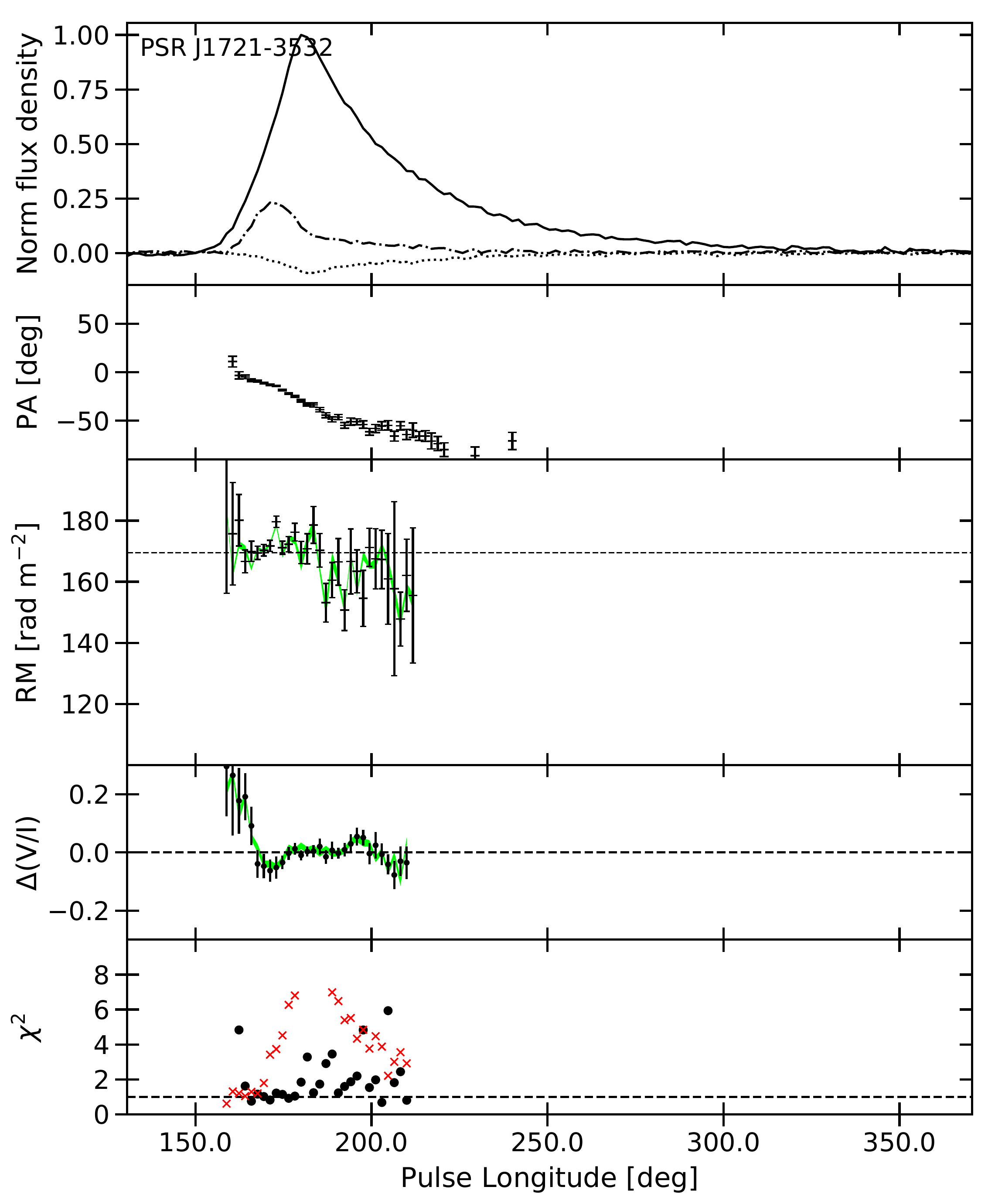} \\
\end{tabular}
\end{center}
\caption{Phase-resolved RM plots for PSR J1709$-$1640, PSR J1709$-$4429, PSR J1717$-$3425 and PSR J1721$-$3532. For more details on what is displayed in the individual panels, see Fig. 1.}
\label{fig:allPSR18}
\end{figure}

\begin{figure}
\begin{center}
\begin{tabular}{cc}
\includegraphics[width=0.45\textwidth,angle=0]{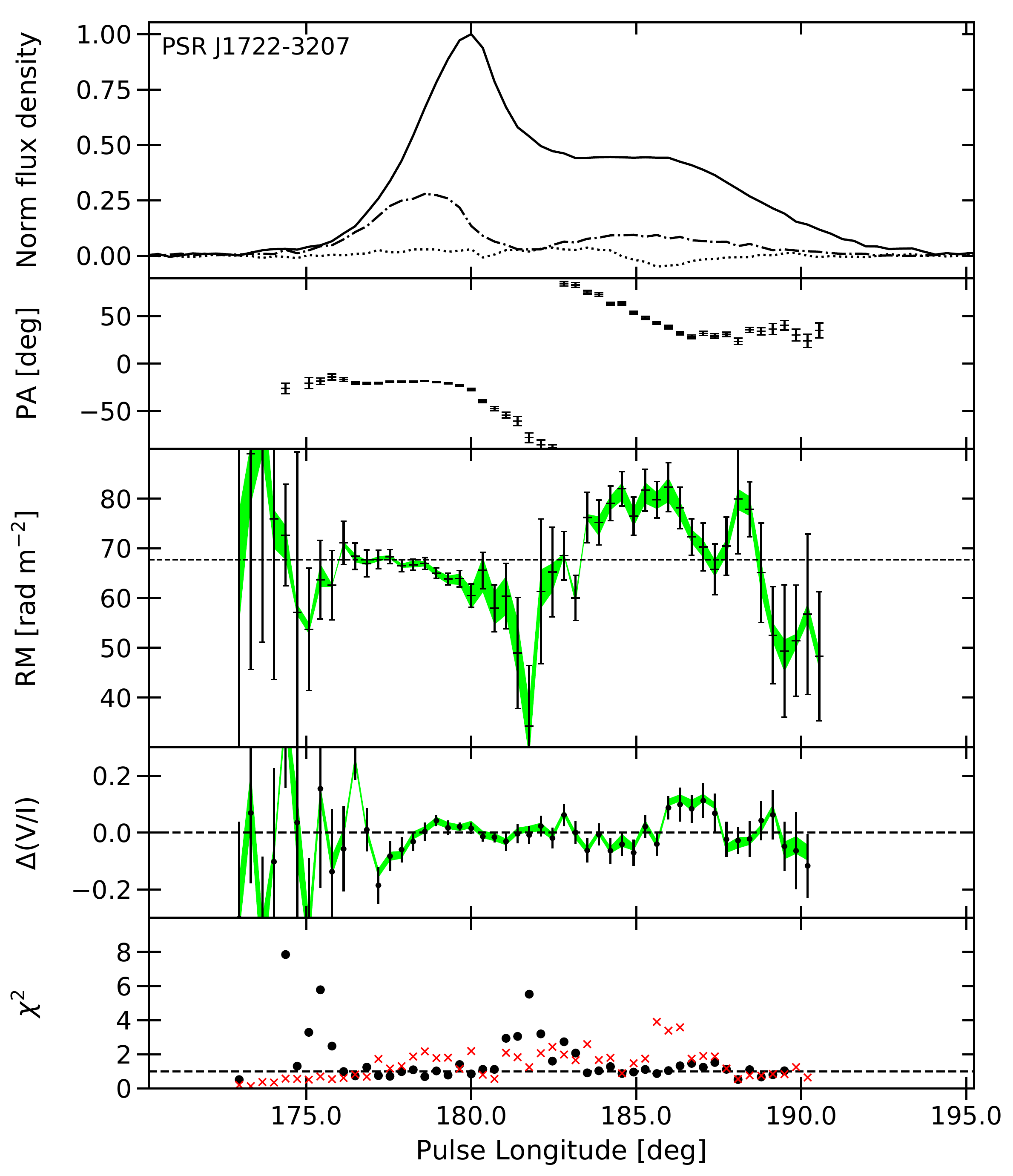} &
\includegraphics[width=0.45\textwidth,angle=0]{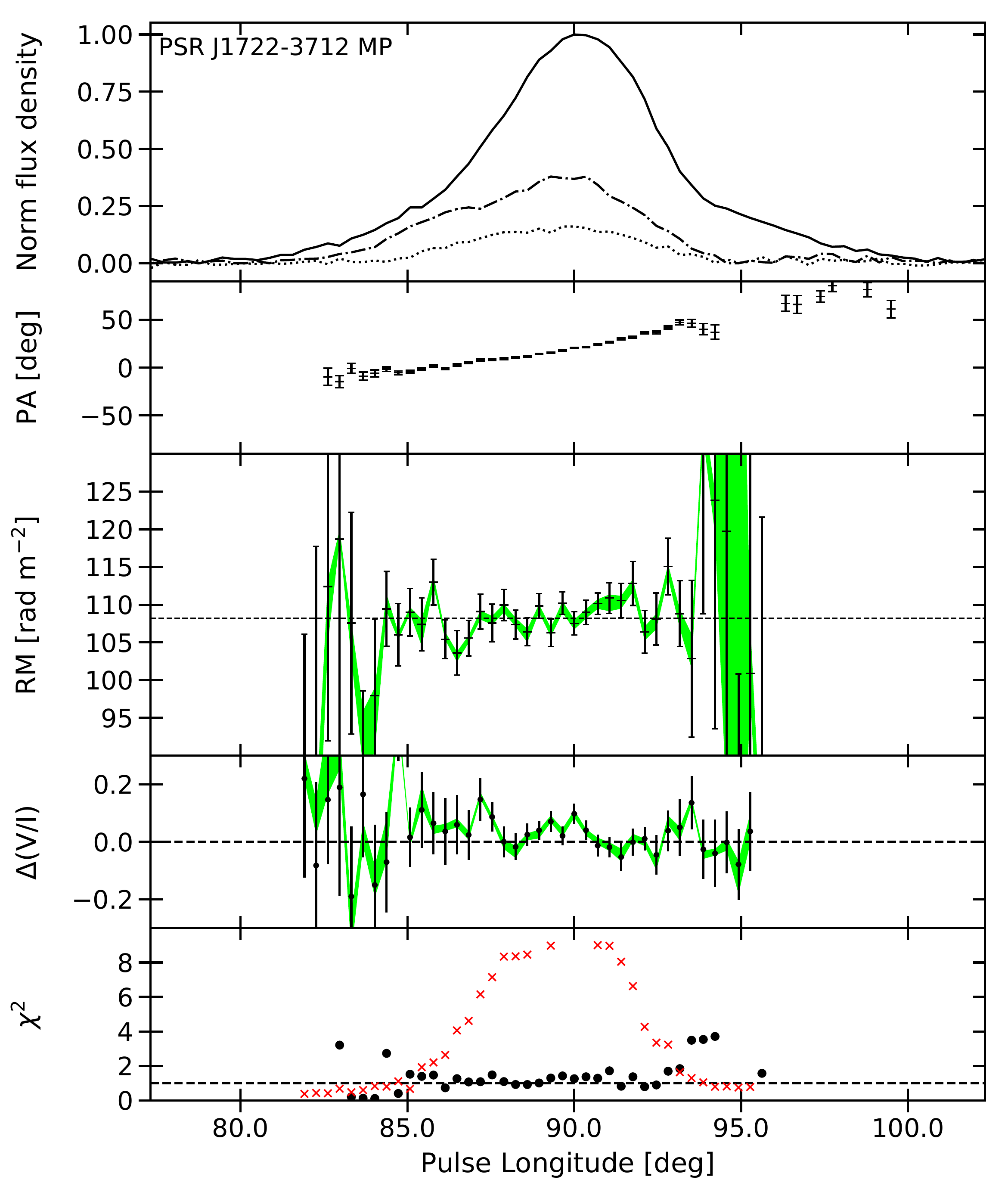} \\
\includegraphics[width=0.45\textwidth,angle=0]{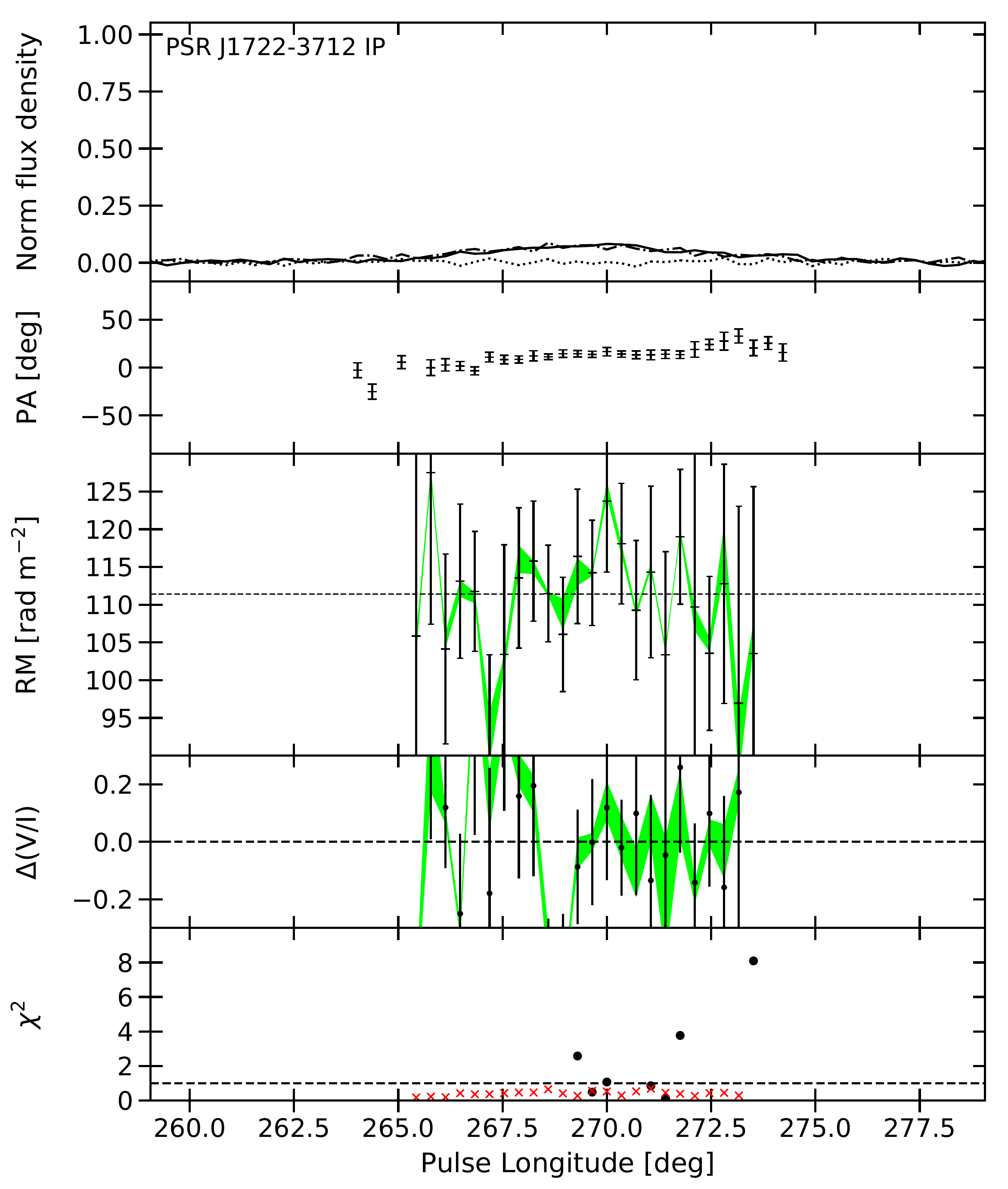} &
\includegraphics[width=0.45\textwidth,angle=0]{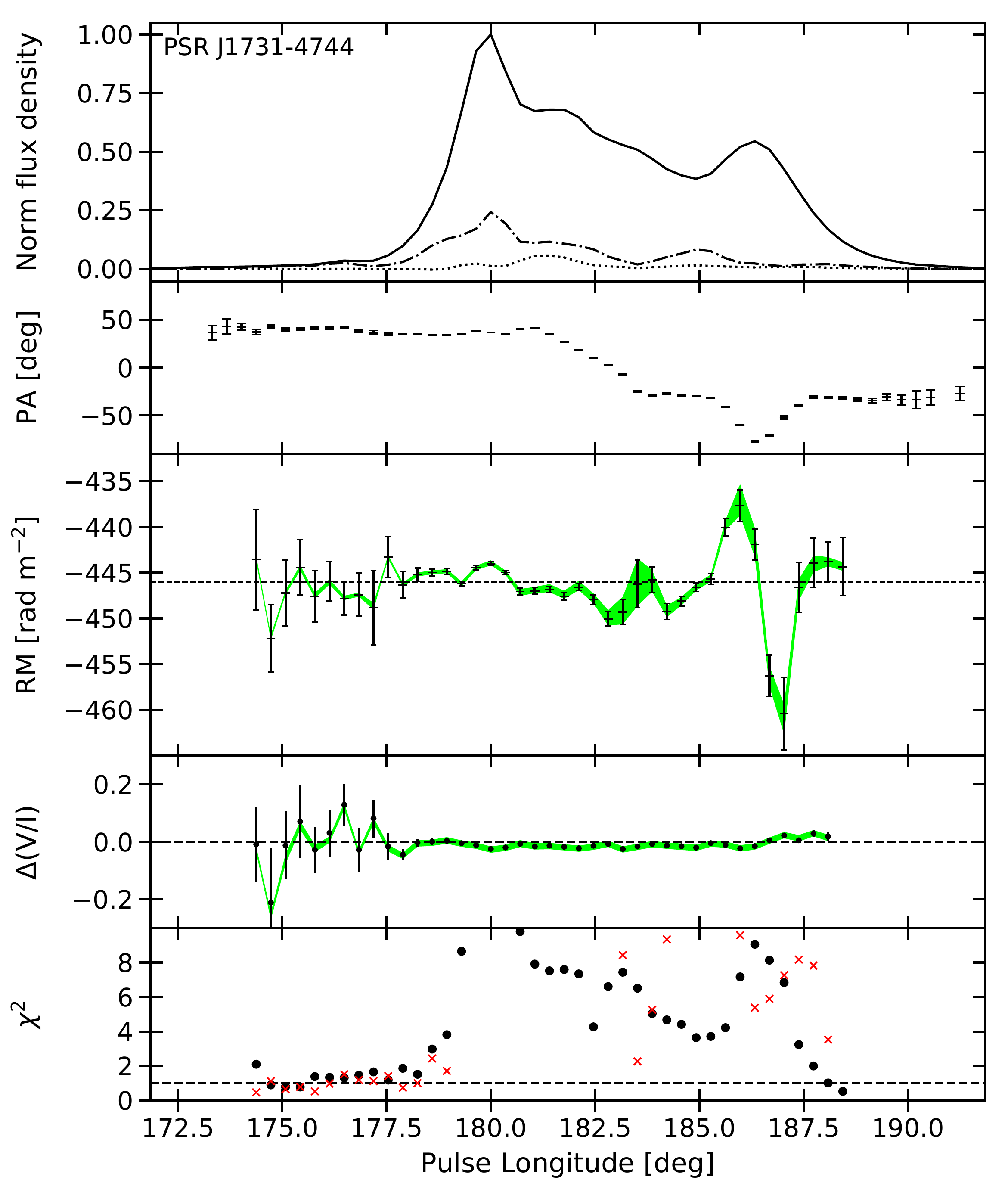} \\
\end{tabular}
\end{center}
\caption{Phase-resolved RM plots for PSR J1722$-$3207, PSR J1722$-$3712 (MP), PSR J1722$-$3712 (IP) and PSR J1731$-$4744. For more details on what is displayed in the individual panels, see Fig. 1.}
\label{fig:allPSR19}
\end{figure}

\begin{figure}
\begin{center}
\begin{tabular}{cc}
\includegraphics[width=0.45\textwidth,angle=0]{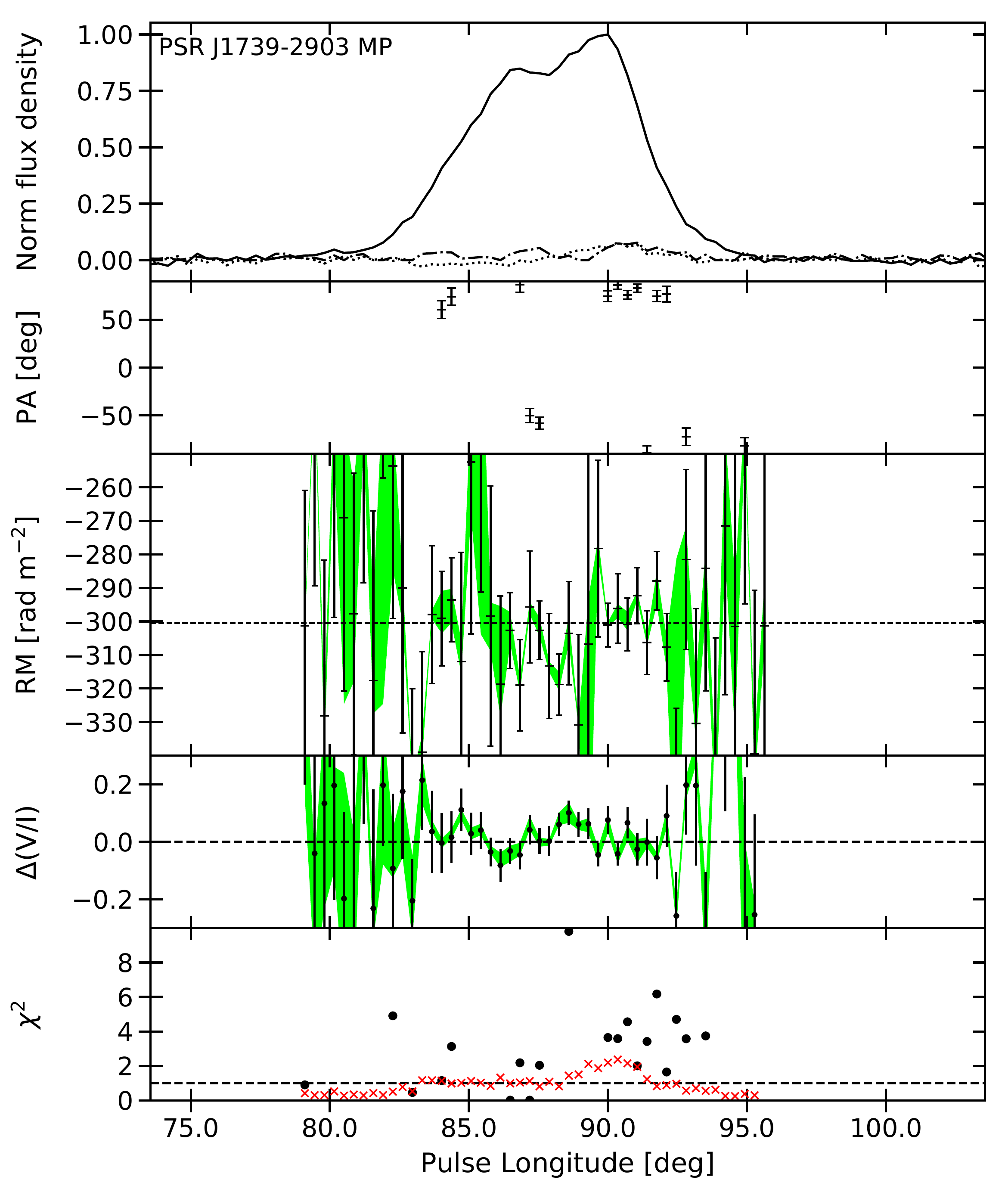} &
\includegraphics[width=0.45\textwidth,angle=0]{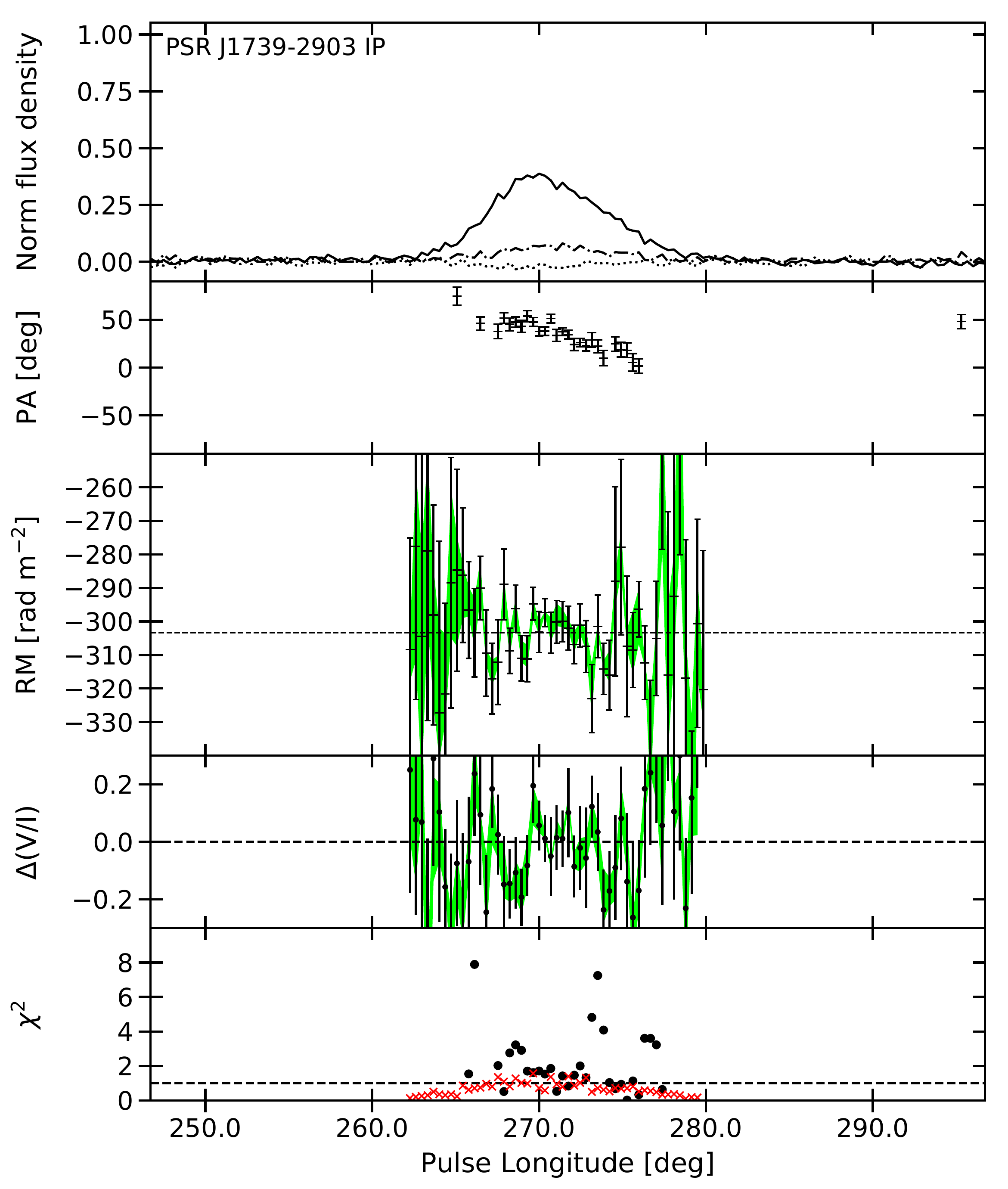} \\
\includegraphics[width=0.45\textwidth,angle=0]{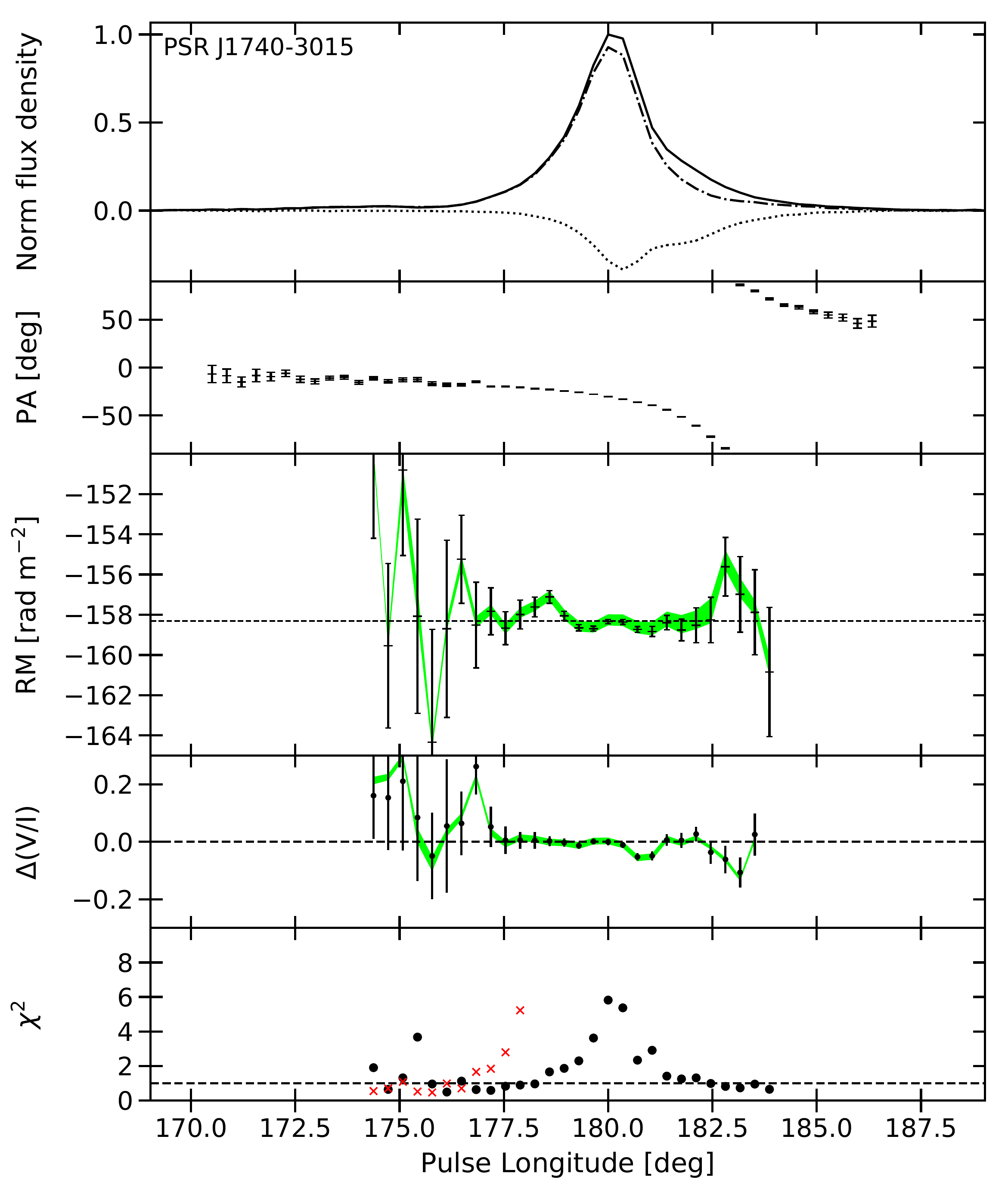} &
\includegraphics[width=0.45\textwidth,angle=0]{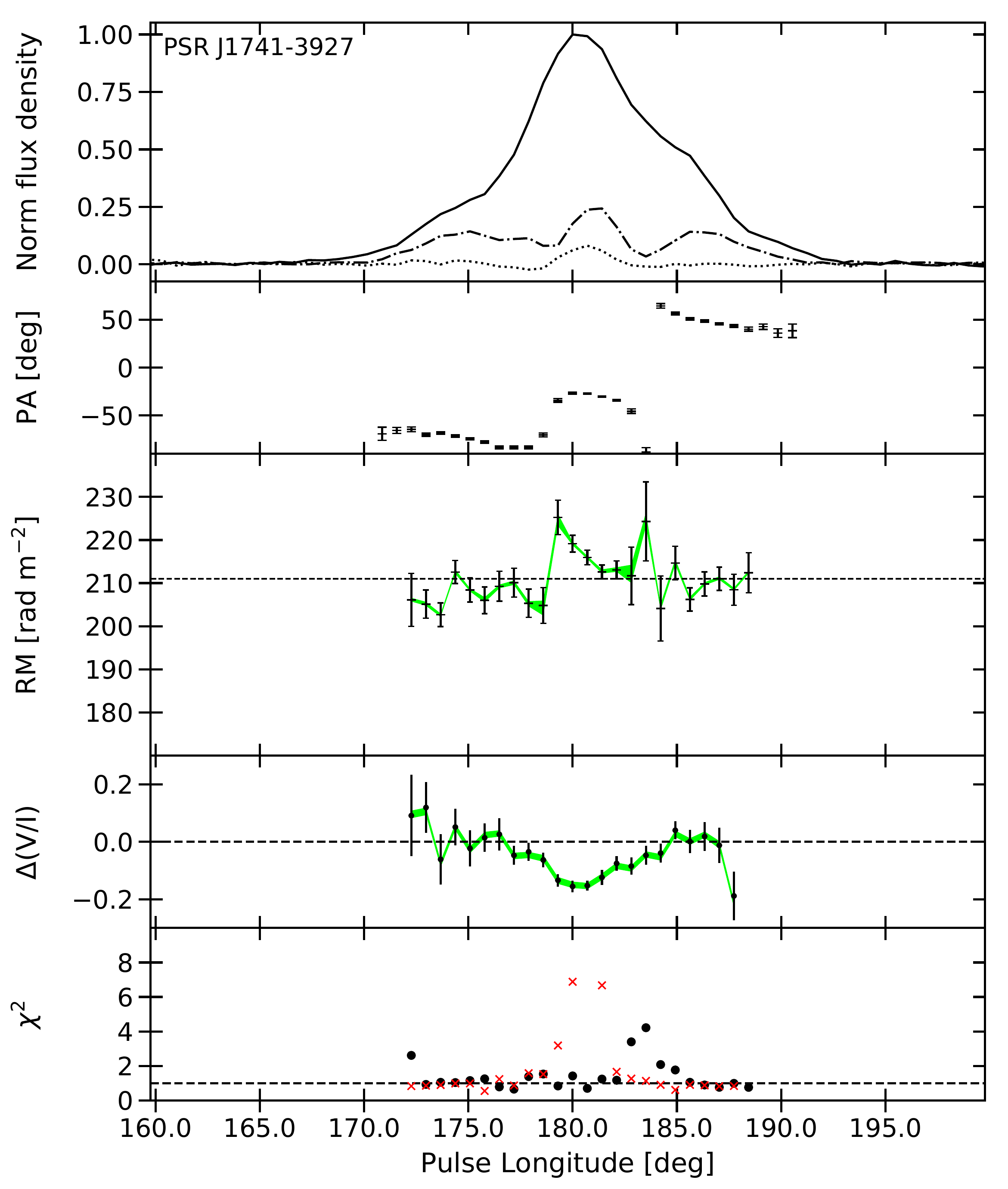} \\
\end{tabular}
\end{center}
\caption{Phase-resolved RM plots for PSR J1739$-$2903 (MP), PSR J1739$-$2903 (IP), PSR J1740$-$3015 and PSR J1741$-$3927. For more details on what is displayed in the individual panels, see Fig. 1.}
\label{fig:allPSR20}
\end{figure}

\begin{figure}
\begin{center}
\begin{tabular}{cc}
\includegraphics[width=0.45\textwidth,angle=0]{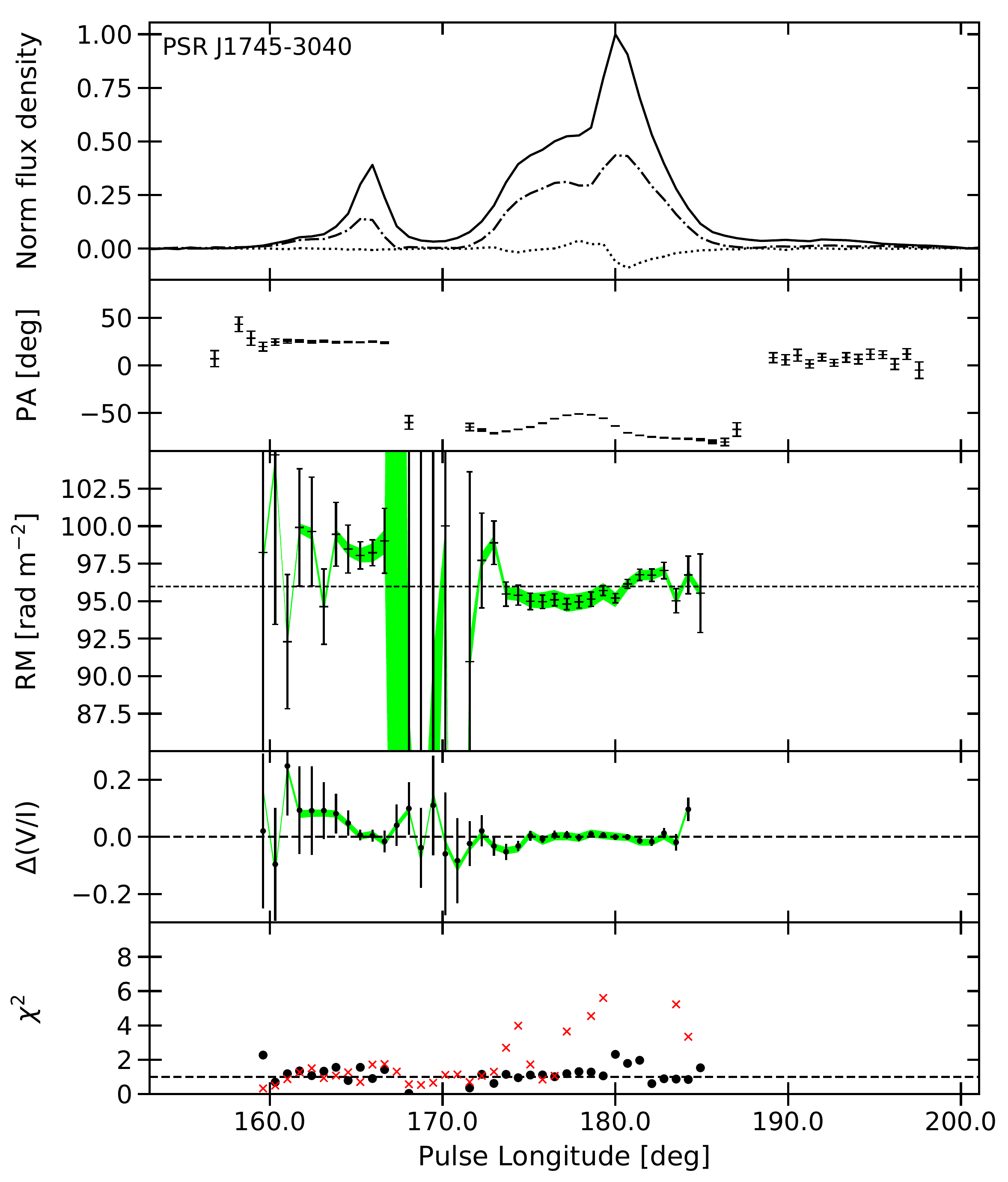}  &
\includegraphics[width=0.45\textwidth,angle=0]{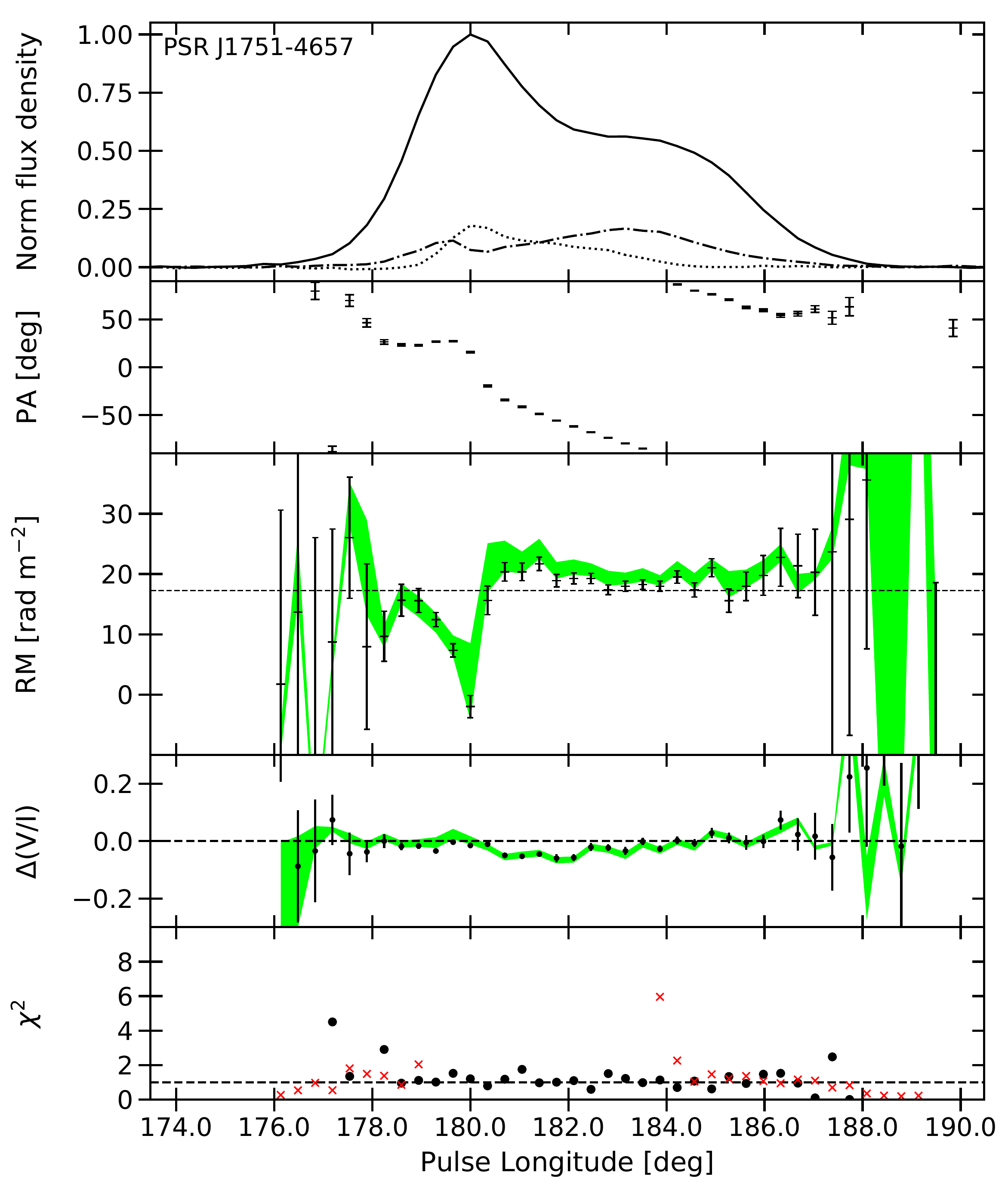} \\
\includegraphics[width=0.45\textwidth,angle=0]{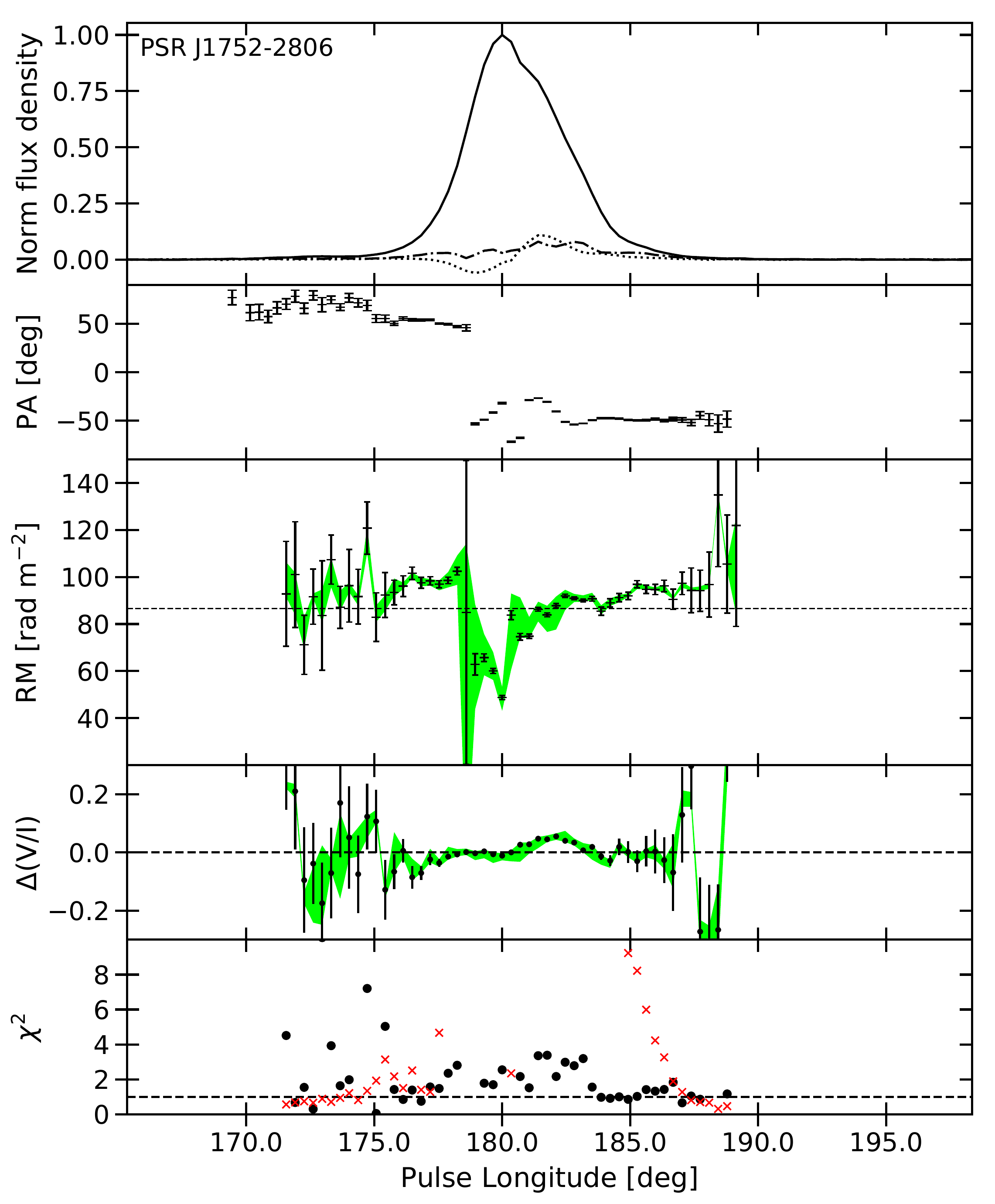} &
\includegraphics[width=0.45\textwidth,angle=0]{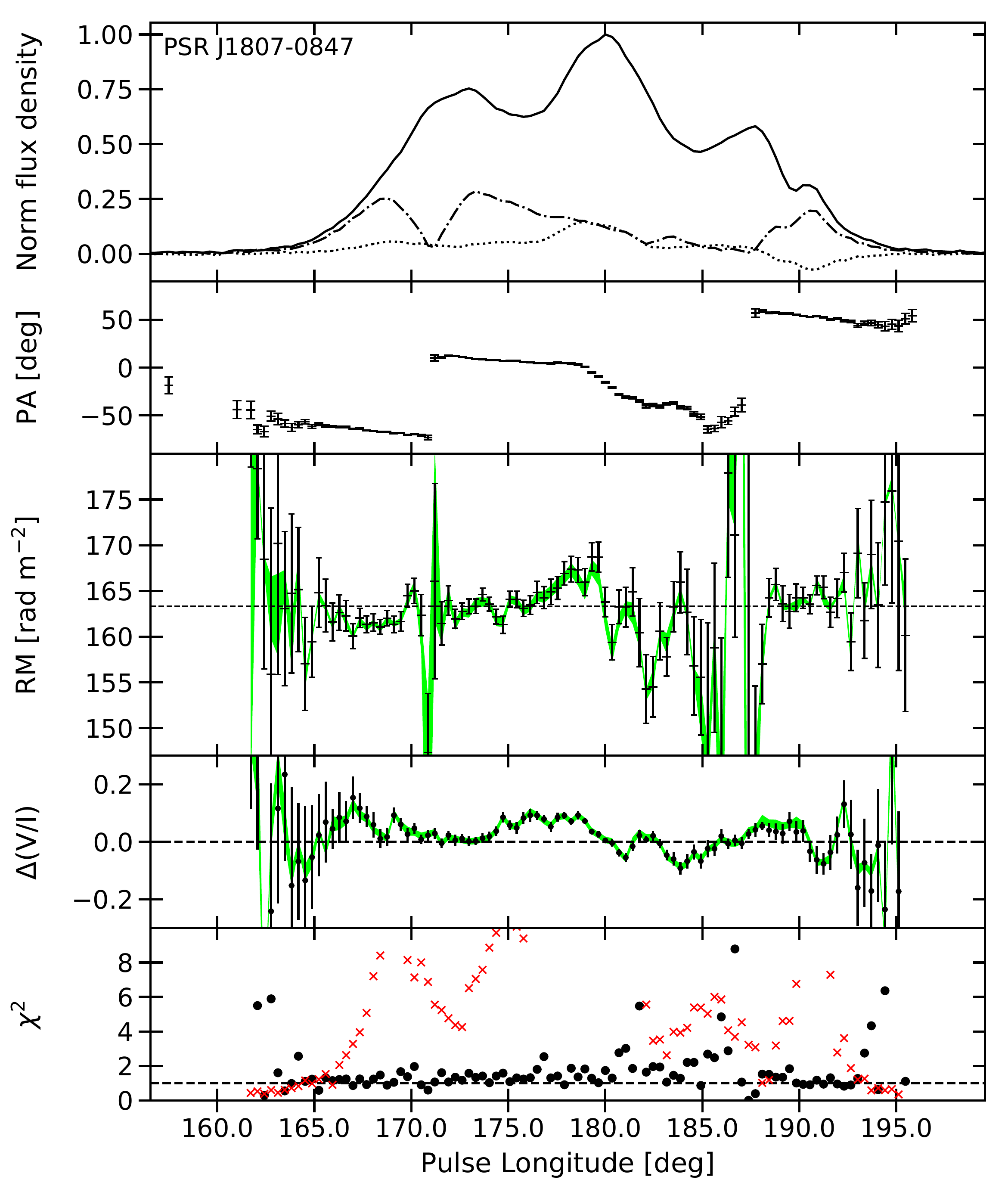} \\
\end{tabular}
\end{center}
\caption{Phase-resolved RM plots for PSR J1745$-$3040, PSR J1751$-$4657, PSR J1752$-$2806, and PSR J1807$-$0847. For more details on what is displayed in the individual panels, see Fig. 1.}
\label{fig:allPSR21}
\end{figure}

\begin{figure}
\begin{center}
\begin{tabular}{cc}
\includegraphics[width=0.45\textwidth,angle=0]{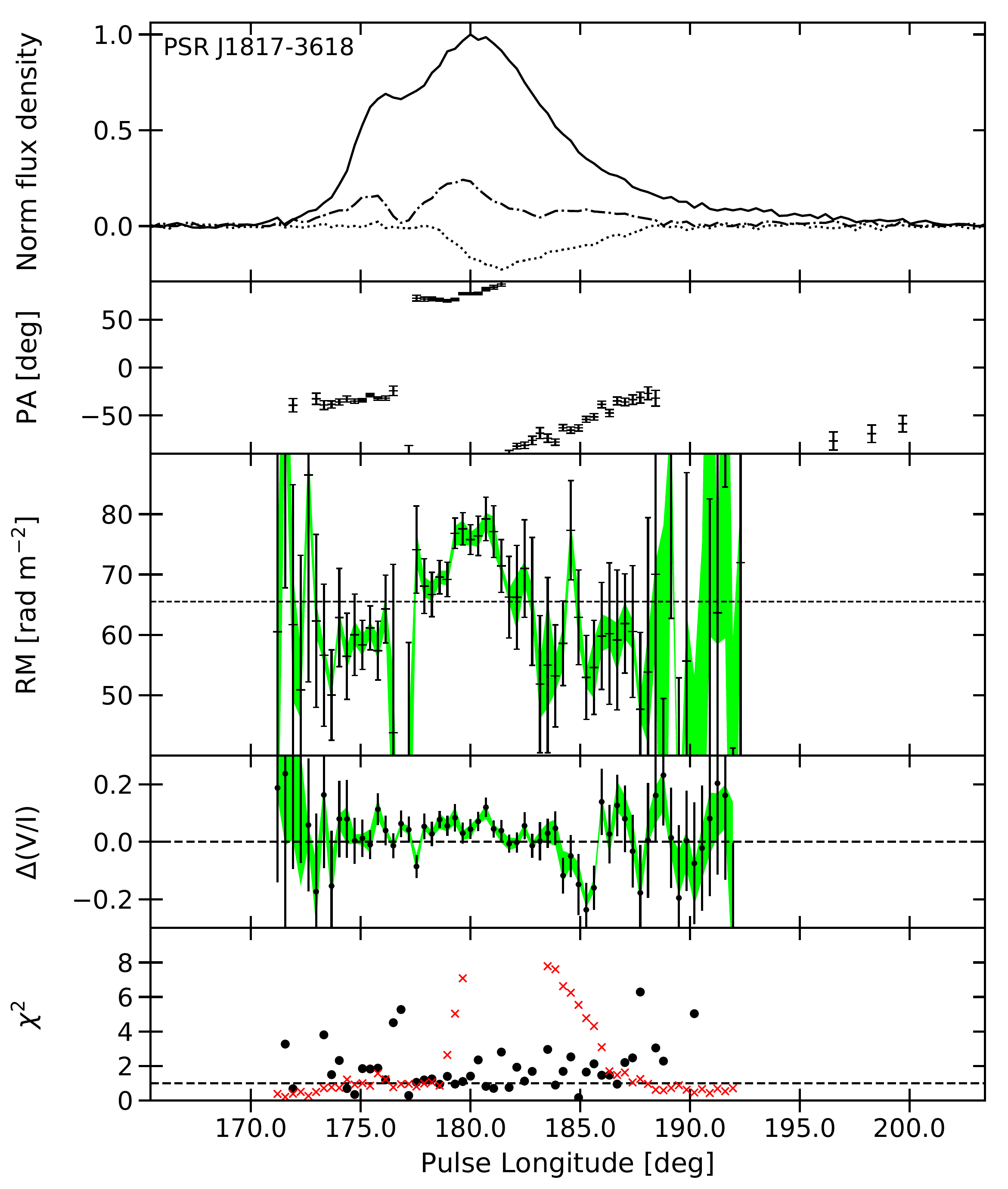} &
\includegraphics[width=0.45\textwidth,angle=0]{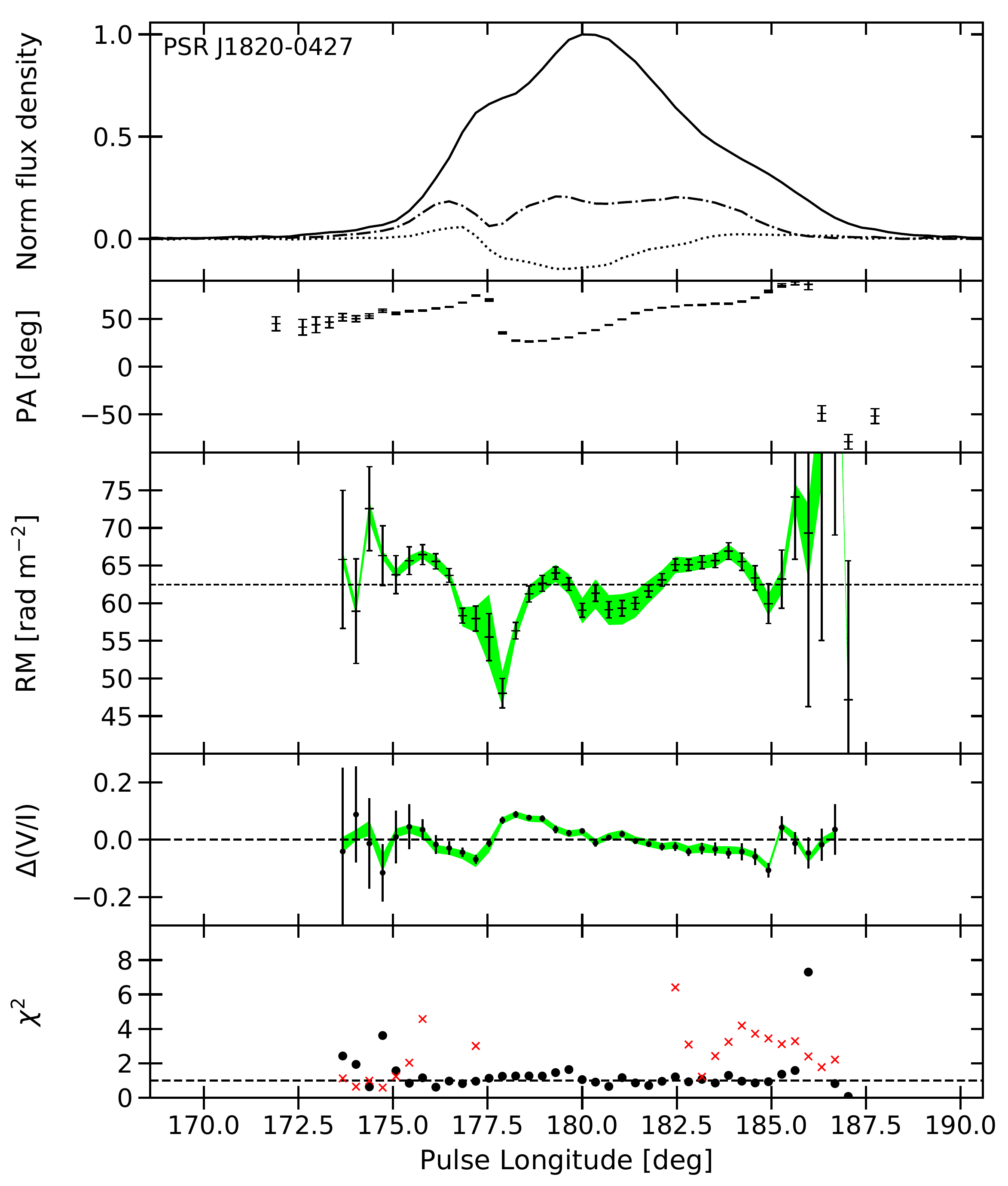} \\
\includegraphics[width=0.45\textwidth,angle=0]{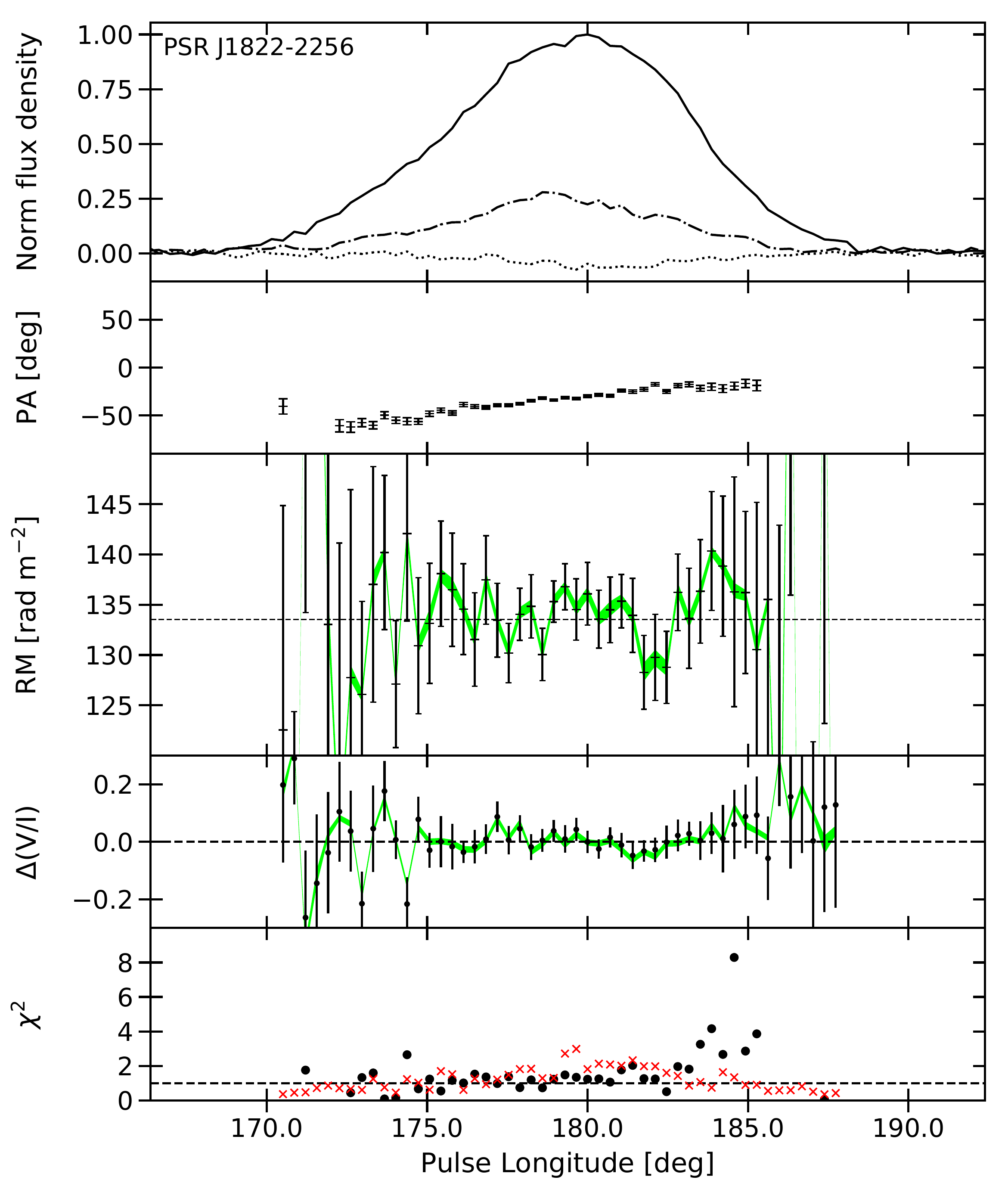} &
\includegraphics[width=0.45\textwidth,angle=0]{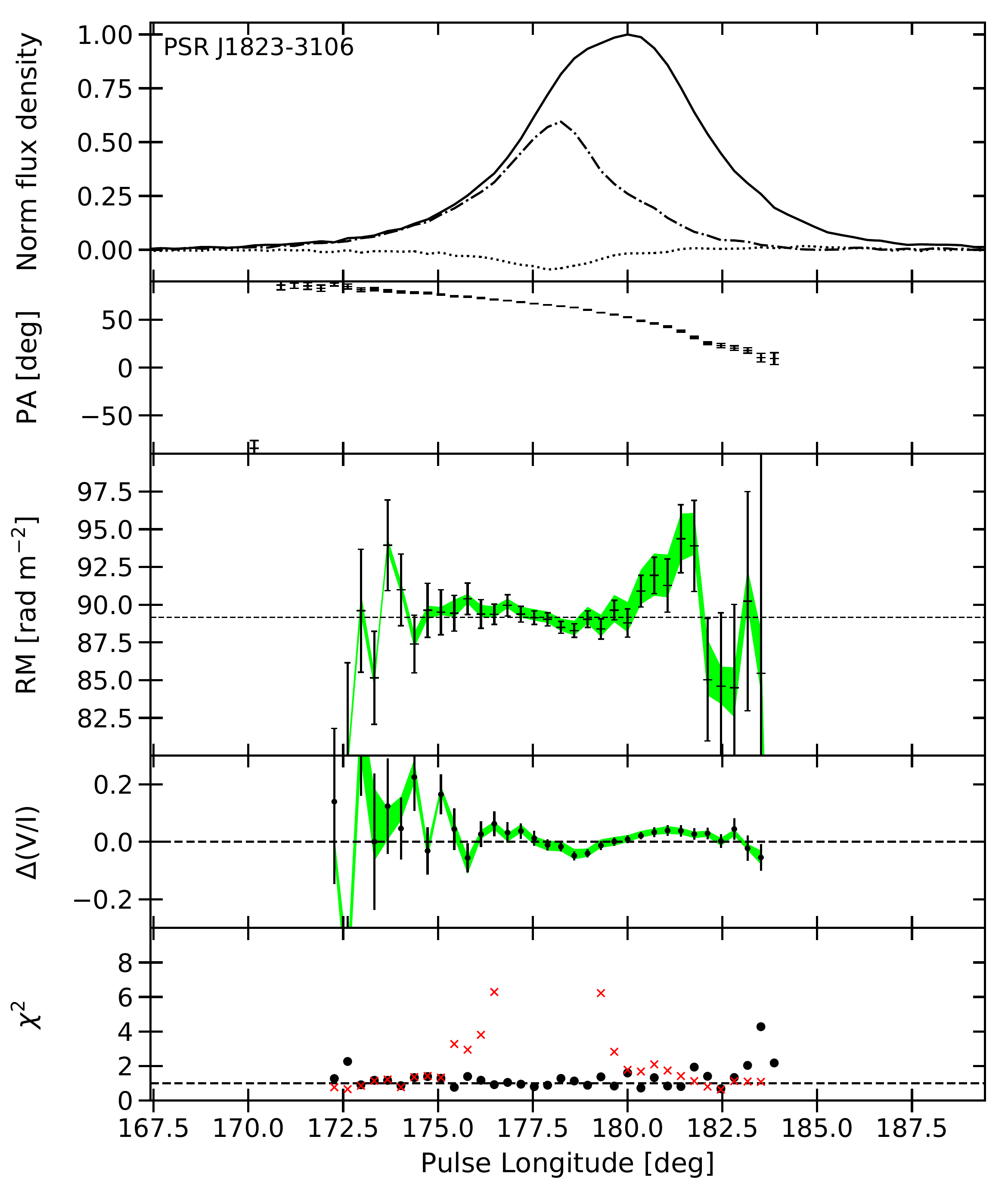} \\
\end{tabular}
\end{center}
\caption{Phase-resolved RM plots for PSR J1817$-$3618, PSR J1820$-$0427, PSR J1822$-$2256 and PSR J1823$-$3106. For more details on what is displayed in the individual panels, see Fig. 1.}
\label{fig:allPSR22}
\end{figure}

\begin{figure}
\begin{center}
\begin{tabular}{cc}
\includegraphics[width=0.45\textwidth,angle=0]{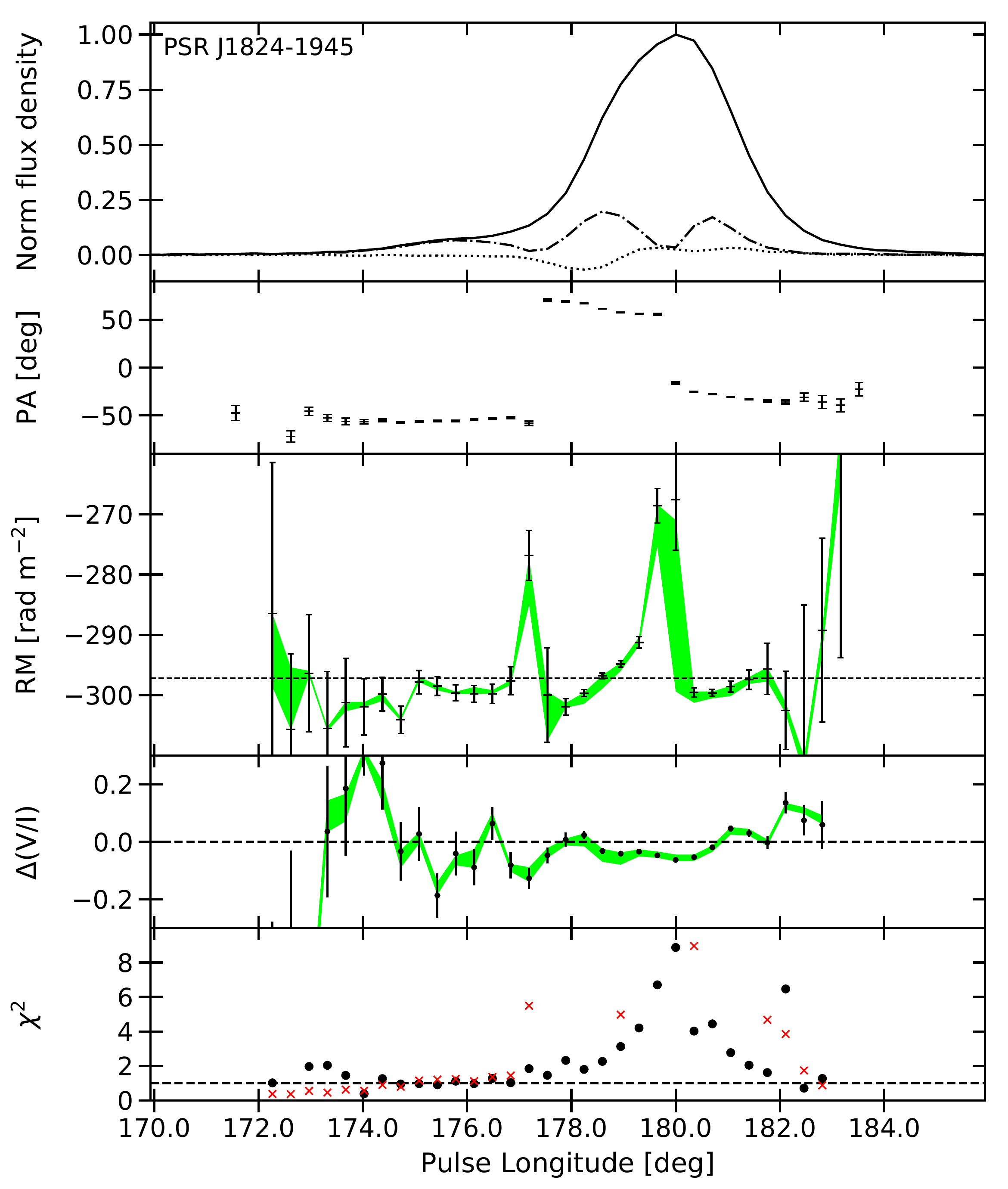} &
\includegraphics[width=0.45\textwidth,angle=0]{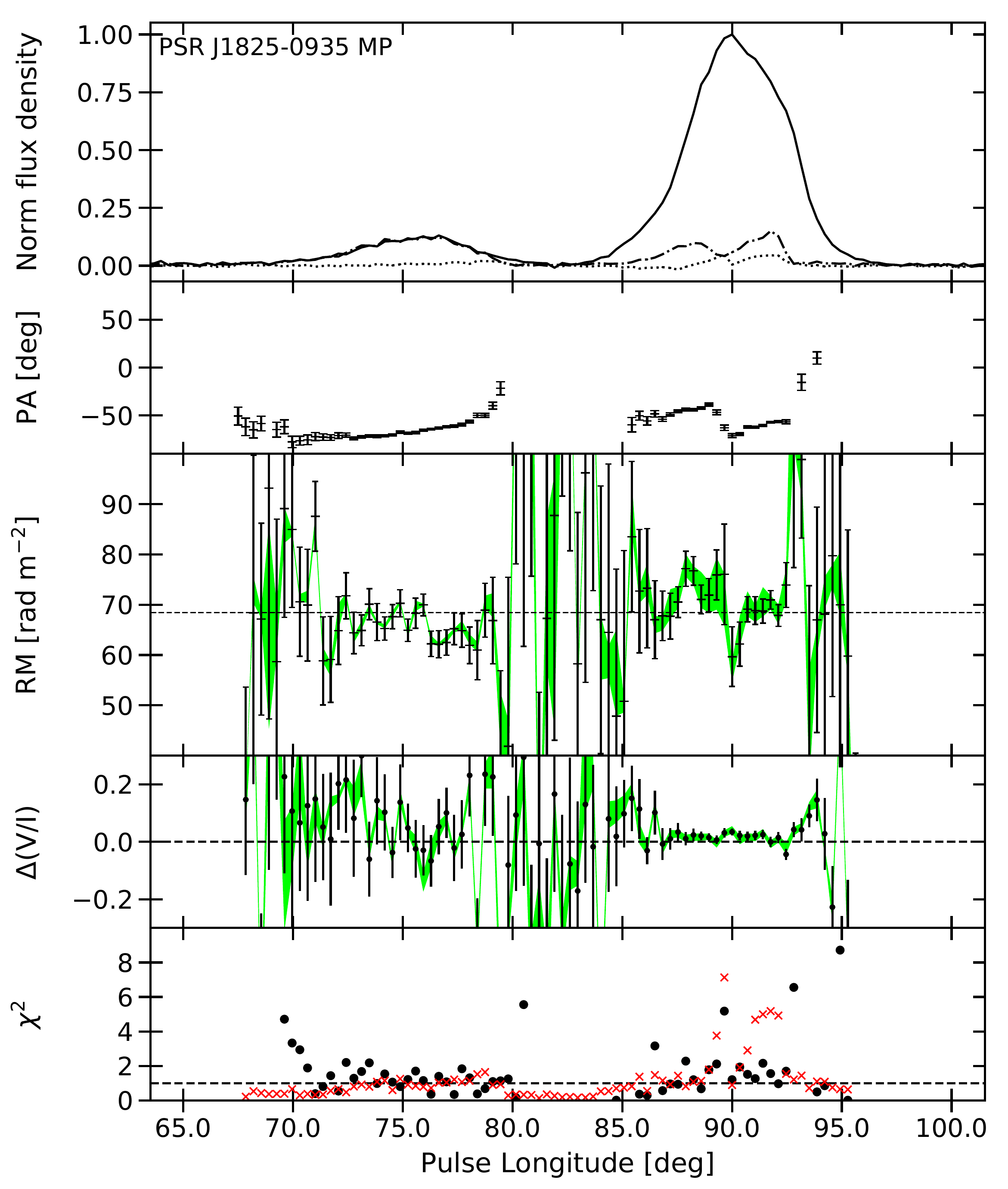} \\
\includegraphics[width=0.45\textwidth,angle=0]{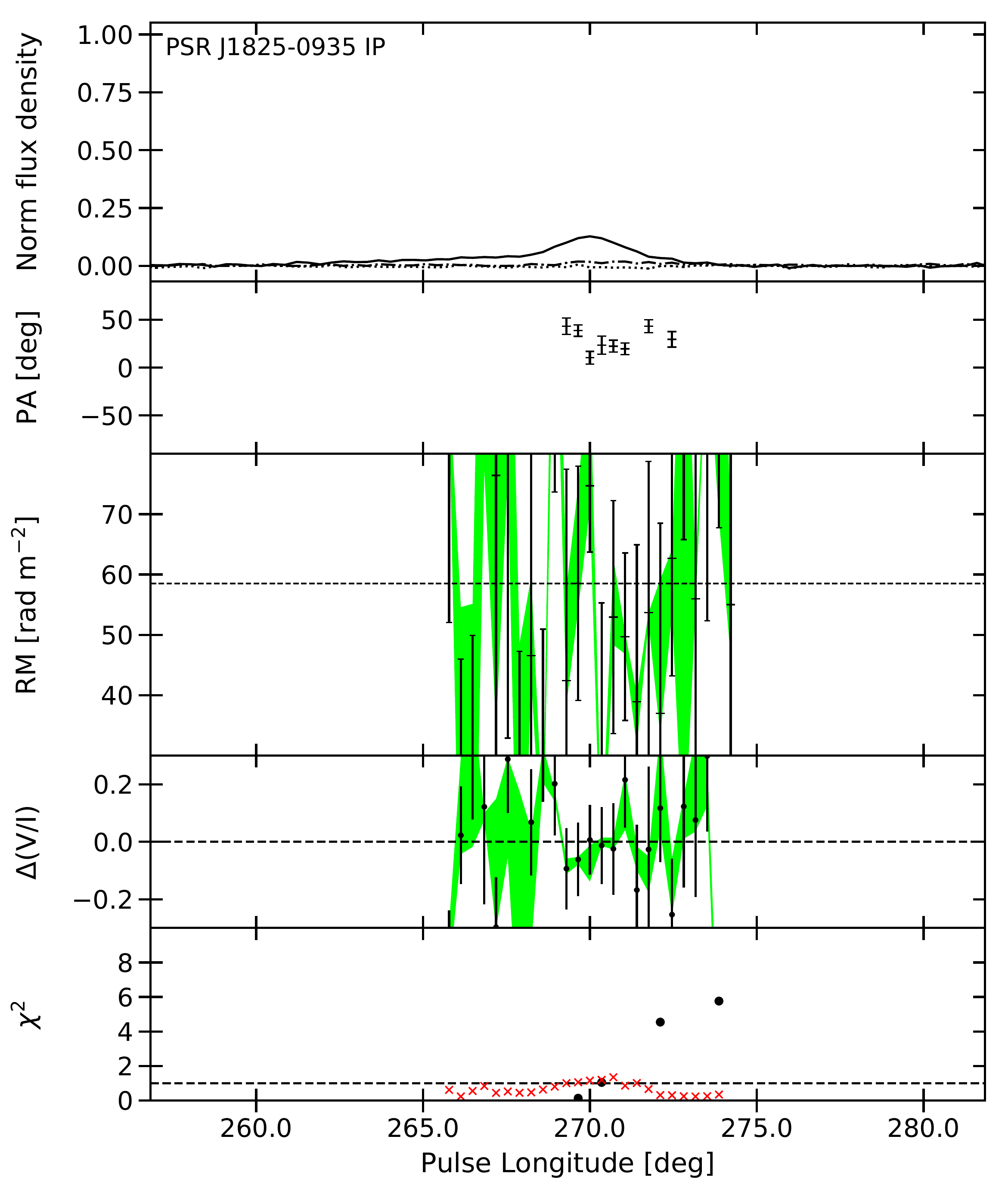} &
\includegraphics[width=0.45\textwidth,angle=0]{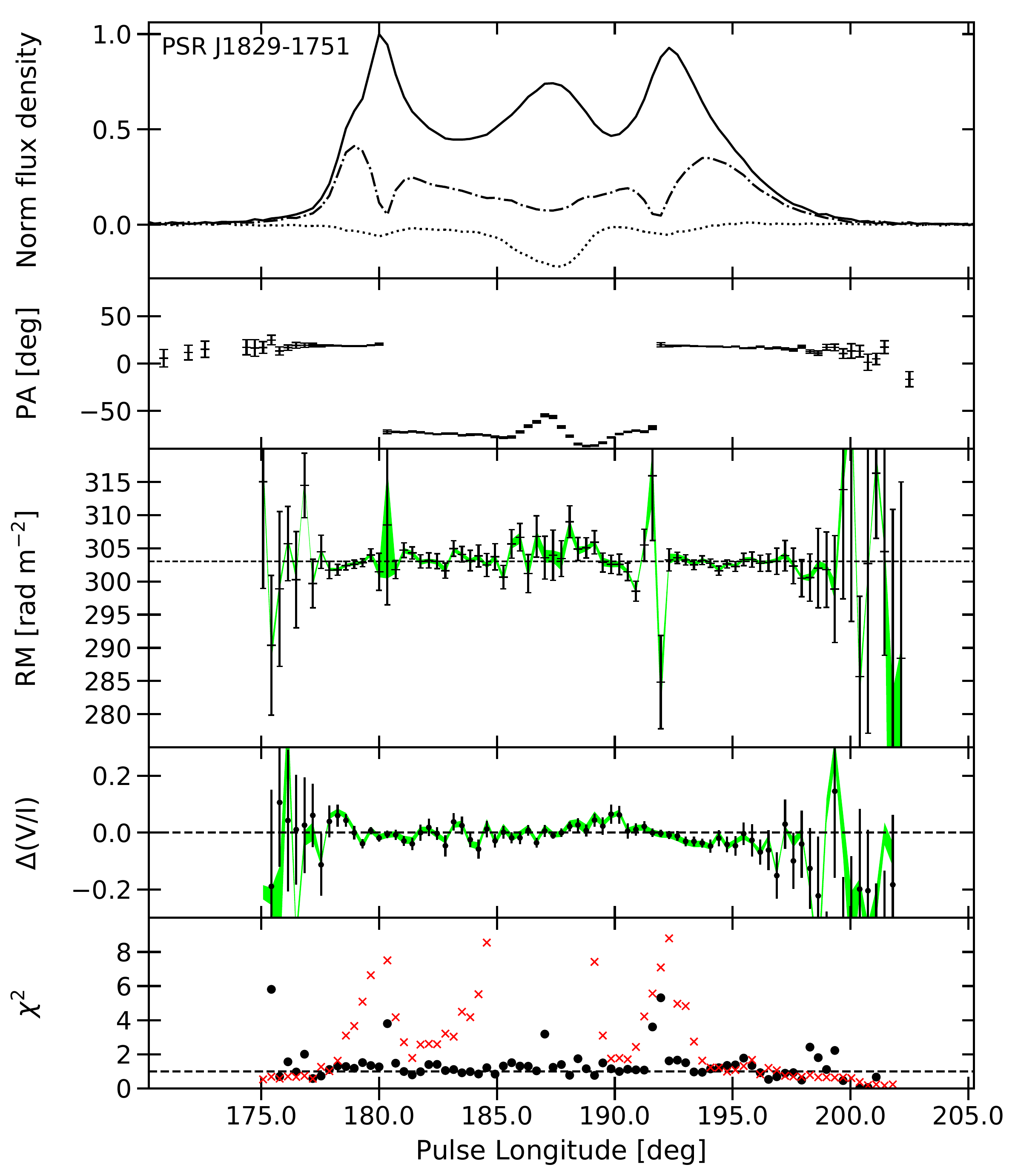} \\
\end{tabular}
\end{center}
\caption{Phase-resolved RM plots for PSR J1824$-$1945, PSR J1825$-$0935 (MP), PSR J1825$-$0935 (IP) and PSR J1829$-$1751. For more details on what is displayed in the individual panels, see Fig. 1.}
\label{fig:allPSR23}
\end{figure}

\begin{figure}
\begin{center}
\begin{tabular}{cc}
\includegraphics[width=0.45\textwidth,angle=0]{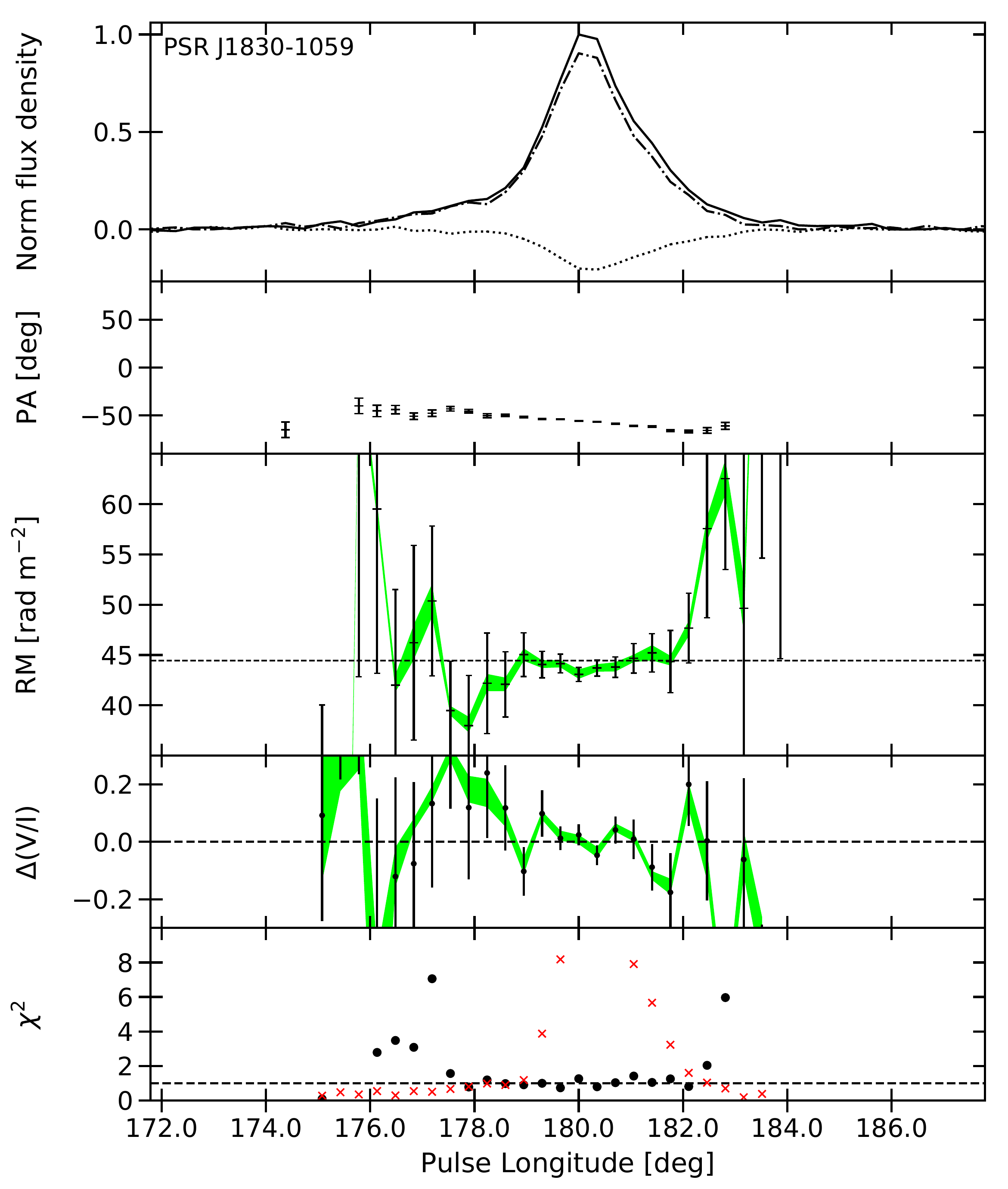} &
\includegraphics[width=0.45\textwidth,angle=0]{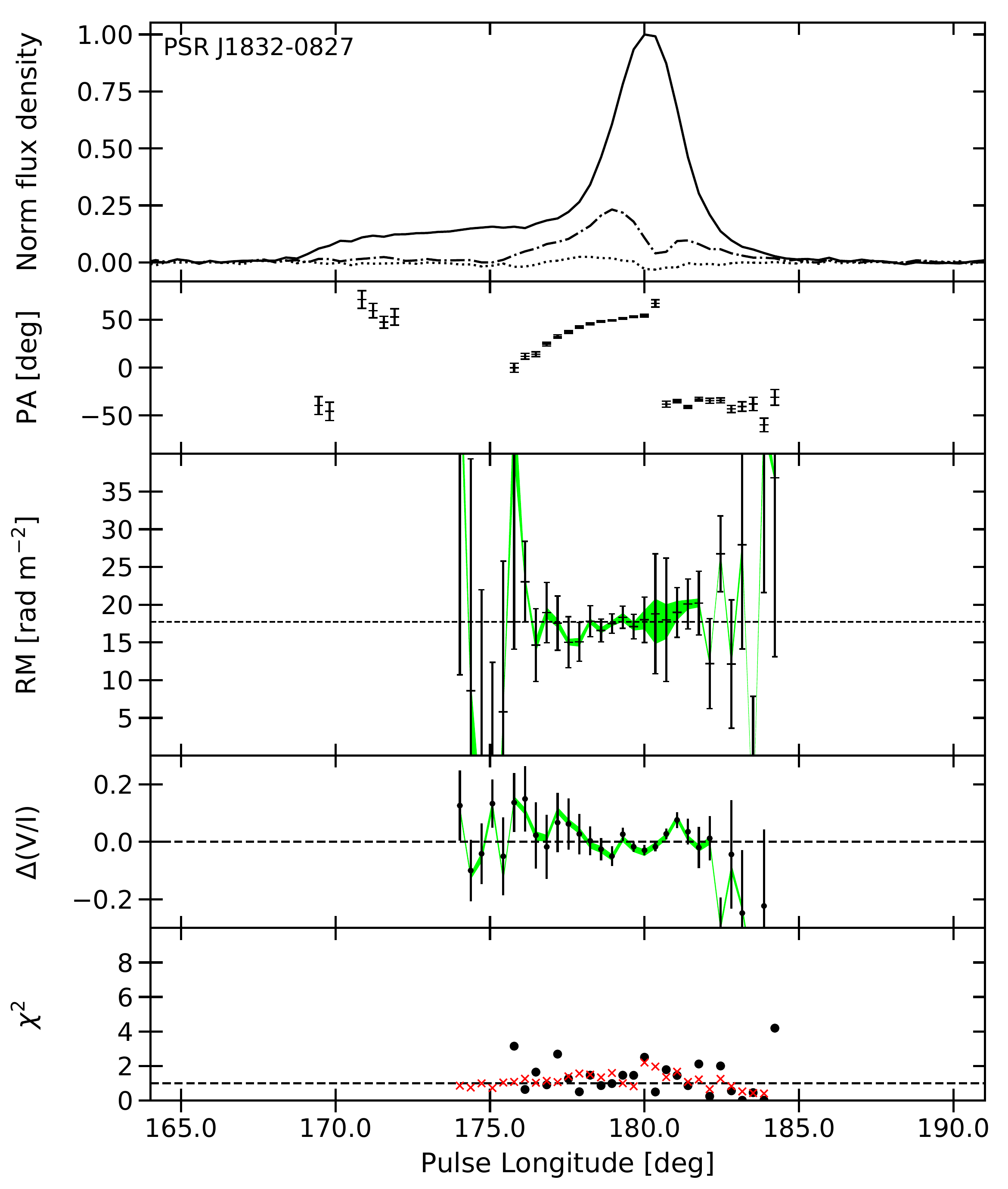} \\
\includegraphics[width=0.45\textwidth,angle=0]{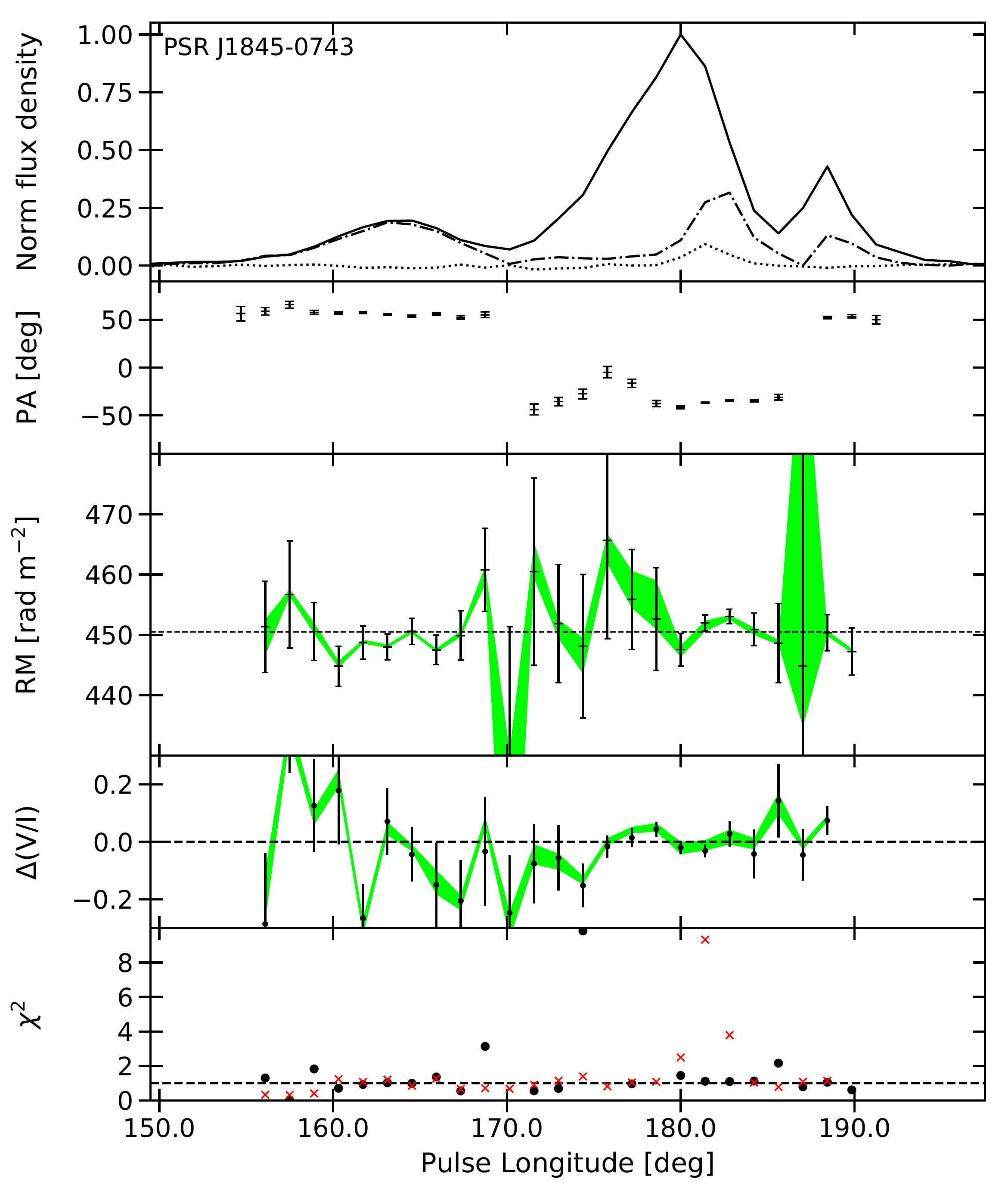} &
\includegraphics[width=0.45\textwidth,angle=0]{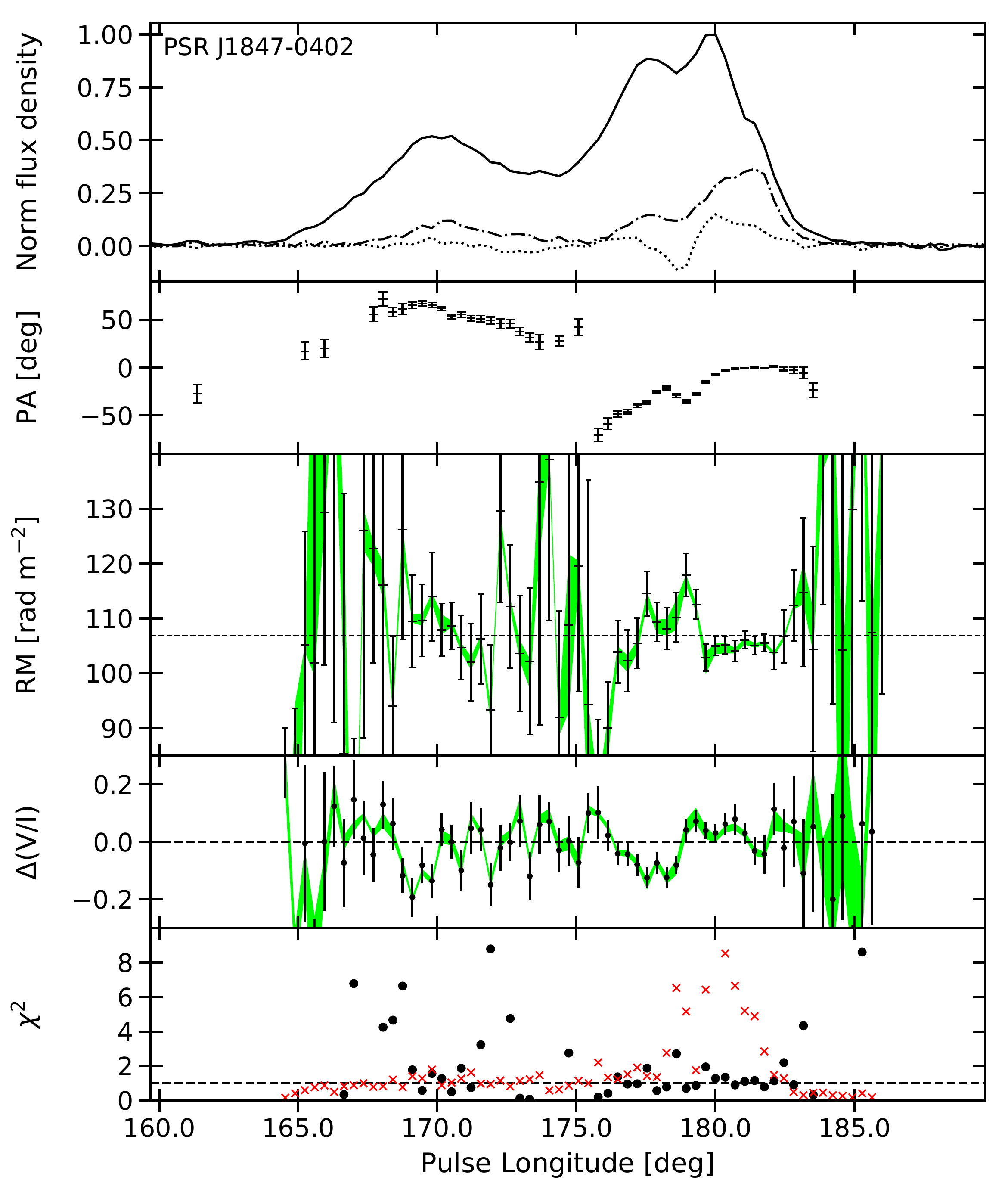} \\
\end{tabular}
\end{center}
\caption{Phase-resolved RM plots for PSR J1830$-$1059, PSR J1832$-$0827, PSR J1845$-$0743 and PSR J1847$-$0402. For more details on what is displayed in the individual panels, see Fig. 1.}
\label{fig:allPSR24}
\end{figure}

\begin{figure}
\begin{center}
\begin{tabular}{cc}
\includegraphics[width=0.45\textwidth,angle=0]{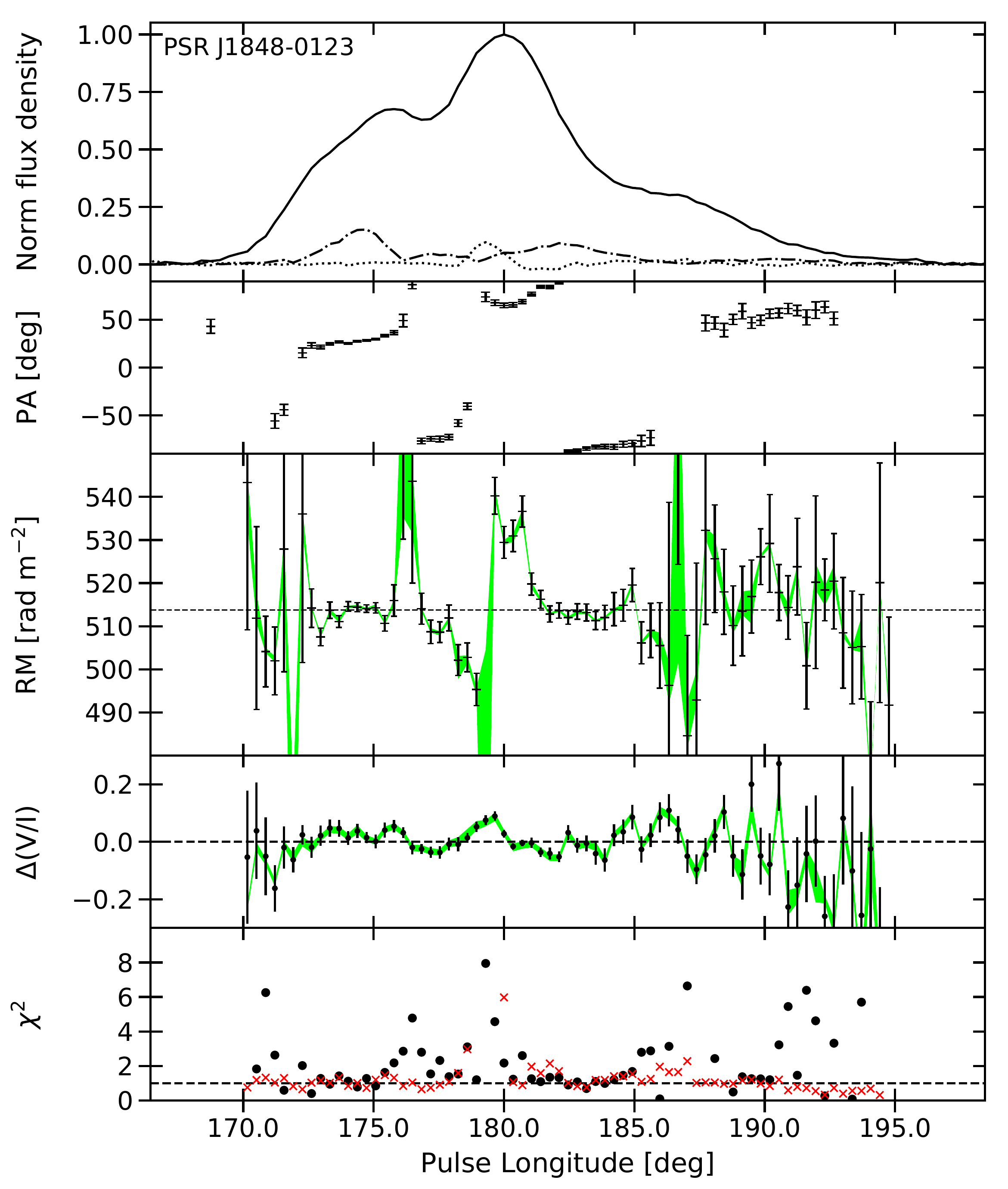} &
\includegraphics[width=0.45\textwidth,angle=0]{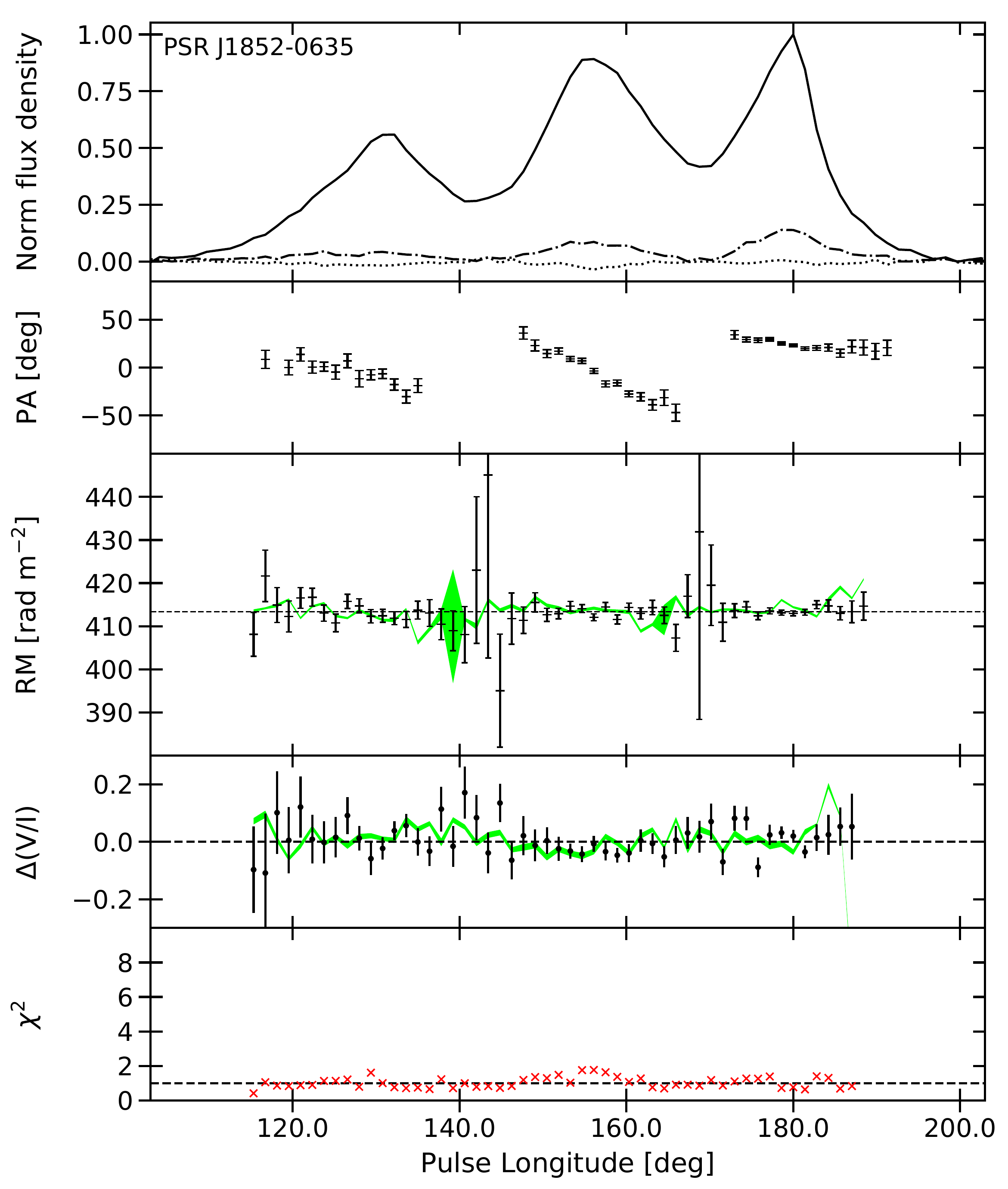} \\
\includegraphics[width=0.45\textwidth,angle=0]{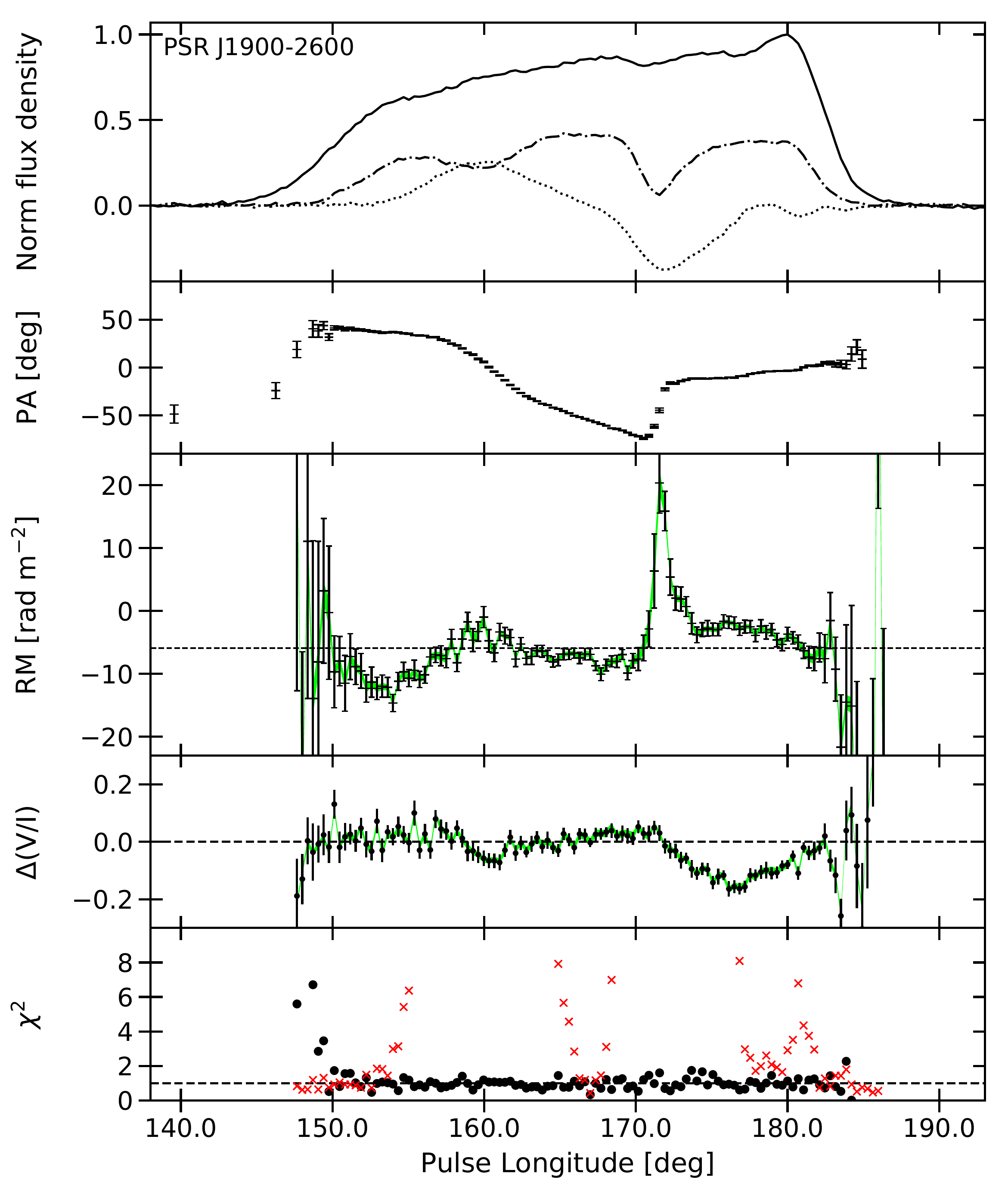} &
\includegraphics[width=0.45\textwidth,angle=0]{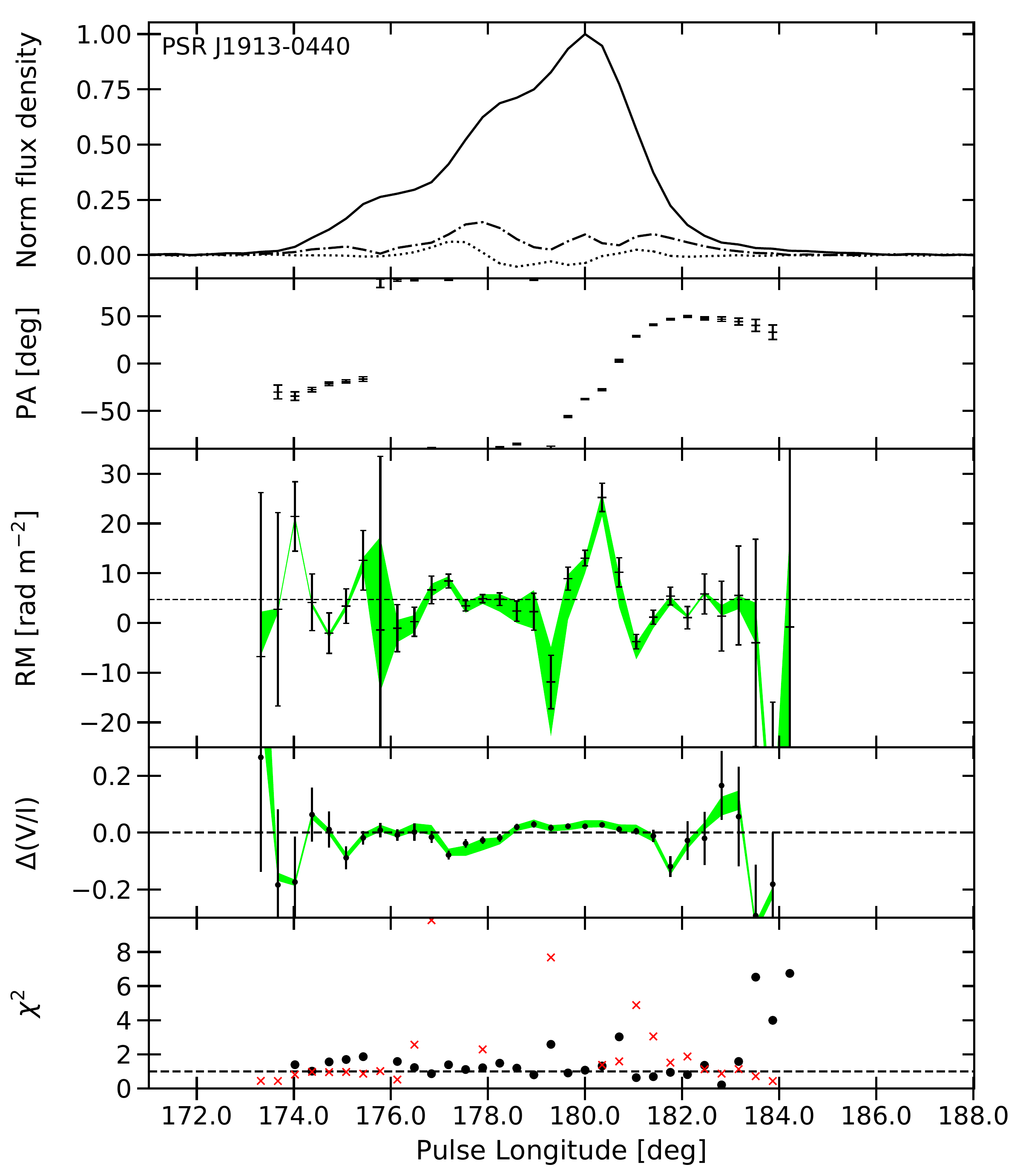} \\
\end{tabular}
\end{center}
\caption{Phase-resolved RM plots for PSR J1848$-$0123, PSR J1852$-$0635, PSR J1900$-$2600 and PSR J1913$-$0440. For more details on what is displayed in the individual panels, see Fig. 1.}
\label{fig:allPSR25}
\end{figure}

\begin{figure}
\begin{center}
\begin{tabular}{cc}
\includegraphics[width=0.45\textwidth,angle=0]{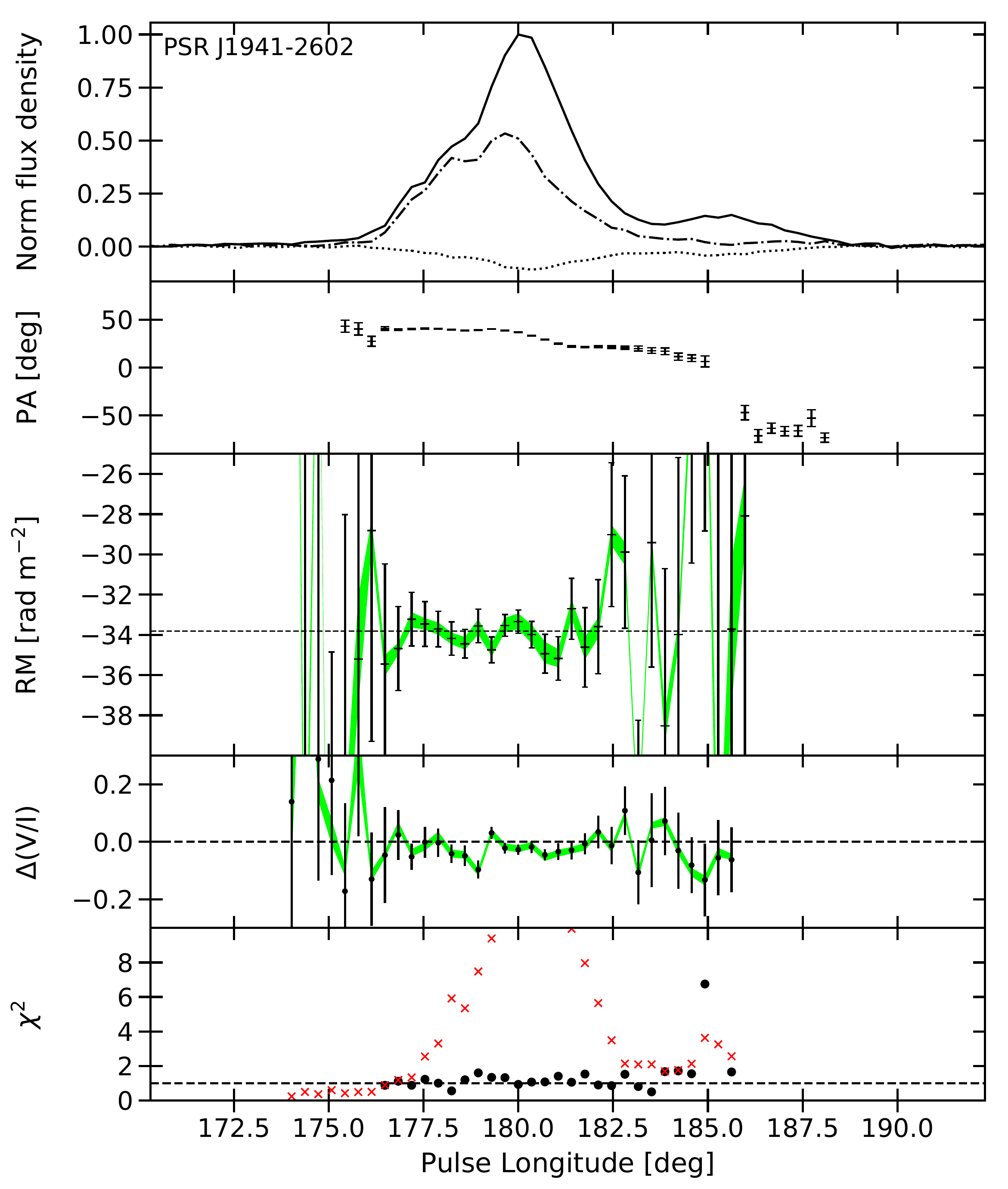} &
\includegraphics[width=0.45\textwidth,angle=0]{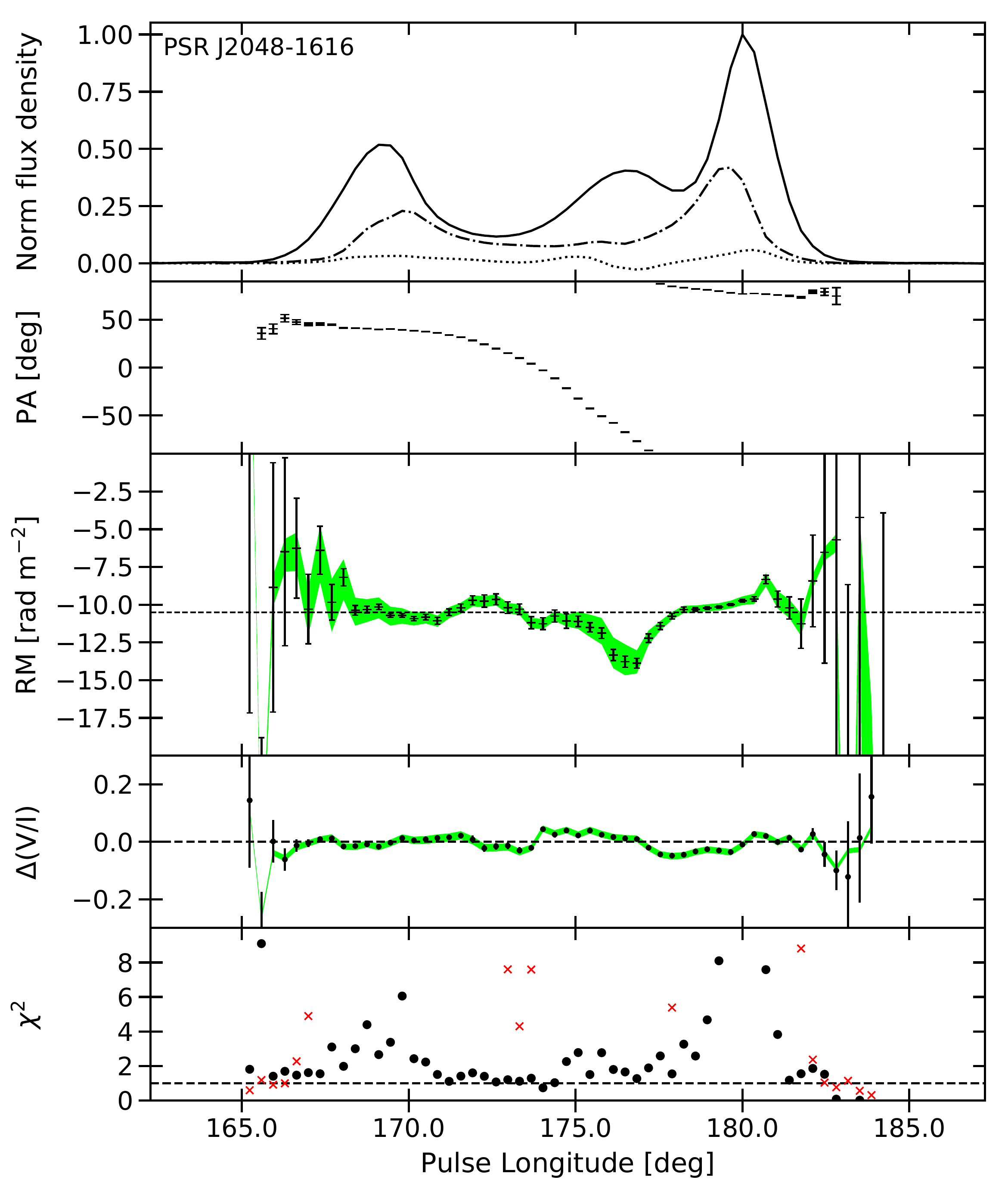} \\
\includegraphics[width=0.45\textwidth,angle=0]{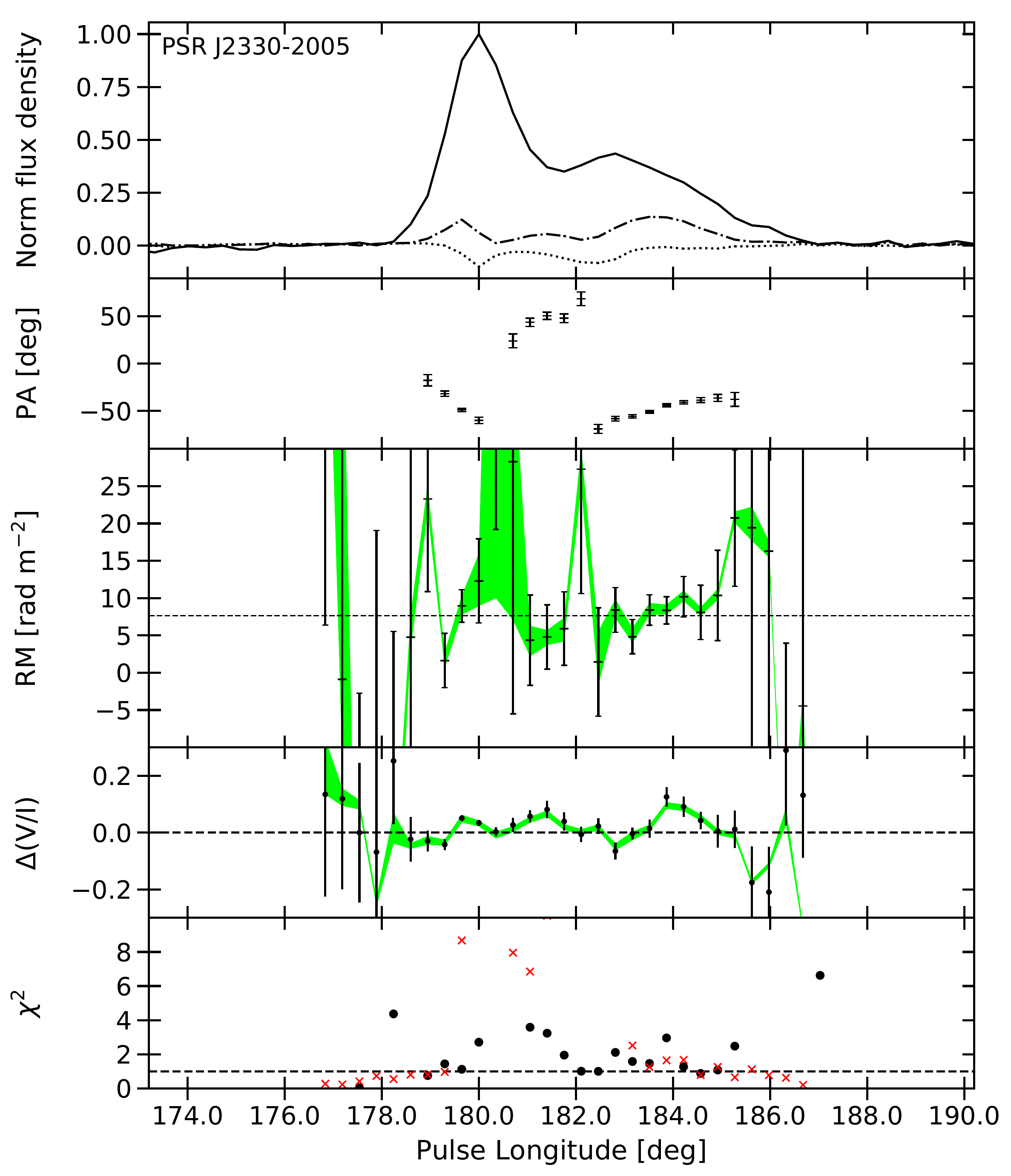} &
\includegraphics[width=0.45\textwidth,angle=0]{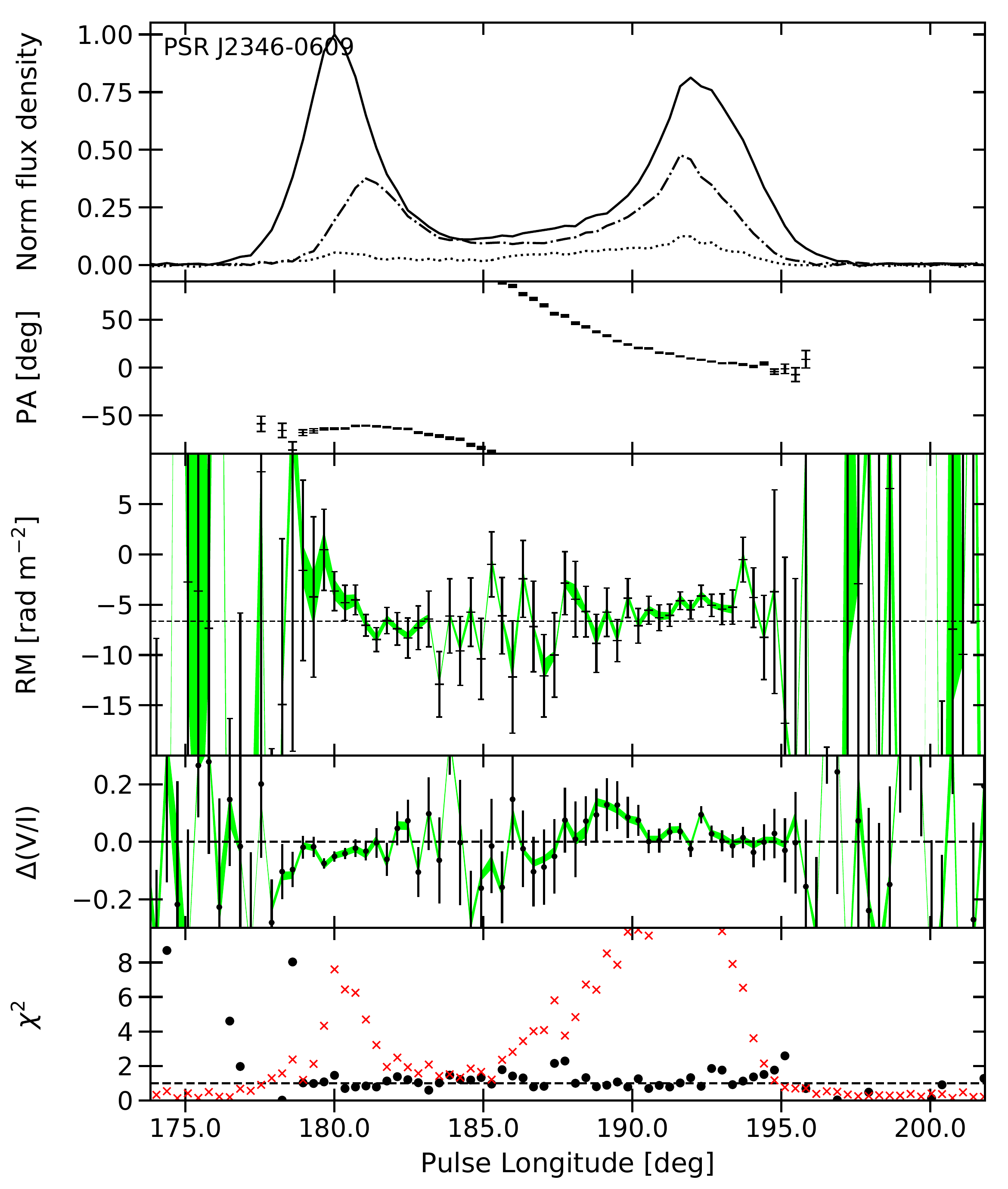} \\
\end{tabular}
\end{center}
\caption{Phase-resolved RM plots for PSR J1941$-$2602, PSR J2048$-$1616, PSR J2330$-$2005 and PSR J2346$-$0609. For more details on what is displayed in the individual panels, see Fig. 1.}
\label{fig:allPSR26}
\end{figure}

In this section, we display the resulting phase-resolved RM plots for the 98 pulsars, in Fig.~\ref{fig:allPSR1}$-$Fig.~\ref{fig:allPSR26}. In each Figure, we display four plots. These plots were aligned so that the total intensity peaked at pulse longitude $180^{\circ}$. For the pulsars which had both a MP and an IP, we aligned them so that the MP peaked at pulse longitude $90^{\circ}$ the IP peaked at longitude $270^{\circ}$.
In each plot, in the top panel, the solid line denotes Stokes $I$, the dashed line shows $L$ and the dotted line Stokes $V$. 
The second panel displays the PA, when $L$ exceeded 2 sigma. 
The third panel shows the RM($\phi$) curve with the associated statistical uncertainties. The shaded region (green in the online version) represents the $1\sigma$ systematic uncertainty contour region. The horizontal dotted line plotted is $\left<\text{RM}(\phi)\right>$. 
The fourth panel shows the phase-resolved $\Delta(V/I)$ values with their associated statistical and systematic uncertainties. 
The bottom panel displays the $\chi^2_{\text{PA}\left(\lambda^2,\phi\right)}$, represented by the black circles  and $\chi^2_{\text{V/I}\left(\nu,\phi\right)}$, represented by the red crosses. The horizontal line corresponds to a reduced $\chi^2$ of 1.

\section{Systematic uncertainties of Vela from RM$_\text{iono}$}

\begin{figure}
\includegraphics[width=0.45\textwidth]{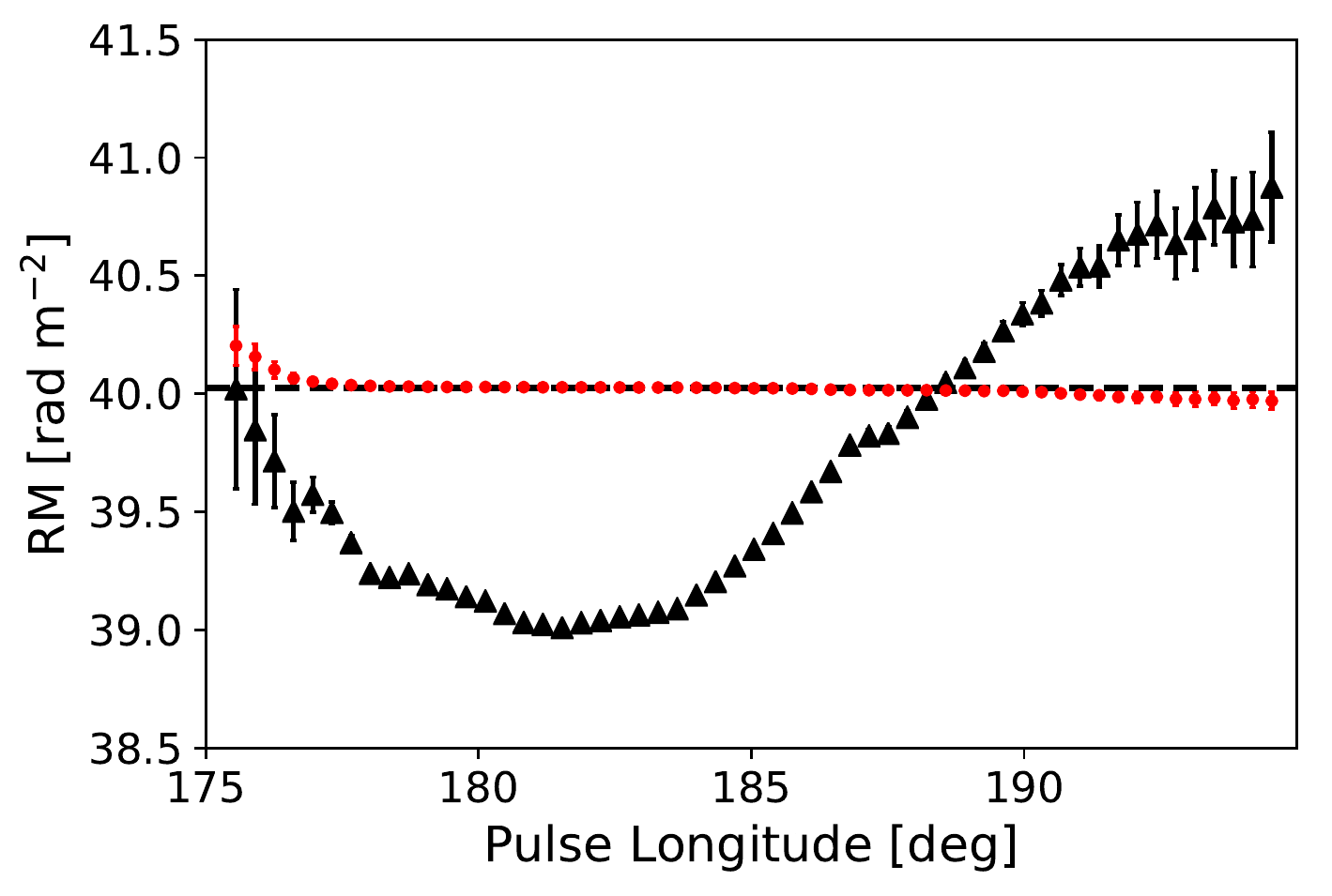}
\centering
\caption{PSR~J0835$-$4510.
The solid black triangles represent the observed RM($\phi$) values with the associated statistical uncertainties. The red circles (in the online version) represent the RM($\phi$) values obtained from simulating the effect of misaligned observations with varying RM$_{\text{iono}}$. The dashed horizontal line represents the mean of these values.}
\label{fig:vela_iono}
\end{figure}

Following the description in Section 3.1, we considered the effect of a varying RM because of changes in the ionosphere. For the Vela pulsar, J0835$-$4510, this leads to systematics effects in the longitude resolved RM curve, as determined in Fig.~\ref{fig:vela_iono}.

